\documentclass[longauth]{aa}  

\pdfoutput=1
\usepackage{longtable}

\usepackage{lscape}

\usepackage{multirow}
\usepackage{natbib}
\usepackage{afterpage}
\usepackage{color}
\usepackage{amssymb,times}
\usepackage[english]{babel}
\usepackage{subfigure}
\usepackage{ifpdf}
\usepackage{morefloats}
\usepackage[rightcaption]{sidecap}
\usepackage{lscape}
\usepackage{multicol}
\usepackage{booktabs}
\usepackage{slashbox, picture}
\usepackage{fixltx2e}
\usepackage{amssymb}
\usepackage{amsmath}
\usepackage[dvipsnames]{xcolor}
\usepackage[colorlinks = true,
            linkcolor = blue,
            urlcolor  = blue,
            citecolor = blue,
            anchorcolor = blue]{hyperref}

\newif\ifpdf \ifx\pdfoutput\undefined\pdffalse\else\pdftrue\fi
\ifpdf\pdfoutput=1
        \usepackage{graphicx}
        \usepackage{epstopdf}
        \DeclareGraphicsExtensions{.pdf,.png,.gif,.jpg}
        \else \usepackage[dvips]{color,graphicx} \fi

\def\Msun{\hbox{M$_{\odot}$}}               %% solar mass
\def\Lsun{\hbox{L$_{\odot}$}}               %% solar luminosity
\def\Rstar{\hbox{R$_{\star}$}}              %% stellar radius
\def\Mdot{\hbox{$\dot{M}$}}               %% Mdot      
\def\arcsec{\hbox{$^{\prime\prime}$}}

\def\deg{\hbox{$^\circ$}}       %% \def for overwriting, \box for math

\def\Al2O3{\hbox{Al$_2$O$_3$}}

\begin{document} 

\selectlanguage{english}
\newcommand{\blue}[1]{\textcolor{blue}{#1}}
\newcommand{\red}[1]{\textcolor{red}{#1}}
\newcommand{\cyan}[1]{\textcolor{cyan}{#1}}
\newcommand{\magenta}[1]{\textcolor{magenta}{#1}}
\newcommand{\pink}[1]{\textcolor{magenta}{#1}}
\newcommand{\gray}[1]{\textcolor{gray}{#1}}
\newcommand{\green}[1]{\textcolor{OliveGreen}{#1}}

\title{{\sc atomium}: ALMA tracing the origins of molecules in dust forming oxygen rich M-type stars}
\subtitle{ Motivation, sample, calibration, and initial results}
 
\author{C.A. Gottlieb\inst{1}
 \and L. Decin\inst{2,3}
 \and A.M.S. Richards\inst{4}
%  +++++++++++++++++++++++++ 
  \and  F. De Ceuster\inst{2,5} 
  \and W. Homan\inst{2}
%  +++++++++++++++++++++++++   
 \and S.H.J. Wallstr{\"o}m\inst{2}
 \and T. Danilovich\inst{2}
 \and T.J. Millar\inst{6}
% ++++++++++++++++++++++++++++++
 \and M. Montarg{\`e}s \inst{2,7} 
% ++++++++++++++++++++++++++++++
 \and K.T. Wong\inst{8}   
 \and I. McDonald\inst{4,9}
 \and A. Baudry\inst{10} 
 \and  J. Bolte\inst{2} 
 \and  E. Cannon\inst{2} 
 \and  E. De Beck\inst{11}  
% \and  F. De Ceuster\inst{2,11} 
 \and  A. de Koter\inst{2,12} 
 \and  I. El Mellah\inst{2,13} 
 \and  S. Etoka\inst{4} 
 \and D. Gobrecht\inst{2} 
 \and M. Gray\inst{4,14} 
 \and F. Herpin\inst{10} 
% \and W. Homan\inst{2}  
 \and M. Jeste\inst{15} 
 \and P. Kervella\inst{16} 
 \and T. Khouri\inst{11} 
 \and E. Lagadec\inst{17} 
 %  +++++++++++++++++++++++++++++
\and S. Maes\inst{2}
\and J. Malfait\inst{2}
%  +++++++++++++++++++++++++++++
 \and K.M. Menten\inst{15}  
 \and H.S.P. M{\"u}ller\inst{18} 
 \and B. Pimpanuwat\inst{4,14}
 \and J.M.C. Plane\inst{3} 
 \and R. Sahai\inst{19}  
 %  ++++++++++++++++++++++++++++=
 \and M. Van de Sande\inst{20,2}
  %  ++++++++++++++++++++++++++++= 
 \and L.B.F.M. Waters\inst{21,22}  
 \and J. Yates\inst{5} 
 \and A. Zijlstra\inst{4,23}    
}

\offprints{Leen.Decin@kuleuven.be}

\institute{
  Harvard-Smithsonian Center for Astrophysics, 60 Garden Street Cambridge, MA 02138, USA  
  \and
Institute of Astronomy, KU Leuven, Celestijnenlaan 200D, 3001 Leuven, Belgium
%  Instituut voor Sterrenkunde, Katholieke Universiteit Leuven, Celestijnenlaan 200D, 3001 Leuven, Belgium  
  \and
  University of Leeds, School of Chemistry, Leeds LS2 9JT, United Kingdom
  \and
  Jodrell Bank Centre for Astrophysics, Department of Physics and Astronomy, University of Manchester, Manchester M13 9PL, 
  United Kingdom 
 \and
University College London, Department of Physics and Astronomy, London WC1E 6BT,  United Kingdom
 \and 
  Astrophysics Research Centre, School of Mathematics and Physics, Queen’s University Belfast, University Road, Belfast BT7 1NN, 
  United Kingdom
% ++++++++++++++++++++++++++++++  
\and
LESIA, Observatoire de Paris, Université PSL, CNRS, Sorbonne Université, Université de Paris, 5 place Jules Janssen, 92195 Meudon, 
France  
% ++++++++++++++++++++++++++++++  
  \and
Institut de Radioastronomie Millim{\'e}trique, 300 rue de la Piscine, 38406 Saint Martin d'H{\`e}res, France
  \and
Open University, Walton Hall, Milton Keynes, MK7 6AA United Kingdom  
\and
Universit{\'e} de Bordeaux, Laboratoire d'Astrophysique de Bordeaux, 33615 Pessac, France
\and
Chalmers University of Technology, Onsala Space Observatory, 43992 Onsala, Sweden
%\and
%University College London, Department of Physics and Astronomy, London WC1E 6BT,  United Kingdom
\and
University of Amsterdam, Anton Pannekoek Institute for Astronomy, 1090 GE Amsterdam, The Netherlands
\and
KU Leuven, Center for mathematical Plasma Astrophysics, 3001 Leuven, Belgium
\and
National Astronomical Research Institute of Thailand, Chiangmai 50180, Thailand
\and
Max-Planck-Institut f{\"u}r Radioastronomie, 53121 Bonn, Germany
\and
Laboratoire d'Etudes Spatiales et d’Instrumentation en Astrophysique, Observatoire de Paris, Universit{\'e} Paris Sciences et Lettres, 
Centre National de la Recherche Scientifique, Sorbonne Universit{\'e}, Universit{\'e} de Paris, 92195 Meudon, France
\and
Universit{\'e} C{\^o}te d'Azur, Laboratoire Lagrange, Observatoire de la C{\^o}te d'Azur, F-06304 Nice
Cedex 4, France
\and
Universit{\"a}t zu K{\"o}ln, I. Physikalisches Institut, 50937 K{\"o}ln, Germany
\and
California Institute of Technology, Jet Propulsion Laboratory, Pasadena CA 91109, USA
\and
School of Physics and Astronomy, University of Leeds, Leeds LS2 9JT, UK
\and 
SRON Netherlands Institute for Space Research, NL-3584 CA Utrecht, The Netherlands
\and
Radboud University, Institute for Mathematics, Astrophysics and Particle Physics (IMAPP), Nijmegen, The Netherlands
\and
University of Hong Kong, Laboratory for Space Research, Pokfulam, Hong Kong
}

\date{Received date 28 January 2021; accepted date 17 November 2021}

\abstract
{This overview paper presents {\sc atomium}, a Large Programme in Cycle~6 with the Atacama Large Millimeter/submillimeter 
Array (ALMA). 
The goal of {\sc atomium} is to understand the dynamics and the gas phase and dust formation chemistry in the winds of evolved 
asymptotic giant branch (AGB) and red supergiant (RSG) stars.
A more general aim is to identify chemical processes applicable to other astrophysical environments.
Seventeen oxygen-rich AGB and RSG stars spanning a range in (circum)stellar parameters and evolutionary phases were observed in a homogeneous observing strategy allowing for an unambiguous comparison.
Data were obtained between 213.83 and 269.71\,GHz at high ($\sim$0\farcs025-0\farcs050), medium ($\sim$0\farcs13--0\farcs24), and low 
($\sim$1\arcsec) angular resolution.
The sensitivity per $\sim$1.3\,km/s channel was 1.5--5\,mJy/beam, and the line-free channels were used to image the millimetre wave continuum. 
Our primary molecules for studying the gas dynamics and dust formation are CO, SiO, AlO, AlOH, TiO, TiO$_2$, and HCN; 
secondary molecules include SO, SO$_2$, SiS, CS, H$_2$O, and NaCl. 
The scientific motivation, survey design, sample properties, data reduction, and an overview of the data products
are described. In addition, we highlight one scientific result --- the wind kinematics of the {\sc atomium} sources. 
Our analysis suggests that the {\sc ATOMIUM} sources often have a slow wind acceleration, and a fraction of the gas reaches 
a velocity which can be up to a factor of two times larger than previously reported terminal velocities assuming isotropic expansion.
Moreover, the wind kinematic profiles establish that the radial velocity described by the momentum equation for a spherical wind structure 
cannot capture the complexity of the velocity field.
In fifteen sources, some molecular transitions other than $^{12}$CO v=0 J=2-1 reach a higher outflow velocity, with a spatial emission zone 
that is often greater than 30 stellar radii, but much less than the extent of CO.  
We propose that a binary interaction with a (sub)stellar companion may (partly) explain the non-monotonic behaviour of the projected velocity field. 
The {\sc atomium} data hence provide a crucial benchmark for the wind dynamics of evolved stars in single and binary star models.}

\keywords{Stars: AGB and post-AGB, Stars: mass-loss, Stars: circumstellar matter, Binaries: general, instrumentation: interferometers, astrochemistry}

\titlerunning{ALMA {\sc atomium} Survey}
\maketitle

%  ############################################################################################# 

\section{Introduction} \label{Sec:Introduction}

A long-standing question in astrophysics is the physicochemical mechanism describing the complex phase transition from small molecules --- containing typically only two or three atoms ---  to larger gas phase clusters, and eventually tiny dust grains, with the first thermochemical computations probably presented in the first half of the 1930s \citep{Wildt1933ZA......6..345W, 1934ApJ....79..317R}.  
We are still struggling to predict how the composition of the gas with specific initial conditions for the thermodynamical and other
physical properties (such as temperature, density, and velocity) will evolve in time.
Aiming to unravel this question, astronomers have focussed their attention on low- and intermediate-mass asymptotic giant branch (AGB) stars and their more massive counterparts, the red supergiants (RSGs). 
The winds of AGB and RSG stars have long been recognised as key chemical laboratories in which more than 90 molecules and 15 dust species have been detected thus far \citep{Habing1996A&ARv...7...97H, Habing2004agbs.book.....H, Heras2005A&A...439..171H, Verhoelst2009A&A...498..127V, Waters2011ASPC..445..227W, Gail2013pccd.book.....G, Hofner2018A&ARv..26....1H}.
Convection-induced dredge-ups in the atmosphere, shocks, nucleation, and stellar and interstellar UV photons in the circumstellar envelope are just a few of the physicochemical processes that determine the chemical fingerprints of AGB and RSG stellar winds (see Sect.~\ref{Sec:chemistry}).
A large variety of chemical reactions occur in the wind, including unimolecular, two- and three-body reactions, cluster growth, and grain formation. 
Through their winds, AGB and RSG stars contribute $\sim$85\% of the gas and $\sim$35\% of the dust from stellar sources to the Galactic ISM \citep{Tielens2005pcim.book.....T}, and are the dominant source of pristine building blocks of interstellar material. 

\citet{Hoyle1962MNRAS.124..417H} were the first to propose that the wind acceleration in AGB stars is caused by radiation pressure on newly formed dust grains. Molecules might carry the analogous potential to launch a RSG wind, with grains taking over farther out in the wind \citep{Gustafsson1992iesh.conf...86G}. It is generally accepted that pulsations are a key ingredient of AGB mass-loss with pulsation-induced shock waves levitating the gas to larger distances where the temperature is low enough for dust to condense 
\citep[][and references therein]{1982ApJ...252..697H, 1997AJ....114.2686H, Bowen1988ApJ...329..299B, McDonald2016ApJ...823L..38M,
Hofner2018A&ARv..26....1H}. 
Convection-induced pulsation amplitudes are, however, much lower for RSG stars and the role of pulsations in triggering the RSG wind is 
thought to be negligible.
The prevailing streamlines in the AGB and RSG winds outside $\sim$5$R_{\star}$ are radial \citep[][and references therein]{Hofner2018A&ARv..26....1H}, although recent observations with ALMA have added structural complexities to this picture (see Sect.~\ref{Sec:dynamics}).
Even so, the dynamical behaviour in the winds is much simpler than in other chemically rich environments, such as high-mass star-forming regions, young stellar objects and protoplanetary disks. If we can disentangle the (thermo)dynamical and chemical processes in the winds, we might be able to lay the foundation for a better understanding of the gas-to-dust phase transition as well as 
some of the physiochemical processes that occur in (pre-biotic) chemistry in these more complex environments. 

The ALMA {\sc atomium}\footnote{{\sc atomium}: ALMA Tracing the Origins of Molecules In dUst-forming oxygen-rich M-type stars; \url{https://fys.kuleuven.be/ster/research-projects/aerosol/atomium/atomium}.
The {\sc atomium}  proposal was selected as a Large Programme in Cycle~6 with 113.2 hr allotted (2018.1.00659.L, PI L.\ Decin), 
and is the first ALMA Large Programme in the field of `Stellar Evolution'.} 
Large Programme has been constructed with the specific aim of understanding the chemistry of dust precursors and dust formation, 
as well as the more general aim of identifying chemical processes applicable to other astrophysical environments (including novae, 
supernovae, protoplanetary nebulae, and interstellar shocks).
The obvious choice of targets for the {\sc atomium} project are \textit{oxygen-rich} AGB and RSG stellar winds (O-rich, C/O$<$1; see Sect.~\ref{Sec:Objects}), because ALMA provides the unique ability to study the many oxide and hydroxide precursors of dust in O-rich winds --- something we cannot do for carbonaceous grains in carbon-rich (C/O$>$1) winds, where the likely precursors such as aromatic molecules and polycyclic aromatic hydrocarbons (PAHs) 
are not observable with ALMA.

In this paper we discuss the scientific motivations for {\sc atomium}, introduce the survey strategy as well as the source 
and spectral line sample (Sect.~\ref{Sec:ATOMIUM}), and describe the calibration process (Sect.~\ref{Sec:Observations}). 
All the data are available in the ALMA Science Archive, but in addition enhanced data products have been prepared.
These are described in Sect.~\ref{Sec:Release}, and they will serve as a legacy for the astronomical community 
and will seed new insights in the dynamical and chemical process in evolved stars and other astronomical media.
The quality and the properties of the data products are illustrated in the example of the OH/IR star IRC$-$10529 
in Sect.~\ref{Sec:IRC-10529} and the accompanying figures.
In Sect.~\ref{Sec:Results} we focus on one scientific result --- the wind kinematics in the circumstellar envelopes of evolved stars.
We discuss the efficiency of the wind initiation and show how the presence of a binary companion can be revealed via a study 
of the wind velocity profile, thereby demonstrating how the {\sc atomium} data provides a crucial benchmark for single and 
binary star models of the wind dynamics of evolved stars.

Other results will be presented in separate papers including:  detailed discussions of the individual sources; 
 a chemical inventory of the molecular species in all 17~stars, observed in the three array configurations with an angular resolution 
that spans 50\,\,mas$-10^{\prime\prime}$; and 
studies of the dust precursors, masers, and the wind morphology \citep{DecinScience, 2020A&A...644A..61H, 2021A&A...651A..82H}. 
In addition, various hydrodynamical, chemical, and radiative transfer models that simulate the wind properties of AGB and RSG stars 
and support the analysis of the {\sc atomium} data, have already been published or are underway 
\citep[see, for example, ][]{2021ARA&A..59..337D, 2020MNRAS.492.1812D}.

%  ###############################################################################################

\vspace{-0.35cm}
\section{The ALMA ATOMIUM Large Programme} \label{Sec:ATOMIUM}

\subsection{Scientific goals} \label{Sec:Goals}
	
The goal of the ALMA {\sc atomium} Programme is:
(1) to derive the morpho-kinematical and chemical properties of the winds; 
(2) to unravel the phase change from gaseous to solid-state species;
(3) to identify the dominant chemical pathways;
(4) to study the role of (un)correlated density structures\footnote{The term `correlated density structures' refer to arcs, spirals, disks, 
bipolar structures, shells, etc.  `Uncorrelated density structures' refer to clump-like morphologies which do not appear to be correlated with 
any other morphology.} on the overall wind structure; and 
(5) to examine the reciprocal effect between various dynamical and chemical phenomena
in 17 oxygen-rich AGB and RSG sources which cover a range of initial stellar masses, pulsations, mass-loss rates, 
and evolutionary phases (see Sect.~\ref{Sec:Objects}).

Summarised in the following paragraphs are key science questions that are addressed in this large programme.
For sake of clarity, we differentiate between physical and chemical phenomena, although both are coupled in an intimate way, 
as for example via the dust extinction efficiency $Q_\lambda$ described in Sect.~\ref{Sec:Results}.

\vspace{-0.25cm}
\subsubsection{Dynamical behaviour of stellar winds} \label{Sec:dynamics}

\noindent\textbf{Wind initiation in the inner wind region} (1\,\Rstar $\la$\,\,r\,\,\la$10$\,--\,30\,\Rstar): 
The winds in O-rich AGB stars can only be predicted theoretically on the premise of pulsation-induced higher density regions close 
to the star where large transparent grains can form 
\citep{1982ApJ...252..697H, 1997AJ....114.2686H, 1985ApJ...299..167B, Bowen1988ApJ...329..299B, Woitke2006A&A...460L...9W, 
Hofner2008A&A...491L...1H, Bladh2019A&A...626A.100B}. 
For RSGs the role of grains close to the star remains unresolved 
\citep{Josselin2007A&A...469..671J, Bennett2010ASPC..425..181B, Scicluna2015A&A...584L..10S,Kervella2018A&A...609A..67K,Montarges2019MNRAS.485.2417M}. 

\citet{Fonfria2008ApJ...673..445F} have used mid-infrared bands of molecules to study the dust formation zone.  
High-resolution ALMA data carry the same diagnostics, and trace the region closer to the star if high-excitation lines are studied.
Recent observational studies have shown that the wind acceleration for O-rich AGB stars is often less efficient  than previous predictions 
obtained by solving the momentum equation 
\citep[][see Eq.~\eqref{Eq:momentum} in Sect.~\ref{Sec:Results}]{Decin2010A&A...521L...4D, Khouri2014A&A...561A...5K, 
VandeSande2018A&A...609A..63V, Decin2018A&A...615A..28D}.
This  behaviour couples directly to the unknown grain composition (see also Sect.~\ref{Sec:chemistry}). 
Moreover, it is not yet known if the wind acceleration profile is different for regular versus irregular pulsators.

As a first step in determining where the wind is initiated, the wind kinematics of 17 oxygen-rich AGB and RSG stars in the {\sc atomium} sample have been 
derived (see Sect.~\ref{Sec:Results}), which in turn allows us to correlate the wind acceleration profile to the specific stellar (and hence pulsation) characteristics and 
chemical properties; and will contribute to recent studies that investigate the role of pulsations as triggers for the onset of the mass loss and in controlling the rate of the mass loss \citep{McDonald2016ApJ...823L..38M, McDonald2019MNRAS.484.4678M}. 

\medskip
\noindent\textbf{Enforced dynamics in the intermediate wind region} ($r\!\sim$30\,--\,400\,\Rstar): 
Accurate measurements of the wind velocities are a major factor in determining the AGB (RSG) mass-loss rate, and thus the 
lifetime and impact on Galactic enrichment.
Recent ALMA data revealed a thought-provoking picture of the wind kinematics in the intermediate wind region \citep{Decin2018A&A...615A..28D}: 
(i)~the wind acceleration appears to continue beyond $\sim$30\,\Rstar, in contradiction to the solution of the momentum equation 
(see also Sect.~\ref{Sec:Results}); and 
(ii)~the line profiles indicate that the maximum wind velocity --- as derived from the primary tracer CO and other molecules ---
is much higher than the previously determined terminal wind velocity, with differences of up to a factor $>$4 in the case 
of R~Dor \citep{Decin2018A&A...615A..28D}.
This surprising behaviour is seen for all AGB  and RSG stars for which the ALMA line sensitivity is greater than a few mJy/beam. 
The reason for these enforced wind dynamics is still unclear, since further grain growth seems implausible owing to the 
low densities in regions far from the star (but see Sect.~\ref{Sec:Interpretation}).
Because the wings of the (low-excitation) lines carry the diagnostic information needed to unravel this science question, 
 a sample of evolved stars was observed at very high sensitivity in {\sc atomium}
(complemented with other data, part of which has already been obtained with ALMA).  
Prior to this, only a handful of evolved stars underwent such observations with ALMA.

\medskip
\noindent\textbf{Wind morphology}:
The first step for identifying the wind-shaping mechanism(s) and retrieving the wind kinematics in AGB and RSG stars, 
was to map the 3D wind morphology. 
The $^{12}$CO v=0 J=1-0 and J=2-1 channel maps observed with single antennas at an angular resolution of $21^{\prime\prime}$ 
and $13^{\prime\prime}$, respectively, indicated that about 80\% of the AGB and RSG winds show a large scale spherical symmetry.
Observations of 24 oxygen rich AGB stars with a  synthesised beam of about $4^{\prime\prime}$ \citep{Neri1998A&AS..130....1N}, 
found that most have an outer circumstellar envelope that is mainly circular and an inner envelope whose shape was not easily discerned 
at the limited resolution. 
However, departures in the spherical symmetry of the CO J=1-0 and J=2-1 emission in the circumstellar envelopes of some oxygen rich 
AGB stars were identified when they were observed at a modest resolution of $1^{\prime\prime}$ or lower  by \citet{2010A&A...523A..59C}.

Data acquired subsequently with ALMA at higher angular resolution revealed that a significant fraction of the winds exhibit structural 
complexities embedded in the smooth radially outflowing wind which include arcs, shells, bipolar structures, clumps, spirals, tori, and rotating discs 
\citep{Maercker2012Natur.490..232M, Ramstedt2014A&A...570L..14R, Ramstedt2017A&A...605A.126R, Ramstedt2018A&A...616A..61R, 
 Kim2015ApJ...814...61K,  Decin2015A&A...574A...5D, Decin2019NatAs...3..408D, 
DecinScience, Cernicharo2015A&A...575A..91C, Wong2016A&A...590A.127W, Kervella2016A&A...596A..92K, Agundez2017A&A...601A...4A, 
Doan2017A&A...605A..28D, Doan2020A&A...633A..13D, Homan2018A&A...616A..34H, Bujarrabal2018A&A...616L...3B, Guelin2018A&A...610A...4G, 
Randall2020A&A...636A.123R, Hoai2020MNRAS.495..943H}. 
For most of these morphologies, the formation mechanism is unknown, although binarity is suspected to play an important role. 
In two particular cases, the ALMA data suggest there is a planetary companion at a disk's inner rim 
\citep{Kervella2016A&A...596A..92K, Homan2018A&A...616A..34H}.   
In addition, hydrodynamical instabilities occurring in a multi-fluid environment and convection-induced activity can lead to the formation of 
overdense clumps  \citep[see for example,][]{Montarges2019MNRAS.485.2417M}. 

To analyse the correlated density structures, high spatial resolution data which sample a range in molecular excitation regime (and hence sample 
the extended wind region) was acquired. 
The key molecule is CO owing to: its high fractional abundance (with respect to H$_2$); its high dissociation energy; its simple energy level structure; 
and its rotational levels are readily excited by collisions.
Other complementary tracers include the rotational transitions of SiO, HCN, and NaCl \citep[see, e.g.,][]{Kervella2016A&A...596A..92K, Decin2016A&A...592A..76D}.
The first observations acquired in the {\sc atomium} project were with an angular resolution of 0\farcs13--0\farcs24 in the mid array configuration 
(see Sect.~\ref{Sec:Observations}). 
The analysis of the $^{12}$CO v=0 J=2-1 and $^{28}$SiO v=0 J=5-4 and J=6-5 rotational lines\footnote{Hereafter, all rotational transitions are in the ground (v=0) vibrational state unless otherwise specified.}
has provided a unique view of the prevailing wind morphology in the {\sc atomium} sources 
\citep{DecinScience}. 
This is  illustrated by the channel maps of $^{12}$CO (Fig.~\ref{Fig:CO_CM_IRC10529}), {SO$_2$ (Fig.~\ref{Fig:SO2_CM_IRC10529}), 
and SiO (Fig.~\ref{Fig:SiO_CM_IRC10529}) in the OH/IR star IRC\,$-$10529 (see also Sect.~\ref{Sec:IRC-10529}).
None of the {\sc atomium} sources display a spherical wind geometry. 
The derived morphologies: (1) correlate with the mass-loss rate; 
(2) yield important insights into the mechanism(s) determining the 
appearance of AGB descendants, post-AGB stars, and planetary nebulae in which cylindrically symmetric and multi-polar morphologies 
are often observed \citep{Guerrero2003AJ....125.3213G, Ercolano2003MNRAS.340.1153E, Ueta2007AJ....133.1345U}; 
and (3) can be explained by binary interaction \citep{DecinScience}.

\vspace{-0.25cm}
\subsubsection{Chemical processes in stellar winds} \label{Sec:chemistry}

Significant advances have been made in the past few years in  characterising the physical and chemical properties of the
dust in the inner wind owing to: 
(1) the polarimetric direct imaging of the dust in the visible at high angular resolution with {\sc vlt/sphere} 
by \citet{Khouri2016A&A...591A..70K, 2018A&A...620A..75K, 2020A&A...635A.200K},
\citet{2016A&A...589A..91O,  2017A&A...597A..20O}, and \citet{ 2019A&A...628A.132A}; 
and (2) parallel observations of the rotational spectra of potential Ti and Al bearing precursors of the dust
\citep{2016A&A...592A..42K, Kaminski2017A&A...599A..59K, Decin2017A&A...608A..55D,  Takigawa2017SciA....3O2149T, Danilovich2020}.} 
However, very little is known about the physicochemical processes in the intermediate wind where dust-gas interactions occur, and tiny dust grains 
formed in the inner wind, grow in size by accretion of small abundant gaseous molecules onto the grains
\citep[for a comprehensive overview see the review by][and references therein]{2021ARA&A..59..337D}.
As noted in the discussion of the enforced dynamics in Sect.~2.1.1, it was unclear why the wind velocity has not yet reached its terminal velocity
in the intermediate wind region.
One of the main emphases of {\sc atomium} is to better understand the chemistry in the intermediate wind. 

To date most chemical models of oxygen-rich AGB stars have been devoted to the study of either the initial stage 
of dust formation in the inner wind at $\lesssim\!10\!-\!30$\,\Rstar\
\citep{2006A&A...456.1001C,Gobrecht2016A&A...585A...6G,Boulangier2019MNRAS.489.4890B}, 
or to the photon dominated chemistry in the outer wind 
\citep{Willacy1997A&A...324..237W,2016A&A...588A...4L}.
Of the 11 parent molecules considered by \citet{VandeSande2019MNRAS.490.2023V} in their chemical kinetics model of the 
intermediate wind region, all but two were observed in {\sc atomium} (N$_2$ and NH$_3$), allowing us: 
(1) to derive the extent of the emission and potential depletion in the outflowing wind of nine of the 11 molecules in 17 sources 
from observations in the three array configurations; and 
(2) to compare the measured depletions with the predictions of the chemical kinetic models that include dust-gas interactions 
in the AGB outflow.

%   +++++++++++++++++++++++++++++++++++++++++++++++++++++++++++++++++++++=

\medskip
\noindent {\textbf{Continuum radiation}}: 
At millimeter wavelengths, the bulk of the continuum emission comes from the extended stellar atmosphere 
\citep{Reid1997ApJ...476..327R}. 
For most of the stars, the {\sc atomium} observations at the highest resolution allow us to either resolve or to fit a disc to the 1.2~mm stellar continuum  
which is known to be 15-50\% greater than the optical size listed in Table~\ref{Table:targets} \citep{2019A&A...626A..81V}.  
On the assumption the star emits as a blackbody, the stellar flux at millimeter wavelengths can be estimated from the stellar effective temperature and luminosity.
The derived stellar flux has been found to agree with fitting a uniform disc to the millimeter-wave visibilities when the S/N is sufficiently high \citep{2021A&A...651A..82H}.
For at least some of the sample,  an excess of the more extended emission that is typically 
up to a few tens of a percent of the stellar emission is detected with ALMA \citep{Decin2018A&A...615A..28D, Dehaes2007MNRAS.377..931D}, which will allow us to subtract the stellar contribution and to measure the dust emission.
Supplemented by data of the spectral energy distribution (SED) at other wavelengths, the dust mass and the (recent) dust mass-loss rate 
can be derived  \citep[][Khouri et al.,  \textit{in prep}]{Decin2018A&A...615A..28D}. 
Combined with the gas mass-loss rate derived from lines of CO acquired previously with single antennas, the gas-to-dust ratio as a 
function of stellar type can be determined  \citet{2015A&A...581A..60D}.
 In addition, the determination of the positions of SiO masers close to the stellar surface  with even finer precision in  {\sc atomium}, 
allow us to investigate the possible connection between dust clumps, particular molecular emission patterns, and stellar characteristics \citep{2020A&A...644A..61H}.

%   +++++++++++++++++++++++++++++++++++++++++++++++++++++++++++++++++++++

\medskip
\noindent{\textbf{Dust nucleation}}:
A major unknown in current wind models concerns the initial dust nucleation process \citep{Gail2013pccd.book.....G}. 
When the {\sc atomium} project was undertaken, it was not  known which molecules form the large gas phase clusters that transition 
into the first solid-state species in oxygen rich winds \citep{Paquette2011ApJ...732...62P, Plane2013RSPTA.37120335P, Bromley2016PCCP...1826913B}. 
Thermodynamic condensation sequences favour alumina (Al$_2$O$_3$) or Fe-free silicates (such as Mg$_2$SiO$_4$), where the Al$_2$O$_3$ is  
formed at slightly higher temperatures \citep{Tielens1998Ap&SS.255..415T, Bladh2012A&A...546A..76B}. 
Grains of this type, however, need to be large enough ($\sim$200\,nm -- 1\,$\mu$m) and close to the star ($r\!\la\!10$\,\Rstar) for photon scattering 
to compensate for their low near-infrared absorption cross sections, and to trigger the onset of a stellar wind \citep{Hofner2008A&A...491L...1H}. 

Recent NACO and SPHERE data support the presence of large transparent grains ($\sim$0.3\,$\mu$m) at $\sim$1.5\,\Rstar\ in some AGB and RSG stars 
\citep{Norris2012Natur.484..220N, Khouri2016A&A...591A..70K, Haubois2019A&A...628A.101H}, but this data cannot pinpoint the 
chemical build-up of the grains. 
As shown in recent publications \citep{Kaminski2017A&A...599A..59K, Decin2017A&A...608A..55D,  Takigawa2017SciA....3O2149T},
ALMA has paved the way for unraveling the composition of the tiny dust seeds via the study of specific small gaseous precursors. 
The synergy between ALMA and (near-)infrared data is allowing us in turn to establish which gas phase clusters [such as (Al$_2$O$_3$)$_n$ 
with $n >1$] might be the intermediate steps in this dust nucleation history \citep{Decin2017A&A...608A..55D}. 

The metal oxides and hydroxides AlO, AlOH, TiO, OH --- and most prominently SiO --- are the key molecules 
we are using to study the impact of higher density clumps and correlated density structures on the time scales for dust growth
in the inner region, and the efficiency of ice deposition in the intermediate region of the 17 stars in the {\sc atomium} survey. 
The abundance structures are being examined with the recent radiative transfer analysis of vibrationally excited AlO and TiO 
in R~Dor which has provided a new view of the formation of Al$_2$O$_3$ dust  \citep{Danilovich2020} ---
and the same approach is also being applied to CO, HCN, SO, SO$_2$, SiS, AlCl, NaCl, and PO which are observed 
in non-maser emission in the ground and the excited vibrational levels within a couple of $R_\star$ of a number of the stars 
in the {\sc atomium} sample.

% +++++++++++++++++++++++++++++++++++++++++++++++++++++++++++++++++++++++++++++++++++++++++++++++++++++++++
\medskip

\noindent{\textbf{Non-equilibrium gas-phase chemistry}}:
 For a long time, the gas-phase composition of stellar winds was believed to be determined solely by the C/O ratio of the stellar photosphere,  
 hence no carbon-bearing molecules except for CO were expected to form in oxygen rich winds. 
 The detection of CO$_2$, CS, and HCN in oxygen rich winds, and H$_2$O, OH, H$_2$CO, and  SiO in carbon rich winds {has caused this picture 
to be amended \citep{1985Natur.317..336D, 1988A&A...205L..15L, Bujarrabal1994A&A...285..247B, Justtanont1998A&A...330L..17J, Ryde1998Ap&SS.255..301R, Melnick2001Natur.412..160M, Ford2003ApJ...589..430F, Ford2004ApJ...614..990F, Schoier2006A&A...454..247S, Cherchneff2008A&A...480..431D, Schoier2013A&A...550A..78S, Velilla2015ApJ...805L..13V}. 
Pulsation-induced shock chemistry, and/or enhanced photochemical activity in a non-homogeneous outflow in which the harsh interstellar UV photon can deeply penetrate, have been proposed as potential explanations 
\citep{Agundez2010ApJ...724L.133A, Cherchneff2011A&A...526L..11C,  Gobrecht2016A&A...585A...6G, Agundez2017A&A...601A...4A, 
2018A&A...616A.106V, Agundez2020A&A...637A..59A}.
Such chemical modelling codes are based on a range of parameters including: velocity shock strength, specific clumpiness, 
and rates in the chemical network. 

The observation of 24 different molecules and the measurement of approximately 290 rotational lines in {\sc atomium} 
(supplemented with ALMA archival data) is described in a comprehensive Molecular Inventory paper (Wallstr{\"o}m et al.\ \textit{in prep.}), 
which includes the complete tabulation of the measured parameters (peak flux, width, and integrated area) of each rotational line 
observed in the 17 stars in the three array configurations.
This homogenous set of measurements provides the fundamental benchmarks for establishing the essential parameters for the 
development of predictive chemical kinetic codes which includes:
the angular size of the emission region in each molecule;
the column densities and abundance distributions with radial extent; and
the comparison of the spatial distributions of the different molecules in each star.
Also being examined in the Molecular Inventory paper is evidence for trends in the distributions of the molecules according to
pulsation type, pulsation period, pulsation phase, C/O ratio, mass-loss rate, and morphology.

%  ++++++++++++++++++++++++++++++++++++++++++++++++++++++++++++++++++++++++++++++++++++++++++++++

The {\sc atomium} observations also serve as a guide for new laboratory kinetic measurements, and quantum chemical calculations of 
accurate theoretical structures and kinetic reaction rates needed to assess the relevant gas phase reaction rates in prior and newly 
developed chemical kinetic codes 
\citep[e.g.,][]{Gobrecht2018CPL...711..138G, West2019ApJ...885..134W, 
McCarthy2019JMoSp.356....7M, Boulangier2019MNRAS.489.4890B, Escatllar2020A&A...634A..77E}.
The first paper resulting from these observations entails a detailed analysis of the rotational spectra of the  aluminium 
halides in W~Aql, augmented with supplementary observations from {\it Herschel} \citep{2021A&A...655A..80D}.
We found that the abundance profiles  calculated with an existing chemical kinetic model \citep{2018A&A...616A.106V} better
reproduces the observations when six new reactions of Al, AlO, and AlOH with HF and HCl were added to the gas phase rates provided 
in the UMIST database by \citet{2013A&A...550A..36M}, 
where the newly incorporated reaction rates in  \citeauthor{2021A&A...655A..80D} were obtained from detailed theoretical quantum chemical
calculations in support of this project.
The revised chemical kinetic code derived by  \citeauthor{2021A&A...655A..80D} should yield more accurate predictions of the abundances 
of these species in other S-type stars.

\noindent{\textbf{New identifications}}:
About 60 unidentified (U) lines have been observed in the {\sc atomium} survey.
Potential carriers of interest include the gaseous oxides, hydroxides, and sulfides of Ca, Fe, Mg, and Zr; HSiO and H$_2$SiO; 
and more complex oxides of Al (e.g., AlOAlO, AlO$_3$, and Al$_2$O$_3$), and of Si (e.g., SiO$_3$, Si$_2$O, and Si$_3$O).	
Relating strengths of rotational lines of unidentified species observed at high sensitivity with ALMA across frequency bands and 
(circum)stellar properties is a crucial step in assigning the molecular carrier, and will  
empower us to build a detailed molecular census which will serve as a legacy for the entire astronomical community.

\subsection{ATOMIUM sample} \label{Sec:Objects}

The {\sc atomium} sample consists of 17 O-rich sources which span a range in (circum)stellar properties of evolved 
AGB and RSG stars.
Our sample was selected so that the stars are observable with ALMA, but had not been previously observed 
at high angular resolution at millimeter wavelengths.
The sources have been selected to cover some of the most important parameters for determing the wind characteristics 
of evolved giant stars such as: mass-loss rate, pulsation behaviour, and red supergiant versus AGB stars.
As commented on above, ensemble studies are not yet possible with ALMA in its high resolution mode. 
Therefore a well-selected, yet small sample is the  best way forward for enhancing our knowledge of these systems.
The sample covers a range in mass-loss rates of $\sim$$10^{-7}$ to $\sim$$10^{-5}$ M$_{\odot}$/{yr}, as inferred from s
ingle antenna observations, and consists of stars that are as close to Earth as possible.
The selection criteria did not take into account prior evidence for possible binary companions. 

%  ------------------------------------
Table~\ref{Table:targets} gives an overview of some of the important (circum)stellar parameters.
More details on how these parameters have been selected and the references to relevant papers, can be found in Sect.~S1 in the Supplementary Materials in \citet{DecinScience}. 
The only changes with respect to the values of the 14 stars cited in \citet{DecinScience} are the newly adopted:
(i)~mass-loss rate of IRC\,$-$10529 from the more recent results of \citet{2015A&A...581A..60D}; and 
(ii)~the distance towards U~Her from the improved maser parallax determination of 266\,pc \citep{Vlemmings2007A&A...472..547V}, 
which also impacts the estimate of the effective temperature. 
Also included in the {\sc atomium} survey are the three red supergiants AH~Sco, KW~Sgr, and VX~Sgr.\footnote{While this paper was 
in the final stage of preparation, the Gaia Early Data Release~3 (Gaia~EDR3) became available.  
In 12 out of 15 {\sc atomium} stars the distances in Gaia~EDR3 are within $\sim$20\% of the distances we have used here (see Table~1); 
two objects have no Gaia measurements (IRC--10529 and IRC+10011); and two of the remaining three stars have maser parallax distances 
that should be more accurate.
We have adopted the distance for  W~Aql  from Gaia~EDR3 which is consistent with that in \citet{2021A&A...655A..80D};
and the maser parallax distance for AH~Sco of 2260~pc \citep{2008IAUS..252..247C}, which is closer (by 21\%) 
than the Gaia~DR2 distance.
Although the maser parallax distance for VX~Sgr that we have adopted is at odds with Gaia~EDR3, it is consistent with all previous 
measurements including Gaia~DR2 \citep{2005A&A...431..773R, 2018Natur.553..310P}.  }

%   +++++++++++++++++++++++++++++++++++++++++++++++++++++++++++++++++++++++++++++++++++++++++++++++

\begin{table*}[htp]
\caption{\textbf{Summary of some (circum)stellar parameters of the {\sc atomium} sample. } }
\label{Table:targets}
\setlength{\tabcolsep}{2.00mm}
\begin{tabular}{l|l|r|l|l|r|l|r|r}
\hline
\multicolumn{1}{l|}{Star$^{(a)}$}	         & \multicolumn{1}{c|}{Variability}               & \multicolumn{1}{c|}{Mass-loss}     
&  \multicolumn{1}{c|}{Pulsation}	         & \multicolumn{1}{c|}{Distance}                & \multicolumn{1}{c|}{Stellar}	          
&  \multicolumn{1}{c|}{$T_{\rm{eff}}$ $^{(b)}$} & \multicolumn{1}{c|}{$L$ $^{(c)}$} & \multicolumn{1}{c}{$v_{\rm{LSR}}^{\rm{new}}$\,$^{(d)}$}
   	\\

\multicolumn{1}{l|}{}                                 & \multicolumn{1}{c|}{type$^{(e)}$}	   & \multicolumn{1}{c|}{rate	}	           
& \multicolumn{1}{c|}{period $P$}           &  \multicolumn{1}{c|}{$D$}	                   & \multicolumn{1}{c|}{diameter$^{(b)}$ }	   
&			                                         &			                                           &		                                                   \\

\multicolumn{1}{l|}{}                                 &		                                                   &	\multicolumn{1}{c|}{(\Msun/yr)}		
&	\multicolumn{1}{c|}{(days)}		 &	\multicolumn{1}{c|}{(pc)}		           &	\multicolumn{1}{c|}{$\theta_d$ (mas)}		
&	\multicolumn{1}{c|}{(K)}		         &	\multicolumn{1}{c|}{(\Lsun)}		   &		\multicolumn{1}{c}{(km/s)}	\\
	\hline
\rule[0mm]{0mm}{4mm}S Pav	&  SRa & $8\times10^{-8}$ $^{(aa)}$ & 381 $^{(aa)}$       &	190 $^{(jj)}$	         &	12.		        &	3100 $^{(xx)}$		&	4900		&	$-$18.2	\\
T Mic	&	SRb	      &	$8\times10^{-8}$ $^{(aa)}$		& 347 $^{(aa)}$          &	210 $^{(jj,ll)}$           &	9.3		        &	3300	 $^{(xx)}$	        &	4700		&	25.5	\\
U Del	&	SRb	      &	$1.5\times10^{-7}$	$^{(aa)}$	        &119 $^{(jj)}$             &	330 $^{(jj,ll,yy)}$	 & 7.9 $^{(uu)}$	&	2800 	                &	4100		&	$-$6.8	\\
RW Sco$^{(f)}$ & Mira  & $2.1\times10^{-7}$ $^{(bb)}$                 & 389 $^{(bb)}$	  &	514 $^{(jj)}$	         &	4.9		        &	3300	 $^{(xx)}$	        &	7700		&	$-$69.7	\\
V PsA	&	SRb	      &	$3\times10^{-7}$ $^{(aa)}$	        &148 $^{(aa)}$		  &	278 $^{(jj)}$		 &	13.		        &	2400	 $^{(aa)}$	        &	4100		&	$-$11.1	\\
SV Aqr	&	LPV	      &	$3\times10^{-7}$ $^{(aa)}$	        & $\cdots$	          &	389 $^{(jj)}$		 &	4.4		        &	3400	 $^{(xx)}$		&	4000		&	6.7	\\
R Hya$^{(g)}$	& Mira    & $4\times10^{-7}$	$^{(cc)}$	                & 366 $^{(ll)}$	          &	165 $^{(pp)}$		 & 23. $^{(uu)}$	&	2100 $^{(cc)}$	        &	7400		&	$-$10.1	\\
U Her	&	Mira	      & $5.9\times10^{-7}$ $^{(dd)}$	                & 402 $^{(ll)}$	          &	266 $^{(qq)}$	         & 11. $^{(uu)}$       &	3100		                &	8000		&	$-$14.9	\\
$\pi^1$ Gru$^{(g,h)}$     & SRb & $7.7\times10^{-7}$ $^{(ee)}$    & 150 $^{(cc)}$	  &	197 $^{(jj,ll)}$	         & 21. $^{(vv)}$	&	2300	 $^{(cc)}$	        &	4700		&	$-$11.7	\\
AH Sco	&	SRc	       &	$1\times10^{-6}$ $^{(ff)}$	        & 738 $^{(mm)}$       &    2260 $^{(rr)}$	         &	5.8 $^{(ss)}$    &	3700		                &  330000        &	$-$2.3	\\
R Aql$^{(f)}$  & Mira      & $1.1\times10^{-6}$ $^{(dd)}$	        & 268 $^{(ll)}$	          &	230 $^{(jj,ll)}$	         & 12. $^{(uu)}$	&	2800 $^{(cc)}$ 	&	4900		&	47.2	\\
W Aql$^{(g,h)}$ & Mira. & $3\times10^{-6}$ $^{(gg)}$                   & 479 $^{(ll)}$           &    375 $^{(jj)}$              & 11. $^{(uu)}$	&	2800		                &	9700		&	$-$23.0	\\
GY Aql	&	Mira	      &	$4.1\times10^{-6}$ $^{(hh)}$	        & 468 $^{(ll)}$	         &	152 $^{(jj)}$		 &	21.		        &	3100 $^{(xx)}$	        &	9600		&	34.0	\\
KW Sgr	&	SRc	      &	$5.6\times10^{-6}$ 	$^{(ii)}$	        & 647 $^{(nn)}$	 &    2400 $^{(ss)}$	         &	3.9 $^{(ss)}$    &	3700		                &  175700	&	$-$4.4	\\
IRC$-$10529$^{(f)}$    &  Mira& $4.5\times10^{-6}$ $^{(cc)}$      & 680 $^{(cc)}$	 &	760 $^{(cc)}$	         &	6.5		        &	2700 $^{(cc)}$		&    14400	&	$-$16.3	\\
IRC+10011$^{(f)}$       &  Mira  & $1.9\times10^{-5}$ $^{(cc)}$    & 660 $^{(cc)}$	         &	740 $^{(cc)}$		 &	6.5		        &	2700	 $^{(cc)}$	        &    13900	&	10.1	\\
VX Sgr     &	SRc	     & $6.1\times10^{-5}$ $^{(jj)}$	               & 732 $^{(oo)}$	         &	1560 $^{(tt)}$		 &	8.8 $^{(ww)}$  &	3500		                &  102300	&	5.7	\\
\hline
\end{tabular} \\
\tablefoot{   
$^{(a)}$~Stars are ordered by increasing mass-loss rate.   \\
$^{(b)}$~For all the stars, either the stellar diameter ($\theta_d$) or $T_{\rm{eff}}$ (or both) are derived from direct measurement; 
there are no objects for which indirect calculations are used for both parameters. 
The references in the footnotes refer to the direct measurements. 
The other parameter is then derived from the relation  $L(R_\star,T_{\rm{eff}})$ with $R_\star$ determined from the stellar diameter and the distance.  \\
  $^{(c)}$~Derived from the $M_{\rm{bol}}(P,L)$ relation in \citet{DeBeck2010A&A...523A..18D} unless indicated otherwise.  \\
$^{(d)}$~Estimate of the local standard of rest velocity derived from a sample of rotational lines with well behaved line profile shapes
and laboratory measured frequencies observed in the ALMA {\sc atomium} survey.  \\
$^{(e)}$~Mira variables have regular, large amplitude variations in the visible with $\delta V\!>\!2.5$\,mag and are thought to be fundamental mode pulsators; semiregular variables (SR) are of smaller amplitude, $\delta V\!<\!2.5$\,mag, with pulsations in the fundamental, first, and even higher overtone modi \citep{Wood2015MNRAS.448.3829W}. Semiregular variables that have stable periodicity are classified as SRa, while variables with different duration of individual cycles are classified as SRb. SRc semiregulars are variable supergiants. 
A source is classified as a long-period variable (LPV) if no regular pulsation period $P$ could be deduced from the observations, in which
case $P$ is indicated by `$\cdots$' in  column~4. \\
$^{(f)}$~OH/IR star -- Mira variables that show strong OH maser emission in the hyperfine split ground state transitions at 18~cm.  \\
$^{(g)}$~Known binary system. \\
$^{(h)}$~S-type AGB star with a carbon to oxygen ratio (C/O) slightly less than 1.   \\
%  +++++++++++++++++++++++++++++++++++++++++++++
{\bf References:} \\
%$\dot{M}$:
$^{aa}$\citet{Olofsson2002A&A...391.1053O}; $^{bb}$\citet{Groenewegen1999A&AS..140..197G}; $^{cc}$\citet{DeBeck2010A&A...523A..18D}; 
$^{dd}$\citet{Young1995ApJ...445..872Y}; $^{ee}$\citet{Doan2017A&A...605A..28D}; 
$^{ff}$\citet{Josselin1998A&AS..129...45J};
$^{gg}$\citet{Ramstedt2017A&A...605A.126R};
$^{hh}$\citet{Loup1993A&AS...99..291L}; 
$^{ii}$\citet{Vogt2016ApJS..227....6V};
$^{jj}$\citet{Gaia2018A&A...616A...1G};  
%  -------------------------------------------------------
%\vspace{0.25cm}   \\
%$P$: 
$^{kk}$\citet{Andronov2012OAP....25..148A}; 
$^{ll}$\citet{Perryman1997A&A...323L..49P};
$^{mm}$\citet{Kiss2006MNRAS.372.1721K};
$^{nn}$\citet{Wittkowski2017A&A...597A...9W};
$^{oo}$\citet{2017ARep...61...80S};  
%  -------------------------------------------------------
%\vspace{0,25cm}   \\
%$D$: 
$^{pp}$\citet{Zijlstra2002MNRAS.334..498Z};
$^{qq}$\citet{Vlemmings2007A&A...472..547V}; 
$^{rr}$\citet{2008ChJAA...8..343S}; 
$^{ss}$\citet{Arroyo2013A&A...554A..76A}; 
$^{tt}$\citet{2007ChJAA...7..531C};   
%  -------------------------------------------------------
%\vspace{0,25cm}   \\
%$\theta_d$: 
$^{uu}$\citet{2005A&A...431..773R}; 
$^{vv}$\citet{2018Natur.553..310P};
$^{ww}$\citet{Chiavassa2010A&A...511A..51C}; 
%  -------------------------------------------------------
%\vspace{0,25cm}   \\
%$T_{\rm{eff}}$:
$^{xx}$\citet{Marigo2008A&A...482..883M}.;
$^{yy}$\citet{2021AJ....161..147B}. \\
}
\end{table*}	
%   +++++++++++++++++++++++++++++++++++++++++++++++++++++++++++++++++++++++++++++++++++++++++++++++

\section{Observations} \label{Sec:Observations}

\subsection{ATOMIUM observing strategy} \label{Sec:Observing_strategy}

A primary requirement for the {\sc atomium} project was homogeneous observations across the sample
that would allow unambiguous comparison among sources.
The most efficient way for ALMA to achieve the science goals described in Sect.~\ref{Sec:Goals} was to target specific 
spectral frequency regions, and to observe all 17 {\sc atomium} sources in the same spectral regions.
We know exactly which molecules to monitor in Band~6 to determine the dynamical behaviour of the winds 
(Sect.~\ref{Sec:dynamics}), and to answer the questions of gas-phase chemistry and dust nucleation (Sect.~\ref{Sec:chemistry}). 
The spectral range was chosen so we automatically had the same appropriate molecules to trace the gas phase chemistry in all 17 stars
in the {\sc atomium} survey, while serendipitous detections came for free.

To spatially resolve the dust condensation region ($r\!\la$10\,--\,30\,\Rstar), an angular resolution (AR) of $\sim$25\,--\,50\,mas was needed 
for our targets, which all have large stellar angular diameters of between 3.9 and 20.5\,mas  (see Table~\ref{Table:targets}).
The finest AR requested was 25\,mas for each target, while we allowed for an upper limit of 35\,mas for stars with stellar angular diameter 
$<$9\,mas and of 50\,mas for the larger stars. 
This was offered in C43-8/C43-9 with maximum recoverable scale (MRS)  $\sim$0\farcs38--0\farcs62  (henceforth referred to as either 
`extended' or `high resolution').
To attain the full line strength of the transitions, we needed to complement these observations with data from a more compact configuration, C43-5/C43-6, 
at an AR of 0\farcs24/0\farcs13 with maximum MRS of 1\farcs5 (henceforth `mid' or `medium resolution'). 
Extended emission of the CO and SiO transitions in the ground-vibrational state  might still be resolved out even with the mid configuration. 
Hence, observations with an even more compact configuration were needed to recover the total fluxes of these transitions. 
For all targets, various single antenna CO line measurements are available to derive the global thermal structure of the wind. 
Hence, the request for the low-resolution observations was primarily based on the estimated extents of the SiO emitting regions of the targets. 
We have estimated the angular size of the SiO photodissociation region for each target using the results of \citet{Gonzalez2003A&A...411..123G}. 
The photodissociation radius of most targets varies between 2\farcs5 and 10\arcsec, except that of AH~Sco and KW~Sgr, which is less than 1\arcsec. 
Hence, we requested C43-2 observations at an AR of $\sim$1\arcsec\ (MRS ranging between 8\arcsec--10\arcsec; 
henceforth `compact' or `low resolution') for 15 out of 17 targets and in the two spectral setups that cover the SiO J=5-4 and J=6-5 lines. 
The CO J=2-1 line is also covered in the same setup as SiO J=5-4.

The 24 molecules identified in {\sc atomium} can be separated into groups according to their chemical properties, or to their utility as probes 
of the wind kinematics and wind shaping mechanisms.  
Five molecules were observed in stars of all six pulsation types (CO, SiO, HCN, SO, and SO$_2$), and three of these (CO, SiO, and HCN) are universal tracers 
of the gas dynamics.  
Four other molecules (AlO, AlOH, TiO, and TiO$_2$) are suspected precursors in the initial dust formation process that occurs in the inner wind 
within a few $R_{\star}$ of the central star.
Three molecules (SiS, H$_2$O, and CS) were observed in all but one of the pulsation types.
Four (SO, SO$_2$, SiS, and CS) inform us about the sulphur budget \citep{Danilovich2017A&A...606A.124D}, 
and one (NaCl) is a probe of the coupling of the chemistry and dynamics \citep{Decin2016A&A...592A..76D}. 
Because of the central role of these 13 molecules in characterizing the physicochemical properties of the inner and intermediate winds, 
we found it useful to designate the 13 molecules as the  `primary’  molecules.
Hereafter CO, SiO, HCN, AlO, AlOH, TiO, and TiO$_2$,
SO, SO$_2$, SiS, H$_2$O, CS, and NaCl are referred to as the `primary molecules’ in the {\sc atomium} survey.

The primary molecules all have principal rotational transitions in spectral Band~6.
Figure~\ref{Fig:spw} shows the frequency coverage between 213.83 and 269.71\,GHz, 
the frequency tunings (a, b, c, d, e, f; see also Table~\ref{Table:vel}), 
and the atmospheric transmission for the range of precipitable water vapour (PWV) recorded during the {\sc atomium} observations. 
The actual bandwidth within the total span of $\sim$56\,GHz is approximately 27\,GHz for the mid and extended configurations
(after trimming the edges), and 13\,GHz for the compact configuration.
To ensure that all the principal transitions of the primary molecules were covered in the {\sc atomium} survey, 
it was necessary to constrain the bandwidths of three of the spectral windows (spw 07, 08, and 13) to 1/2 the width of the 13 other 
spws because of: 
(1) the constraints of the ALMA Local Oscillator system on fitting the spws within the basebands; and
(2) the need to  minimise  the total number of the local oscillator tunings for efficient use of observing time.\footnote{See the ALMA Cycle~6 
Technical Handbook at \url{https://almascience.nrao.edu/documents-and-tools/cycle6/alma-technical-handbook}.}
In all three array configurations, the line free channels (or about one half the total bandwidth) are available to image the millimeter-wave continuum.

A spectral resolution of $\sim$1.3\,km/s provided sufficient resolution elements per line with typical full width at half maximum (FWHM) line widths 
ranging between 5--60\,km/s, where the smaller line widths probe the wind acceleration region. 
The velocity widths of our spectral windows (spw) and channels are shown in Table~\ref{Table:vel}. 

\begin{figure*}[t]
   \centering
\includegraphics[width=1.0\textwidth]{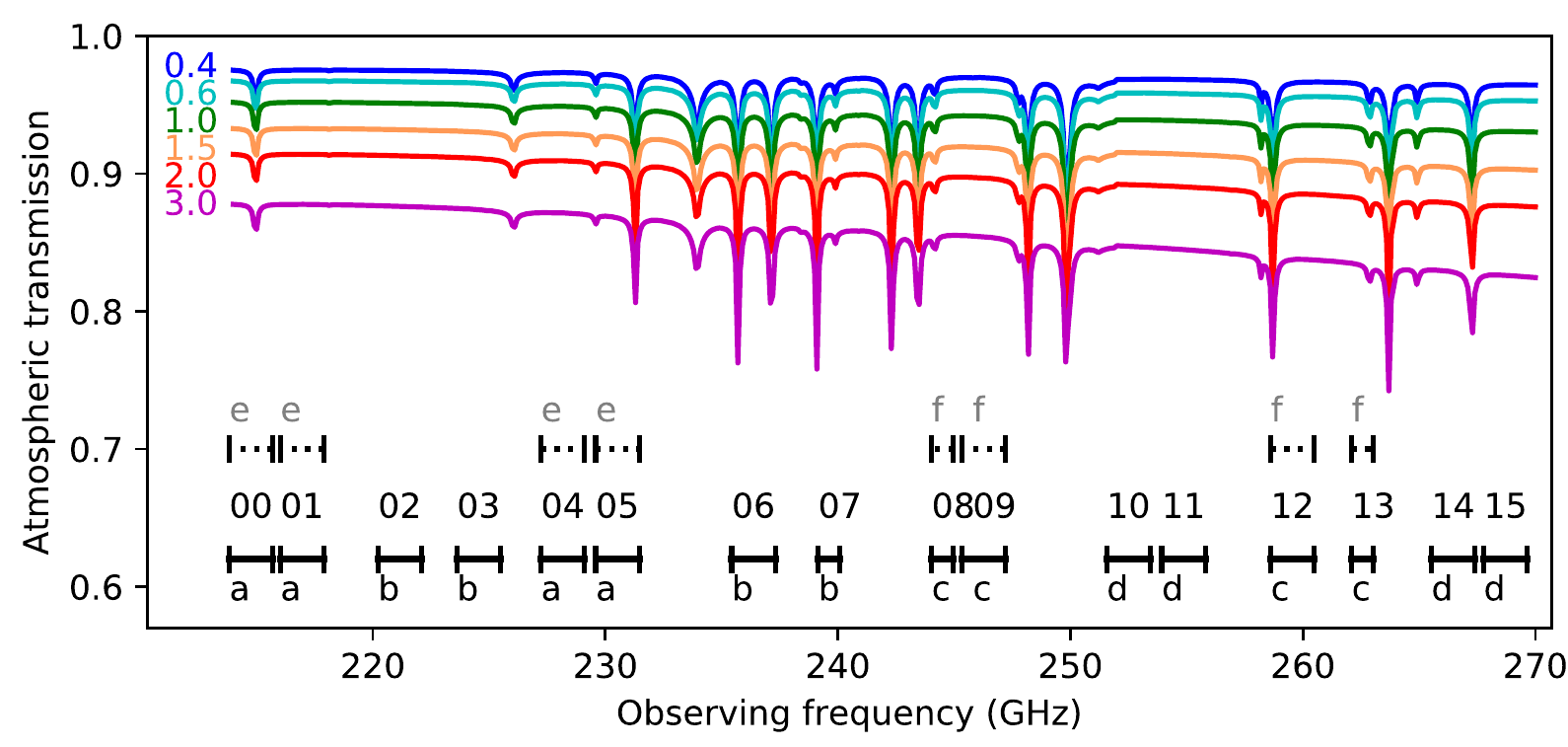}
\caption{\textbf {Frequency coverage of the {\sc atomium} project in each array configuration (see Sect.~\ref{Sec:Observing_strategy} 
and Table~\ref{Table:vel})}.       
Each black bar represents the frequency coverage of a spectral window (spw), labelled with the same index number as in our final released 
data products. 
The solid lines and letters a, b, c, d represent the frequency tunings for the medium and extended configurations; 
the dotted lines and grey letters e, f represent the frequency tunings for the compact configuration. 
The exact spectral coverage for each target depends on the adjustment to the assumed $v_{\rm{LSR}}$ on the dates of observation 
(see Table~\ref{Table:targets}). 
Each frequency tuning covered 4 spws grouped as follows: [00,01,04,05], [02,03,06,07], [08,09,12,13], and [10,11,14,15]. 
The first (second) pair of spw in each frequency tuning corresponds to the lower (upper) sideband in which the channel numbering is in 
descending (ascending) frequency order. 
The coloured lines represent the atmospheric percentage transmission labelled by the precipitable water vapour (PWV), in mm.}
\label{Fig:spw}%
\end{figure*}

\begin{table}[htp]
\caption{{\bf Velocity widths $\Delta v$ and resolutions $\delta v$ of the {\sc atomium} observations.}  }
\centering \begin{tabular}{lllll}
\hline
spw         & $\nu_{\mathrm{central}}$ & $\Delta v$    & $\delta v$       &  {Frequency}    \\  
                & (GHz)                                & (km/s)           & (km/s)              &  {tuning}          \\
\hline
00&214.8& 2598 &1.36 &     a/e  \\
01&217.0& 2572 &1.35 &     a/e  \\
02&221.2& 2523 &1.32 &     b  \\
03&224.6& 2485 &1.30 &     b  \\
04&228.2& 2445 &1.28 &     a/e  \\
05&230.5& 2420 &1.27 &     a/e  \\
06&236.4& 2360 &1.24 &     b  \\
07&239.7& 1164 &1.22$^{*}$ &     b  \\
08&244.5& 1141 &1.20$^{*}$ &     c/f  \\
09&246.3& 2266 &1.19 &     c/f  \\
10&252.6& 2209 &1.16 &     d  \\
11&254.9& 2189 &1.15 &     d  \\
12&259.6& 2149 &1.13 &     c/f  \\
13&262.6& 1062 &1.12$^{*}$ &     c/f \\
14&266.5& 2093 &1.10 &     d  \\
15&268.7& 2076 &1.09 &     d  \\
\hline
\end{tabular}
\tablefoot{
The exact central frequency  $\nu_{\mathrm{central}}$  depends on  the adjustment to the assumed $v_{\rm{LSR}}$ on the dates of 
observation.
Velocity widths are given using  $\nu_{\mathrm{central}}$. 
Due to Hanning smoothing in the correlator, the velocity resolution is 15\% broader except for spectral windows (spw) marked $^*$. 
These were observed in half the maximum bandwidth in order to fit within the frequency sidebands; the original velocity channel width 
was half that shown, and the additional averaging gives a final velocity resolution which is only 1\% broader than the channel spacing.     
The letters a, b, c, d represent the frequency tunings for the medium and extended configurations; e, f represent the 
frequency tunings for the compact configuration.     
}
\label{Table:vel}
\end{table}

To diagnose the (wide) velocity tails and hence extract the kinematical behaviour, a sensitivity of a few mJy/beam was needed 
\citep{Decin2018A&A...615A..28D}. 
The most stringent constraint on the sensitivity was set by the metal oxides, most especially by AlO
 --- the gaseous precursor of aluminum oxide (Al$_2$O$_3$) grains
\citep{2016A&A...592A..42K, Decin2017A&A...608A..55D,  Takigawa2017SciA....3O2149T}.
We calculated the expected AlO line strength for each target, and aimed for a signal-to-noise ratio of $>$3.
The sensitivity ranged from 1.5\,mJy/beam to 5\,mJy/beam for C43-8/C43-9 and C43-5/C43-6. 
For the SiO observations in C43-2, the sensitivity was 5\,mJy/beam. 

Standard ALMA observing procedures were followed, including system temperature and PWV monitoring.  
Bright, compact quasi stellar objects (QSOs) were used for calibration of the bandpass and flux scale; the latter 
was determined with respect to approximately fortnightly monitoring of Neptune or Uranus.
Phase referencing was used with a nearby, compact quasar. 
A check source --- that is to say, a known, compact source at a similar angular separation from the phase reference as the target 
--- was also observed in the extended configuration. 
Table~\ref{tab:obs_prop} summarises the observations, including the phase reference sources used; see the ALMA Science Archive\footnote{\url{http://almascience.eso.org/aq/}} for more details.   

\subsection{ATOMIUM data reduction} \label{Sec:data_reduction}

The {\sc atomium} project is among the first to collect a large volume of ALMA data for a set of three different baseline configurations, 
including long baselines. 
A substantial effort was made to explore various calibration strategies to enhance the data quality. 
In this section, we describe the standard data reduction methodology. 
The details on the calibration of the specific datasets for individual stars will be available in the 
{\sc atomium} data release (see Sect.~\ref{Sec:Release}) and where needed, additional information will be provided in separate papers.

\subsubsection{Processing each configuration}

Each fully observed Scheduling Block (SB) was processed using the ALMA
calibration and imaging Pipelines\footnote{\url{https://almascience.eso.org/processing/science-pipeline}}
\citep{Humphreys2016alma.confE...1H} implemented in CASA{\footnote{\url{https://casa.nrao.edu/}}} (the Common Astronomy Software Applications package), 
or in a few cases with manually steered
scripts, where the end result was equivalent in the two procedures. 
The calibration pipeline applies all instrumentally derived calibration (e.g., from PWV measurements) as well as corrections derived from observations of calibration and phase-reference sources.
The line free channels were initially identified from the visibility data and a linear fit to these was subtracted from the data. 
Data cubes were then made for each subtracted spw, and the line-free continuum was also imaged.

We inspected the web logs; occasionally a few instances of over- or under-flagging\footnote{`Flagging' is a term used in radio astronomy 
which refers to the process of identifying faulty or questionable portions of data that are not used in further steps of the data analysis and 
imaging.} were identified, but the former were too trivial to affect sensitivity significantly and the latter were remedied during our processing.  

For each star, each full set of tunings in each configuration was processed by the following steps:

\begin{enumerate}
\item{Two copies of the pipeline calibrated target data were split out: 
one at a `continuum' spectral resolution of 15.625\,MHz, and the other at a `line' spectral resolution of 0.9765625\,MHz 
which ranges from 1.09 to 1.37 km/s in velocity units.
These were then concatenated to make continuum and line datasets containing the full spectral coverage for each star and array configuration. 
The concatenation task aligns the phase centre of each input visibility dataset with that of the one measured at the earliest date.
The extended configuration data were all taken within 5 weeks, so any errors in the predicted proper motions would cause $<$1\,mas discrepancy  
(see Sect~\ref{sec:comb}) and the self-calibration (see step 5) takes care of relative alignment. 
}\vspace{.5mm}

\item{The Lumberjack\footnote{\url{https://github.com/adam-avison/LumberJack}} package was used to
  identify line free channels from the pipeline image cubes.  
  The selection was adjusted to correspond to the channelisation of the
  continuum and line datasets, and checked interactively using the
  visibility data.}
  
  \item{The continuum-only channels of the dataset were imaged. 
In most cases the continuum emission distribution was dominated by a compact peak, but at the highest resolution some stars 
were slightly resolved.
Nonetheless the peak signal-to-noise ratio (S/N) was $\gtrsim$100 for all the stars except SV~Aqr where it was $\sim$50.
}

\item{The stellar peaks were offset by up to a few hundred mas from the
  predicted continuum positions (see Sect.~\ref{sec:comb}). The measured position was
  used as the imaging field centre for further extended configuration
  images as the displacement could be a significant fraction of the
  chosen image size. Mid and compact configuration images were made
  using the observing phase centres.}
  
\item{The clean components from the first continuum image  were used as a starting model for self-calibration. 
This removes any small offsets between SBs due to differences in calibration or proper motion uncertainty, and improves 
the image quality. 
If the signal-to-noise ratio was sufficient,  an image using a first-order spectral index provided a model for more cycles of 
self-calibration, including amplitude self-calibration. 
Fortunately, amplitude offsets are only significant above the noise in sources bright enough for self-calibration.  
In the case of continuum sources with complex structure, we checked that the apparent complex structure was not due to an incorrect model 
or inappropriate imaging parameters --- for example, if a secondary, compact component was present, we investigated whether it remained 
after using a single point model.  
Once the optimum level of calibration was achieved, images with and without the primary beam correction were made.}

\item{The corrections were also applied to the line data set, and we then checked the selection of line-free channels and subtracted the 
continuum using a first-order fit.}

\item{A spectral image cube for each spw and configuration was made large enough to encompass all detectable circumstellar 
emission at that resolution. 
Weighting for the optimum balance between sensitivity and the required resolution resulted in a synthesized beam 
$\theta_{\mathrm B}$ that varied slightly depending on target elevation and exact antenna positions. 
Cubes were made with and without the primary beam correction. 
Automasking was used for the mid and compact configurations; the masks derived for mid were also used for the extended configuration.}

\item{Spectra were extracted for a range of circular apertures (as appropriate for the configuration resolution and image size), centred on the 
stellar peak.}
  
\end{enumerate}

The properties of each continuum and cube image are listed in Table~\ref{tab:cont} and Table~\ref{tab:cube}, respectively. 
The values for the maximum recoverable scale (MRS) apply to both line and continuum for a given configuration and target, 
although the imaging fidelity for cubes is slightly worse due to the narrower coverage of the visibility plane per channel as compared 
to the broadband continuum.

\subsubsection{Accuracy}

In this section, we cover the overall accuracy of the {\sc atomium} observations. 

\medskip

Astrometric position uncertainties in the {\sc atomium} data arise from several factors: 
\begin{itemize}
\item Transferring phase corrections from the reference source to the target is affected by the difference in the angular separation 
and in the time between the observations of the phase reference and the target. 
Following expressions from \citet{Taylor1999ASPC..180.....T},
it can be estimated from the magnitude of the initial target self-calibration phase corrections
and for the 43 antennas in use, that the position error is roughly equal to: (synthesised beam) $\times$ (phase error in degrees/1450). 
The phase corrections are typically $\sim$40$^{\circ}$, and the uncertainty is  $\sim\theta_{\mathrm B}/40$ which corresponds to
$\sim$0.7, $\sim$6, and $\sim$25\,mas for the extended, mid, and compact configurations.

\item The phase reference position is
usually accurate to $<$1 mas as most phase reference source positions are taken from  the Very Long Baseline Interferometer (VLBI) calibrator catalogues: see the ALMA Calibrator Source Catalogue.\footnote{ \url{https://almascience.eso.org/sc/}}

\item Position measurement
accuracy for a compact source such as the star is given by  $f\times \theta_{\mathrm B}$/(S/N) where $f\!=\!0.5$ is appropriate 
for a well-filled array, tending to $f\!=\!1$ for the extended configuration. 
S/N is the signal-to-noise ratio, typically $\ga\!100$, leading to stochastic
position errors of no more than $\sim$0.25, $\sim$2, and $\sim$5\,mas for the extended, mid, and compact configurations.  

\item Antenna r.m.s. position errors now contribute $\sim$1\,mas astrometric errors (for typical target-calibrator separations of around 6~degrees). 
The measurement technique involving QSO observations described in \citet{ALMA2015ApJ...808L...1A} has since been improved by the addition of more weather stations across the ALMA tracks which refine the measurements of the atmospheric delay.
\end{itemize}

Thus, the total astrometric uncertainty has typical values of 2, 7, and 26\,mas for the extended, mid, and compact configurations. 
This is consistent with the typical extended-configuration check-source position errors of 1\,--\,5\,mas. 
The stellar positions used for astrometry were measured before any self-calibration as this cannot improve the astrometry.  

The faintest stars are the most difficult case for self-calibration, and self-calibration was only performed for the phase with a solution 
interval of a single scan. 
This removes errors due to the phase-reference-target angular separation  and inconsistencies between antennas --- at least 
halving the phase errors. 
A residual 20$^{\circ}$ phase error would give a  4\% amplitude error. 
The direct causes of amplitude errors fluctuate more slowly than for phase errors, so the solution transfer has a smaller uncertainty.

The ALMA flux density scale has an uncertainty of up to 5\% in Band~6 due to the variability of QSOs in between monitoring intervals. 
After the data reduction was completed, a problem was identified with the $T_{\rm{sys}}$  normalisation that might affect the 
flux scales in a channel
%  ++++++++++++++++++++++++++++++++++++++++++++++++++++++++++++++++++++++++++++++++++++++++++
dependent way,\footnote{\url {https://almascience.eso.org/news/amplitude} \break \url{-calibration-issue-affecting-some-alma-data}} 
%  ++++++++++++++++++++++++++++++++++++++++++++++++++++++++++++++++++++++++++++++++++++++++++ 
Using scripts provided by ESO, we confirmed that --- in the example of the CO J=2-1 line observed in KW~Sgr in the mid configuration 
--- the magnitude of the effect is only $\sim$2\%.

Each of our targets was observed several times separated by months, with uncertainties at each epoch, 
so the total amplitude scale error is at least 10\% (to be analysed in more detail in future papers).

Each channel is labelled with its central frequency and the only significant uncertainties in $v_{\rm{LSR}}$ arise from poorly known 
rest frequencies for a few little-studied species.

\subsubsection{Combining configurations}
\label{sec:comb}

In order to combine the mid, extended and (if used) compact configurations, the data sets have to be aligned in position and 
flux density scale.  
The observing schedules were prepared using \emph{Hipparcos}-based positions and proper motions.  
The highest proper motions are $\sim$70 mas yr$^{-1}$, referenced to epoch 2000, and by the time of the observations, the offsets from the predicted positions were up to a few hundred mas, suggesting errors of up to 10~mas~yr$^{-1}$. 
The mid and extended configurations completed so far were taken up to 9 months apart, so the alignment error could be 
a significant fraction of the extended configuration resolution, requiring position corrections before combination. 
Future comparisons between positions derived from {\sc atomium} and GAIA might lead to improved accuracy.

After applying all self-calibration to the data taken in each configuration, the continuum visibility data were split out 
and the line channels flagged (to allow later averaging). 
We used task {\sc fixvis} to rotate the phase centre of each visibility data set to the position of its continuum image peak.  
We took the peak position measured from the extended configuration image (with the highest astrometric accuracy), 
as the reference position and used task {\sc fixplanets} to re-label the centre of the mid and compact configuration data sets 
to this position.  
All positions are given in ICRS.

We then plotted amplitude against $uv$ distance (i.e., projected baseline lengths) for each configuration,
using the data sets with the peak at the phase centre, averaging all continuum channels, to investigate whether the amplitude scaling 
was consistent on baseline lengths common to all configurations.
The situation is complicated because, as well as possible flux scale errors of order 10\%, the photospheric pulsation or the formation 
of dust could cause a flux variation of a few percent, and the extended line emission is likely to be much less affected on the same
time scale. 
In three cases the emission in the extended configuration was >10\%  brighter, probably due to a known bias in the flux scale
calibration of long baselines with phase noise, and we rescaled these data to be consistent with the flux densities in the other
configurations. 
The position-corrected continuum data were then concatenated giving each data set equal weight. 
We imaged the combined data applying a $uv$ taper, equivalent to a Gaussian beam of 20\,mas at the FWHM,  
in order to avoid artefacts owing to the relatively sparse coverage on the longest baselines, giving $\theta_{\mathrm B} \sim 50$~mas.

The calibrated line data were then similarly split out, the position corrections applied, and the data concatenated and image cubes made.
All spw were imaged using an image size of 4\arcsec\ and multi-scale clean, and giving higher weight to the largest scales. 
This maintained high resolution whilst ensuring that all scales in the data were imaged smoothly, 
avoiding over-emphasied, spotty small scales owing to the higher sensitivity of the extended observations.
The emission of a few lines in the ground vibrational state was extended over more than 4\arcsec. 
In the example of the $^{12}$CO  J=2-1 line we made a 40\arcsec\ image to the 0.2~primary beam sensitivity level.
The size (8192$\times$8192 pixels$^2$) and time taken to clean made it impractical to make such images for more than a few hundred channels
for each target. 

\vspace{-0.25cm}

\section{ATOMIUM data release} \label{Sec:Release}

An important motivation for the {\sc atomium} survey was to provide the community with a set of accurately calibrated ALMA data of evolved stars, which --- on the grounds of its homogeneous setup --- can advance our insights into dynamical and astrochemical processes in various astrophysical media, and spark related research. 
To that end, we will release a suite of data products which go beyond the normal standard contents in the ALMA Archive
where all the {\sc atomium} data are now available.

\subsection{Data products}
\label{Sec:Products}

The enhanced data products for each star will include: 
(1) the visibility data self-calibrated as described in Sect.~\ref{Sec:Observations}, with all tunings aligned per configuration, 
and the data sets from the three configurations combined; 
(2) consistent image data cubes of manageable size, covering the full spectral range; 
(3) continuum images; and 
(4) spectra extracted at a range of apertures. 
%  +++++++++++++++++++++++++++++++++++++++++++++++++++++++++++++++++=
Documentation describing the data products will be provided, and all the principal data poducts 
will be available in the ALMA Archive standard format via the ALMA Large Programme web pages in 2022.
%{\red {within the next few months of 2021}}.
In addition, we will provide the parameters of all the spectral lines observed in the three array configurations, 
and a Table with the parameters of all the unidentified lines. 
The spectral and imaging templates that will be created will allow the astronomical community to explore the
entire dataset, and to exploit these libraries in other research domains which will in turn serve as a legacy 
for the community.\footnote{For announcements and links to products see:  \hfill\break  \url {https://fys.kuleuven.be/ster/research-projects/}   
 \url{aerosol/atomium/atomium } }

\subsection{Example of the OH/IR star IRC$-$10529}
\label{Sec:IRC-10529}
     
In the following discussion we refer to the example of the OH/IR-star IRC$-$10529 for each of the data products included in the {\sc ATOMIUM} 
data release. 
This target was chosen because the morphology is not too complex \citep{DecinScience}, it is rich in molecular spectral lines, 
and the data can be used for a straightforward demonstration of the {\sc atomium} data products and their role for scientific 
inference.\footnote{{{ Many OH/IR stars were first observed in the two micron Caltech (IRC) sky survey \citep{1969tmss.book.....N}, 
and were subsequently shown by radio astronomers to have intense lines from OH and H$_2$O masers.  Most of the stars in the IRC catalog 
are M-type stars which have high mass-loss rates.}}} 

\begin{itemize}

\item{{\sc continuum image:} The low, medium, and high spatial resolution continuum maps of IRC$-$10529 are displayed in Fig.~\ref{Fig:cont_IRC10529}. 
For each resolution, the emission is spatially resolved with deconvolved sizes of $0\farcs348\times0\farcs310$, $0\farcs085\times0\farcs046$ 
and $0\farcs015\times0\farcs011$ for the compact, mid, and extended configuration respectively.
The peak continuum flux densities are  6--7\,mJy/beam in all configurations.
For a star with effective temperature of 2700\,K and angular diameter of 6.47\,mas at a distance of 760\,pc, the stellar blackbody contribution in the selected spectral windows is $\sim$3.7\,mJy. Hence roughly 50--60\% of the continuum flux can be attributed to dust emission and the radio photosphere.}

\item{{\sc channel maps:} 
Figure~\ref{Fig:CO_CM_IRC10529} shows the low resolution channel map of $^{12}$CO  J=2-1 at 230.538\,GHz and a lower state 
energy (E$_{\rm{low}}$) of 5.53\,K. 
%  ----------------------------------------------
{ {We could not find any measurements with single antennas  of the $^{12}$CO J=2-1 v=0 line in IRC-10529 in the literature, 
although there are such measurements for other sources in the {\sc atomium} sample.  
Therefore to estimate the total amount of the CO flux recovered for this source, we referred to the observations of the J=1-0 line with 
SEST \citep{1992A&AS...93..121N} and the J=3-2 line observed with APEX-2a \citep{DeBeck2010A&A...523A..18D}, and  
to the Jy/K factors from the APEX/SEST web site\footnote{\url{https://www.apex-telescope.org/telescope/efficiency/index.php.old} 
\hfill\break \url{https://www.apex-telescope.org/sest/html/telescope-instruments/telescope/index.html}}
and R.~Laing (personal communication). 
We estimate an average flux density of $\sim$44~Jy for the J=2-1 transition on  the basis of  a peak flux density of 16.2~Jy 
for the J=1-0 line and 71~Jy for the J=3-2 line, on the assumption that a  flat-topped approximation to the profiles is adequate
owing to the other uncertainties.  
ALMA recovered a peak flux density of  25~Jy for the J=2-1 line, implying that we recovered $>$55\% of the most extended emission 
(and probably all the emission of the compact front and back caps).}}
%  ----------------------------------------------
\hfill\break
\indent As discussed by \citet{DecinScience}, the data show the prevalence of a broken spiral-like structure which can be explained by binary interaction caused by an as yet undetected (sub-)stellar companion. 
As an example of a medium-resolution channel map, we show the emission of the $11_{1,11} - 10_{0,10}$ transition of SO$_2$ (at 221.965\,GHz 
and E$_{\rm{low}}$\,=\,49.71\,K), where  the brightness distribution is composed of a hollow shell structure located at a radius of $\sim$2\farcs5 (Fig.~\ref{Fig:SO2_CM_IRC10529}). 
A similar shell-like structure was previously seen for SO in oxygen-rich AGB stars with a high mass-loss rate by \citet{Danilovich2016A&A...588A.119D}, but the limitations of their data did not allow these authors to study the spatial distribution of SO$_2$. 
The current ALMA data now confirm the emission of both SO and SO$_2$ can have a shell-like structure, in accord with recent chemical model predictions
\citep{2018A&A...616A.106V, Danilovich2020MNRAS.tmp..647D}.
The high-resolution channel map of the SiO J=5-4 emission (at 217.105\,GHz) is displayed in Fig.~\ref{Fig:SiO_CM_IRC10529}. 
Although the channel map is challenging to interpret at face value, the moment1-map has proven to be very valuable for understanding 
the velocity vector field in the inner wind region of various {\sc atomium} sources \citep{DecinScience}, 
which (as shown in the next item) is illustrated very nicely in IRC$-$10529.}

\item{{\sc moment1-map:} First moment (or \textit{moment1}) maps are  utilised as a tool for  visualising structures in the velocity fields. The maps are obtained by 
\begin{equation}
M_1 = \frac{\sum_{\nu_{\rm{blue}}}^{\nu_{\rm{red}}}I_\nu \, v_\nu \, d\,\nu}{\sum_{\nu_{\rm{blue}}}^{\nu_{\rm{red}}}I_\nu \, d\nu}\,,
\end{equation}
with the velocity channels centred around $v_{\rm{LSR}}$. We illustrate the strength of this visualisation for the low, medium, and high spatial resolution data of the SiO J=5-4 emission of IRC\,$-$10529 (see Fig.~\ref{Fig:mom1_IRC10529}). The line velocity map exhibits distinct red-shifted and blue-shifted components, which is the classical signature of rotation or a bipolar outflow \citep{Kervella2016A&A...596A..92K, 
DecinScience}.}

\item{{\sc spectra and line identifications:}
Shown in Fig.~\ref{Fig:spectra_IRC1} and Figs.~\ref{Fig:spectra_IRC2}--\ref{Fig:spectra_IRC4} are
the 16 spectral windows (spw) of IRC$-$10529 observed with medium resolution (mid configuration).
The spectra were extracted with an aperture radius of 1\farcs8 in the upper part of the panels 
and 0\farcs2 in the lower part, and are plotted on a common frequency scale. 
The line identifications shown in the upper panel were made using the spectral line  catalogues of the Cologne Database for 
Molecular Spectroscopy \citep[CDMS,][]{Muller2001A&A...370L..49M, Muller2005JMoSt.742..215M, Endres2016JMoSp.327...95E} 
and the Jet Propulsion Laboratory \citep[JPL,][]{Pickett1998JQSRT..60..883P}, and by referring to prior spectral line surveys.  
In all, 60 lines from 12 molecules are observed in IRC$-$10529 in the mid configuration.  
These include CO, SiO, HCN, SO, SO$_2$, SiS, CS,  H$_2$S, H$_2$O, NaCl, KCl, and PO.
Some molecular features, such as those of AlOH and OH (although not shown here), are only visible in the extended configuration, 
because the longer on-source observing time provides higher sensitivity to compact emission.
In addition, there are a couple of weak features whose carriers have not yet been identified.
The parameters of the molecular lines in IRC-10529 will be presented in the Molecular Inventory paper 
(Wallstr{\"o}m et al.\ \textit{in prep}, see Sect.~\ref{Sec:chemistry}).}

\end{itemize}

%   ####################################### CONTINUUM MAPS ##############################################

\begin{figure*}[htp]
                                \centering\includegraphics[angle=0,width=.33\textwidth]{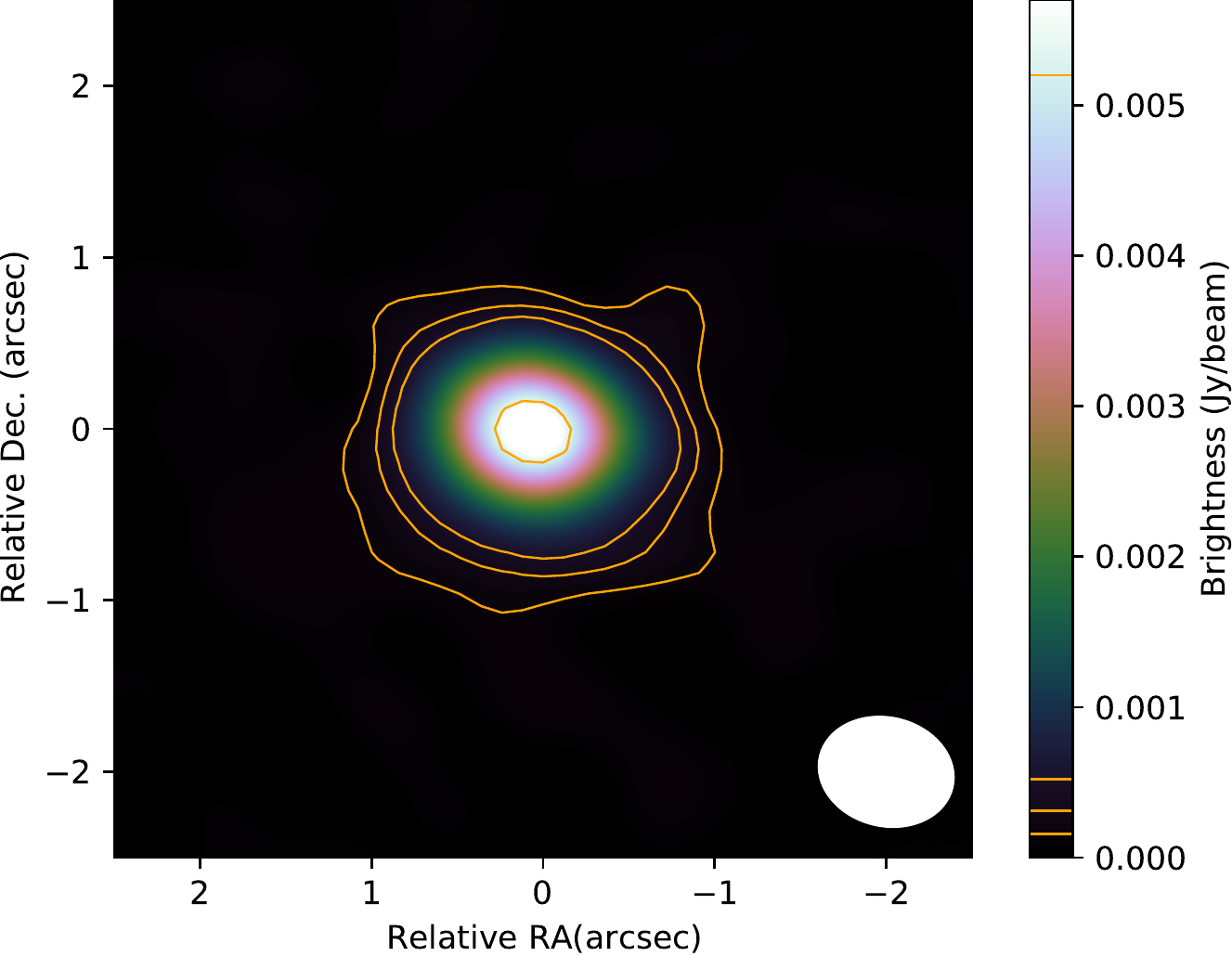}
    \hfill
        \centering\includegraphics[angle=0,width=.33\textwidth]{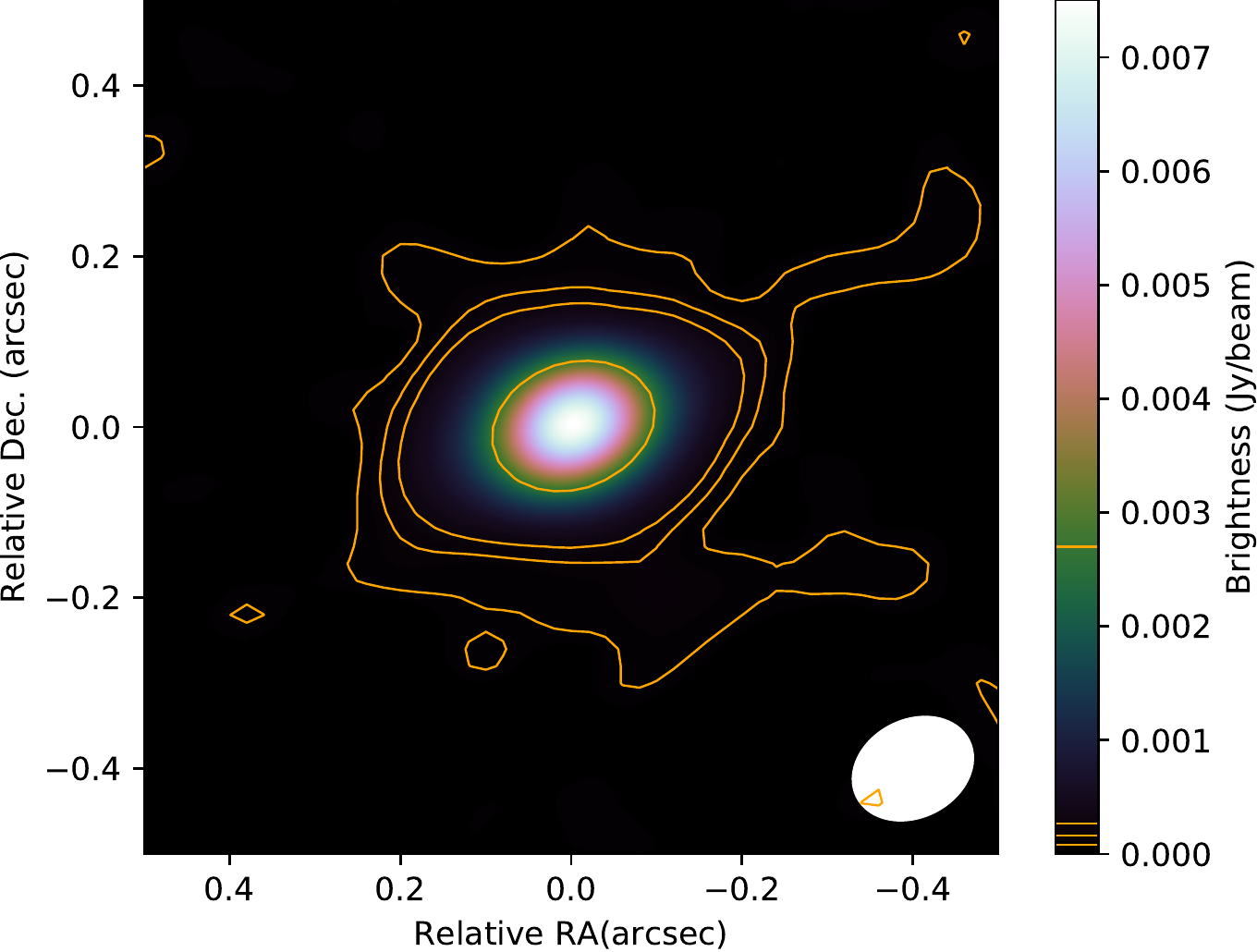}
    \hfill
        \centering\includegraphics[angle=0,width=.33\textwidth]{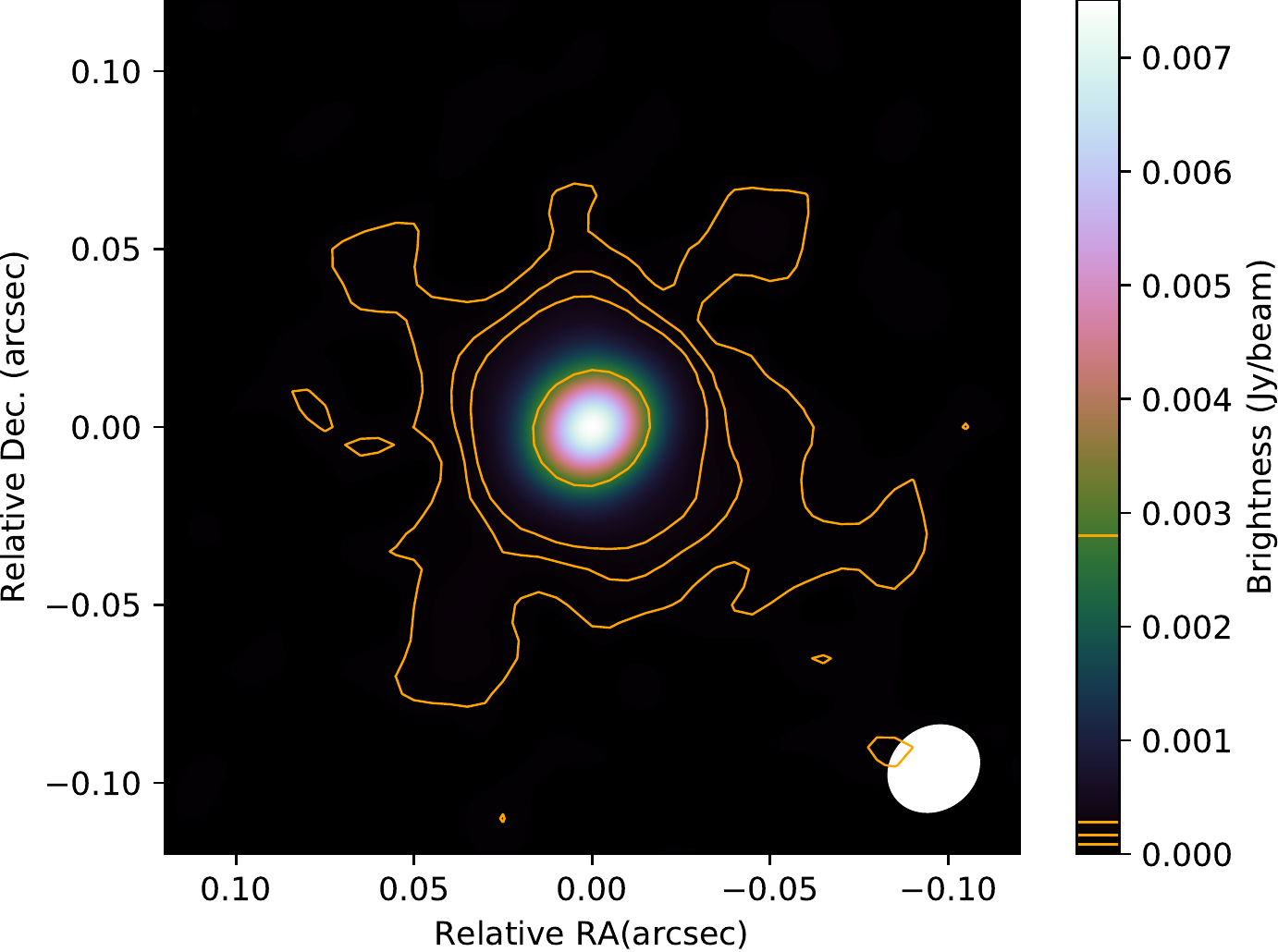}
\caption{\textbf{Continuum-maps of IRC\,$-$10529.} Low (left panel), medium (middle panel), and high (right panel) spatial 
resolution continuum map. 
Contours (in orange) are indicated in steps of (3, 6, 10, 100)$\times \sigma_{\rm{rms}}^{\rm{cont}}$ (see Table~\ref{tab:cont}).
The ALMA synthesized beam is shown as a white ellipse in the lower right corner of each panel (see Table~\ref{tab:cont}).
}
\label{Fig:cont_IRC10529}
\end{figure*}

%   ####################################### CHANNEL MAPS ##############################################

%  +++++++++++++++  CO   ++++++++++++++++++

\begin{figure*}[!htbp]
\vspace*{1ex}
\centering\includegraphics[angle=0, width=185mm]{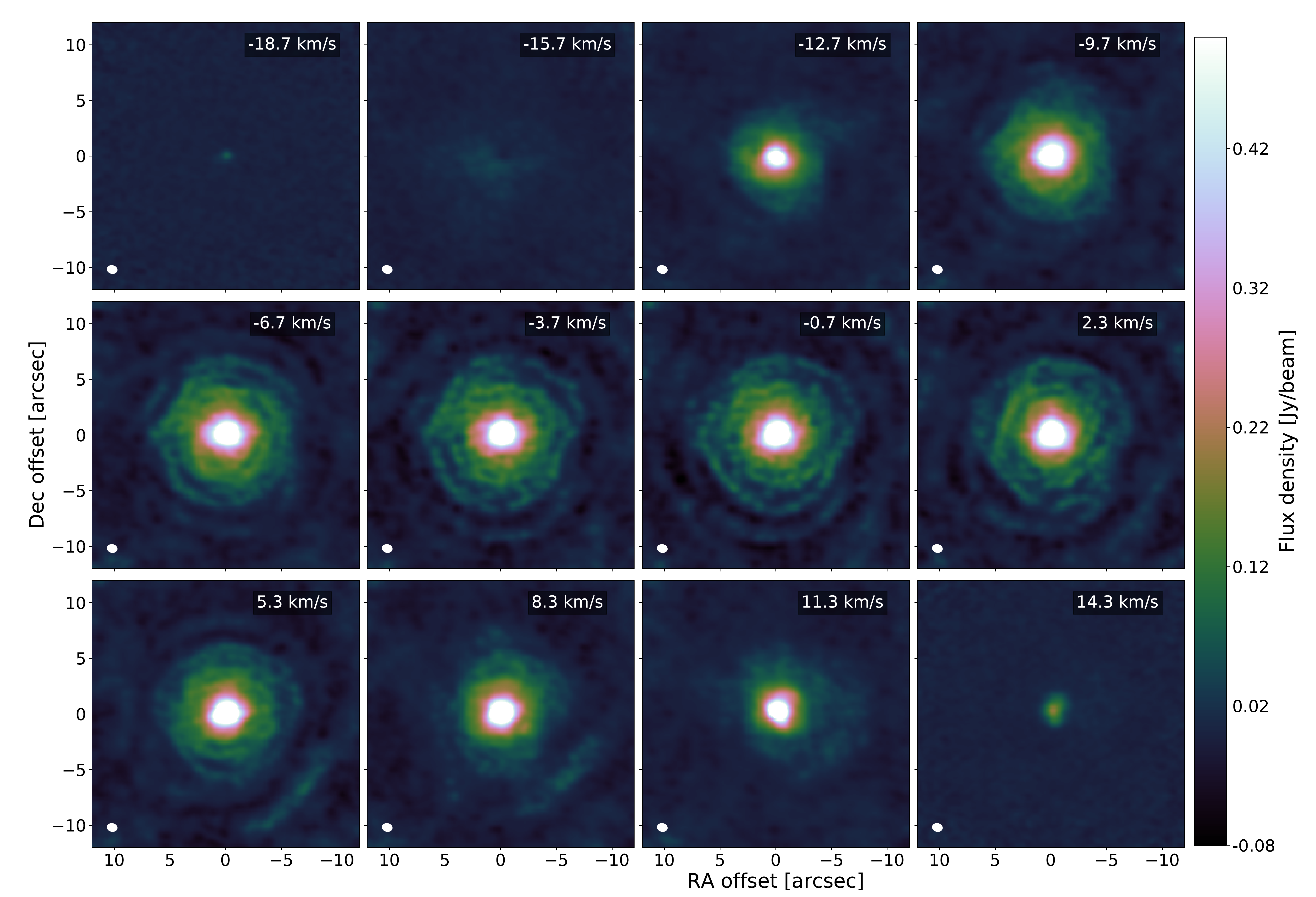}
\caption{\textbf{Low resolution channel map of  $^{12}$CO v=0 J=2-1 in IRC\,$-$10529.} 
The peak of the continuum emission is at (0,0). 
The velocity (in km~s$^{-1}$) is with respect to the stellar velocity of $-16.3$~km~s$^{-1}$ (see the last column in 
Table~\ref{Table:targets}), and is indicated in the upper right corner of each panel. 
The ALMA synthesized beam is shown as a white ellipse in the lower left corner of each panel (see Table~\ref{tab:cont}).
The offsets in right ascension and declination are with respect to the peak of the continuum emission. 
%(where north is up and east is to the  left). 
The channel maps are best viewed in the electronic version.
}  
\label{Fig:CO_CM_IRC10529}
\end{figure*}

%  +++++++++++++++  SO2   ++++++++++++++++++

\begin{figure*}[!htbp]
\centering\includegraphics[angle=0, width=170mm]{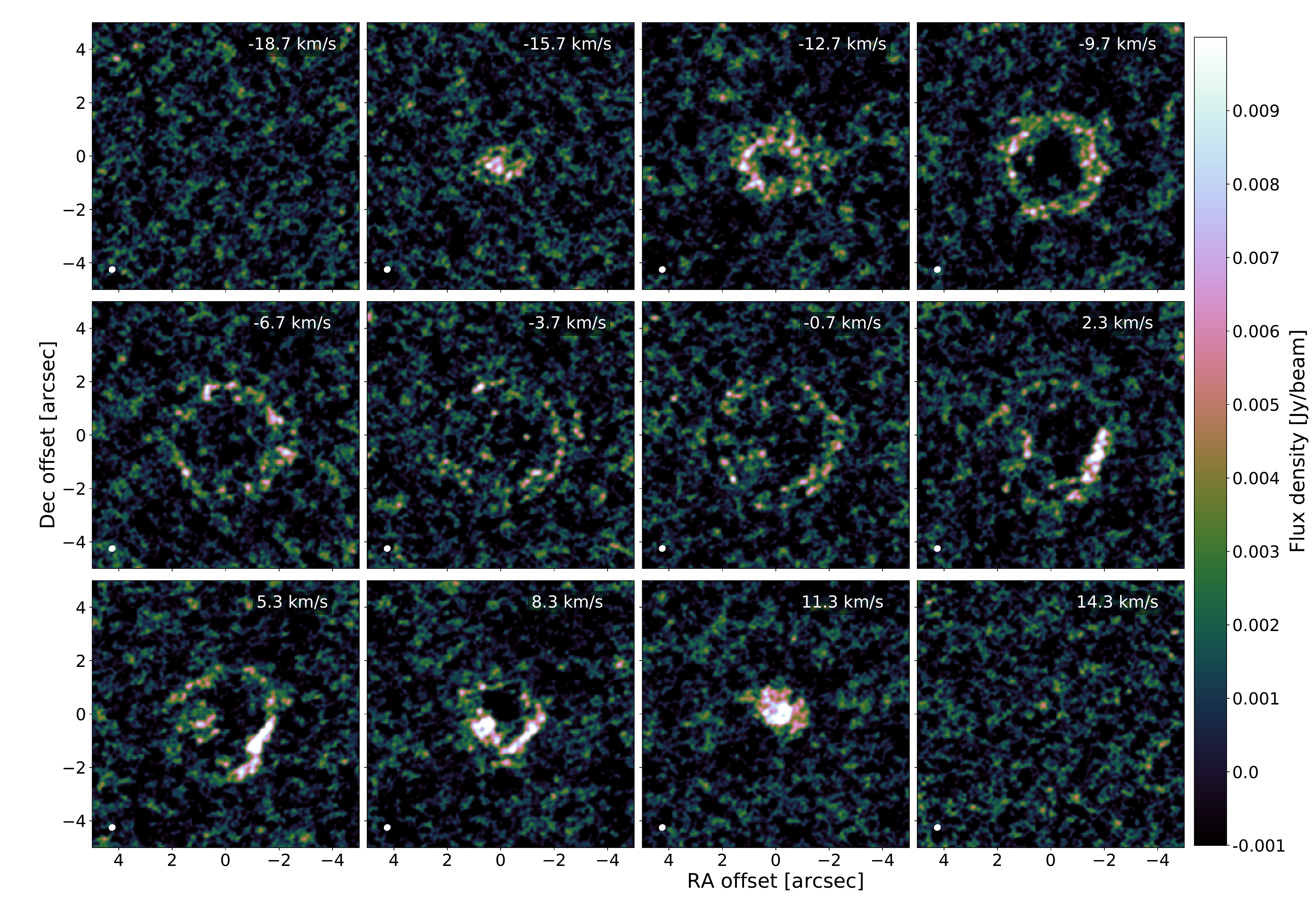}
\caption{\textbf{Medium resolution channel map of SO$_2$ v=0  11(1,11)-10(0,10)  in IRC\,$-$10529.} 
See Fig.~\ref{Fig:CO_CM_IRC10529} caption.
The emission shows a hollow shell structure located at a radius of $\sim$2\farcs5.}
\label{Fig:SO2_CM_IRC10529}
\end{figure*}

%  +++++++++++++++  SiO    ++++++++++++++++++

\begin{figure*}[!htbp]
\centering\includegraphics[angle=0, width=165mm]{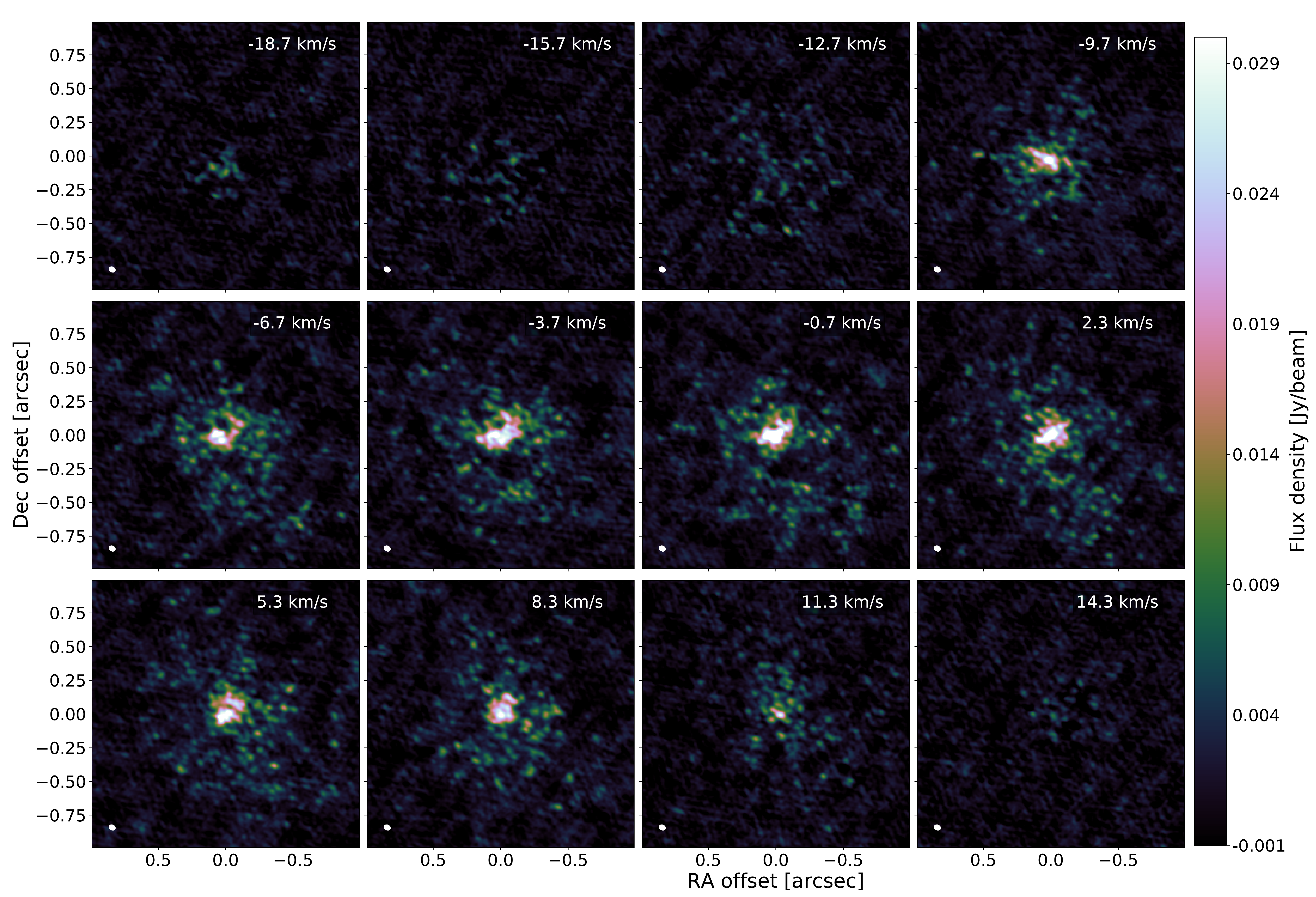}
\caption{\textbf{High resolution channel map of SiO v=0 J=5-4 in IRC\,$-$10529.} 
See Fig.~\ref{Fig:CO_CM_IRC10529} caption.}
\label{Fig:SiO_CM_IRC10529}
\end{figure*}

%   ########################################     MOMENT 1 MAPS. ##############################################

\begin{figure*}[htp]
 \centering\includegraphics[angle=0,width=.33\textwidth]{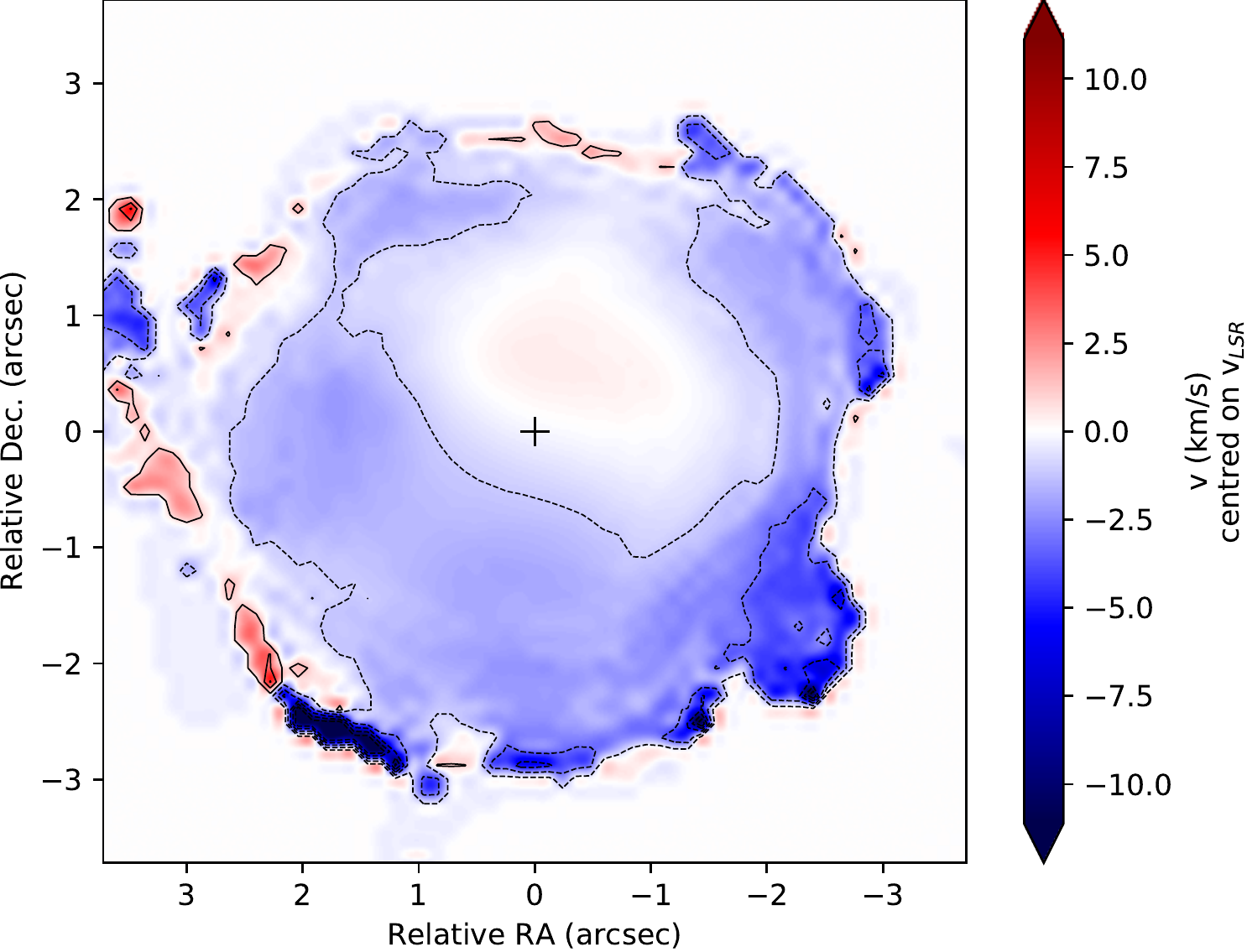}
\hfill
\centering\includegraphics[angle=0,width=.33\textwidth]{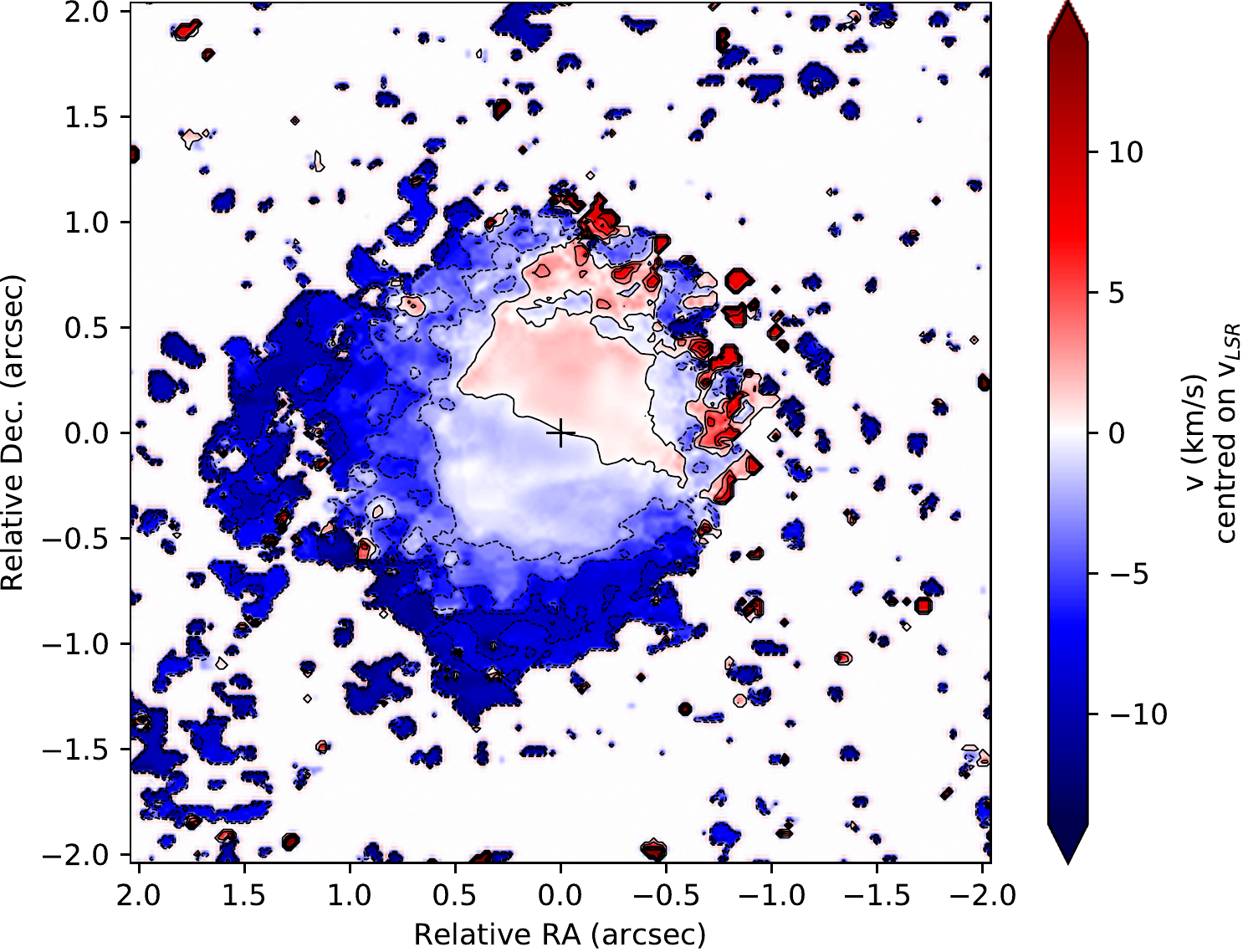}
    \hfill
\centering\includegraphics[angle=0,width=.33\textwidth]{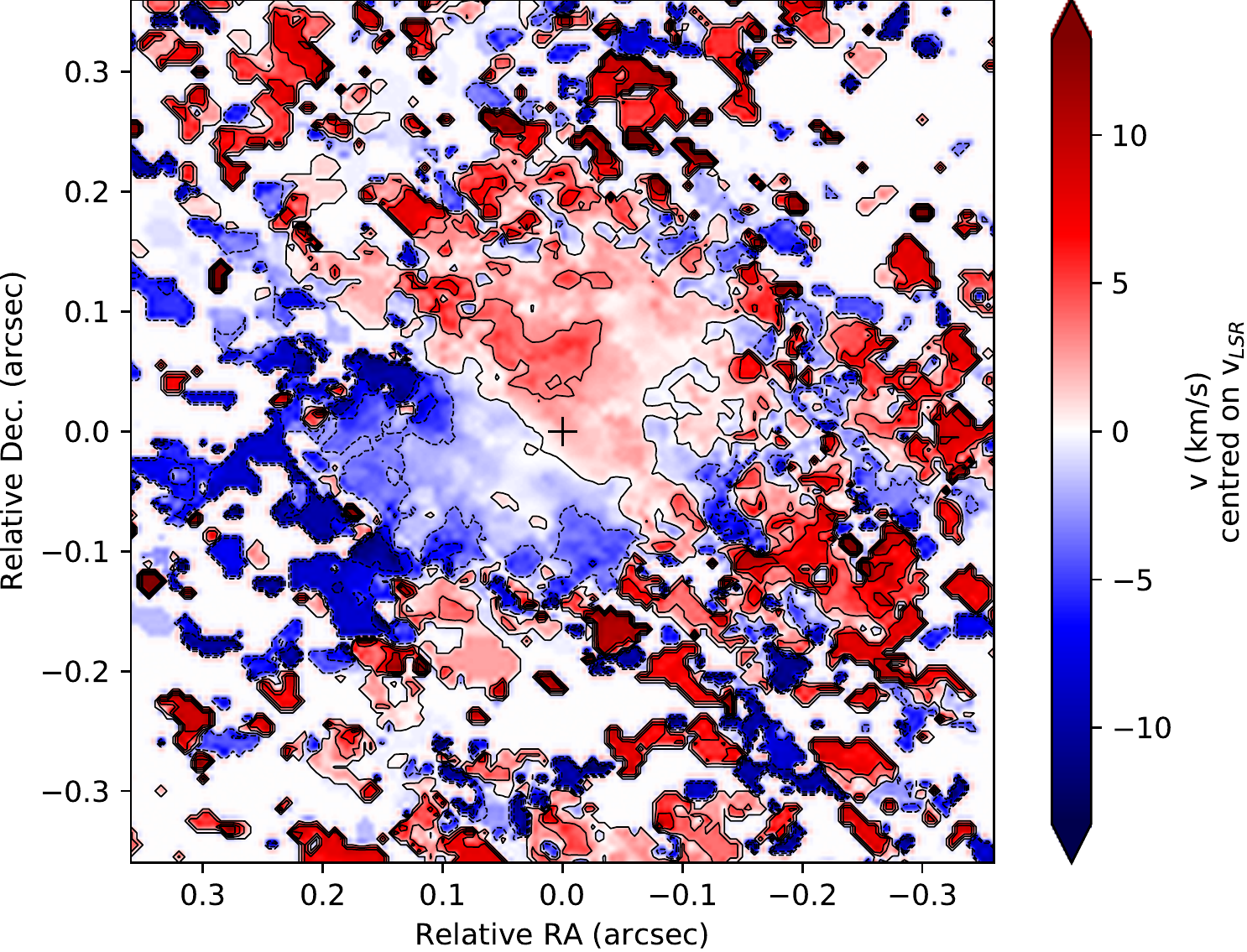}
\caption{\textbf{SiO moment1-maps of IRC\,$-$10529.} Low (left panel), medium (middle panel), and high (right panel) spatial resolution moment1-map of SiO v=0 J=5-4. The black cross indicates the position of the AGB star. The distinct spatial difference between red and blue-shifted velocity components indicates signs of rotation or bipolarity in the inner $\sim$0\farcs5 region of IRC\,$-$10529.}
\label{Fig:mom1_IRC10529}
\end{figure*}

%   #########################################################################################################
% The figures were made with /Users/leen/leda/latex/ALMA/ATOMIUM/overview_paper/programmes/plot_spectrum.pro
\begin{figure*}[htp]
\centering\includegraphics[angle=0,width=.48\textwidth]{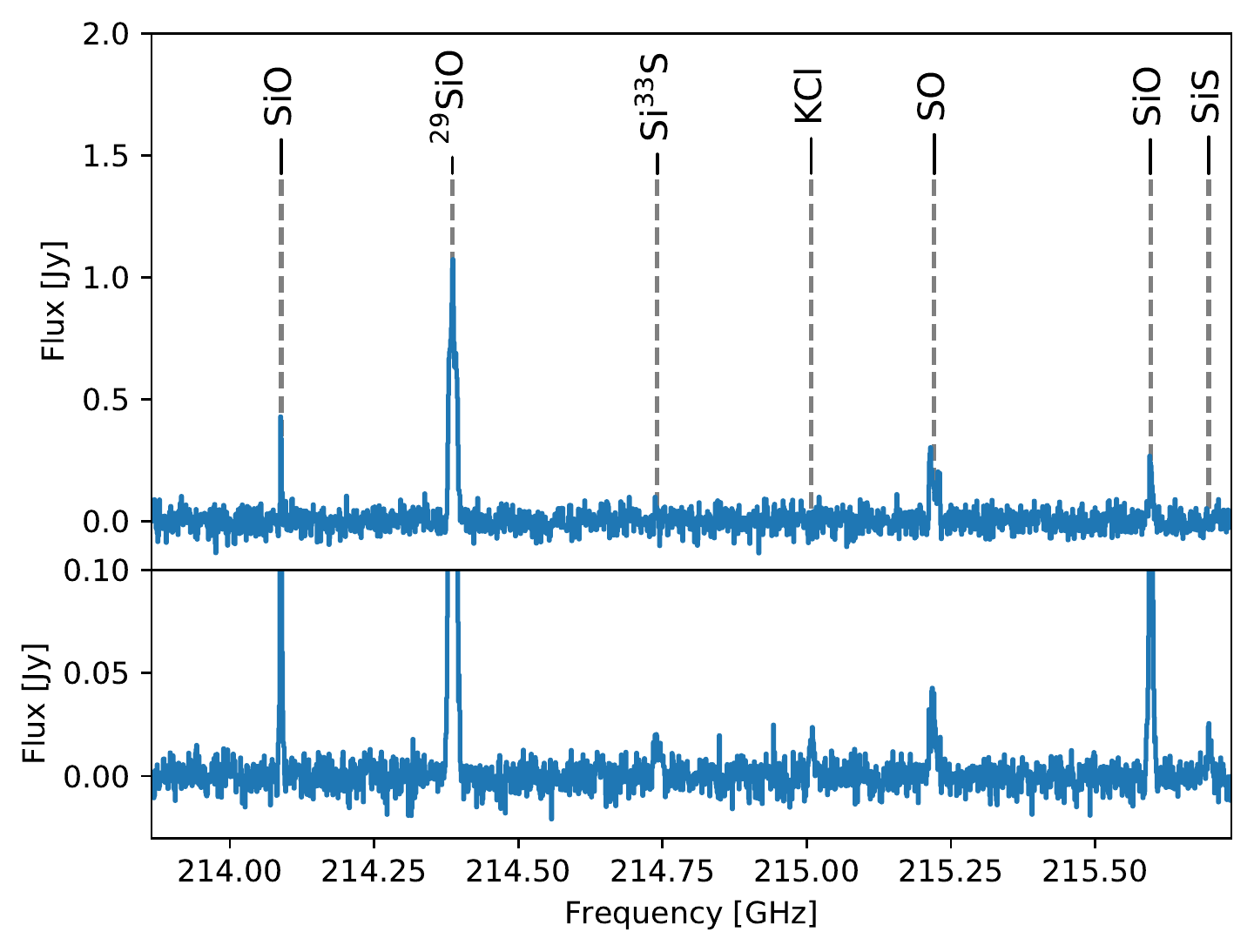}
\hfill
\centering\includegraphics[angle=0,width=.48\textwidth]{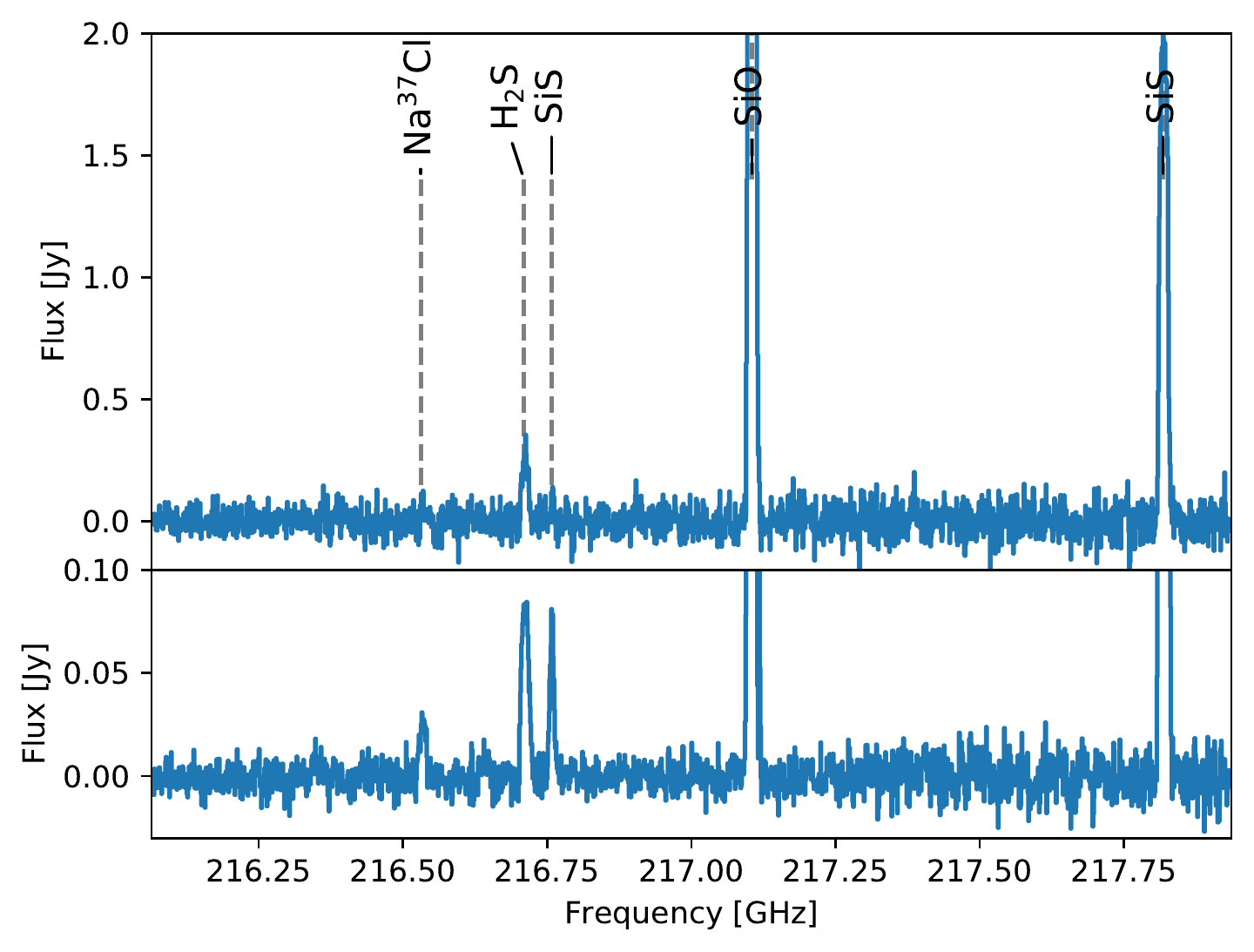}
\caption{\textbf{Spectra of IRC$-$10529.} 
Spectra of IRC\,$-$10529 for cubes 00 (left panel) and 01 (right panel) observed in the medium-resolution configuration with ALMA. 
The spectra were extracted with an aperture whose radius is 1\farcs8 (upper part) and 0\farcs2 (lower part).
The frequency scale refers to the rest frequency adjusted to the $v_{\rm{LSR}}$ of the star indicated in Table~\ref{Table:targets}.
Plots for the remaining cubes 02--15 are shown in Figs.~\ref{Fig:spectra_IRC2}--Figs.~\ref{Fig:spectra_IRC4} in the Appendix.}
\label{Fig:spectra_IRC1}        
\end{figure*}
%  ++++++++++++++++++++++++++++++++++++++++++++++++++++++++++++++++++++++++++++++++++++++++++

\section{Result --- Wind kinematics of the {\sc{atomium}} AGB and RSG sources} \label{Sec:Results}

\subsection{Background} \label{Sec:background}

The {\sc atomium} data introduced in Sect.~\ref{Sec:Goals} provides a unique opportunity for studying the wind kinematics in the circumstellar envelope of the 17 AGB and RSG sources. 
Here we use the data to understand where the wind is initiated, how fast it is accelerated, and if a terminal velocity is reached at some
distance from the central star (Sect.~\ref{Sec:dynamics}). 
These questions can be answered by retrieving the wind velocity profile by analysing  the extent of the emission from an ensemble of  molecular transitions \citep[for examples see][]{Decin2015A&A...574A...5D, Decin2018A&A...615A..28D}.

When the ALMA proposal was submitted, it was generally expected that most of the {\sc atomium} sources,
with the exception of W~Aql, $\pi^1$~Gru, and R~Hya \citep{Danilovich2015A&A...574A..23D, Feast1953MNRAS.113..510F, Mason2001AJ....122.3466M}, were single stars. 
However, even for these three AGB sources, the known companion resides at a separation $>$150\,au so its gravitational field should not disturb the wind kinematics in the inner wind region ($r\la10-30$\,\Rstar) where the wind is initiated. Hence, even for these three sources, the {\sc atomium} data should allow us to study the efficiency of the wind initiation.

A first highlight of the {\sc atomium} programme, however, was that no source displays a smooth spherical wind. 
Instead, the observed morphologies include bipolar geometries with a central waist, equatorial density enhancements (EDE) 
and disk-like geometries, spiral-like structures, arcs, and `eye'-like shapes. 
These morphologies, supported by a population synthesis approach, led to the conclusion that most {\sc atomium} sources are part of a binary system, 
although the stellar or planetary properties and the orbital parameters of the companion remain unknown \citep{DecinScience}.
It is expected that very-low-mass objects, including brown dwarfs and large planets, play a larger role than previously assumed.
Therefore the {\sc atomium} data renders a crucial observational benchmark for  both binary-star and single-star  
theoretical simulations of the wind dynamics of AGB and RSG sources.
Even though (sub-)stellar  companions might be omnipresent, if the mass of the companion is low or the separation is large, there will be
little departure of the velocity streaming lines from radial motion and the observed wind kinematics can guide single-star models. 
Moreover, even if a companion disturbs the radial velocity pattern substantially, the effect is `localised' and the velocity pattern 
retains its radial character farther out in the wind. 
As discussed by \citet{2020A&A...637A..91E}, any density structure imprinted in the wind will then expand in a self-similar way. 

For the single-star and binary-star models, the question about the impact of resolved-out flux on the observables needs to be assessed. 
This mainly affects the low-excitation CO emission, {\red {however}} measurements of the velocity measure are not affected. 
Our observations are sensitive to MRS $\gtrsim$8$^{\prime\prime}$  (Sect.~\ref{sec:comb}), therefore all but the smoothest emission 
is detected. 
The perturbations and anomalous velocities we examine occur within $4^{\prime\prime}$ of the star (as do the extreme velocities 
from a spherical shell). so resolved-out flux does not affect an investigation of the cause of perturbations.
A next step entails assessing the mass fraction of the wind that is diverted for the binary-star models. 
To answer that question, single antenna observations are currently being acquired and analysed (Jeste et al.\ \textit{in prep).}

%  ---------------------------------------------------------------------------------------------------------------------------------------------

\paragraph{Single star models:}
It is generally accepted that the winds of AGB stars are radiation driven. 
Pulsations lift material to greater heights where the temperature is  
$\la$1800\,K, allowing gas to condense into grains \citep{Hoyle1962MNRAS.124..417H, Gail2013pccd.book.....G}. 
The absorption of stellar radiation by these newly formed dust grains creates a net force that can overcome gravity \citep{Hofner2018A&ARv..26....1H}. 
The gas is then accelerated beyond the escape velocity. This is expressed in the radial momentum transfer equation 
\citep{Goldreich1976ApJ...205..144G}
\begin{equation}
  v(r) \frac{dv(r)}{dr} = (\Gamma(r) -1) \frac{G M_{\star}}{r^2}\,,
  \label{Eq:momentum}
\end{equation}
where $v(r)$ refers to the gas velocity at a radial distance
$r$ from the star, $M_{\star}$ the stellar mass, $G$ the gravitional constant, and $\Gamma(r)$ the ratio of the radiation pressure force on the dust to the gravitational force that can be written as 
\citep{Decin2006A&A...456..549D}
\begin{equation}\label{Gamma}
  \Gamma(r) = \frac {3 v(r)}{16 \pi \rho_s c G M_{\star}
    \dot{M}(r)}\int\!\!\!{\int{\frac{Q_\lambda(a,r) L_{\lambda} \dot{M}_d(a,r)}{a
        [v(r)+v_{\rm{drift}}(a,r)]}}} \,d\lambda\,da \,,
\end{equation}
with $\rho_s$ the specific density of dust, $c$ the speed of light, $\dot{M}$ the gas mass-loss rate, $\dot{M}_d$ the dust mass-loss rate,  $v_{\rm{drift}}(a,r)$ the drift velocity of a grain of size $a$, $Q_{\lambda}(a)$ the dust extinction efficiency, $L_{\lambda}$ the monochromatic stellar luminosity at wavelength $\lambda$. 
A solution for the gas velocity as derived from solving the momentum equation (Eq.~\eqref{Eq:momentum}) for IK~Tau is shown as the full black line in Fig.~\ref{Fig:velocities}.
If the grain properties change with radial distance so that, for example,  $Q_\lambda(a,r)$ increases, a gradual wind acceleration at larger distances can arise \citep{Chapman1986MNRAS.220..513C}. In general, the particular behaviour of $Q_\lambda(a,r)$ has a strong influence on the wind acceleration, as discussed in detail by \cite{Elitzur1993ApJ...410..701N}. 

%  ++++++++++++++++++++++++++++++++++++++++++++++++++++++++++++++++++++++++++++++++++++++++++++++++
% figures made with /Users/leen/leda/latex/ALMA/ATOMIUM/overview_paper/programmes/plot_vel_structures.pro
\begin{figure}[htp]
\centering
\includegraphics[width=0.5\textwidth]{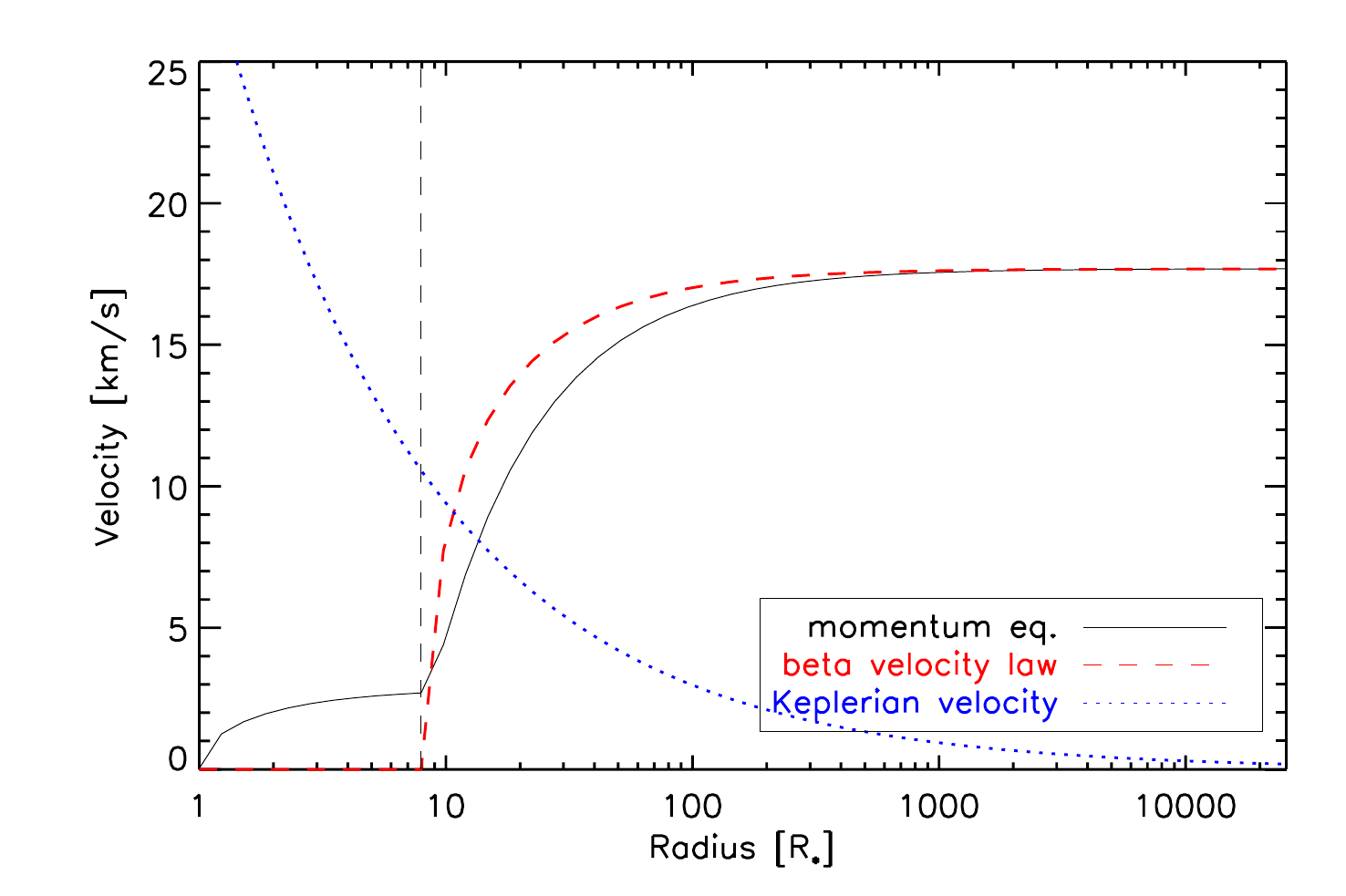}
\caption{\textbf{Illustration of different velocity laws.} 
The full black line at radii beyond 8\,\Rstar represents the solution of the momentum equation (Eq.~\eqref{Eq:momentum}) derived for the wind 
velocity profile of the oxygen-rich AGB star IK~Tau \citep{Decin2010A&A...516A..69D}; for the region between 1--8\,\Rstar\ a beta-velocity law 
with $\beta\,=\,0.5$ is used \citep{Decin2006A&A...456..549D}. 
The red dashed line illustrates the $\beta$-velocity law (Eq.~\eqref{Eq:velocity}) for $\beta\!=\!0.5$, $v_0\!=\!2.7$\,km/s, 
$v_\infty\!=\!17.68$\,km/s, and the vertical black dashed line indicates $R_{\rm{dust}}$ at 8.6\,\Rstar. 
An almost perfect fit to the velocity as derived from the momentum equation (Eq.~\eqref{Eq:momentum}) would be obtained for $\beta\,=\,1$. 
The dotted {\blue{blue}} line represents the Keplerian velocity law for material bound to a star of mass 1\,\Msun. }
\label{Fig:velocities}
\end{figure}

%  ++++++++++++++++++++++++++++++++++++++++++++++++++++++++++++++++++++++++++++++++++++++++++++++++

The wind initiation mechanism for RSG stars is less well understood. Mechanisms based on turbulent pressure in combination 
with radiation pressure on molecular lines or freshly synthesized dust grains and magneto-accoustic waves are invoked, 
or a combination of the above  \citep{Josselin2007A&A...469..671J, Thirumalai2012MNRAS.422.1272T}. 
In general, these alternative processes might also support the AGB stellar wind, although their role in driving the wind is still
very much debated \citep{Wood1990ASPC...11..355W, Gustafsson2003agbs.conf..149G}.
Solutions for the momentum equation (Eq.~\eqref{Eq:momentum}) indicate that the velocity profile of AGB and RSG winds 
can be approximated by the so-called $\beta$-type velocity law \citep{Lamers1999isw..book.....L} 
\begin{equation}
v(r) = v_0 + (v_\infty - v_0) \left( 1 - \frac{R_{\rm{dust}}}{r}\right)^\beta\,,
\label{Eq:velocity}
\end{equation}
with $r$ the distance to the star, $v_0$ the velocity at the dust condensation radius $R_{\rm{dust}}$, and $v_\infty$ 
the terminal wind velocity (see red dashed  line in Fig.~\ref{Fig:velocities}).

%  ------------------------------------------------------------
The beta velocity law assumes that the CSE is physically homogenous, apart from a decrease in number density and temperature 
as a function of distance from the star.  
Low values for $\beta$ describe a situation with a high wind acceleration.
For carbon-rich AGB stars, $\beta$ is around 0.5 \citep{Decin2015A&A...574A...5D} owing to the very opaque carbon dust grains 
that facilitate photon momentum transfer. 
Recent observational studies indicate that the wind acceleration for oxygen-rich AGB stars might be much lower than for carbon-stars, and values of $\beta$ between $1-5$ have been derived 
\citep{Decin2010A&A...521L...4D, Khouri2014A&A...561A...5K, VandeSande2018A&A...609A..63V, Decin2018A&A...615A..28D}. 
The cause for this slow wind acceleration is not yet fully understood.
The fact that oxygen-rich dust grains (such as aluminium oxides and silicates) are more transparent than carbon-rich grains offers part of the solution.
Using colour-dependent absorption, it has been shown that silicates become progressively more iron-rich (hence opaque) as the material gets farther from the star \citep{Woitke2006A&A...460L...9W, Bladh2012A&A...546A..76B}.
However, even then we cannot explain why for some sources the observed wind acceleration continues beyond $\sim$50 stellar radii where the densities are too low for efficient momentum exchange between the gas and dust particles \citep[see, for example, IK~Tau in Fig.~9 of ][]{Decin2018A&A...615A..28D}. 
Fractal grains within an inhomogeneous clumpy wind increase the radiation pressure efficiency and can potentially explain the more gradual but ultimately more forceful acceleration \citep{Decin2018A&A...615A..28D}.

\paragraph{Binary star models:}
As discussed in 
\citet{DecinScience}, 
we expect that most of the {\sc atomium} sources are part of a binary system.
Binary interaction with a (sub-)stellar companion results in distinct non-spherical wind geometries which are readily probed in CO and SiO channel maps.
Observationally derived wind profiles can also provide a means for constraining the presence and properties of a companion.
Compared with a single-star model, the companion will perturb the radial character of the velocity vector field as expressed, for example, in the momentum equation (Eq.~\eqref{Eq:momentum}) or in the $\beta$-velocity law (Eq.~\eqref{Eq:velocity}). 
For example, in the case that the companion induces the formation of a Keplerian disk-like structure, the tangential velocity component is given by 
$\sqrt{G\,M_\star/r}$, with $G$ the gravitational constant, $M_\star$ the mass of the AGB star to which the disk is gravitationally bound, and $r$ the radial distance \citep[][see blue dotted line in Fig.~\ref{Fig:velocities}]{Kervella2016A&A...596A..92K}. 
In addition, the companion's gravity lowers the effective gravity felt by a particle driven from the primary AGB star, which can lead to a local enhancement of the velocity amplitude. 
Based on a 3D hydrodynamical simulation for a binary system containing a mass-losing AGB star \citep[Fig.~\ref{Fig:Ileyk};][]{2020A&A...637A..91E}, we illustrate this effect in Fig.~\ref{Fig:velocities_binaries}. 
In this simulation, the wind is initially accelerated from the primary star following a $\beta$-velocity profile with $\beta$\,=\,5 (see black dashed curve in Fig.~\ref{Fig:velocities_binaries}, illustrating the single-star model). 
The presence of the secondary object impacts the velocity profile; see dotted lines in Fig.~\ref{Fig:velocities_binaries}.
The first up/down peak in the radial velocity profile in the direction of the secondary (blue dotted line in Fig.~\ref{Fig:velocities_binaries}) is due to the wind being first accelerated, but then dissipating most of its radial kinetic energy in the spiral shock, and having to be re-accelerated from scratch. 
To a lesser extent the same phenomena are apparent each time the radial ray crosses the spiral shock, hence the oscillating motion around the mean isotropic profile. 
Notice also that, as expected, the blue and red profiles oscillate with almost opposite phases. 
The isotropic velocity profile (black dotted line in Fig.~\ref{Fig:velocities_binaries}) is then the average over all azimuthal and longitudinal angles of the velocity profile. 
In the example shown here, the isotropic velocity profile has a wave-like character, in which the first peak indicates the orbital separation and the higher harmonics are linked to the spiral-arm crossing.  
As we discuss below, the velocity profile of W~Aql might be an example of the binary-induced effect described here.

%  ++++++++++++++++++++++++++++++++++++++++++++++++++++++++++++++++++++++++++++++++++++++++++++++++

\begin{figure}[htp]
\centering
\includegraphics[angle=0,width=0.48\textwidth]{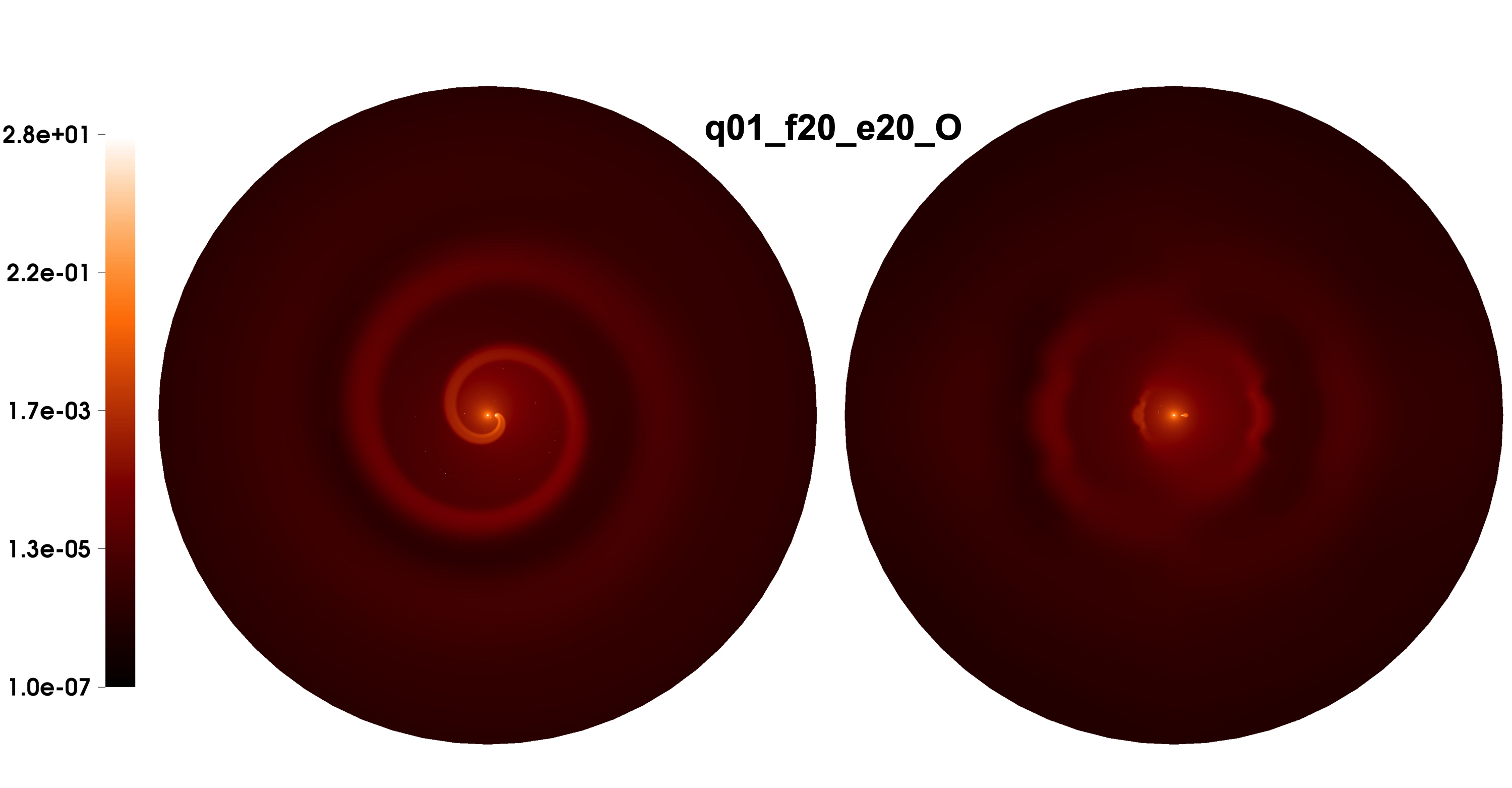}
\caption{\textbf{3D hydrodynamical simulation for a binary system containing a mass-losing AGB star.} 
Slices of density are shown, in units of the density at the sonic point, in the orbital plane (left column) and in the plane containing the orbital axis 
and the line joining the two bodies (right column).
The dimensionless parameters for this simulation are the mass ratio, $q\,=\,M_1/M_2\,=\,1$; the ratio of the terminal to orbital speed $\eta\,=\,v_\infty/v_{\rm{orb}}\,=\,2$, the dust condensation radius filling factor $f = R_d/R_{R,1}\,=20\,\%$ (with $R_{R,1}$ the Roche lobe radius of the primary), and the $\beta$ exponent setting the steepness of the velocity profile here being 5 \citep{2020A&A...637A..91E}. For a dust condensation radius set to 3\,\Rstar, the dimensionless parameters translate into an orbital separation of $\sim$35\,\Rstar. Due to binary interaction, a spiral shock is created in the circumstellar envelope. 
The spiral structure is readily  recognised in the density slice in the orbital plane and the width of the successive spiral windings can be deduced from the density arcs in the edge-on view.}
\label{Fig:Ileyk}
\end{figure}

\begin{figure}[htp]
\centering
\includegraphics[angle=0, width=0.48\textwidth]{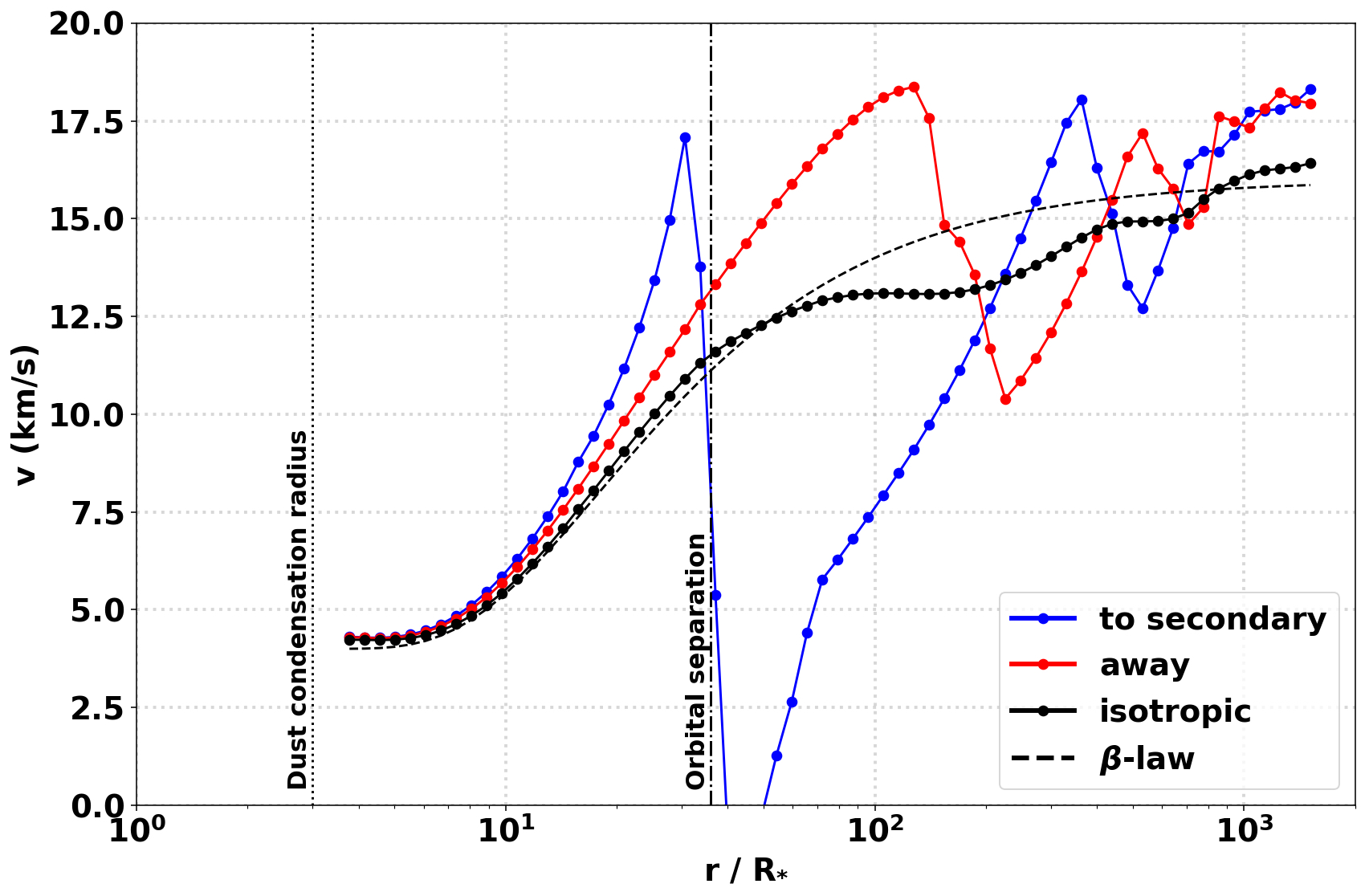}
\caption{\textbf{Illustration of the impact of a binary companion on the velocity field.} Given the simulations shown in Fig.~\ref{Fig:Ileyk} 
\citep{2020A&A...637A..91E}, the dotted black line represents the 1D isotropic radial velocity profile (w.r.t.\ the star). The dotted blue line 
is the radial velocity profile in the direction of the secondary, and the dotted red line is the radial velocity profile in the direction opposite 
to the secondary. For comparison, the dashed black curve illustrates the $\beta$-velocity law (Eq.~\eqref{Eq:velocity}) for $\beta\!=\!5$ representing the single-star situation. For better comparison with the observed velocity profiles (Sect.~\ref{Sec:Methodology}), the same 
figure but with a linear x-axis is shown in App.~\ref{App:add_figures}, Fig.~\ref{Fig:velocities_binaries_linear}. See text for more details.}
\label{Fig:velocities_binaries}
\end{figure}

%  ++++++++++++++++++++++++++++++++++++++++++++++++++++++++++++++++++++++++++++++++++++++++++++++++

\subsection{Methodology} \label{Sec:Methodology}

To constrain the wind kinematics of the {\sc atomium} sources, we followed the same methodology described in 
\citet{Decin2015A&A...574A...5D} and \citet{Decin2018A&A...615A..28D}, which was augmented with several additional steps.
Figures~\ref{Fig:deconvolution_SiO}--\ref{Fig:deconvolution_NaCl} in the Appendix illustrate the methodology for the SiO J=6-5 
and NaCl J=20-19 transition observed in the medium-resolution configuration of IRC\,$-$10529.

\begin{enumerate}
\item In the first step, the spectrum of each molecular transition was extracted for a range of circular apertures.
The minimum diameter of the extraction aperture was the major axis of the synthesized beam (`$b_\text{maj}$', see Table~\ref{tab:cube}),
the maximum diameter was the MRS (see Table~\ref{tab:cube}), where the step size was $2\!\times\!b_\text{maj}$. 

\item  
The velocity of the blue and red wings\footnote{Negative velocities (i.e., blue shifted) with respect to the $v_{\rm{LSR}}$ represent 
material coming towards the observer, and positive velocities (i.e., red shifted) with respect to the $v_{\rm{LSR}}$ represent material 
receding from the observer.} 
was determined for all extraction apertures.
Accounting for the noise around the line spectrum ($\sigma_{\rm{line}}$), the blue and red wing velocity were taken as the closest points 
to the line centre for which the flux was less than $3\times\sigma_{\rm{line}}$. These sensitivity-limited velocity widths are likely the lower bounds.
The maximum of these numbers in absolute values was retained as the wind `velocity measure' for the transition. 
The uncertainty in the velocity measure was taken from the spectral resolution of the data.

\item We then computed the zeroth moment map (integrated intensity or moment0 map) of each line (between the measured 
red and blue wing velocities) and measured the angular FWHM by fitting a 2D Gaussian profile to the moment\,0 map using the 
Levenberg–Marquardt algorithm.  
If the least-squares  minimisation was unsuccessful, the molecular transition was not retained for further analysis.
This led to a significant reduction in the number of transitions retained for the analysis of the kinematical behaviour.  
This particularly affects transitions with low signal to noise ratios and/or high upper energy levels which are associated with 
having a small angular extent. 
Transitions whose moment\,0 maps differ significantly from a 2D Gaussian profile  --- such as the SO$_2$ 
distribution shown in Fig.~\ref{Fig:SO2_CM_IRC10529} --- were not retained after this step.

\item  If the FWHM of the fitted 2D Gaussian was comparable to the axes of the synthesised beam ($b_\text{min}$, $b_\text{maj}$; 
see Table~\ref{tab:cube}), the non-deconvolved extent was taken as an upper limit to the  distribution of the species. 
For transitions for which the FWHM of the fitted 2D Gaussian was larger than the synthesised beam, we deconvolved the beam. 
As in \citet{Decin2018A&A...615A..28D}, we assumed that the spatial FWHM of the molecular emission zone represents the 
dominant line formation region.

\item In the case of successful least-square minimization, the covariance matrix was used to estimate the variance of the FWHM 
($\sigma_{\rm{FWHM}}$).  
Accounting for the interferometric capabilities of ALMA, the total accuracy of the measured size of the emission is then given by 
\begin{equation}
\sigma_{\rm{ext}} = \sqrt{\sigma_{\rm{FWHM}}^2 + \left(\sqrt{2} * K * \theta_{\mathrm B} / {\rm{S/N}} \right)^2}\,,
\end{equation}
where the S/N is the signal-to-noise ratio of the moment\,0-map, $\theta_{\mathrm B}$ the beamsize as given by  
($b_\text{min}$, $b_\text{maj}$), and $K$ is 0.5, 1, or 1.5 for the low, medium, and high spatial resolution data
\citep{Condon1997PASP..109..166C, Taylor1999ASPC..180.....T}.

\item In the final step, the outcomes of the analysis of the velocity versus aperture information extracted from
the low, medium, and high spatial resolution data were merged to produce a single output:
for transitions in which the emission zone could be deconvolved in observations at various spatial resolution, the largest velocity measure 
and largest extent (often measured by the observation at the lowest spatial resolution) were retained. 
This should ensure that the impact of resolving out flux was kept to a minimum. 
In addition, blended lines were removed from the sample.
\end{enumerate}

Our analysis provides a unique view of the wind kinematics of 17 oxygen-rich AGB and RSG sources. 
The outcome of this analysis for IRC\,$-$10529 is shown in Fig.~\ref{Fig:IRC10529_kinematics}, and that of the prototypical
source W~Aql is shown in Fig.~\ref{Fig:W_Aql_temporary}. 
The wind dynamics for the 15 other sources are in Appendix Sect.~\ref{Sect:kinematics_other_sources} 
(see Figs.~\ref{Fig:S_Pav_kinematics} -- \ref{Fig:RW_Sco_kinematics}).
Until now similar velocity profiles were only obtained for the carbon-rich AGB star CW~Leo (\Mdot\,=\,1.5$\times$10$^{-5}$\,\Msun/yr),
and the two oxygen-rich AGB stars R~Dor (\Mdot\,=\,1$\times$10$^{-7}$\,\Msun/yr) and IK~Tau (\Mdot\,=\,5$\times$10$^{-6}$\,\Msun/yr) 
\citep{Decin2015A&A...574A...5D, Decin2018A&A...615A..28D}. 

One of the obvious limitations of the method is that we use Gaussian fits to the zeroth moment maps to determine the size of the emission zone.
Inspection of individual channel maps and zeroth moment maps shows a variety of intensity distributions, depending on excitation effects, formation 
and depletion mechanisms, and the potential influence of a companion (see Sect.~\ref{Sec:atomium_as_binary}). 
In future work we will investigate different techniques such as fitting  a modified power law 
\citep[as in][]{Sahai1993AJ....105..595S}, an azimuthal average, or a cutoff such as $3\!\times\sigma_{\rm{rms}}$. 
These methods for determining the angular size are likely to strengthen the inference drawn from this analysis (see Sect.~\ref{Sec:Interpretation}), 
because fewer transitions would be rejected owing to the lack of convergence of the Gaussian fitting method.
We also will need to consider whether we are detecting essentially all the emission of each transition; 
or whether we are bounded by missing extended flux and/or sensitivity which varies between species. 
In general, the estimates of the maximum extent of any one transition might be biased if
the CSE is not truly spherical, and depends on the orientation to the line of sight of any elongation.

%  +++++++++++++++++++++++++++++++++++++++++++++++++++++++++++++++++++++++++++++++++++++++++++
%  +++++++++++++++++++++++++  FIGURE 11 WIND KINEMATICS OF IRC-10529  +++++++++++++++++++++++++++++
%  +++++++++++++++++++++++++++++++++++++++++++++++++++++++++++++++++++++++++++++++++++++++++++

\begin{figure*}[!htpb]
\vspace{1ex}

\begin{minipage}[t]{.495\textwidth}
        \centerline{\resizebox{\textwidth}{!}{\includegraphics[angle=0]{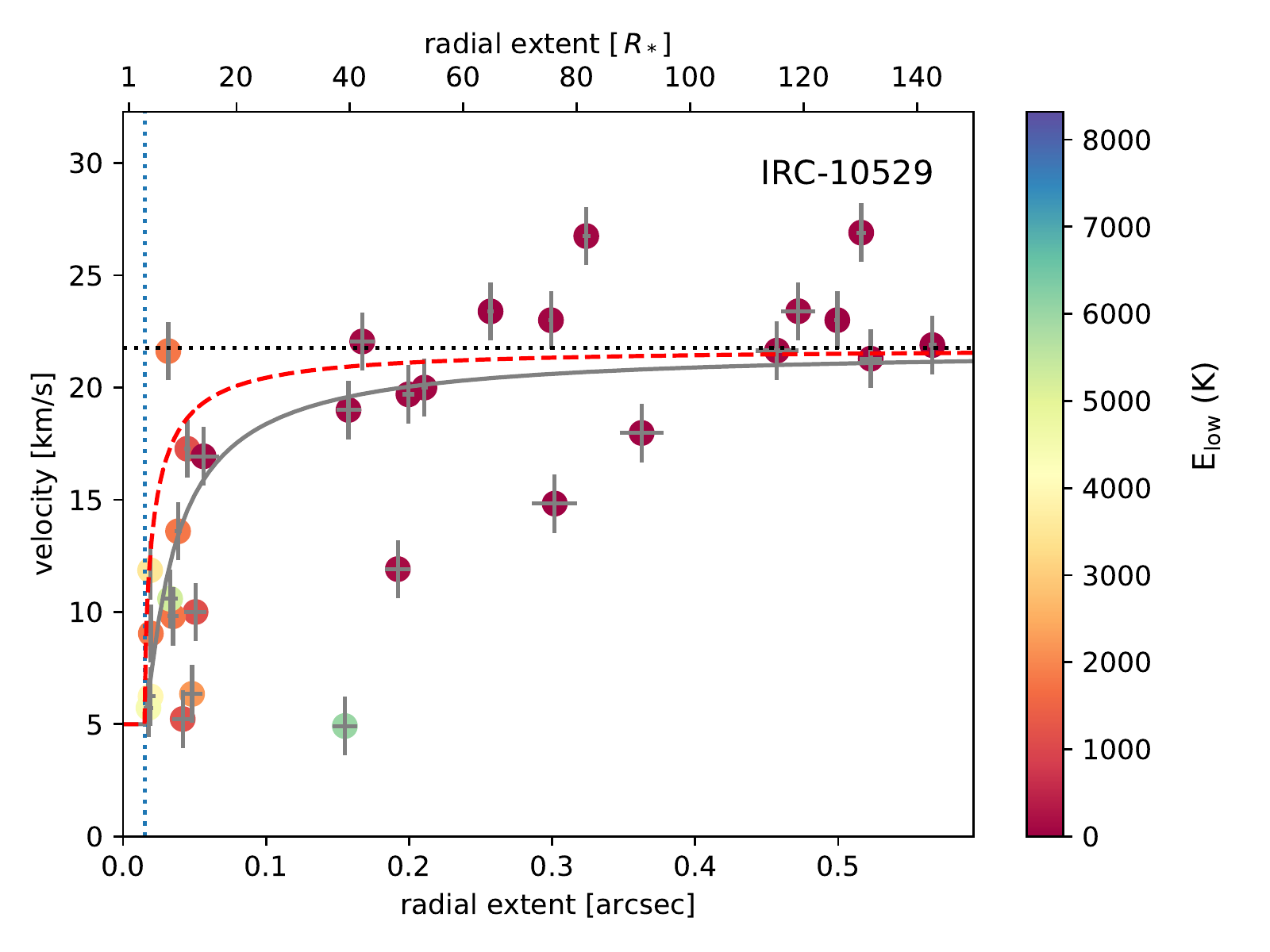}}}
\end{minipage}
    \hfill
\begin{minipage}[t]{.495\textwidth}
        \centerline{\resizebox{\textwidth}{!}{\includegraphics[angle=0]{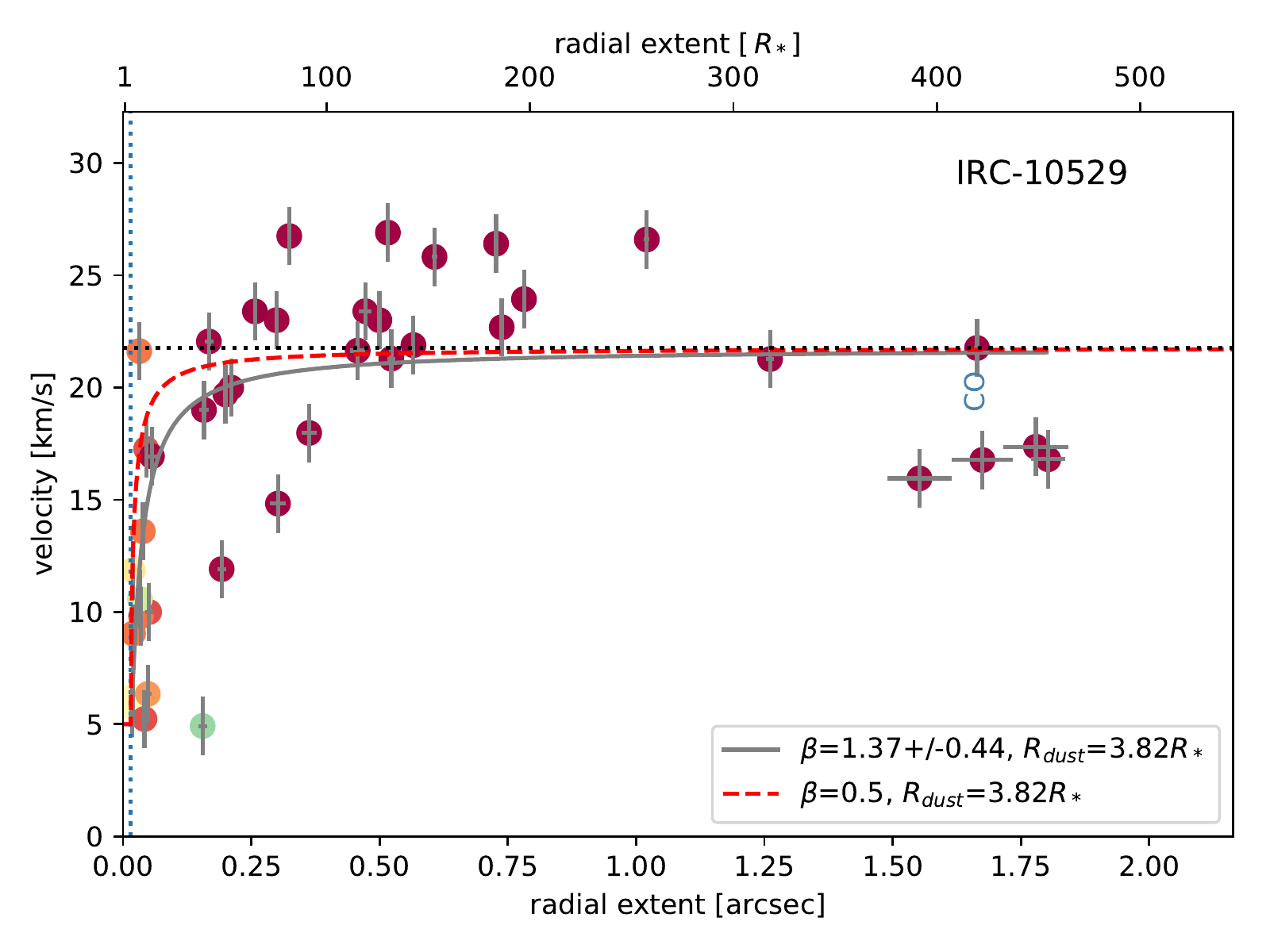}}}
\end{minipage}
\caption{\textbf{Wind kinematics for IRC\,$-$10529.} 
The wind velocities for all the molecular transitions observed in the low, medium, 
and high spatial resolution observations of IRC\,$-$10529 derived by the methodology described in Sect.~\ref{Sec:Methodology}. 
The velocities are plotted versus half of the spatial full-width-half-maximum (FWHM) of the molecular emission zone, and represent 
the dominant line formation region \citep{Decin2018A&A...615A..28D}. 
The dotted blue vertical line indicates the radius at which the winds begin  being accelerated ($R_{\rm{dust}}$ in Eq.~\eqref{Eq:velocity}; 
see Sect.~\ref{Sec:Interpretation}),
and the dotted black horizontal line is the velocity measure of the $^{12}$CO v=0 J=2-1 line. 
Only emission zones which could be spatially deconvolved are plotted (as dots) ---   the colour of the symbols are related to the energy 
of the lower state as indicated by the colour code bar.
The grey cross on top of each coloured dot indicates the error bar in the derived velocity and in the FWHM, and is often smaller than the 
size of the dots. 
The error bars represent the fitting margin.  
Not accounted for here are the uncertainties due to the Gaussian approximation which might result in a systematic underestimate 
of the angular extent (see Sect.~\ref{Sec:Methodology}).
%\hfill\break
%\hfill\break
The bottom axis is in units of arcseconds and the top axis is in units of the stellar radius. 
The {\it left hand panel} zooms into the 150 stellar radii of the circumstellar envelope, and the right hand panel shows the full extent of the 
detected wind emission.
The $^{12}$CO v=0 J=2-1 transition (indicated in blue) only appears in the {\it right hand panel} owing to its large angular extent. 
The deconvolved data with a velocity less than that of the $^{12}$CO J=2-1 transition, were fitted with a $\beta$-velocity law 
(Eq.~\eqref{Eq:velocity}). 
Indicated in the legend in the {\it right hand panel} is $\beta$.
The fit is represented by the full  grey line which can be compared with a $\beta$-velocity law for $\beta$\,=\,0.5 indicated by the 
red dashed line. 
}
\label{Fig:IRC10529_kinematics}
\end{figure*}

%  ++++++++++++++++++++++++++++++++++++++++++++++++++++++++++++++++++++++++++++++++++++++++++++++

Arguably, this is a simplified view of the wind kinematics since:
(i)~we only obtain a projected 1D view of the velocity vector field;  and
(ii)~the spatial extents plotted in these figures might not always reflect the region where some of the  extreme velocities arise from, 
for example in the situation where the extreme velocities are only reached at close distances from the AGB star while the bulk of the 
emission further out has lower velocities.
Nevertheless, interesting conclusions can already be derived from these results (see Sect.~\ref{Sec:Interpretation}).

\subsection{Combining different angular resolution data and the issue of resolved-out flux} \label{Sec:Resolved-out flux}

To date, there are two stars for which the three individual spatial resolution data and the combined dataset 
have been analysed in detail ---  $\pi^1$~Gru \citep{2020A&A...644A..61H} and R~Hya   \citep{2021A&A...651A..82H} --- 
thereby offering an opportunity to study the effect of resolved-out flux at a maximum recoverable scale for the lowest resolution 
ALMA {\sc atomium} observations of $\sim$10\arcsec.
Resolving out flux might be an issue for the measurement of the angular extent of the molecular emission, but this does not hinder the
determination of the velocity measure, 
because resolving out flux mainly reduces the measured line flux of extended emission around the central velocities and not that of the 
line wings which represent the more compact front and back caps of the circumstellar envelope.

The combined data are significantly more sensitive to emission which has structure on sub-arcsec scales, than the individual configurations.
For an angular resolution of between $\sim$0.1 and $\sim$8~arcsec, the combined data sets contain at least  twice as much data as an 
individual configuration. 
The combined visibility data can be weighted as a function of baseline length to provide greater sensitivity for any desired resolution 
in this range, 
and this will be done in the detailed studies of each star. 
On smaller scales of tens of mas, the combined data set may also give a better image in the presence of extended structure although 'clean' 
stability can then be an issue.
The combined data therefore allows for a more accurate measure of the total extent of emission from some lines at intermediate distances 
from the star.

The three observing dates of the low, medium, and high spatial resolution data differ by up to $\sim$9 months,
while AGB stars have typical pulsation periods of 
$\sim$0.5–2~yr  \citep[see Table~1;][]{Cernicharo2014ApJ...796L..21C, Fonfria2018ApJ...860..162F}. 
Therefore, the higher-excitation molecular transitions which are excited close to the stellar surface, can vary between epochs, 
and while this may lead to a more complete picture in the combined data, extreme variability (e.g., SiO masers) causes  
artefacts that result in combined images which might not be reliable. For these lines, the velocity measurement is more accurate 
for the merged dataset 
(see Step~6 in Sect.~\ref{Sec:Methodology}) than for the combined dataset. 

For some molecular transitions the Levenberg--Marquardt minimization used for fitting the 2D Gaussian profiles failed for the combined dataset, 
but was successful for (one of) the individual observations. 
An example is the $^{12}$CO J=2-1 line where the emission zone in low spatial resolution data often better resembles a 2D Gaussian profile, 
but in the combined dataset the individual substructures are more prominent and  a 2D Gaussian profile does not reproduce the observed emission.
In general, an excellent correspondence is found for those molecular transitions that could be deconvolved in both the merged outcomes 
(of the low, medium, and high spatial resolution data; see step~6 in Sect.~\ref{Sec:Methodology}) and the combined dataset. 
The differences between extents measured in the two ways have a dispersion of 0\farcs06.

Given these outcomes for $\pi^1$~Gru, we have used the merged results from the datasets of individual spatial resolution for the study 
of the kinematics in all the {\sc atomium} AGB and RSG envelopes.

The only question still remaining is whether some flux is resolved out beyond the $\sim$10\arcsec\ scale.
The two molecules most affected by resolved-out flux are CO and SiO, owing to their high abundance in these oxygen-rich sources, 
and hence potentially extended molecular envelopes.
For each target, we have estimated the CO and SiO photodissocation region using the formula given by \citet{Mamon1988ApJ...328..797M} 
and \citet{Gonzalez2003A&A...411..123G}, respectively (see Sect.~\ref{Sec:Observing_strategy}).
For all the {\sc atomium} sources, the SiO emission zone --- as estimated from the photodissociation radius following 
\citet{Gonzalez2003A&A...411..123G} --- is less than $\sim$3\farcs5, with the exception of GY~Aql where owing to its proximity to Earth
the estimated SiO emission zone is $\sim$9\arcsec. 
Hence, resolving out flux should not be an issue for the SiO measurements. 
However, the CO emission zone for the {\sc atomium} sources ranges between $\sim$20--300\,\arcsec, although the CO J=2-1 transition 
is not excited up to the very outer boundary. 
This implies that the measured sizes of the $^{12}$CO and $^{13}$CO J=2-1 transitions will be lower limits.
This, however, does not hinder our present study on the wind kinematics, since we focus  here on the wind initiation efficiency in the inner wind region, the maximum and minimum velocities deduced from the {\sc atomium} data, and the imprint of a binary companion on the observationally derived wind velocity profiles.
For all sources, additional observations of the $^{12}$CO J=2-1 and $^{13}$CO J=2-1 transitions have been or will be acquired with the 
APEX 12~m single antenna, allowing us to constrain the CO emission region with higher precision.

\subsection{Interpretation} \label{Sec:Interpretation}

In this section, we aim for an interpretation of the observationally derived wind kinematic profiles displayed in Figs.~\ref{Fig:IRC10529_kinematics}, \ref{Fig:W_Aql_temporary}, and \ref{Fig:S_Pav_kinematics} -- \ref{Fig:RW_Sco_kinematics}.
We first examine the wind acceleration in the {\sc atomium} sources. 
As will be discussed, the majority of the {\sc atomium} sources display a slow wind acceleration  characterised by quite high values of $\beta$. 
However, the wind kinematic profiles  also make it readily clear that the radial velocity description as provided by the momentum equation (Eq.~\eqref{Eq:momentum}) or the
$\beta$-velocity law (Eq.~\eqref{Eq:velocity}) for single-star models cannot capture the complexity of the velocity field in the 
{\sc atomium} sources. 
We therefore extend this discussion with a more detailed examination of some of the observationally derived velocity profiles in the context 
of binary-star models.

%\vspace{10.0cm}
\subsubsection{Single-star models: wind initiation and terminal wind velocities}\label{Sec:atomium_as_single}

The first theoretical studies discussing analytical approximations \citep{Gehrz1971ApJ...165..285G, Gilman1972ApJ...178..423G} and numerical solutions of the equation of motion \citep{Kwok1975ApJ...198..583K, Goldreich1976ApJ...205..144G}, resulted in gas velocity profiles having a characteristic sharp rise and reaching a constant `terminal' wind velocity, $v_\infty$, within the first few stellar radii (see also Fig.~\ref{Fig:velocities}). 
Values for $v_\infty$ have been observationally derived from the half-line width of low-excitation rotational CO lines \citep[see, for example,][]{Loup1993A&AS...99..291L}. 
The predicted fast acceleration of the wind velocity also motivated various modellers to assume a constant wind velocity 
\citep[for example,][]{Ramstedt2009A&A...499..515R, Sargent2011ApJ...728...93S, Groenewegen2014A&A...561L..11G}. 
However, the {\sc atomium} data show that the notion of a constant wind velocity being reached within $\sim$10\,\Rstar\ is not consistent with the observations. 
For all {\sc atomium} sources, large variations in the derived velocity measures are seen, even at radial distances greater than 50\,\Rstar, 
where (1D) spherical symmetric wind models with a steadily increasing velocity profile predict the velocity measure to have asymptotically reached the terminal wind velocity  (see Appendix~\ref{Sect:terminal_velocity}, Fig.~\ref{Fig:theory_vwind}).

In the approach adopted here we apply the same classical methodology in which a low excitation line of CO is used to determine the 
terminal wind velocity.
However our approach differs from most earlier studies because:
(1) it relies on a quantifiable metric linked to the 3$\sigma$ rms noise in spectra that were observed at very high angular resolution and sensitivity; and 
(2) it does not rely on subjective judgement derived from visual inspection of less sensitive observations, or by referring to a model.

%  ++++++++++++++++++++++++++++++++++++++++++++++++++++++++++++++++++++++++++++++++++++++++++++++
\begin{table*}[htpb]
\caption{\textbf{Velocity parameters of the {\sc atomium} sample. }}
%
%\small
%
\label{Table:beta}
\setlength{\tabcolsep}{1.1mm}
\begin{tabular}{l|rcc|cc|cc|cc}
\hline
\multicolumn{1}{c|}{(1)} & \multicolumn{1}{c}{(2)} & \multicolumn{1}{c}{(3)} & \multicolumn{1}{c|}{(4)} & \multicolumn{1}{c}{(5)} & \multicolumn{1}{c|}{(6)} & \multicolumn{1}{c}{(7)} & \multicolumn{1}{c|}{(8)} & \multicolumn{1}{c}{(9)} & \multicolumn{1}{c}{(10)}\\ 
\multicolumn{1}{c|}{Target}& $v_\infty^{\rm{old}}$(CO) & $v_\infty^{\rm{com}}$(CO) &$v$(CO) & $v_{\rm{max}}$\,$^{(a)}$ &Transition$^{(b)}$ & $v_{\rm{min}}$\,$^{(a)}$ &Transition$^{(b)}$ &$R_{\rm{dust}}$ &$\beta$\\
	& (km/s) & (km/s) & (km/s) & (km/s) & ($v_{\rm{max}}$) & (km/s) & ($v_{\rm{min}}$) & (\Rstar) & \\
\hline
S Pav&9.0 (1)&13.0 & 15.5&21.2&SiO J=5-4&3.9&SO$_2$ $20_{2,18}$--$19_{3,17}$&2.0&0.7$\pm$0.2 \\
T Mic&6.1 (2)&12.7 & 16.0&21.8&SiO J=5-4&4.0&CS J=5-4&2.0&0.6$\pm$0.3 \\
U Del$^{(c)}$&7.5 (1)&14.6 & 18.4&19.4&SiO J=6-5&9.5&SiO v=1 J=5-4&5.1&$-$ \\
RW Sco&11.0 (3)& 18.5 & 18.5&18.8&SO$_2$ $11_{1,11}$--$10_{0,10}$&4.8&CS J=5-4&6.7&9.1$\pm$2.3 \\
V PsA$^{(c)}$&14.4 (1)&18.8 & 23.1&28.4&SiO J=6-5&5.9&SiO v=4 J=6-5&3.5&$-$ \\
SV Aqr$^{(c)}$&7.9 (4)& 15.9 & 17.0&23.8&SiO J=6-5&4.9&Si$^{34}$S J=14-13&5.2&$-$ \\
R Hya&12.5 (5)&22.2 & 22.2&24.8&SiO J=6-5&3.9&SO$_2$  $44_{6,38}$--$43_{7,37}$&2.0& 0.6$\pm$0.1 \\
U Her&11.5 (6)&19.7 & 19.7&23.0&SiO v=1 J=5-4&4.4&SO$_2$  $4_{3,1}-4_{2,2}$&2.6&2.0$\pm$0.5 \\
$\pi^1$ Gru&30.0 (7)&64.5 & 64.5&64.5&CO J=2-1&3.9&SiO v=3 J=6-5&2.0&2.6$\pm$0.6 \\
AH Sco$^{(d)}$ & 23.0   (8) & $-$ & 35.4 & 52.0 & HCN J=3-2 & 5.8  &  TiO v=1 $\Omega$=1 J=7-6 & 3.8 & 5.0$\pm$0.9\\
R Aql&9.5 (6)& 12.8 & 15.8&21.4&SiO J=5-4&4.3&SO$_2$   $\nu_2$=1 $30_{4,26}$--$30_{3,27}$&2.3&1.0$\pm$0.3 \\
W Aql&20.0 (5)& 24.6 & 27.1&42.5&SiO J=6-5&4.0&Si$^{34}$S v=1 J=14-13&2.2&2.9$\pm$0.4 \\
GY Aql&16.2 (9)& 15.0 & 18.1&22.9&SiO J=5-4&4.6&$^{13}$CS J=5-4&3.1&3.2$\pm$1.1 \\
IRC\,$-$10529&16.5 (5)&21.8 & 21.8&26.9&SiS J=12-11&4.3&CO v=1 J=2-1&3.8&1.4$\pm$0.4 \\
KW Sgr$^{(c,d)}$ & 27.0   (10) & $-$ & 27.7 & 34.0 & SiO J=5-4 & 3.9  &  $^{29}$Si$^{34}$S  $J=13-12$ & 7.9 & $-$\\
IRC\,+10011&19.8 (5)&23.1 & 23.1&34.9&Si$^{34}$S J=14-13&4.1&PO\,$^2\Pi_{1/2}$\,$J,F$=$5.5,6$-$4.5,5$&6.5&2.8$\pm$0.6 \\
VX Sgr&24.3 (5)&32.9 & 34.4&66.5&HCN J=3-2&4.0&$^{34}$SO$_2$  $24_{2,22}$--$24_{1,23}$&3.9&2.2$\pm$0.5 \\
\hline
\end{tabular}\\
\tablefoot{\hspace{0.5cm} 
The target name is in column {\sl (1)}\,; 
the terminal wind velocity  $v_\infty^{\rm{old}}$(CO) in column {\sl (2)}\, was obtained from observations of CO listed in the references
indicated in parentheses; the wind velocity $v_\infty^{\rm{com}}$(CO) determined from the compact {\sc atomium} $^{12}$CO v=0 J=2-1 observation is in column {\sl (3)}\,;
the velocity $v$(CO) determined from all {\sc atomium} configurations of the  $^{12}$CO v=0 J=2-1 transition is in column {\sl (4)}\,, 
following Step~2 in Sect.~\ref{Sec:Methodology};
the maximum velocity $v_{\rm{max}}$ derived from the {\sc atomium} observations is in column {\sl (5)}\, with the corresponding molecular transition  in  column {\sl (6)}\,;
the minimum velocity $v_{\rm{min}}$  derived from the {\sc atomium} observations is in column {\sl (7)}\, with the corresponding molecular transition  in column {\sl (8)}\,;
the two parameters derived from fitting the $\beta$-velocity law (Eq.~\eqref{Eq:velocity}) are in columns {\sl (9)} and {\sl (10)}  
(see text for more details).\\
$^{(a)}$ Includes both deconvolved and non-deconvolved data obtained after step~3 as described in Sect.~\ref{Sec:Methodology}.\\
$^{(b)}$ All rotational transitions are in the ground ($v=0$) vibrational state unless otherwise noted.     \\
$^{(c)}$ Not enough data are available to determine $\beta$.   \\
$^{(d)}$ Targets only observed at medium and high spatial resolution.    \\
%$^{(d)}$ Fast, bipolar outflow with velocity increasing from 14\,km\,s up to 100\,km/s.\\
References: (1) \citet{Olofsson2002A&A...391.1053O}; (2) \citet{Kerschbaum1999A&AS..138..299K}; (3) \citet{Groenewegen1999A&AS..140..197G}; (4) \citet{Kerschbaum1998A&A...336..654K}; (5) \citet{DeBeck2010A&A...523A..18D}; (6) \citet{Young1995ApJ...445..872Y}; (7) \citet{Doan2017A&A...605A..28D}; (8) \citet{Josselin1998A&AS..129...45J}; (9) \citet{Loup1993A&AS...99..291L}; (10) \citet{Mauron2011A&A...526A.156M}
}
\end{table*}

%  ++++++++++++++++++++++++++++++++++++++++++++++++++++++++++++++++++++++++++++++++++++++++++++++

Various sources such as S~Pav show some substantial changes in the velocity amplitude in the innermost few stellar radii. 
This behaviour might be reminiscent of pulsation-induced shocks for which 
hydrodynamical simulations  show that they can lead to time-variable velocity characteristics within the first $\sim$4 stellar radii and with amplitudes of around 5--10\,km/s \citep[][see Fig.~\ref{Fig:theory_vwind_shock}]{Bladh2019A&A...626A.100B, Hoai2020MNRAS.495..943H}. 
The strongest effect is within the first 2\,\Rstar, but rapidly fades out at greater distances \citep{Liljegren2018A&A...619A..47L, Bladh2019A&A...626A.100B}. 
While the shock in the 3D hydrodynamical models is global in scale, the maximum velocity reached by the gas in the shock front is not uniform but 
rather clumpy. The medium and low-resolution data are less sensitive to compact emission, and hence the high-resolution and combined datasets 
should be used to diagnose this region that is disturbed by the shocks.
Analogous to \citet{Khouri2016MNRAS.463L..74K, Khouri2019A&A...623L...1K}, we derive a first estimate of the velocity amplitude in this complex region from the high-resolution observations of the highly excited OH and CO v=1 transitions.
These data indicate velocities of around 6\,--\,10\,km/s. 
For three source (R~Aql, R~Hya, and S~Pav) the OH and CO v=1 transition both have an inverse P-Cygni profile which	
--- in the framework of a 1D single-star model --- can be interpreted as a sign of infall of material with velocities of around 10\,--\,15\,km/s. 
As discussed in Appendix~\ref{Sect:terminal_velocity} for the case of the `normal shock' model, these shock characteristics cannot be traced 
in the medium and low-resolution data in the theoretical simulations of the $^{12}$CO v=0 J=2-1 and $^{28}$SiO v=0 J=5-4 line,
and this is a conclusion which we extrapolate to the other molecular lines.

% --------------------------------------------------------------------
To compare our results here with the prior literature, we use the velocity retrieved from the ALMA CO v=0 J=2-1 line  $v$(CO) 
to define the terminal wind velocity ($v_\infty$), 
and thereby assume CO traces the velocity of the bulk material at large distances from the star.
% ---------------------------------------------------------------------
To avoid potential impact from pulsation-induced shocks (see above and Sect.~\ref{Sect:terminal_velocity}),
we opt to use the (low-resolution) data from the compact configuration [$v_\infty^{\rm{com}}$(CO), see column~3 in Table~\ref{Table:beta}].

As discussed in the Appendix~\ref{Sect:terminal_velocity}:
(1) the effects of thermal and turbulent broadening in the wings of the line profile can be distinguished when CO is observed 
at high sensitivity; and 
(2) the low resolution CO data is indeed a good diagnostic for the terminal wind velocity.
At a temperature of 2\,500\,K, thermal broadening amounts to $\sim$1.2\,km/s. 
The turbulent broadening is difficult to estimate, but the example of \citet{DeBeck2012A&A...539A.108D} indicates a value of around 
1.5\,km/s, so the combined effect yields a Gaussian broadening of the line wings with HWHM of around 2.2\,km/s. 
Excluding the fast, bipolar outflow traced in the $^{12}$CO J=2-1 data of $\pi^1$~Gru 
\citep[see Sect.~\ref{Sec:atomium_as_binary};][]{Doan2017A&A...605A..28D}, 
the three red supergiants in the sample (AH~Sco, KW~Sgr, and VX~Sgr) have the largest CO velocities.
Even when we account for this broadening effect and for the spectral resolution of our ALMA data of $\sim$1.3\,km/s, 
the results in Table~\ref{Table:beta} still indicate higher terminal velocities than were derived previously for most {\sc atomium} sources.}  
% ---------------------------------------------------------------------
With the exception of GY~Aql, the compact configuration {\sc ATOMIUM} data of CO yield a larger terminal wind velocity than previous values (see columns~2 and 3 in Table~\ref{Table:beta}), where the maximum difference is a factor of 2.1. 
The reason for this difference is the higher sensitivity of the {\sc atomium} data which allows us to trace the broad CO wings whose intensity 
is low. 

In their survey of 42 mostly southern AGB stars that includes 21 M-type stars,
\citet{2020A&A...640A.133R} observed the J=2--1 line of $^{12}$CO in five of the {\sc atomium} sources 
(T~Mic, IRC--10529, IRC+10011, SV~Aqr, and R~Hya), 
but their observations were done with the ALMA 7\,m Compact Array (ACA) --- rather than with the ALMA 12\,m Array which we used here.  
The synthesized beam of the ACA is about 5~arcsec and the $1\sigma$ rms of the data is about 40--110~mJy/beam in Band~6
\citep[see Table B.1 in][]{2020A&A...640A.133R}.
When the {\sc atomium} observations were made at low angular resolution in the compact configuration, the synthesized beam was about 
7 times smaller (750~mas) and the $1\sigma$ rms (5~mJy/beam) was about 15 times smaller than in \citeauthor{2020A&A...640A.133R}.
The method for determining the CO velocity measure in \citeauthor{2020A&A...640A.133R}  and {\sc atomium} are similar: 
(1)  the CO line profile is extracted for a circular aperture, which in \citeauthor{2020A&A...640A.133R} consists of a fixed circular 
aperture of $18^{\prime\prime}$ that is close to the maximum recoverable scale for these observations; and 
(2) the total velocity width (divided by a factor 2) is used to determine the CO velocity measure. 
Dividing the full width of the CO J=2--1 line in Table~B1 of \citeauthor{2020A&A...640A.133R} by a factor of 2, 
the $\upsilon_{\rm{max}}$(CO) is 6.25~km/s for T~Mic, 15.25~km/s for IRC-10529, 19.5~km/s for IRC+10011, 
9.4~km/s for SV~Aqr, and 10.5~km/s for R~Hya.
Comparing these with {\boldmath  $v_\infty^{\rm{com}}$(CO)} in Table~\ref{Table:beta},
we find that $v_\infty^{\rm{com}}$(CO) determined in {\sc atomium} is $1.5 - 2$ times higher in four stars 
and 1.2 times higher in IRC+10011 than in  \citeauthor{2020A&A...640A.133R}
As illustrated  in Fig.~\ref{Fig:TMic_CO}, the 15 times higher sensitivity in the {\sc atomium} observation of T~Mic 
yields a much more sensitive diagnostic of the low level emission from the wings of the CO line profile and an estimation of 
$\upsilon_{\rm{max}}$(CO).

Owing to resolved-out flux (see Sect.~\ref{Sec:Resolved-out flux}), in five out of the 17 sources a transition other than the 
$^{12}$CO J=2-1 line has the largest apparent emission zone.
In W~Aql and GY~Aql, the $^{13}$CO J=2-1 line has a larger extent, while in IRC\,$-$10529 and IRC\,+10011 some  transitions 
of SO$_2$ probe larger regions. 
KW~Sgr is in various ways an exception, because we have not acquired the low-resolution data and the $^{12}$CO J=2-1 line remains undetected in the medium-resolution data --- i.e.,  the data are sensitivity limited, owing to its large distance of 2\,400\,pc. 
Hence the CO extent plotted in Fig.~\ref{Fig:KW_Sgr_kinematics} is deduced from the high resolution data and some rotational transitions 
of SiO in the vibrational ground state which trace larger emission zones.

For all sources, except for $\pi^1$~Gru, the value of $\upsilon$(CO) is, however, lower than the velocity measure from other transitions with 
spatial emission zones lower than the extent of CO (see columns 4--6 in Table~\ref{Table:beta}), implying the wind profile will never be 
captured by the solution of the momentum equation~(Eq. (2)) and cannot be adequately reproduced using a $\beta$-velocity law. 
Nevertheless, we have tried to quantify approximately the region of the wind initiation and the wind acceleration efficiency by fitting 
the $\beta$-velocity law (Eq.~\eqref{Eq:velocity}) to the data in which the velocity measure is lower than the velocity determined 
from the $^{12}$CO J=2-1 line.
The fits account for the variance on the measure of the velocity, and the emission extent. 
Eq.~\eqref{Eq:velocity} has three free parameters ($v_0, R_{\rm{dust}}$, and $\beta$). 
The parameters $v_0$ and $R_{\rm{dust}}$ are not straightforward to quantify, because of the pulsation-induced shocks (see above). 
We therefore empirically estimate these two parameters.\footnote{The acceleration of the gas and dust begins at $R_0$, 
however --- consistent with the prior work by \citet[][and references therein]{2016A&A...591A..44M, Decin2018A&A...615A..28D} --- 
we do not distinguish between $R_{\rm{dust}}$ and $R_0$ here, because the difference between $R_{\rm{dust}}$ and $R_0$ of 
$\lesssim 10$\,\Rstar\ is small compared with the large scale description of the wind velocities that extend up to about 
$100-200$\,\Rstar\ in many of the {\sc atomium} stars.}   
 
To be a reliable measure of the velocity, the line profile should encompass at least three spectral resolution elements. 
Given the spectral resolution of $\sim$1.3\,km/s, this results in the minimum measurable velocity of around 4\,km/s 
(as can be seen in column~7 of Table~\ref{Table:beta}). 
Often the corresponding transitions are high-excitation transitions (see column~8 in Table~\ref{Table:beta}), although in some cases 
the lowest velocity measure is derived for weak, low-excitation transitions with restricted signal-to-noise ratios.
The parameter $v_0$ is determined as the minimum of the derived velocity measures (for both non-deconvolved and deconvolved emission; 
see Step 4 in Sect.~\ref{Sec:Methodology}), but should be larger than the local sound speed of $\sim$4\,--\,5\,km/s. 

The parameter $R_{\rm{dust}}$ is then quantified as the radius at which $v_0$ is reached, but should be larger than 2\,\Rstar. 
The derived values of $R_{\rm{dust}}$  vary between 2--6.7\,\Rstar (see Table~\ref{Table:beta}). 
We caution against over interpreting $R_{\rm{dust}}$ as the radius where the dust formation is starting, owing to the unknown effects 
of pulsation-induced shocks.

Using a Levenberg--Marquardt minimization routine, $\beta$ is derived with its variance as given by the covariance matrix 
(see last two columns in Table~\ref{Table:beta}) and --- as can be seen in Table~\ref{Table:beta} --- $\beta$ varies between 0.6--9.1. 
In four sources (U~Del, V~PsA, SV~Aqr, and KW~Sgr) not enough data points with $v<v$(CO) are available to determine $\beta$ 
(see Figs.~\ref{Fig:U_Del_kinematics}, \ref{Fig:V_PsA_kinematics}, \ref{Fig:SV_Aqr_kinematics}, \ref{Fig:KW_Sgr_kinematics}). 
For sources with \Mdot$\ga$5$\times$10$^{-7}$\,\Msun/yr in which the spatial emission zone of a significant fraction of the molecular 
transitions could be deconvolved, we often see that the wind acceleration continues up to $\sim$100\,\Rstar\,  and is represented by high 
values of $\beta$.
Some examples include RW~Sco, AH~Sco and GY~Aql.

The {\sc atomium} data provides some insight into the wind initiation efficiency, particularly on the frequently observed low wind acceleration. 
But Figs.~\ref{Fig:IRC10529_kinematics}, \ref{Fig:W_Aql_temporary}, and \ref{Fig:S_Pav_kinematics} -- \ref{Fig:RW_Sco_kinematics} 
confirm that the velocity vector field for all the {\sc atomium} sources is more complex than is captured by current 1D hydrodynamical models.
 {\it Pulsation-induced shocks can explain some of the velocity variation in the innermost few stellar radii, but as discussed 
 in the next section this scenario cannot explain the complex kinematic behaviour seen in most {\sc atomium} sources.} 
 
 %  +++++++++++++++++++++++++++++++++++++++++++++++++++++++++++++++++++++++++++++++++++++++++++++++++++

\begin{table*}[t]
\caption{\textbf{Velocity measures of SiO versus CO.}   }
\centering
\label{Table:large_vel}
\begin{tabular}{l|lrl|lrl}
\hline
\multicolumn{1}{c|}{(1)} & \multicolumn{3}{c|}{(2)} & \multicolumn{3}{c}{(3)} \\
$v$(SiO)$<v$(CO) & \multicolumn{3}{c|}{$v$(SiO)$>v$(CO)} & \multicolumn{3}{c}{$v$(SiO)$>v$(CO)}\\
 & \multicolumn{3}{c|}{No other molecules with large $v$} & \multicolumn{3}{c}{Other molecules with large $v$}\\
 \hline
 Target & Target & \multicolumn{2}{c|}{Radial extent of SiO } & Target & \multicolumn{2}{c}{Radial extent of SiO }\\
 \hline
 RW Sco & S~Pav & 0\farcs20\ \ \ -- & 21\,\Rstar & AH Sco & 0\farcs40\ \ \ -- & 136\,\Rstar\\
 $\pi^1$~Gru & T~Mic & 0\farcs40\ \ \ -- & 42\,\Rstar &  W~Aql & 0\farcs74\ \ \ --  & 133\,\Rstar\\
 	  		  & U~Del & 0\farcs34\ \ \ -- & 85\,\Rstar & GY Aql & 0\farcs82\ \ \ -- & 149\,\Rstar\\
 	  		  & V~PsA & 0\farcs05\ \ \ -- & 9\,\Rstar & IRC\,$-$10529 & 0\farcs78\ \ \ -- & 197\,\Rstar\\
	  		  & SV Aqr & 0\farcs52\ \ \ -- & 263\,\Rstar &  IRC\,+10011 & 0\farcs67\ \ \ -- & 336\,\Rstar\\		  
 	  		  &	R~Hya & 0\farcs97\ \ \ -- & 85\,\Rstar&  VX~Sgr & 0\farcs67\ \ \ -- & 153\,\Rstar\\
 	  	      & U~Her & 0\farcs54\ \ \ -- & 98\,\Rstar &	  	& \\	  
              & R~Aql & 1\farcs48\ \ \ -- & 248\,\Rstar & 	  & \\	
              & KW Sgr & 0\farcs06\ \ \ -- & 29\,\Rstar & & \\              
\hline
\end{tabular}
\tablefoot{  
Listed in the first column are the {\sc atomium} sources in which the velocity of all transitions of SiO is lower than that of CO v=0 J=2-1, 
$v$(SiO)$<v$(CO)). 
The second and third column list the {\sc atomium} sources in which at least one SiO transition has a larger line width than the CO v=0 J=2-1 line, 
$v$(SiO)$>v$(CO). 
In the case where only SiO lines trace velocities larger than the CO v=0 J=2-1 line, the source is listed in the second column; 
if other molecules also trace a larger velocity, the source name is indicated in the last column. 
The largest radial extent probed by the high velocity lines of SiO in the 15 sources with $v$(SiO)$>v$(CO) is listed in columns (2) and (3).
The maximum extent can be probed by SiO lines other than the one listed in Table~\ref{Table:beta}.
}
\end{table*}

%  +++++++++++++++++++++++++++++++++++++++++++++++++++++++++++++++++++++++++++++++++++++++++++++++++++ 

\subsubsection{{\sc atomium} wind profiles interpreted within the context of binary-star models}\label{Sec:atomium_as_binary}

Two conclusions can be drawn from the plots of the wind kinematics: 
(1)  the velocity measures in Table~\ref{Table:large_vel} which were derived from different rotational transitions and from 
different molecules at the same distance from the star (see columns 6 and 8 in Table~\ref{Table:beta}),
differ by more than the 3$\sigma$ of the velocity resolution; and 
(2)  the differences in the velocity measures correspond to real anomalies in the behaviour of the species.
The wind kinematic plots also show that the SiO velocities are greater than than those of CO in 15 of the 17 {\sc atomium} sources 
(i.e., all sources except $\pi^1$ Gru and RW~Sco) by up to a factor $\sim$1.6, and in six sources the velocity measure  
derived from other molecular lines are also greater than that of CO (see Tables~\ref{Table:beta}--\ref{Table:large_vel}).
However, CO has a larger extent than SiO for all sources except KW~Sgr (see Fig.~\ref{Fig:terminal_velocities1}, and Sect.~\ref{Sec:atomium_as_single}). 
In general, the  CO emission is expected to extend roughly an order of magnitude farther than the SiO emission
(see Sect.~\ref{Sec:Resolved-out flux}).
The low-excitation CO line is predominantly collisionally excited, and is a reliable tracer of the density in the outer wind region.
The Einstein $A$ coefficients of the rotational lines of SiO in the ground-vibrational state are three orders of magnitude higher 
than those of CO, and as a result the rotational lines of SiO are sensitive to radiative (de-)excitation effects implying that these 
lines are key diagnostics for tracing the complex kinematics in the inner wind regions.

Given our current physical understanding, pulsation-induced shocks are not a viable mechanism to explain the wind kinematic 
profiles of most of the {\sc atomium} sources, in particular for the case where larger SiO velocities are traced in the compact and medium 
configuration data (see the Appendix~\ref{Sect:terminal_velocity}). 
Hence other mechanisms should be considered.
An obvious candidate is binary interaction --- see our results published in \citet{DecinScience}.
Simulations for binary systems with a mass-losing AGB star as primary indicate that circumbinary disks are dynamically formed for systems 
with orbital separation $\la$6\,au and mass-ratios in the order of 0.5--1 \citep{Chen2017MNRAS.468.4465C}. A 
Keplerian (or in general rotational) velocity field (see dotted blue line in Fig.~\ref{Fig:velocities}) can lead to velocities projected along the 
line-of-sight in excess of the terminal velocity in the innermost few stellar radii; see for example the case of L$_2$~Pup \citep{Kervella2016A&A...596A..92K}.  
However, such a disk cannot explain the high velocity indicated for most sources in the  centre and right hand columns of Table~\ref{Table:large_vel}, 
which show high velocity SiO emission extends by more than $\sim$30\,\Rstar\ (in diameter). 
{\it We propose here that the gravitational influence of the binary companion (residing at a wide separation) is the cause for the latter 
behaviour, in particular for those sources in the last column of Table~\ref{Table:large_vel} for which various molecules trace large velocities.}
For example, if there is a (binary-induced) density contrast between an equatorial density enhancement and a biconical outflow or lobes, the velocities in these directions might differ, and they might not always favour the lobes if a denser equatorial region is more efficiently dust driven.
As the result, the density differential will favour different species in the kinematical plots which are at the same distance from the star, 
but which  have different velocities.

%  ----------------------------------------------------------------------------------------------------------------------------------------------

\begin{figure*}[htpb]
\vspace{1ex}
\begin{minipage}[t]{.495\textwidth}
        \centerline{\resizebox{\textwidth}{!}{\includegraphics[angle=0]{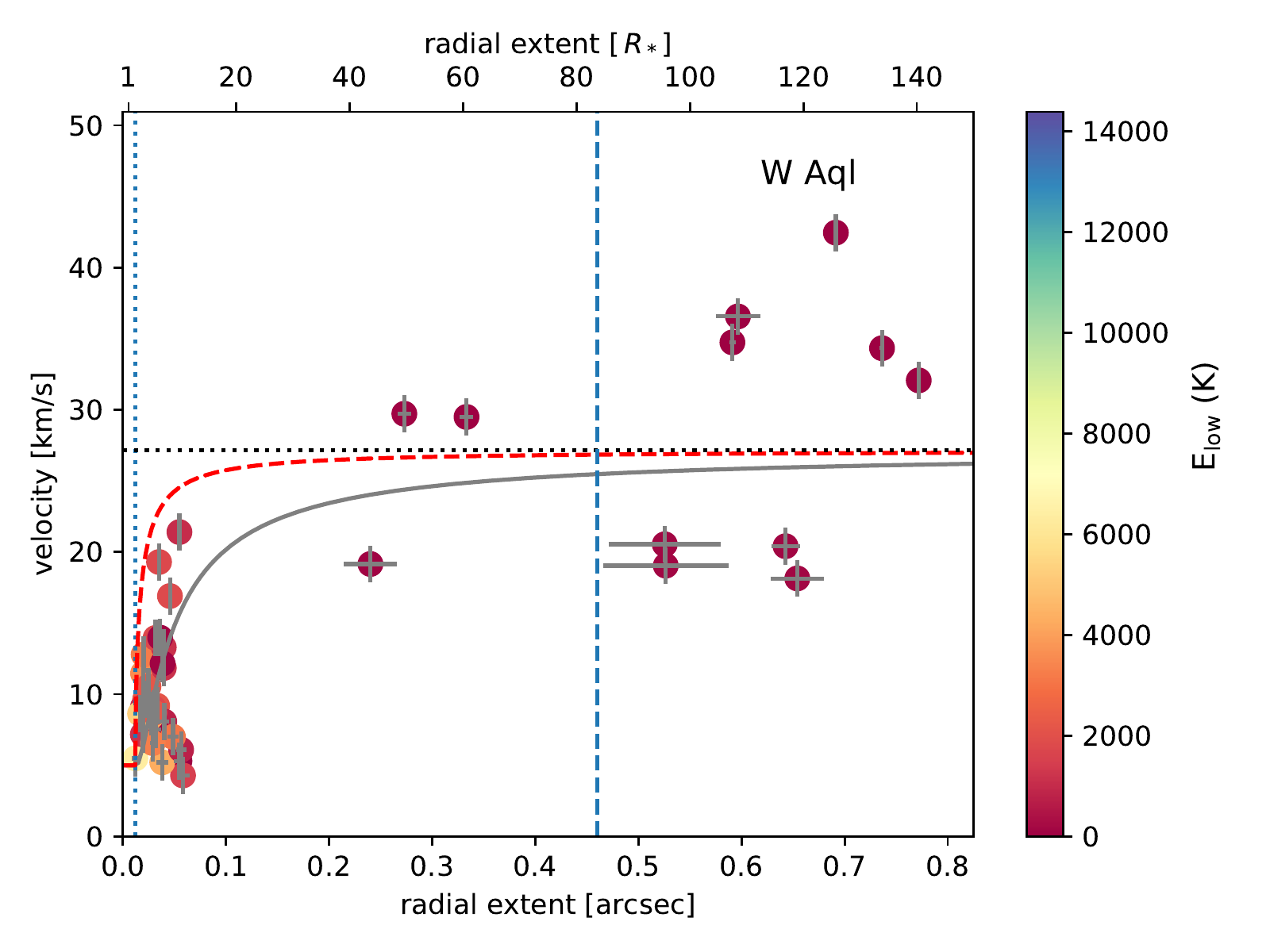}}}
\end{minipage}
    \hfill
\begin{minipage}[t]{.495\textwidth}
        \centerline{\resizebox{\textwidth}{!}{\includegraphics[angle=0]{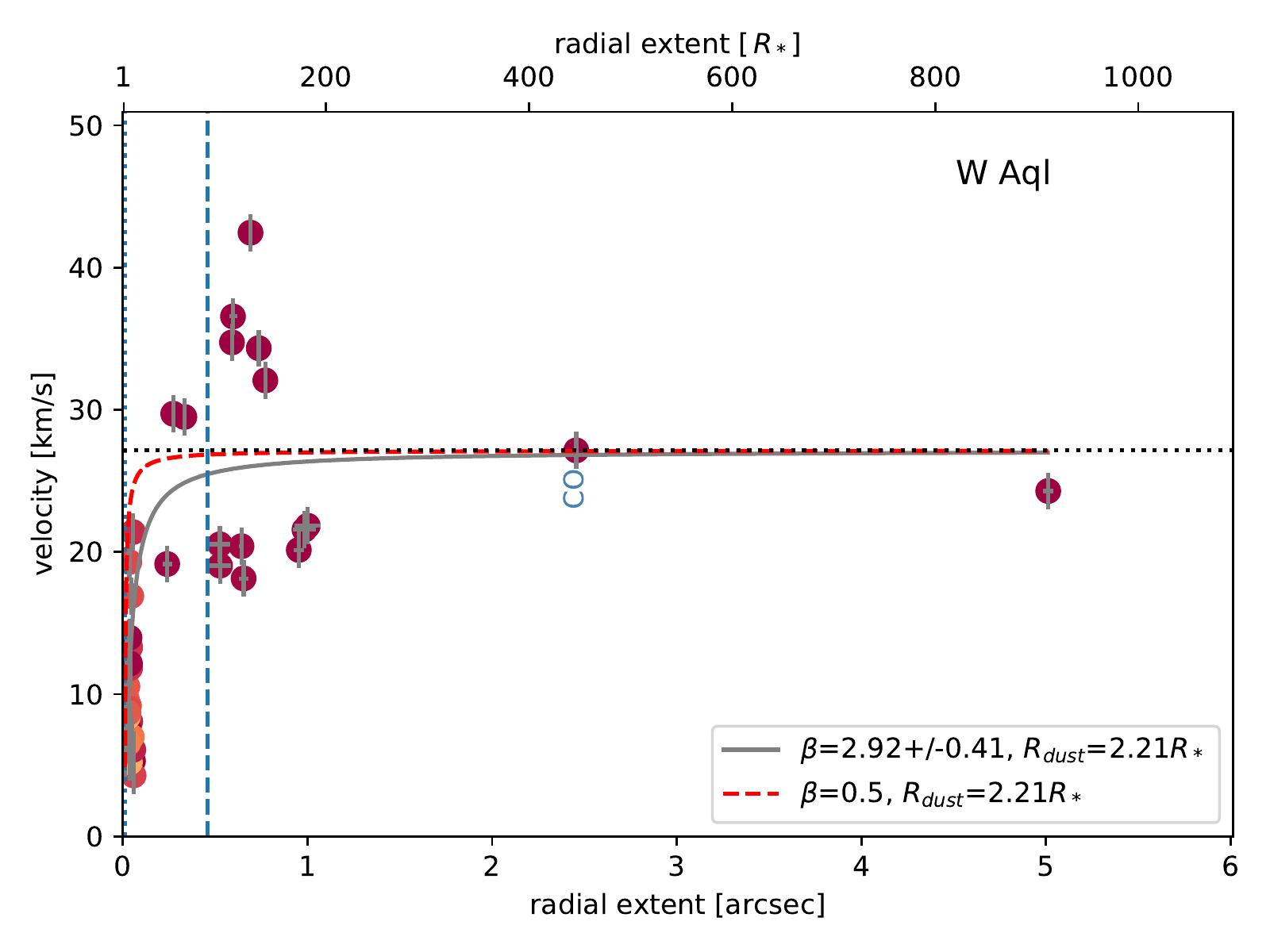}}}
\end{minipage}
\caption{\textbf{Wind kinematics for W~Aql.} See Fig.~\ref{Fig:IRC10529_kinematics}  caption. The vertical blue dashed line at 0\farcs46 
indicates the projected separation of the binary companion as deduced by \citet{Ramstedt2011A&A...531A.148R}.}
\label{Fig:W_Aql_temporary}
\end{figure*}

%  ----------------------------------------------------------------------------------------------------------------------------------------------

Checking the low-, medium-, and high-resolution data of the high-velocity lines, it becomes clear that in most cases  the highest velocity value  
is observed in the high resolution data.
This points towards excess high velocity emission arising from the shocked inner wind region, often within  $\sim$2--10\,\Rstar\ 
\citep[][see the Appendix~\ref{Sect:terminal_velocity}]{Cernicharo1997A&A...319..607C, Herpin1998A&A...334.1037H, Vlemmings2017NatAs...1..848V}.
For the binary hypothesis to hold in the situation of a separation above $\sim$10 stellar radii, we need to know if there is high velocity emission 
at other locations in the wind farther away from the central star, which we refer to here as {\it the persistence test}.  
We therefore check if high velocity emission (from molecules other than CO) can be detected in the low and medium resolution data. 
Because the lower resolution data were observed for a shorter time, they are less sensitive to compact emission. 
If the emission only arises from the innermost few stellar radii, no maximum radius or Gaussian fit can be determined from the lower resolution 
data following the procedure outlined in Sect.~\ref{Sec:Methodology}, and we then categorise the emission as {\it  non-persistent}.
The persistence test reaches two levels:  {\it level~1} applies to the sources in the second column of Table~\ref{Table:large_vel}, in which only 
SiO transitions reach velocities above the one deduced from the CO v=0 J=2-1 line; 
{\it level~2} applies to sources in the last column of Table~\ref{Table:large_vel}, in which molecules other than SiO also reach velocities above CO. 
The persistence test fails for five sources (U~Del, V~PsA, U~Her, R~Aql, and KW~Sgr), but is successful for four sources at 
level~1 (S~Pav, T~Mic, SV~Aqr, and R~Hya), and six sources at level~2 (AH~Sco, W~Aql, GY~Aql, IRC\,$-$10529, IRC\,+10011, and VX~Sgr). 
Obviously, all targets in the last column of Table~\ref{Table:large_vel} pass, which is not unexpected given the argument that molecules other 
than SiO are also diagnostics for the binary hypothesis. 
Hence, for the ten sources that pass the persistence test at either level~1 or level~2, we investigated whether we could deduce 
an approximate orbital separation from the kinematic information in 
%Fig.~\ref{Fig:IRC10529_kinematics} and Figs.~\ref{Fig:S_Pav_kinematics}--\ref{Fig:RW_Sco_kinematics}.}}
Figs.~\ref{Fig:IRC10529_kinematics}, 
\ref{Fig:W_Aql_temporary}, 
\ref{Fig:S_Pav_kinematics},
\ref{Fig:T_Mic_kinematics}, 
\ref{Fig:SV_Aqr_kinematics}, 
\ref{Fig:R_Hya_kinematics},  
\ref{Fig:AH_Sco_kinematics}, 
\ref{Fig:GY_Aql_kinematics}, 
\ref{Fig:IRC10011_kinematics}, 
and \ref{Fig:VX_Sgr_kinematics}.

W~Aql was first identified as a spectroscopic binary by \citet{Herbig1965VeBam..27..164H}.
\citet{Ramstedt2011A&A...531A.148R} used {\sl Hubble Space Telescope} (HST) data, with a spatial resolution of 0\farcs12, to deduce the 
projected separation of 0\farcs46 ($\sim$150\,au or $\sim$85\,\Rstar). The inclination of the orbit is unknown, and therefore the orientation 
of both sources relative to each other could not be deduced from these data, although \citet{Danilovich2015A&A...574A..23D} finds that the 
F8-G0 companion, with mass around 1\,\Msun, cannot be in front of the AGB star. 

%  ----------------------------------------------------------------------------------------------------------------------------------------------
We note that a close companion can increase the terminal wind velocity as compared to a single star model owing to the
slingshot mechanism \citep{2021A&A...653A..25M}.
Morever as illustrated in Fig.~\ref{Fig:velocities_binaries}, the gravitational attraction by a binary companion can induce  an increase in the velocity 
amplitude at radii smaller than the orbital separation, eventually leading to a wave-like velocity profile beyond the orbital separation. 
These predictions are roughly consistent with the velocity profile of W~Aql derived from the {\sc atomium} data 
in Fig.~\ref{Fig:W_Aql_temporary}, 
in which around 80\,\Rstar we see both an increase and a decrease of the velocity measures with respect to the beta law.
However the observed pattern is not as sharp as in the theoretical simulation in Fig.~\ref{Fig:velocities_binaries} 
(see also Fig.~\ref{Fig:velocities_binaries_linear}) because: 
(1) the observations correspond to projected velocities; and
(2) the binary system in W~Aql might be more complex than the simulations.  
Given this first-order agreement, we use the closest location beyond 10\,\Rstar\ where the velocity gradient turns negative 
as a proxy for a tentative indication of the upper limit on the orbital separation for those sources that pass the persistence test. 
Admittedly, this is not deduced straightforwardly for all sources.
For those cases in which there are only a few points to guide us in this exercise, we have opted to list the extent of the molecule 
with the highest velocity measure ($v_{\rm{max}}$) which for W~Aql would yield 0\farcs69 (or 125\,\Rstar). 
This difference between the first estimate of $\sim$80\,\Rstar\ --- as deduced from the negative 
gradient of the velocity pattern, and $\sim$125\,\Rstar\ as retrieved from $v_{\rm{max}}$ --- also marks the limitations of the method
proposed here for estimating the orbital separation.

%  +++++++++++++++++++++++++++++++++++++++++++++++++++++++++++++++++++++++++++++++++++++++++++++++++++

Two other {\sc atomium} sources in addition to W~Aql are confirmed binaries (R~Hya and $\pi^1$Gru). 
The companion of R~Hya is thought to have a very wide orbital separation of 21\arcsec\ \citep{Mason2001AJ....122.3466M} 
which is beyond the field of view of the {\sc atomium} data.
In R~Hya there is also evidence of dramatic perturbations in the CSE within a few 100$R_\star$ of the central star, 
possibly owing to a second companion \citep{2021A&A...651A..82H}. 
$\pi^1$Gru has a companion of spectral type G0V at a separation of 2\farcs7 
\citep[$\sim$500 au,][]{Feast1953MNRAS.113..510F}, 
but there is no signature of the known companion in the line or continuum data of {\sc atomium}.

Following a similar approach as for W~Aql in Fig.~\ref{Fig:W_Aql_temporary}, we derived an orbital separation for IRC--10529 
from the velocity profile in Fig.~\ref{Fig:IRC10529_kinematics}, and the orbital separations for the eight other stars 
that pass the persistence test from the velocity profiles in Appendix Sect.~\ref{Sect:kinematics_other_sources} of: 
$0.20^{\prime\prime} - 0.30^{\prime\prime}$ for S~Pav, T~Mic, SV~Aqr, GY~Aql, and IRC+10011;
and between $0.45^{\prime\prime} - 1.00^{\prime\prime}$ for AH~Sco, R~Hya, and VX~Sgr.

In $\pi^1$Gru, the width of the $^{12}$CO  J=2-1 line is more than a factor of 2 larger than that of any other molecular transition. 
The large line width --- which had also been seen in previous ALMA $^{12}$CO J=3-2 data --- was interpreted as an indication that the  
envelope structure of $\pi^1$~Gru includes a radially expanding equatorial torus (with a velocity of 8--13\,km/s);
and a fast bipolar outflow (with a linear velocity increase from 14\,km/s at the base up to 100\,km/s at the tip), with an angle between 
the line of sight and the equatorial plane of 40\deg\, \citep{Doan2017A&A...605A..28D}.  
However, a spiral pattern has emerged in more recent higher spatial resolution ALMA $^{12}$CO J=3-2 data, and the spiral-arm separation 
hints towards the presence of a companion with a separation of less than 70\,AU \citep[or 34\,\Rstar;][]{Doan2020A&A...633A..13D}.
The {\sc atomium} data will now further refine this picture, since various other molecules have line wings up to $\sim$40\,km/s and exhibit 
clear signs of rotation or bipolarity in their moment~1 maps --- see for example the SiO J=5-4 and J=6-5 lines in \citet{2020A&A...644A..61H}. 
The separation between the arc-like structures observed in the {\sc atomium} $^{12}$CO J=2-1 channel maps 
\citep{DecinScience}, 
indicates the presence of a second companion with an orbital separation of around 0\farcs04; 
and the dynamics traced by the SiO masers, 
suggest a (tentative) upper limit of the companion mass of $\sim$1.1\,\Msun\, 
\citep[][Montarg\`es et al., \textit{in prep}]{2020A&A...644A..61H}.
 
\subsection{{\bf Discussion and implications}}  
\label{Sec:implications}

Putting these results in the context of the overall goals of the {\sc atomium} project, it is clear that important science questions 
posed in Sect.~\ref{Sec:dynamics} can be addressed by the {\sc atomium} data. 
The wind acceleration efficiency, as expressed by the quite large values of $\beta$, seems quite low in general. 
The slow wind acceleration in turn yields constraints on the composition, size, and formation radius of dust grains, expressed for a single-star 
model by the dust extinction efficiency $Q_\lambda(a,r)$ in Eqs.~\ref{Eq:momentum} and \ref{Gamma}. 
In addition, the question of the enforced wind dynamics in the intermediate wind region {needs to be reformulated and should incorporate 
 a search for the impact of binary companions on the wind dynamics of AGB and RSG stars.

Within both a single-star and a binary-star context, the results derived here have an impact on our understanding of the mass-loss rate  
for the following two reasons:

\begin{itemize}
\item[(i)]{When comparing prior results to those obtained in the {\sc atomium} project, 
the velocities $v$(CO) derived from the low excitation line of $^{12}$CO J=2-1 are systematically higher when derived from
the {\sc atomium} data. 
Since $v$(CO) is often used as a measure for the terminal wind velocity in single-star models, a direct implication is that the mass-loss
rate for these sources will be underestimated. 
The high sensitivity of the current ALMA data was the key for deriving these higher wind velocities. 
As noted in Sect.~\ref{Sec:atomium_as_single},  under the condition of single star models the larger CO velocities might imply larger 
terminal velocities and larger values of $\dot{M}$, because the random scatter from the thermal line broadening and turbulence cannot 
explain these large velocities.
As a result, we surmise that the terminal wind velocities and hence gas mass-loss rates will be underestimated for other AGB and RSG 
sources as well. }

\vspace{0.125cm}
It might also be the case that the higher CO velocities (and those of other molecules) indicate exceptional motions owing to binary interaction. 
Under the condition of binary companions, the larger CO velocities might not be an indication of  larger terminal velocities 
of the bulk material, but we are currently unable to estimate the relative effect in the single versus binary model,
because assessing the impact of the companion on the Lagrangian
\citep[see][]{Gregory_R_Douglas2006-04-17} would require extensive hydrodynamical simulations.

\vspace{0.25cm}
\item[(ii)]{The current results support the conclusion in \citet{DecinScience} 
that (sub-)stellar binary interaction is the prime wind shaping agent of the majority of AGB/RSG stars, including the {\sc atomium} sources
whose mass-loss rates exceed the nuclear burning rate of around  1$\times$10$^{-7}$\,\Msun/yr. 
In the majority of the {\sc atomium} sources, molecular transitions other than the $^{12}$CO v=0 J=2-1 line trace a larger velocity amplitude 
than CO, and have a spatial emission zone that is often greater than 30 stellar radii, but is much less than the extent of CO. 
This result has a two-fold repercussion on our historical insight of mass-loss rates in AGB and RSG which were derived within the context 
of single-star models: \\
}

\begin{enumerate}
\item{For close binary systems, a massive planet or stellar companion can enhance the AGB/RSG mass-loss rate by depositing angular momentum into the envelope and by reducing the effective gravity of the mass-losing star. 
Single stars or binary stars isolated from angular momentum deposition hence might suffer from a lower mass-loss rate during the AGB/RSG phase than stars prone to angular momentum deposition \citep{Sabach2018MNRAS.479.2249S}. 
Hydro-chemical simulations stimulated by the results of the {\sc atomium} survey, indicate this difference might be up to almost an order of magnitude  \citep[Bolte et al. {\it{in prep}},][]{2021ARA&A..59..337D}.   \\
} 

\item{There are profound implications for the classical measures of the AGB/RSG mass-loss rate derived under the assumption of a single star
with a spherical wind. For a mass-losing AGB/RSG star in a binary system, the material flow will have a directional preference towards the orbital 
plane; and spiral arcs, circumbinary and accretion disks, etc. can be created \citep[see, for example][]{Mastrodemos1999ApJ...523..357M,
Mohamed2012BaltA..21...88M, Kim2012ApJ...759...59K, Liu2017ApJ...846..117L, Chen2017MNRAS.468.4465C, Saladino2019A&A...626A..68S, 
2020A&A...637A..91E}. 
As such, previous mass-loss rate estimates based on the assumption of spherical symmetry should be interpreted with care since 
systematic errors might occur, 
as shown recently by \citet{Homan2015A&A...579A.118H}, \citet{Homan2016A&A...596A..91H}, and \citet{Decin2019NatAs...3..408D}.  
We conjecture that in general, mass-loss rates hitherto derived from \textit{dust spectral features} will be systematically overestimated. 
This conjecture is based on the fact that the companion’s gravitational attraction can create an equatorial density enhancement (EDE) with a 
density contrast that can be up to an order of magnitude higher than the background wind density \citep{2020A&A...637A..91E}. 
The spectral energy distribution (SED) in the near- and mid-infrared mainly traces warm dust residing close to the star, hence in the EDE. 
Therefore, depending on the inclination of the EDE, the analysis of dust spectral features using a simplified 1D approach  reflects the higher
density in the EDE created by the binary interaction, but not the actual mass-loss rate which will be lower.  
This conjecture is in line with previous observations which indicated that mass-loss rates derived from dust features are about an order of 
magnitude larger than mass-loss rates from CO observations \citep[e.g.,][]{Heske1990A&A...239..173H}.  \\
}

\end{enumerate}

\end{itemize}

These implications have a profound impact on several aspects of stellar evolution. 
Because the mass is the prime parameter determining the evolution and lifetime of a star, any modification to the stellar mass-loss over time has large repercussions on its evolutionary path.

If we would only account for (i), then a higher mass-loss rate implies a shortening of the AGB (RSG) phase. 
However, recent studies indicate that most stars in the universe will have one or more stellar or gas-giant planetary mass companions 
\citep{Moe2017ApJS..230...15M, Nielsen2019AJ....158...13N, Fulton2018AJ....156..264F, Fulton2019ESS.....440101F}. 
Hence, most empirically derived mass-loss rates are from samples containing a large fraction of stars that experience binary interaction with 
a (sub-)stellar companion.
As such, our knowledge of the mass-loss rate will be biased by the impact that companions can have both on the magnitude  
of the mass-loss and on the observed diagnostics from which mass-loss rates are retrieved. 
Given (ii) implies that the mass-loss rate for these cool evolved stars can be seriously overestimated in current stellar evolution models for 
single stars. 
These models use mass-loss rate prescriptions to calculate the change of mass during the AGB and RSG phase, 
whose parametric relations for the mass-loss rate are often based on fitting infrared colours or the dusty SEDs \citep{Reimers1975MSRSL...8..369R, deJager1988A&AS...72..259D, vanLoon1999A&A...351..559V, vanLoon2005A&A...438..273V}.
The impact of the effects discussed in (ii) are somewhat countered by the increase in mass-loss rate prescribed by (i), but the amplitude of the effect discussed in (i) is lower than those in (ii).
The white-dwarf initial-final mass function \citep[see for example][]{Cummings2018ApJ...866...21C} limits the total mass-loss occurring during 
the AGB phase. 
Hence, this result implies that the AGB phase will be longer for single stars. For single stars with initial mass greater 
than $\sim$8\,\Msun, the mass before exploding as supernovae will be higher implying a larger fraction of more massive neutron stars can form.
In addition, this implies that the contribution of cool evolved stars to the (extra)galactic dust budget will be lower than currently stipulated 
\citep[see for example][]{Matsuura2009MNRAS.396..918M, Matsuura2013MNRAS.429.2527M}, and the issue of the `missing dust-mass problem'\footnote{The missing dust-mass problem refers to the Large Magellanic Cloud and other high-z galaxies whose accumulated dust mass 
from AGB and RSG stars (and possibly supernovae) over the dust lifetime is significantly less than the dust mass 
in the ISM.} is far from solved \citep{Matsuura2009MNRAS.396..918M}
 
Dust-to-gas mass ratios for M-type AGB stars retrieved empirically are on average 5.8$\times$10$^{-3}$, while for carbon and S-type AGB stars 
they are around 2.5$\times$10$^{-3}$ and  2.8$\times$10$^{-3}$, respectively
\citep{Groenewegen1998A&A...337..797G, Groenewegen1999A&AS..140..197G, Ramstedt2008A&A...487..645R, Ramstedt2009A&A...499..515R,
2015A&A...581A..60D}.
Combining (i) and (ii) implies the dust-to-gas mass ratio for these samples, with derived gas mass-loss rates $\la\!1\!\times\!10^{-5}$\,\Msun/yr 
for the carbon and S-type AGB stars and $\la\!7\!\times\!10^{-5}$\,\Msun/yr for the M-type AGB stars, 
will be lower than current empirically derived values indicate. 
This conclusion impacts all studies that are (or have been) using a dust-to-gas mass ratio to compute total mass-loss rates from retrieved 
dust masses for which most often the canonical dust-to-gas ratio of 1/200 (as derived from galactic ISM studies) is used \citep[see for example][]{Matsuura2009MNRAS.396..918M, Matsuura2013MNRAS.429.2527M}.

%. ################################################################################################################

\section{Acknowledgements}

\begin{acknowledgements}

The authors are grateful to a referee for the close reading of the manuscript and the constructive comments.
This paper makes use of the following ALMA data: ADS/JAO.ALMA\#2018.1.00659.L, `ATOMIUM: ALMA tracing the origins of molecules forming dust in oxygen-rich M-type stars’. 
ALMA is a partnership of ESO (representing its member states), NSF (USA) and NINS (Japan), together with NRC (Canada) and NSC and ASIAA (Taiwan), in cooperation with the 
Republic of Chile. 
The Joint ALMA Observatory is operated by ESO, AUI/NRAO and NAOJ. 
This paper makes use of the CASA data reduction package: \url{http://casa.nra.edu} \citep{McMullin2007ASPC..376..127M}.
- Credit: International consortium of scientists based at the National Radio Astronomical Observatory (NRAO), 
the European Southern Observatory (ESO), the National Astronomical Observatory of Japan (NAOJ), the CSIRO Australia Telescope National Facility (CSIRO/ATNF), and the Netherlands Institute for Radio Astronomy (ASTRON) under the guidance of NRAO. 
The National Radio Astronomy Observatory is a facility of the National Science Foundation operated under cooperative agreement by 
Associated Universities, Inc.
The authors thank the Quality Assurance team at ESO for  customising the imaging pipeline. 
We thank IRIS for provision of high-performance computing facilities. STFC IRIS is investing in the UK's Radio and mm/sub-mm Interferometry Services in order to improve the data quality 
and allow much more data to be processed. 
This paper makes use of the Cologne Database for Molecular Spectroscopy (CDMS; \url{https://cdms.astro.uni-koeln.de/classic/}) 
and the spectral line catalogs of the Jet Propulsion Laboratory (JPL, \url{https://spec.jpl.nasa.gov});
{\bf Funding:}
We acknowledge travel support for consortium meetings from the RadioNet MARCUs programme under the European Community Framework 
Programme~7, Advanced Radio Astronomy in Europe, grant agreement no.: 227290, 
L.D., D.G., W.H., J.B., J.M.C.P., and S.H.J.W.\ acknowledge support from the ERC consolidator grant 646758 AEROSOL, 
L.D. and E.C.\ acknowledge support from the KU Leuven under the C1 MAESTRO grant C16/17/007, 
W.H.\ acknowledges support from the FWO Flemish Fund of Scientific Research under grant G086217N, 
F.H.\ acknowledges support from the "Programme National de Physique Stellaire" (PNPS) of CNRS/INSU co-funded by CEA and CNES,
J.M.C.P.\ acknowledges support from the UK STFC grant ST/P00041X/1, 
J.Y.\ acknowledges support from the UK STFC grant ST/R001049/1, 
M.V.d.S.\ acknowledges support from the FWO through grant 12X6419N, 
T.D.\ acknowledges support from the FWO through grants 12N9917N \& 12N9920N, 
A.B.\ acknowledges support from the "Programme National de Physique Stellaire" (PNPS), 
I.M.\ acknowledges funding by the UK STFC grant ST/P000649/1, EDB acknowledges financial support from the Swedish National Space Agency, 
I.E.M.\ acknowledges support from the FWO and the European Union's Horizon 2020 research and innovation program under the Marie 
Skłodowska-Curie grant agreement No 665501, 
S.E.\ acknowledges funding from the UK STFC as part of the consolidated grant ST/P000649/1 to the Jodrell Bank Centre for Astrophysics at the University of Manchester, 
P.K.\ acknowledges support from the French PNPS of CNRS/INSU, 
C.A.G.\ acknowledges support from NSF grant AST-1615847, 
T.J.M.\ is grateful to the STFC for support under grant ST/P000312/1, 
A.A.Z.\ was supported by the STFC under grants ST/T000414/1 and ST/P000649/1, 
M.D.G.\ thanks the STFC for support under consolidated grant ST/P000649/1 to the JBCA.

\end{acknowledgements}

%++++++++++++++++++++++++++++++++++++++++++++

\bibliographystyle{aa}
\bibliography{ATOMIUM_overview}

%++++++++++++++++++++++++++++++++++++++++++++
 
 \newpage
 \onecolumn
 \noindent    \hfill\break
 \vspace{-1.5cm}
\begin{appendix}
 
 \section{Additional figures} \label{App:add_figures}
 
 %  ############################################################################################################ 
%  ##########################################   APPENDIX A  ##################################################### 
%  ############################################################################################################ 

 \begin{figure*}[htpb]
\centering
\includegraphics[angle=0,width=1.0\textwidth]{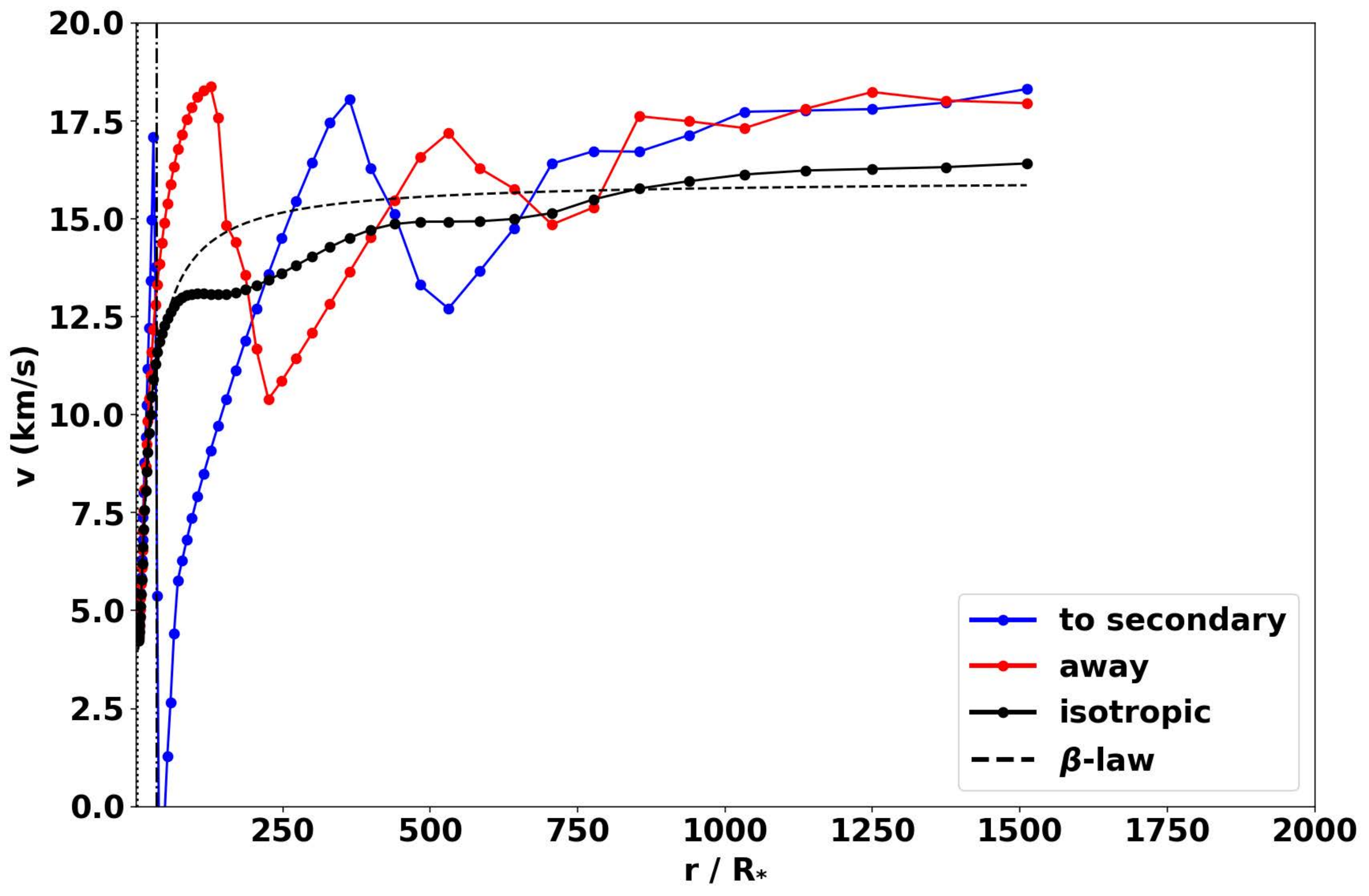}
\caption{  {\bf Illustration of the impact of a binary companion on the velocity field.} 
This is the same as Fig.~\ref{Fig:velocities_binaries}, except it has been replotted with a linear x-axis.} 
\label{Fig:velocities_binaries_linear}
\end{figure*}

%  #################################################################

\afterpage{\clearpage}
\newpage

\begin{figure*}[t]
\centering
\includegraphics[width=0.75\textwidth]{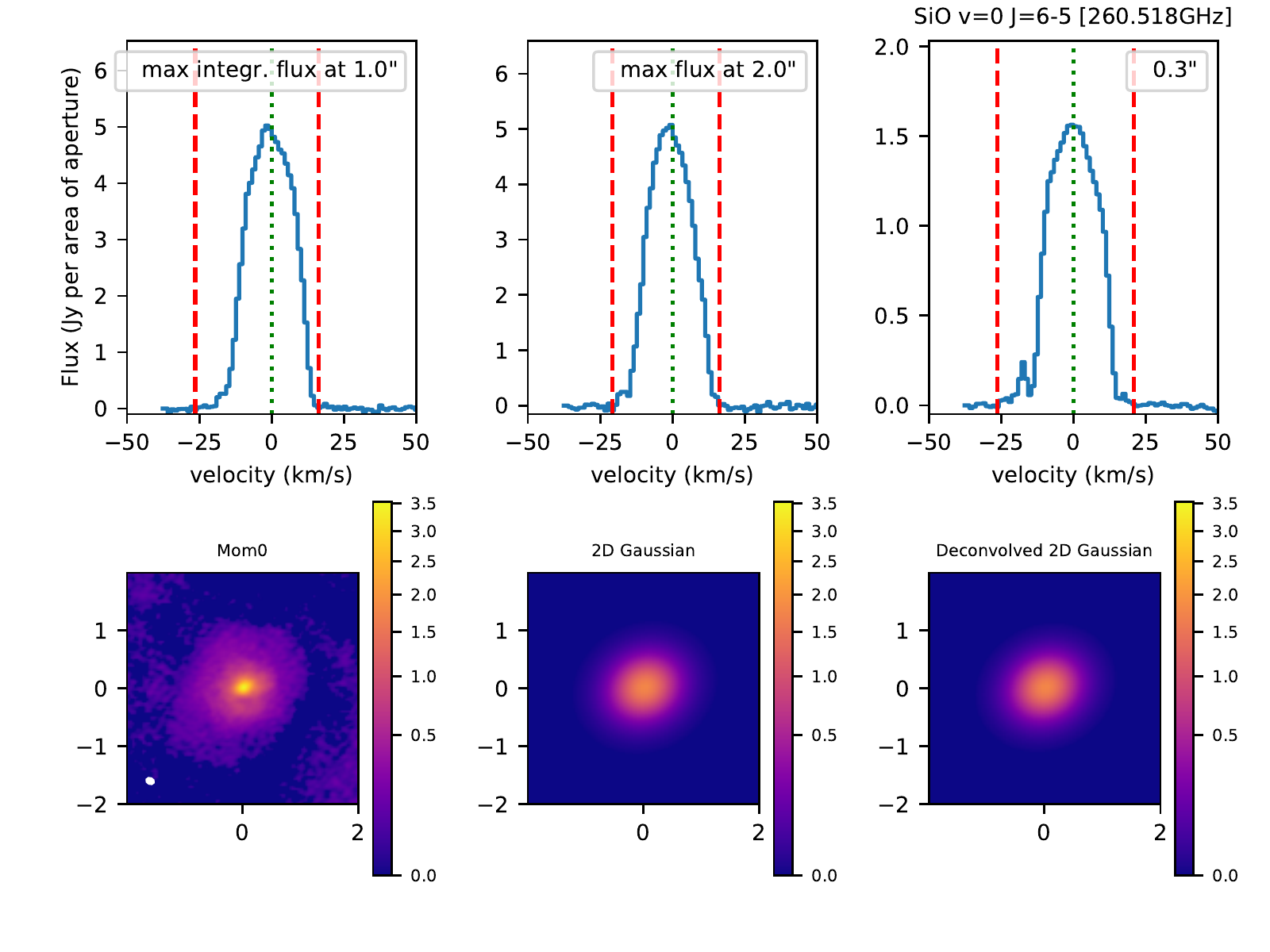}
\vspace{-1.00cm}
\caption{\textbf{Determination of the velocity measure and angular emission zone for the SiO v=0 J=6-5 transition of the medium  
resolution data of IRC\,$-$10529.}
{\it Top row:} 
The maximum integrated flux is attained for an extraction aperture of 1\farcs0 (left panel), the peak flux is maximal for an extraction 
aperture of 2\farcs0 (middle panel), and for each transition a reference spectrum at 0\farcs3 is plotted (right panel). 
The velocity (in km~s$^{-1}$) is with respect to the stellar velocity of $-16.3$~km~s$^{-1}$ (see the last column in Table~\ref{Table:targets}).
The dotted {\green{green}} vertical line in the spectra indicates the central frequency, 
and the two dashed red lines indicate the determination of the velocities of the red and blue wings.
In the example here, the velocity derived for the SiO line (v(SiO)=26.4~km~s$^{-1}$) is larger than the velocity determined 
from the low resolution CO v=0 J=2-1 line (v(CO)=21.8~km~s$^{-1}$).
{\it Bottom row:}  The first image is the moment0 map, the second image the 2D Gaussian fit to the moment0-map, and the last plot 
the image for the deconvolved 2D Gaussian profile.
The  colour scales in the moment0 maps are in units of Jy/beam km/s.
The ALMA synthesized beam is shown as a white ellipse in the lower left corner of the moment0 map.
}
\label{Fig:deconvolution_SiO}
\vspace{-1.5cm}
\end{figure*} 

%   ######################################################################################
%\noident   \hfill\break
%\vspace{-2.0cm}

\begin{figure}[b]
%\begin{figure*}[!htbp]
\centering
\includegraphics[width=.75\textwidth]{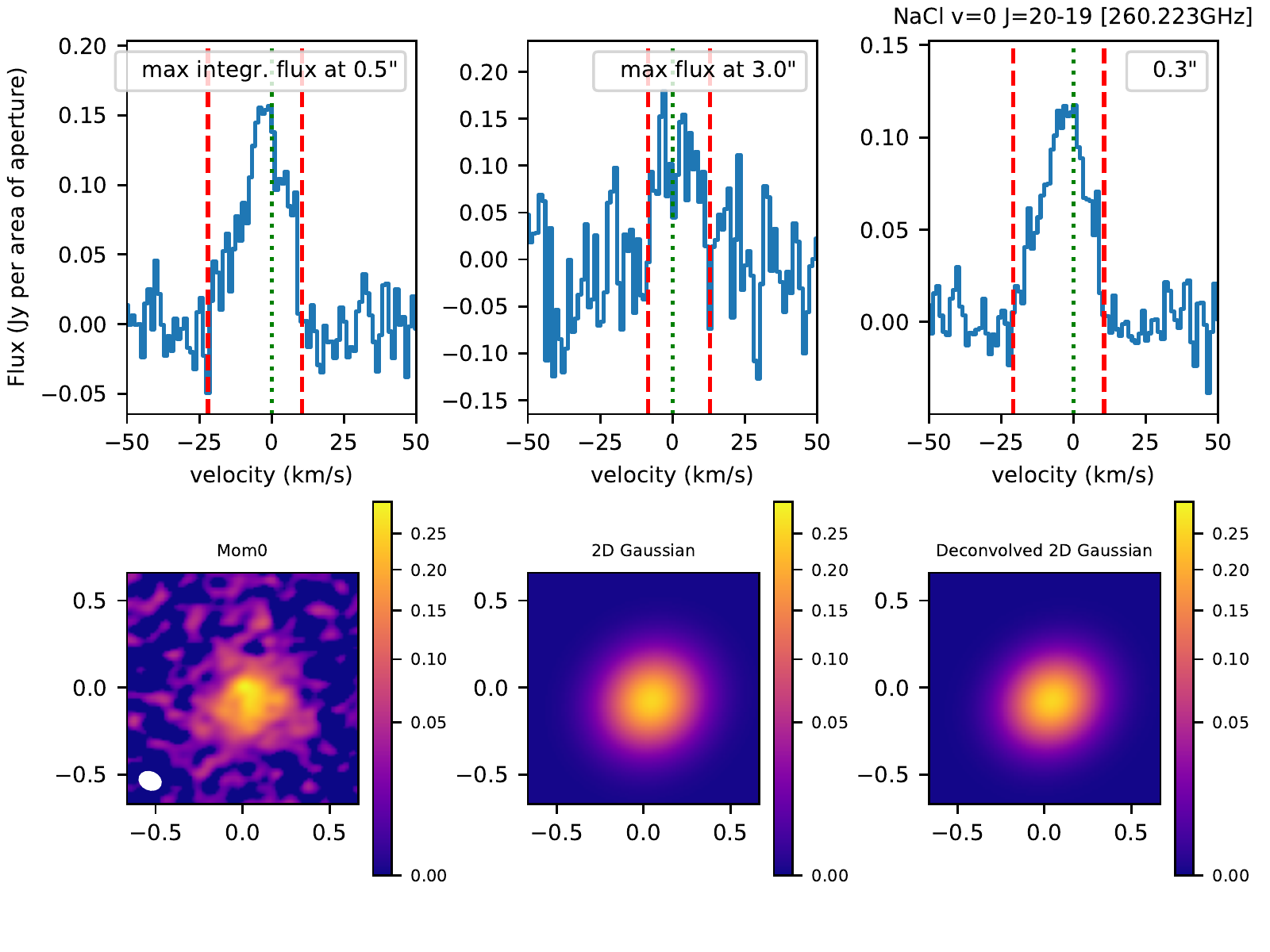}
\vspace{-.75cm}
\caption{\textbf{Determination of the velocity measure and angular emission zone for the NaCl v=0 J=20-19 transition of the medium  
resolution data of IRC\,$-$10529.}
The velocity (in km~s$^{-1}$) is with respect to the stellar velocity of $-16.3$~km~s$^{-1}$ (see the last column in Table~\ref{Table:targets}).
The  colour scales in the moment0 maps are in units of Jy/beam km/s (see Fig.~\ref{Fig:deconvolution_SiO} caption).
}
\label{Fig:deconvolution_NaCl}
\end{figure} 

%  ###################################  T Mic CO  ##########################################################
\noindent    \hfill\break
\vspace{-1.0cm}

\begin{figure}[htp]
   \centering 
 \vspace{0.5cm}
{\includegraphics[width=1.0\textwidth]{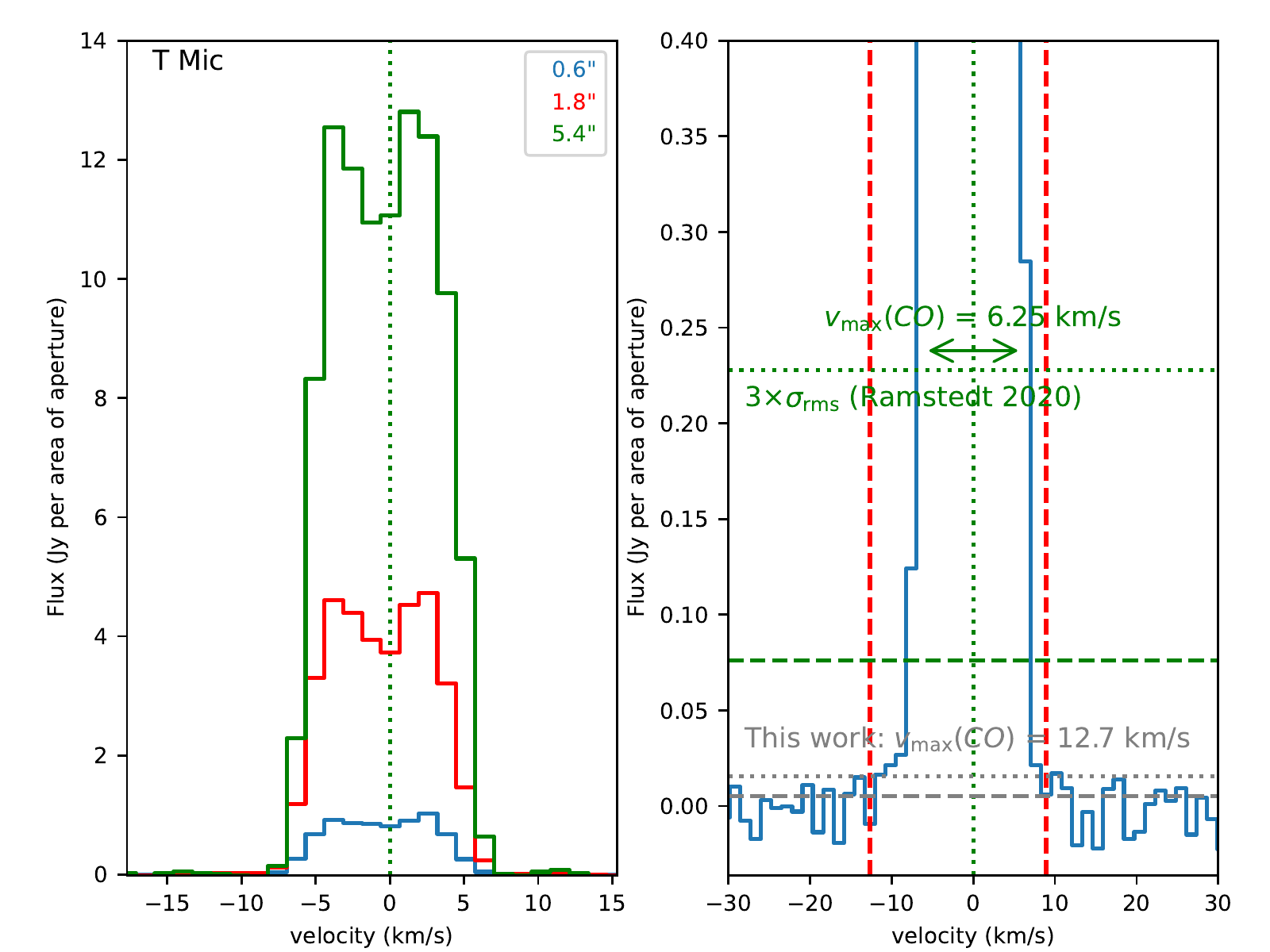}}
\vspace{0.5cm}
\caption{  {\bf The J=2-1 line of CO at 230,538.000~MHz observed in T~Mic in the compact configuration.}  
The plots show how (in the example of T Mic) the velocity {\boldmath  ($v_{\rm{max}}$)} was derived from the molecular lines observed in the ATOMIUM survey.
{\it Left:} The CO spectra extracted with apertures of radius {\boldmath  $0.6^{\prime\prime}$ (blue), $1.8^{\prime\prime}$} (red), 
and {\boldmath $5.4^{\prime\prime}$} (green).
{\it Right:} The CO profile extracted with an aperture of radius {\boldmath $2.06^{\prime\prime}$} is plotted on an expanded flux density scale with a full scale amplitude of 0.4\,Jy/beam to better discern the red and blue wings. 
The horizontal dashed grey line at 5\,mJy/beam corresponds to the 1$\sigma$ peak rms noise (see point \#2 in Sect. 5.2);
the horizontal dotted grey line at 15\,mJy/beam corresponds to the 3$\sigma$ peak rms noise; the green dotted vertical line denotes the line center 
of CO at the {\boldmath  $v_{\rm{LSR}}$ of $+25.5$}~km/sec; 
and the red dashed vertical lines indicate the blue wing velocity of {\boldmath $-12.7$}~km/s, and the red wing velocity of 8.9~km/s. 
The value of {\boldmath $v_{\rm{max}}{\rm{(CO)}}$} with the largest magnitude --- i.e., the blue wing velocity of 12.7~km/s --- is designated 
as the `velocity measure' of the CO J=2-1 transition in T~Mic (see Sect.~\ref{Sec:Methodology}). 
Similarly, the dashed green horizontal line at 76\,mJy/beam corresponds to the 1$\sigma_{\rm{rms}}$ noise in the Band~6 spectrum 
of CO observed in T~Mic by \citet[][see Table~B.1]{2020A&A...640A.133R}. 
The dotted green horizontal line at 228\,mJy/beam corresponds to the 3$\sigma_{\rm{rms}}$ noise and to the point where the full width 
of the CO J=2-1 line profile is equal to 12.5~km/s, whereby \citeauthor{2020A&A...640A.133R} determined 
{\boldmath $v_{\rm{max}}{\rm{(CO)}}= 6.25$}~km/s.
}
\label{Fig:TMic_CO}
\vspace{4.0cm}
\end{figure}

%  #########################################################################################
%  ###################################  APPENDIX B  #########################################
%  #########################################################################################

\afterpage{\clearpage}
\newpage

\section{Medium angular resolution spectra of IRC\,$-$10529} 

In this section, we provide the additional spectra extracted from the medium resolution data of IRC\,$-$10529 for cubes 02--15. 
The spectra for cubes 00 and 01 are shown in Fig.~\ref{Fig:spectra_IRC1}.

\noindent   \hfill\break
\vspace{1.0cm}

\begin{figure*}[htbp]
\centering\includegraphics[angle=0,width=.48\textwidth]{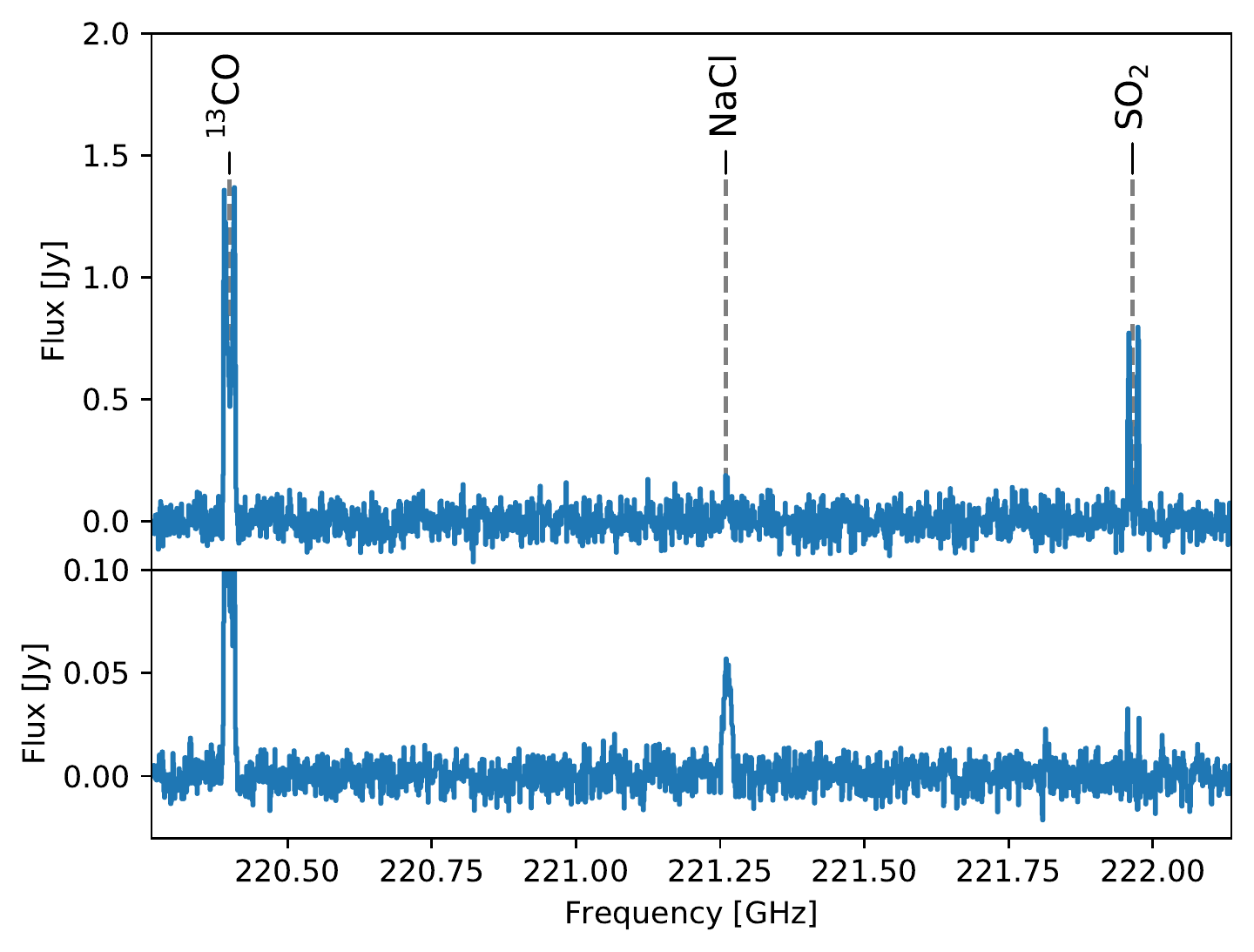}       
            \hfill
\centering\includegraphics[angle=0,width=.48\textwidth]{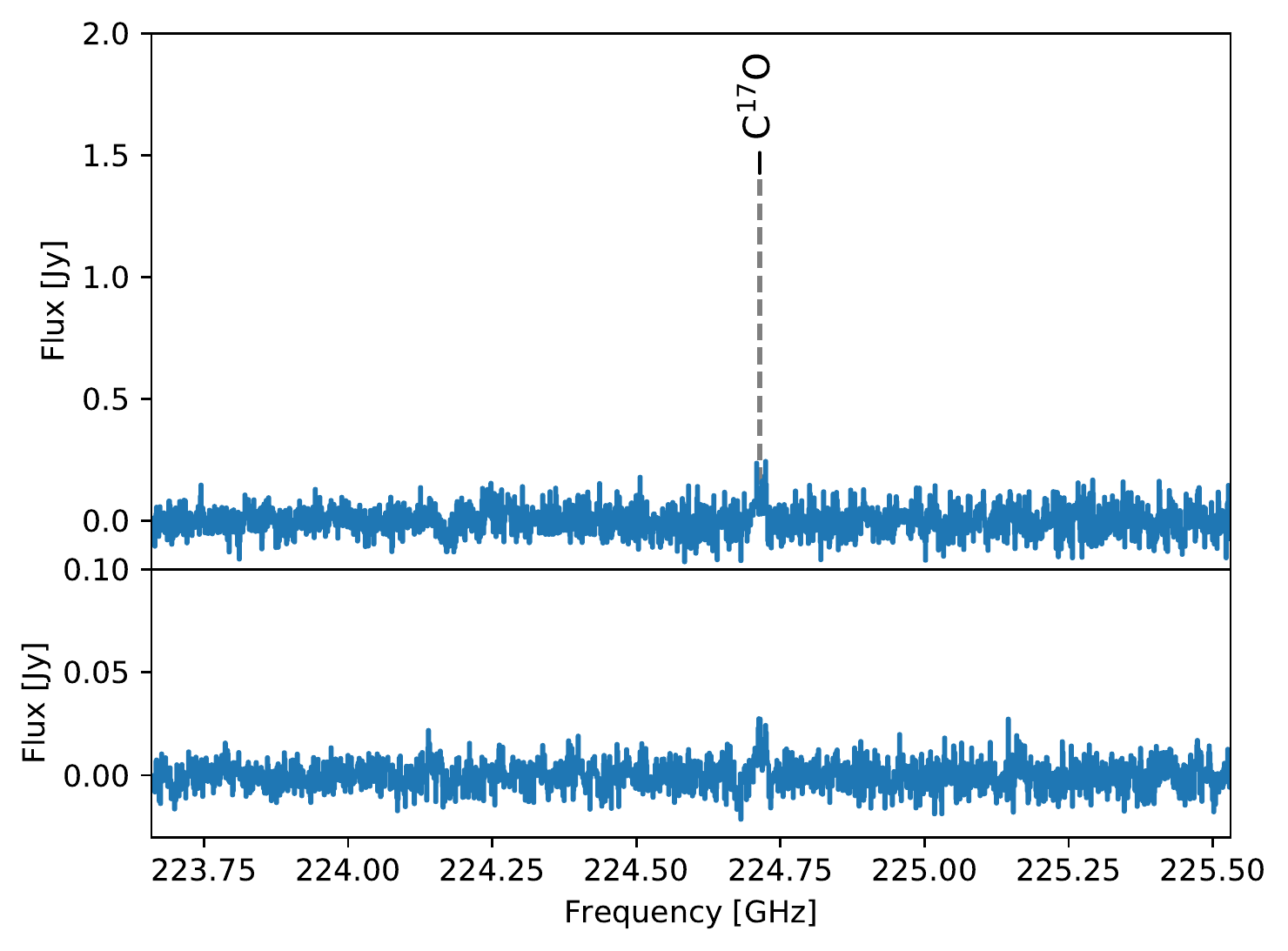}          
\caption{\textbf{ALMA spectra of IRC\,$-$10529.} 
Caption; see Fig.~\ref{Fig:spectra_IRC1}. Data are displayed for cubes 02 to 03.}
\label{Fig:spectra_IRC2} 
\end{figure*}

%.  ++++++++++++++++++++++++++++++++++++++++++++++++++++++++

\afterpage{\clearpage}
\newpage

\begin{figure*}[htbp]
\centering\includegraphics[angle=0,width=.48\textwidth]{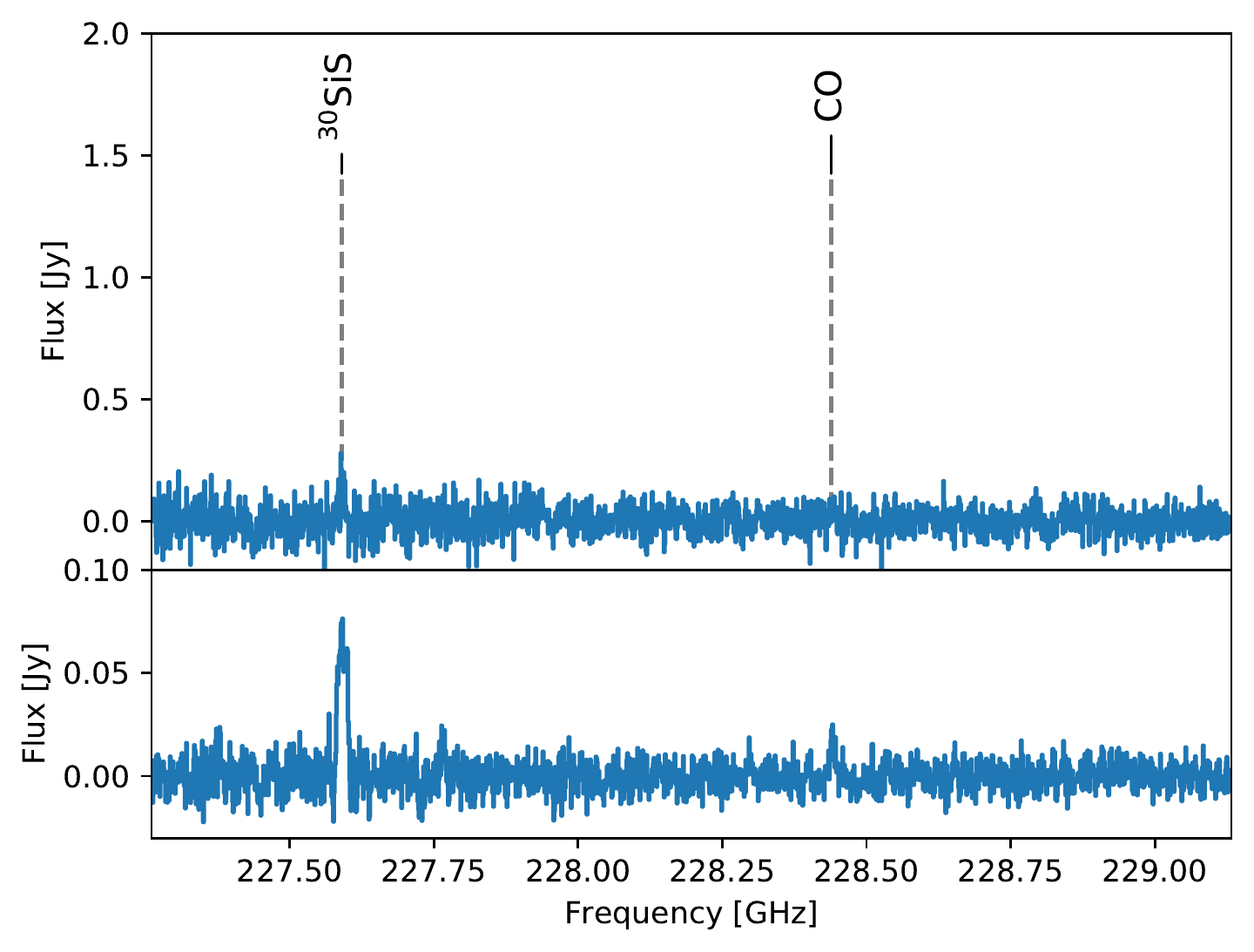}         
           \hfill
\centering\includegraphics[angle=0,width=.48\textwidth]{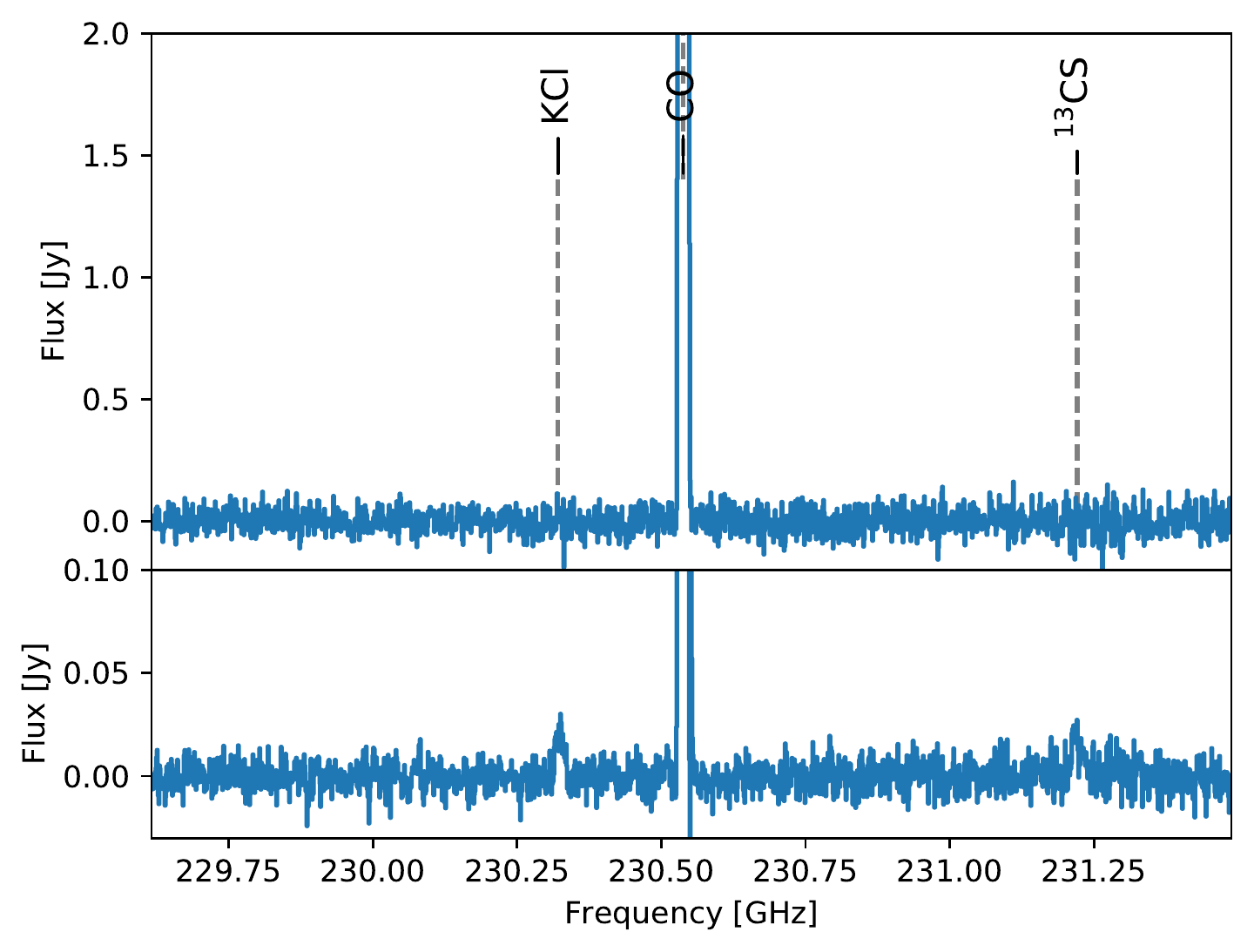}      
\centering\includegraphics[angle=0,width=.48\textwidth]{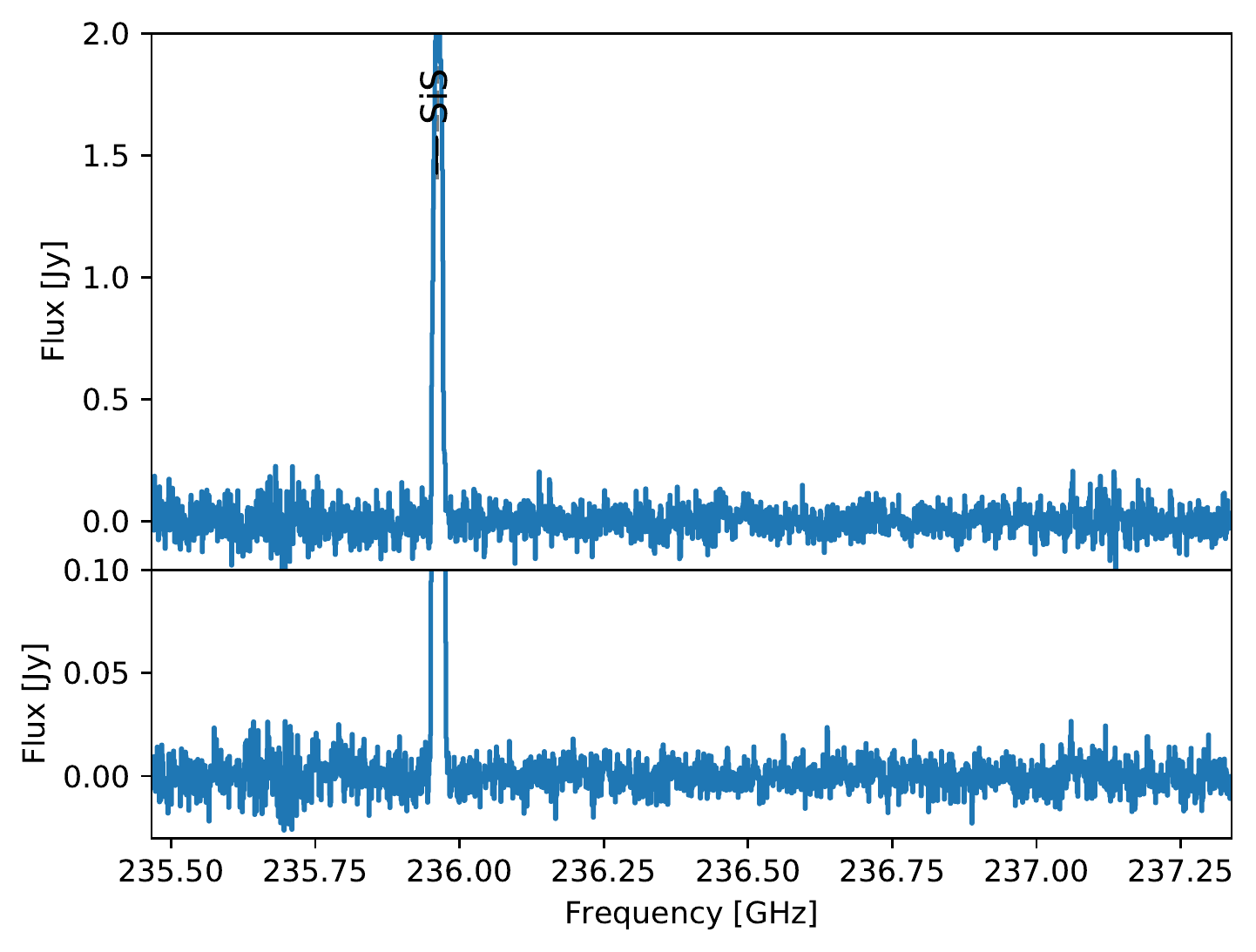}       
            \hfill
\centering\includegraphics[angle=0,width=.48\textwidth]{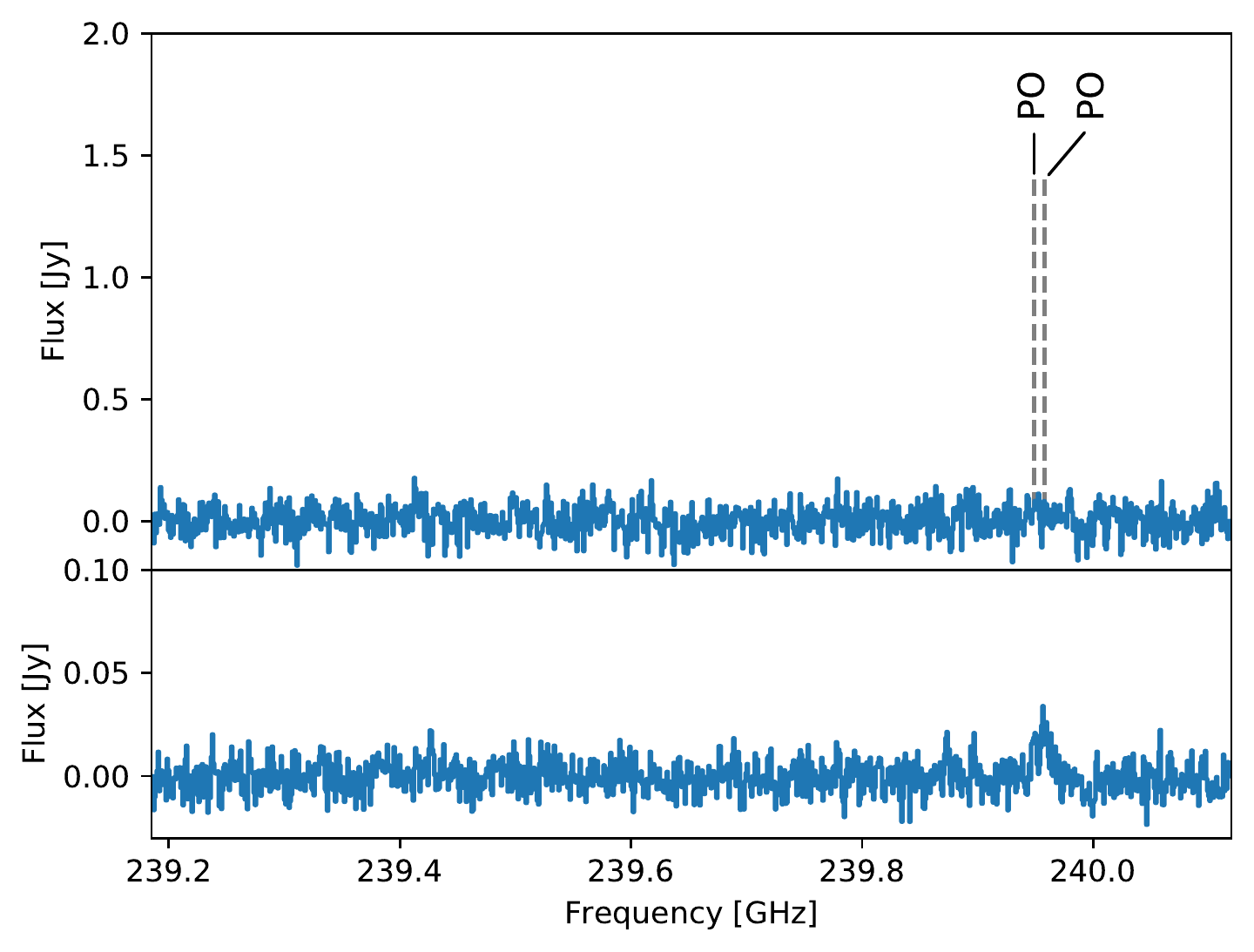}          
\centering\includegraphics[angle=0,width=.48\textwidth]{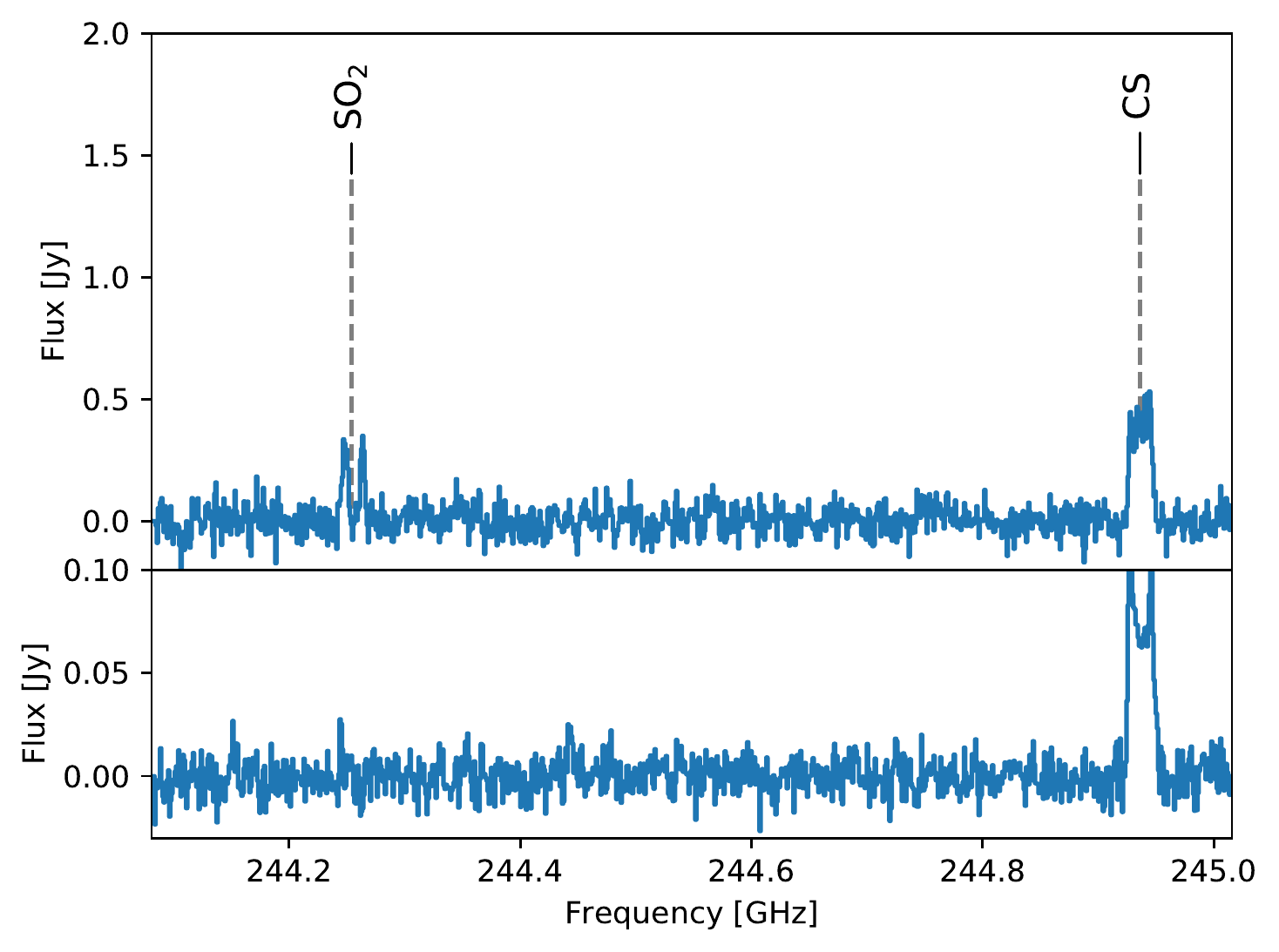}         
           \hfill
\centering\includegraphics[angle=0,width=.48\textwidth]{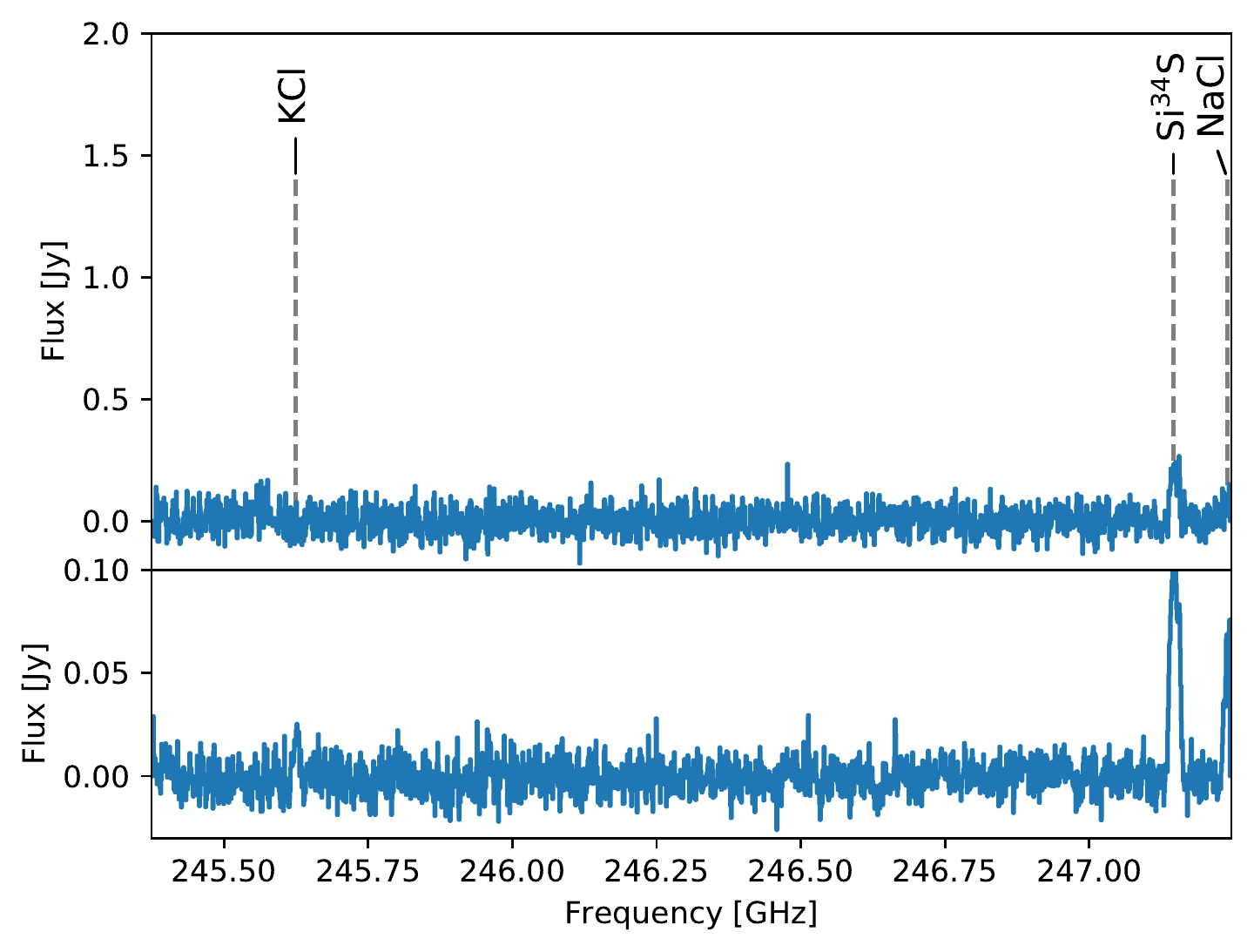}          
\caption{\textbf{ALMA spectra of IRC\,$-$10529.} 
Caption; see Fig.~\ref{Fig:spectra_IRC1}. Data are displayed for cubes 04 to 09.}
\label{Fig:spectra_IRC3} 
\end{figure*}

%.  ++++++++++++++++++++++++++++++++++++++++++++++++++++++++
\begin{figure*}[htbp]

\centering\includegraphics[angle=0,width=.48\textwidth]{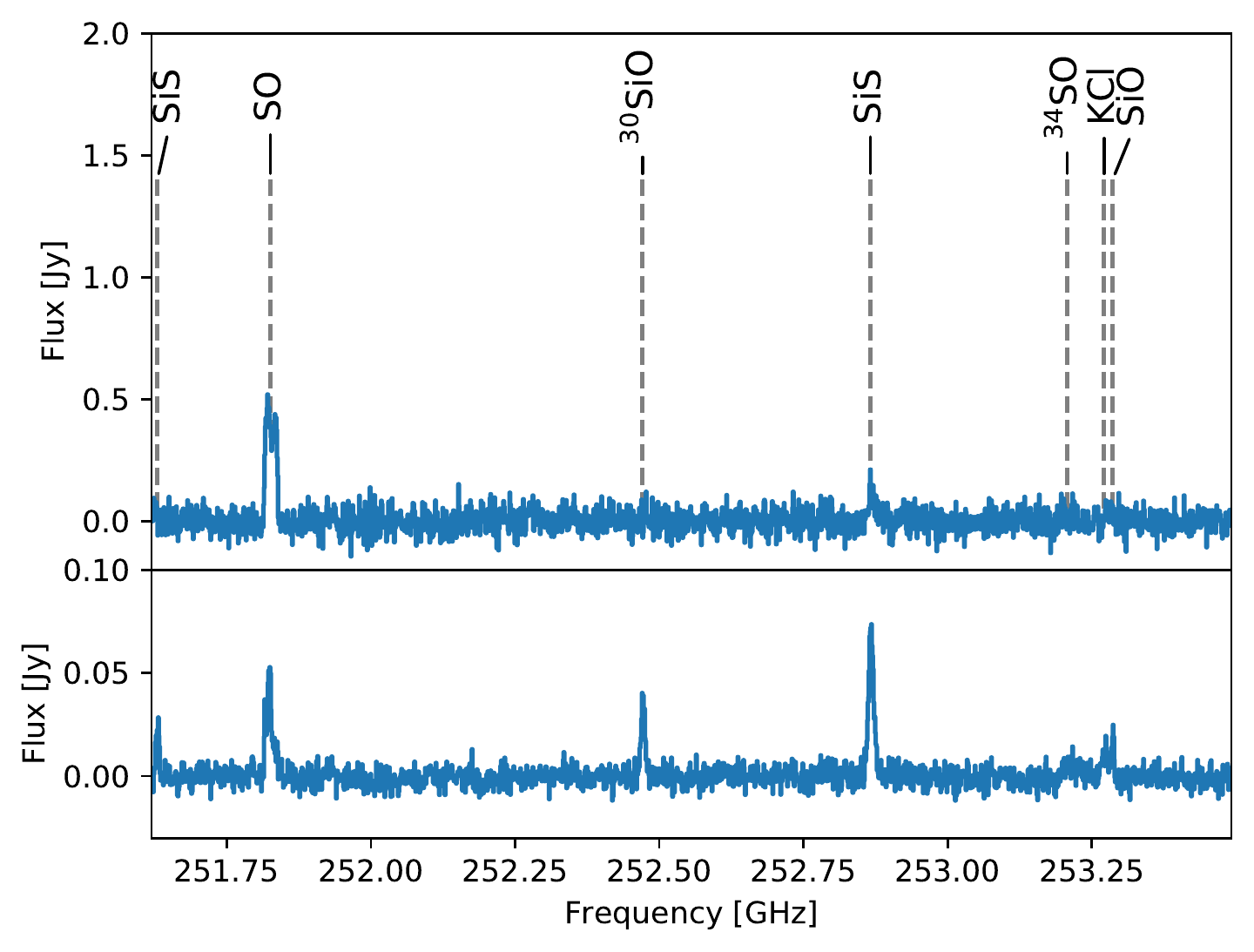}       
\hfill            
\centering\includegraphics[angle=0,width=.48\textwidth]{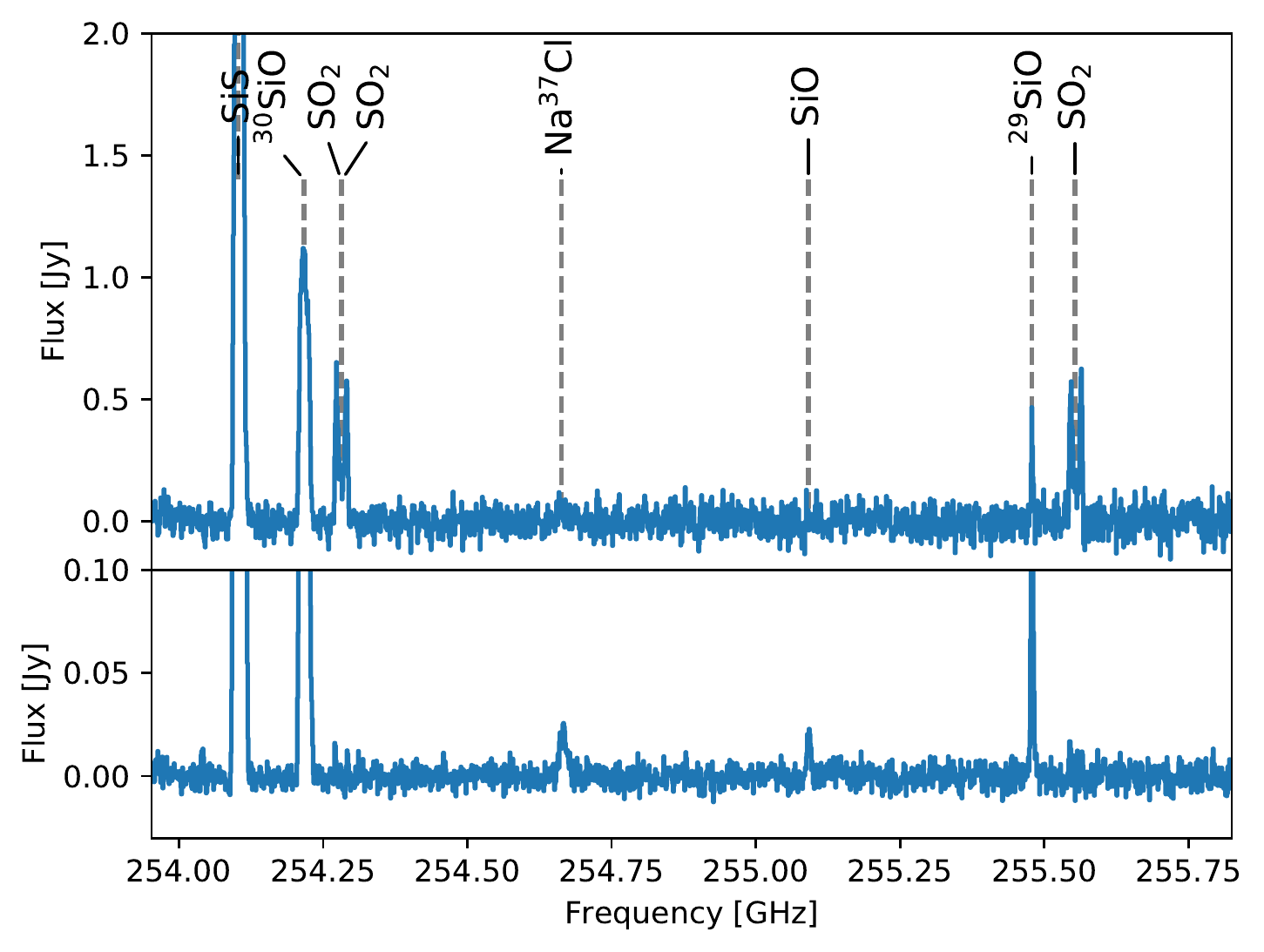}          
\centering\includegraphics[angle=0,width=.48\textwidth]{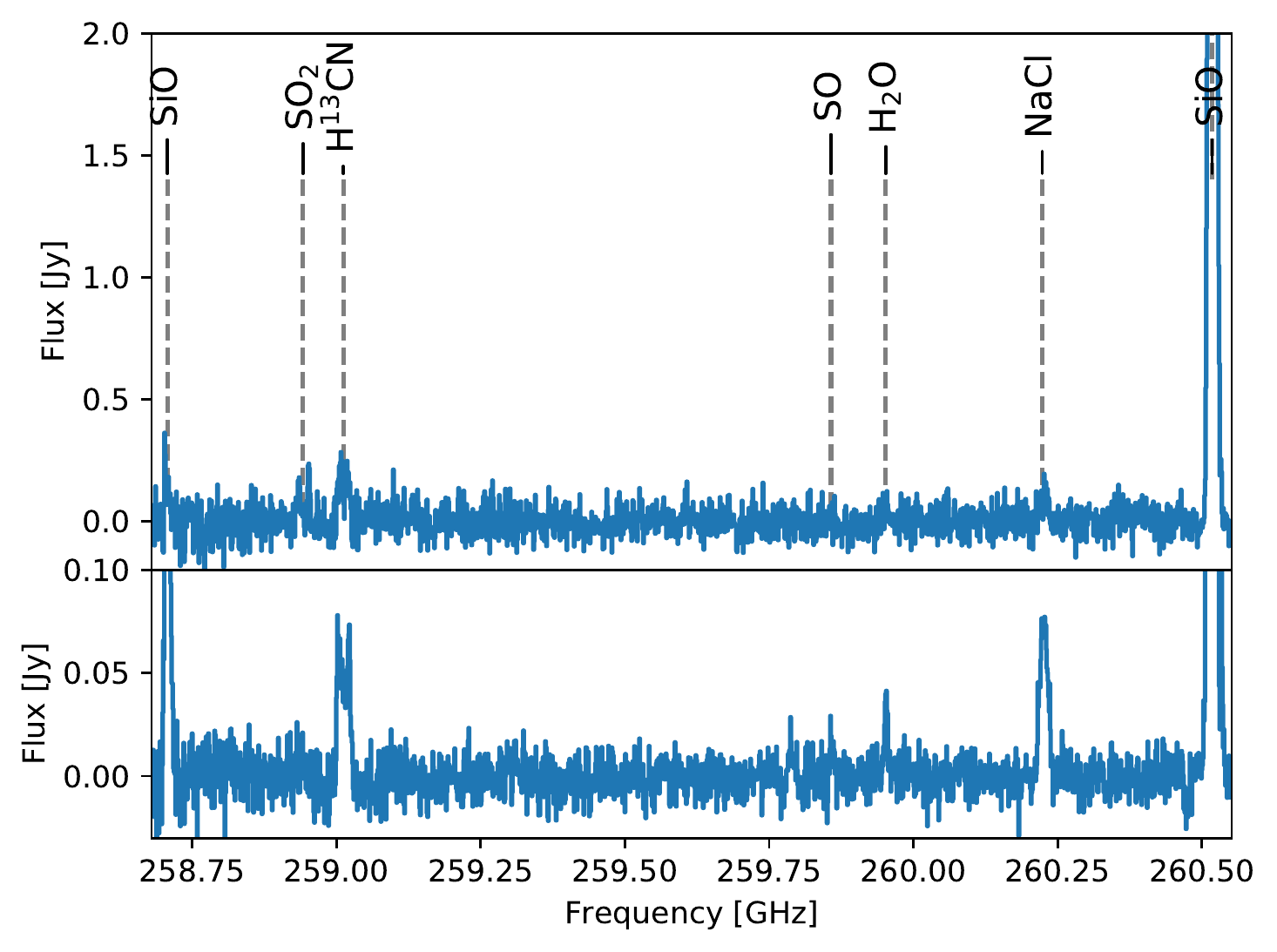}         
           \hfill
\centering\includegraphics[angle=0,width=.48\textwidth]{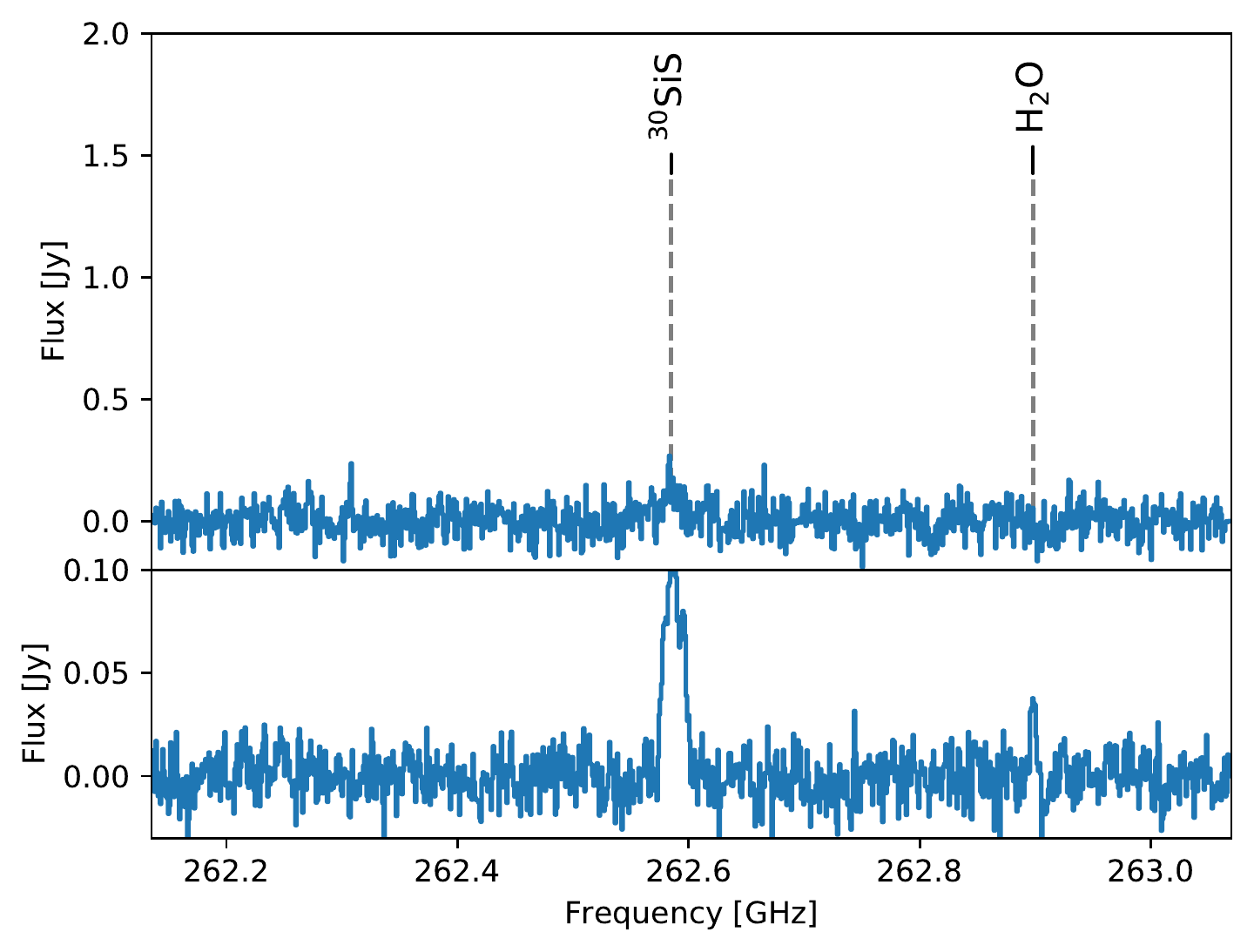}          
\centering\includegraphics[angle=0,width=.48\textwidth]{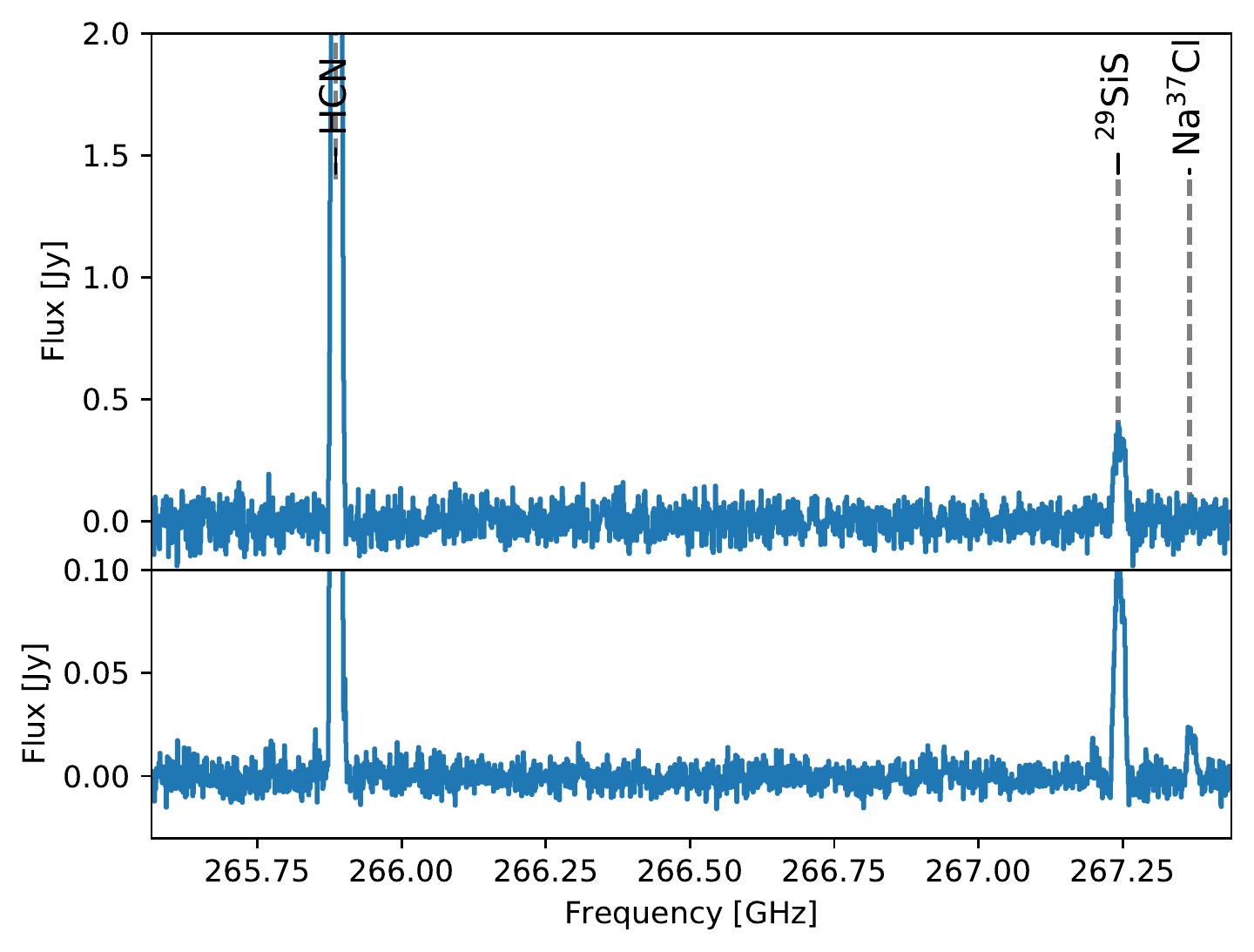}       
            \hfill
\centering\includegraphics[angle=0,width=.48\textwidth]{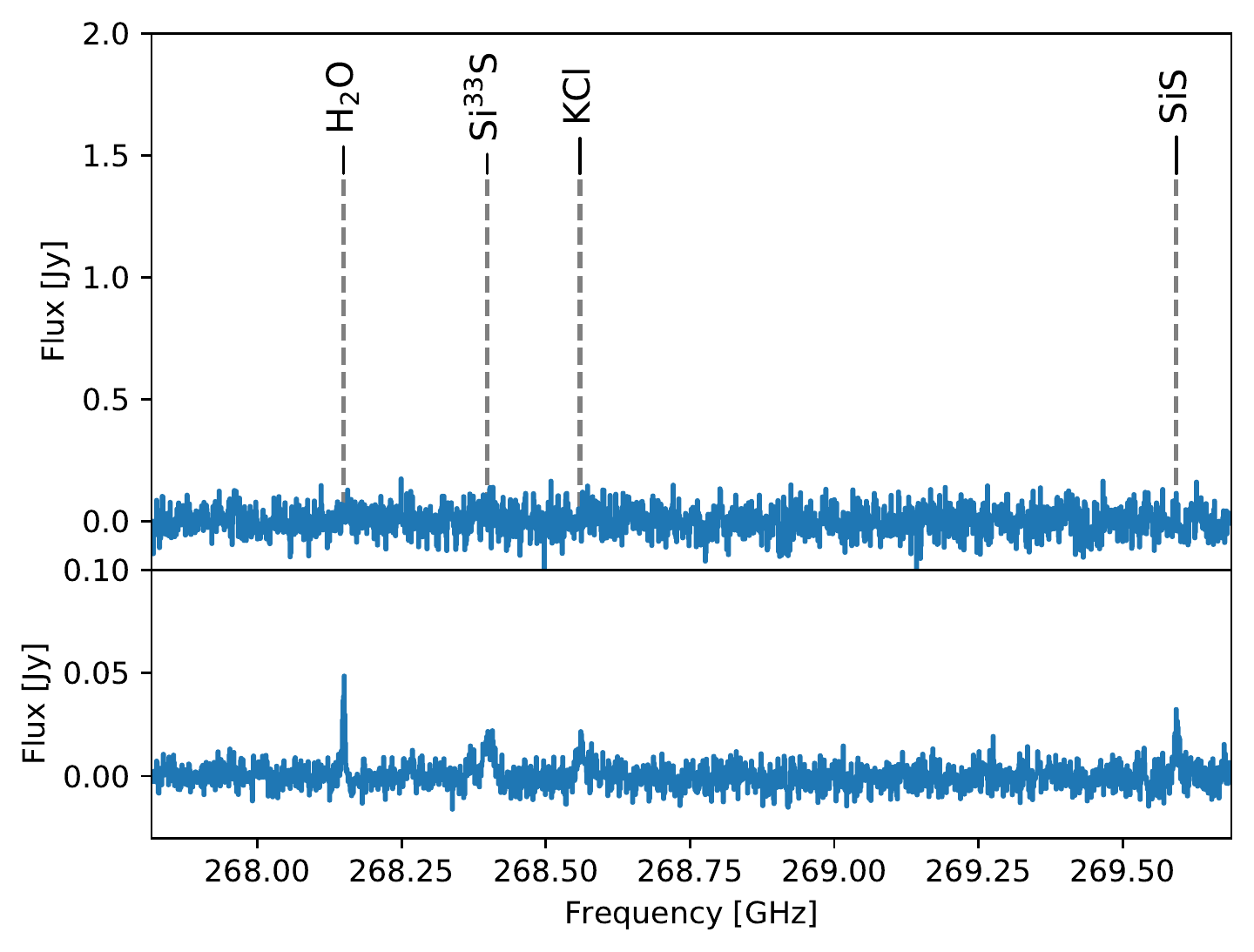}     
\caption{\textbf{ALMA spectra of IRC\,$-$10529.} 
Caption; see Fig.~\ref{Fig:spectra_IRC1}. Data are displayed for cubes 10 to 15.}
\label{Fig:spectra_IRC4} 
\end{figure*}

%  #############################################################################################
%  #######################################. APPENDIX C  #########################################
%  ############################################################################################# 
\afterpage{\clearpage}
\newpage
%\mbox{}

\section{Wind kinematics for 16 {\sc atomium} sources}\label{Sect:kinematics_other_sources}

%  +++++++++++++++++++++++++++++++++++++++++++++++++++++++++++++++++++++++++++++++++++++++++++++++
\mbox{}

\begin{figure*}[htbp]
%\begin{figure*}[!htpb]
\begin{minipage}[t]{.495\textwidth}
        \centerline{\resizebox{\textwidth}{6truecm}{\includegraphics[angle=0]{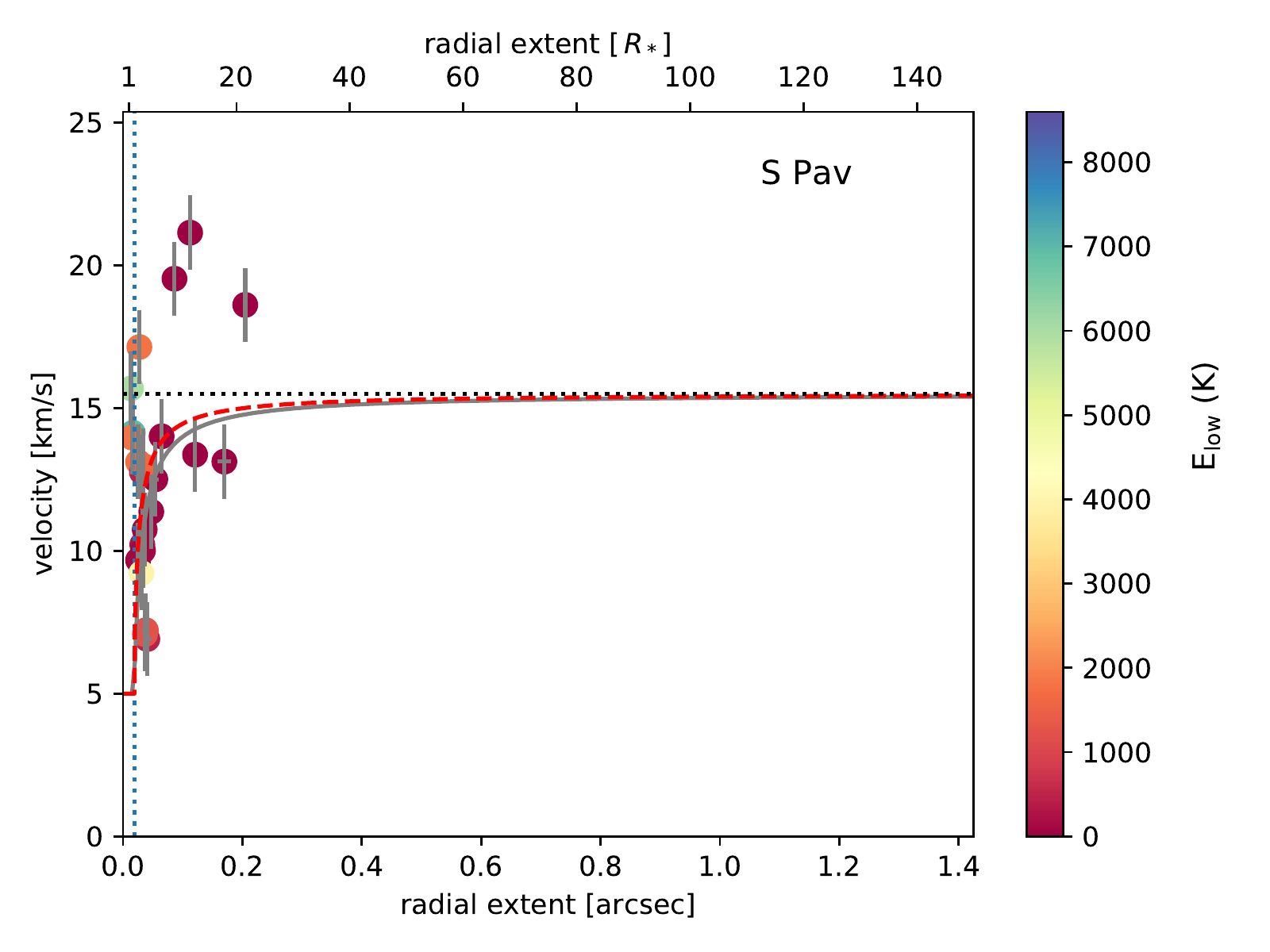}}}
\end{minipage}
    \hfill
\begin{minipage}[t]{.495\textwidth}
        \centerline{\resizebox{\textwidth}{6truecm}{\includegraphics[angle=0]{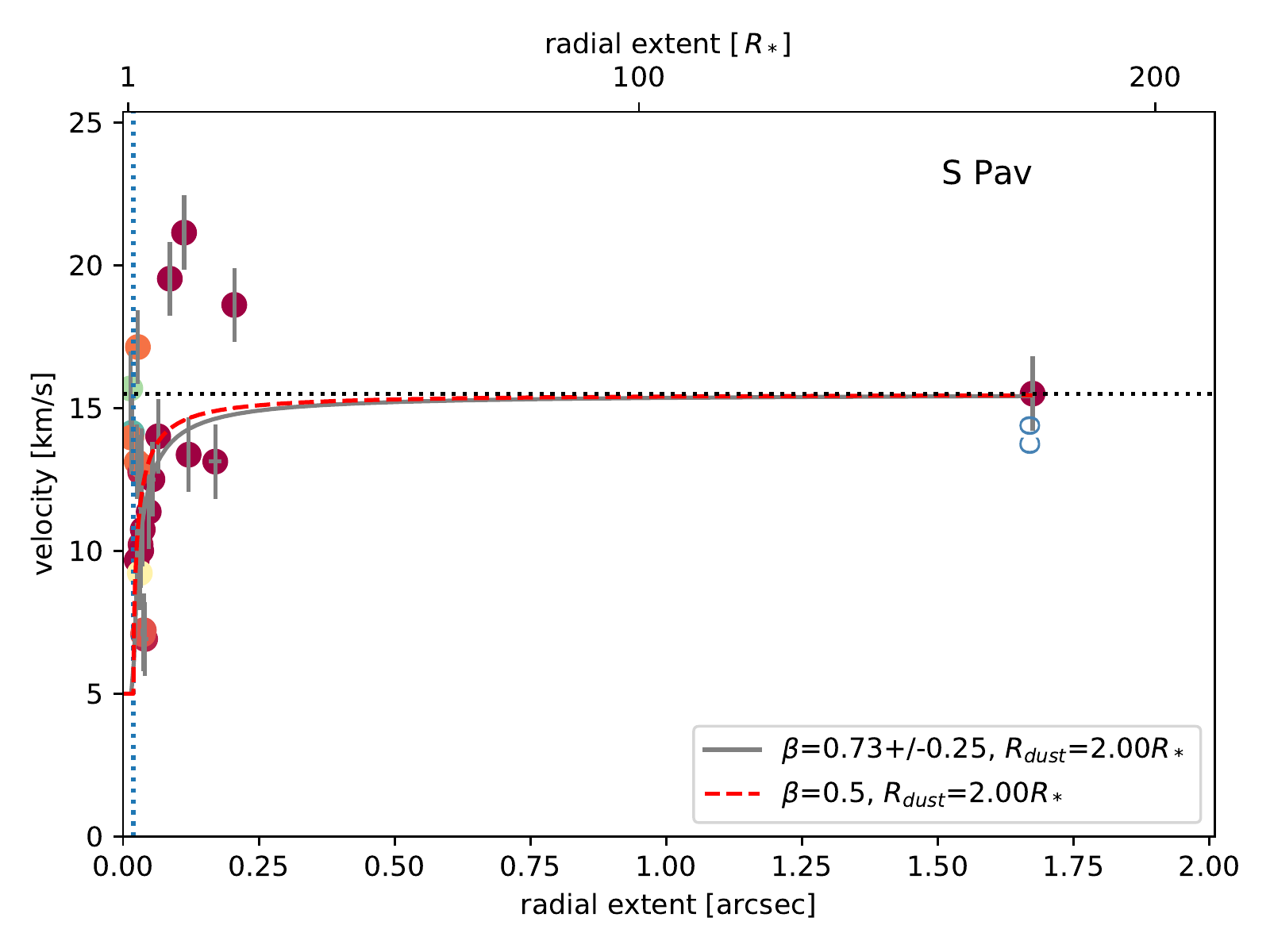}}}
\end{minipage}
\caption{\textbf{Wind kinematics for S~Pav.} See Fig.~\ref{Fig:IRC10529_kinematics} caption.}
\label{Fig:S_Pav_kinematics}
\end{figure*}

\begin{figure*}[htpb]
\begin{minipage}[t]{.495\textwidth}
        \centerline{\resizebox{\textwidth}{6truecm}{\includegraphics[angle=0]{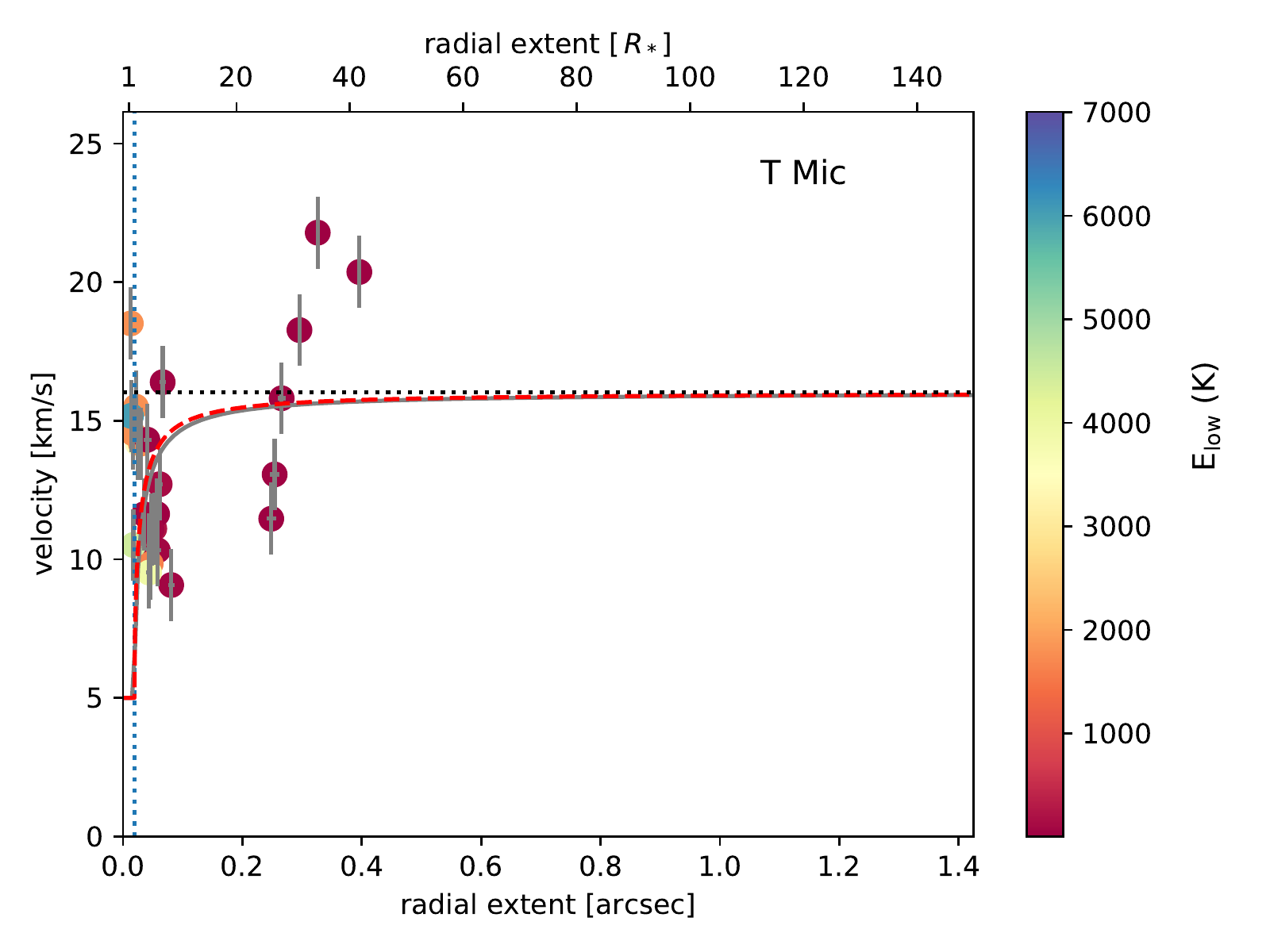}}}
\end{minipage}
    \hfill
\begin{minipage}[t]{.495\textwidth}
        \centerline{\resizebox{\textwidth}{6truecm}{\includegraphics[angle=0]{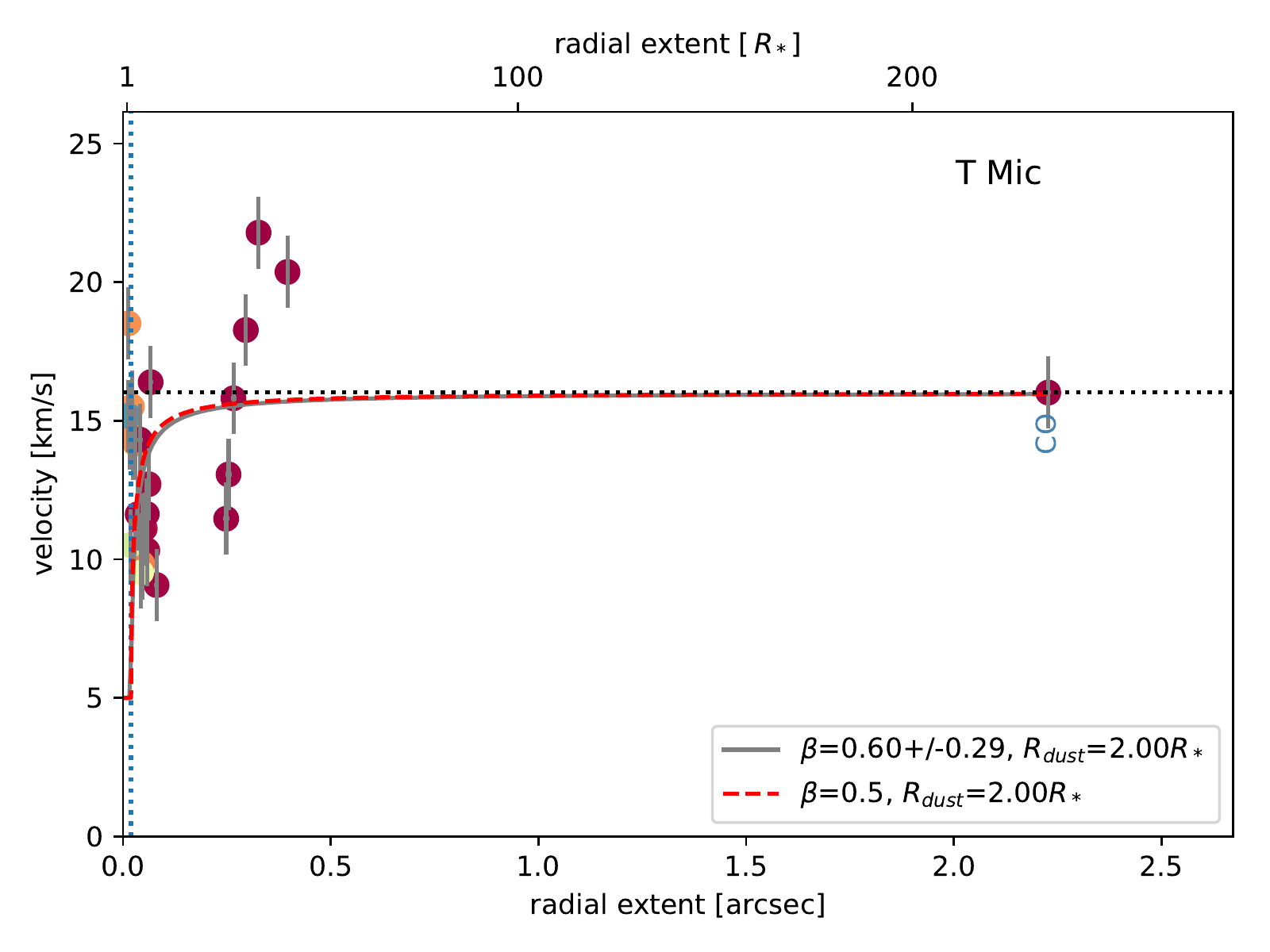}}}
\end{minipage}
\caption{\textbf{Wind kinematics for T~Mic.} See Fig.~\ref{Fig:IRC10529_kinematics}  caption.}
\label{Fig:T_Mic_kinematics}
\end{figure*}

\begin{figure*}[htpb]
\begin{minipage}[t]{.495\textwidth}
        \centerline{\resizebox{\textwidth}{6truecm}{\includegraphics[angle=0]{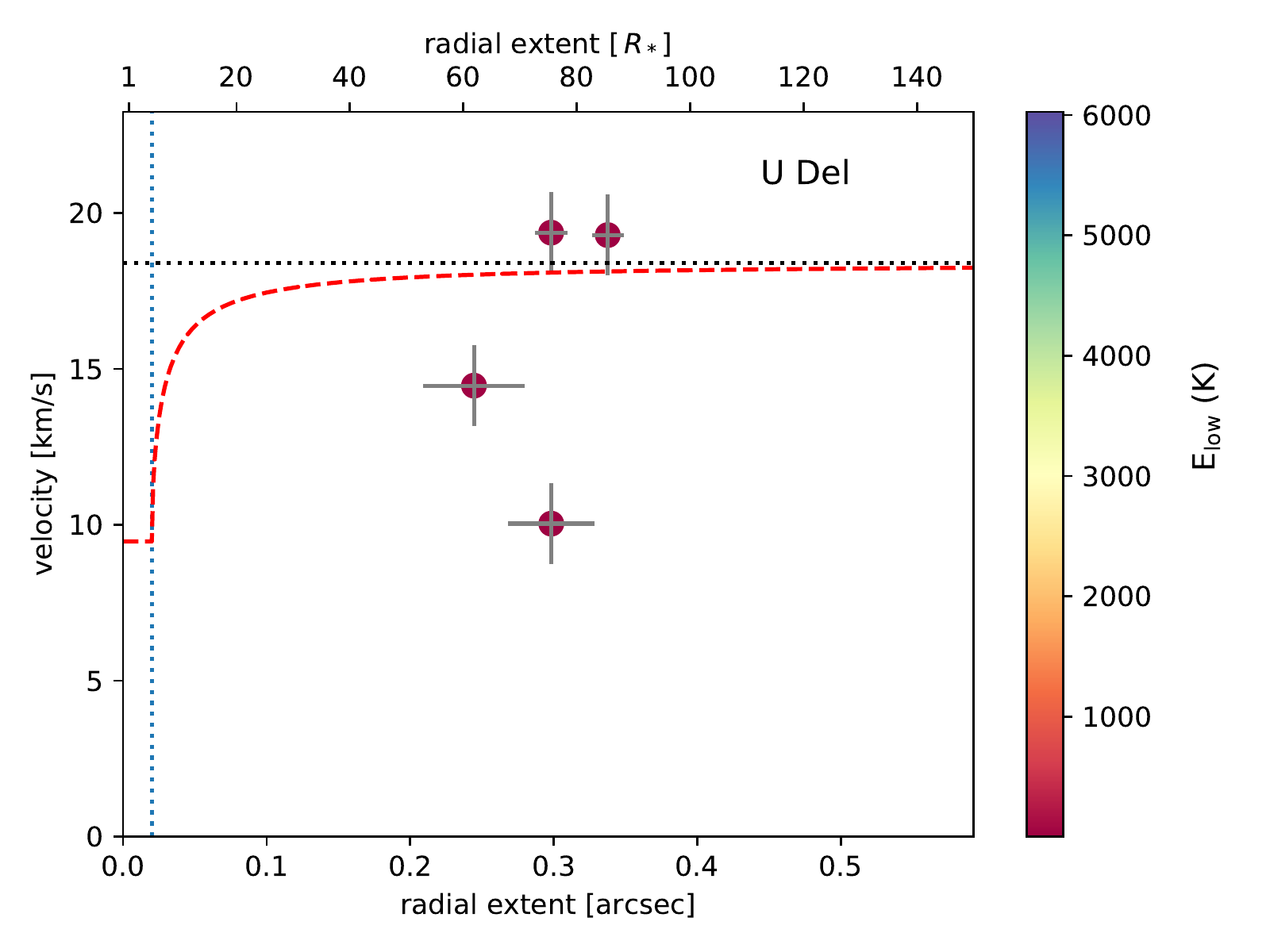}}}
\end{minipage}
    \hfill
\begin{minipage}[t]{.495\textwidth}
        \centerline{\resizebox{\textwidth}{6truecm}{\includegraphics[angle=0]{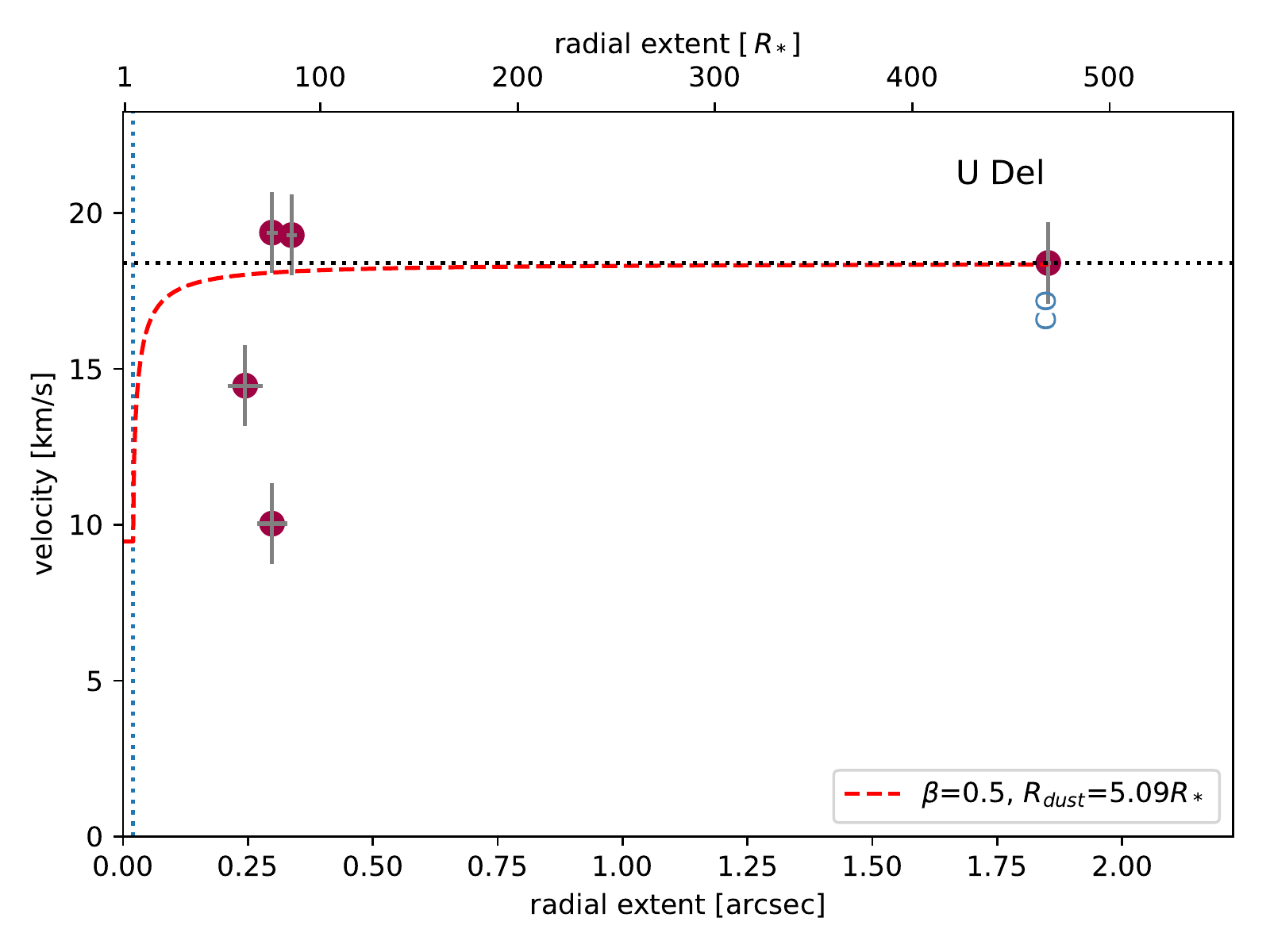}}}
\end{minipage}
\caption{\textbf{Wind kinematics for U~Del.} See Fig.~\ref{Fig:IRC10529_kinematics}  caption.
Not enough data are available for a reliable determination of the $\beta$ parameter.}
\label{Fig:U_Del_kinematics}
\end{figure*}

\begin{figure*}[htpb]
\begin{minipage}[t]{.495\textwidth}
        \centerline{\resizebox{\textwidth}{!}{\includegraphics[angle=0]{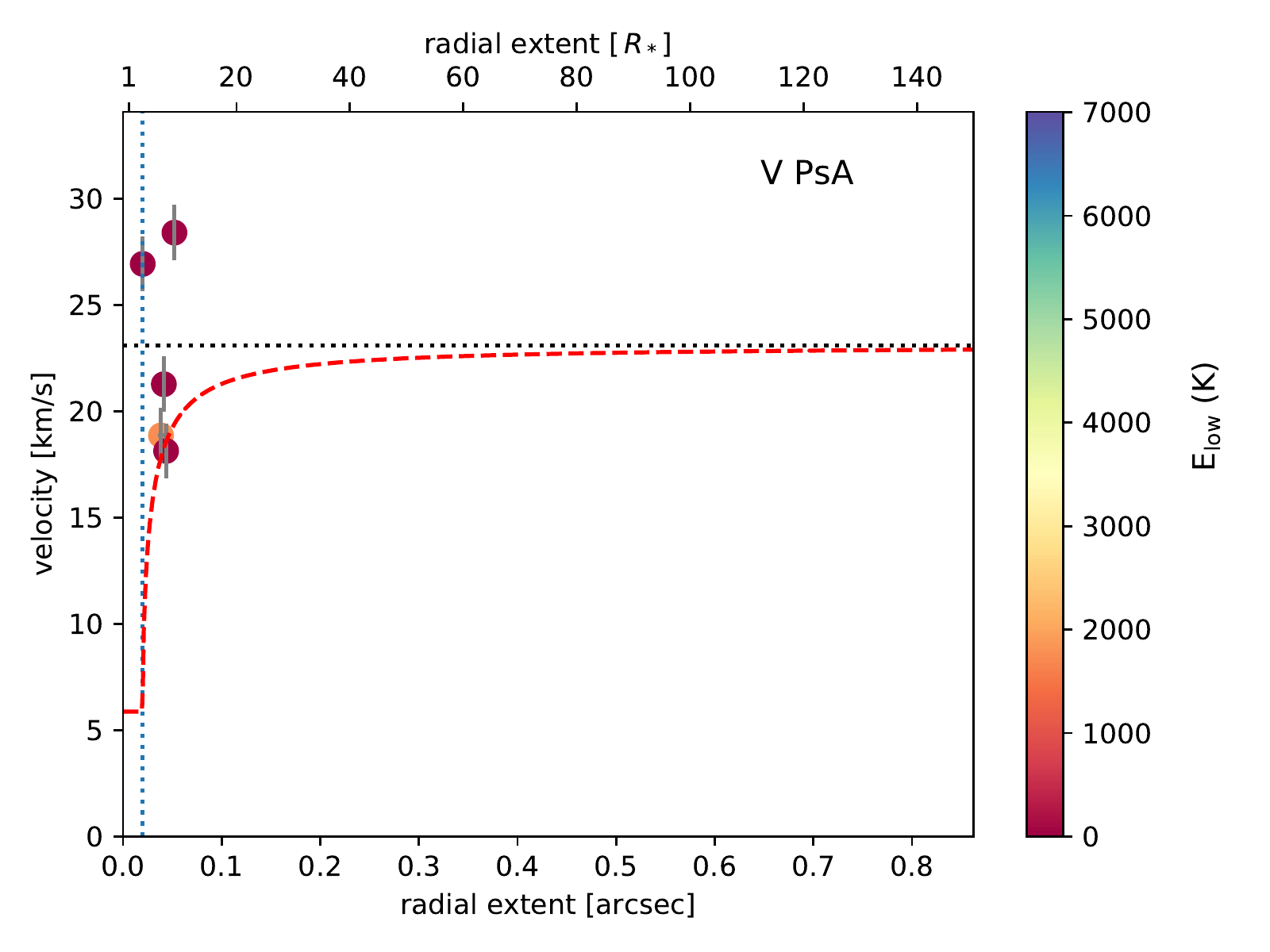}}}
\end{minipage}
    \hfill
\begin{minipage}[t]{.495\textwidth}
        \centerline{\resizebox{\textwidth}{!}{\includegraphics[angle=0]{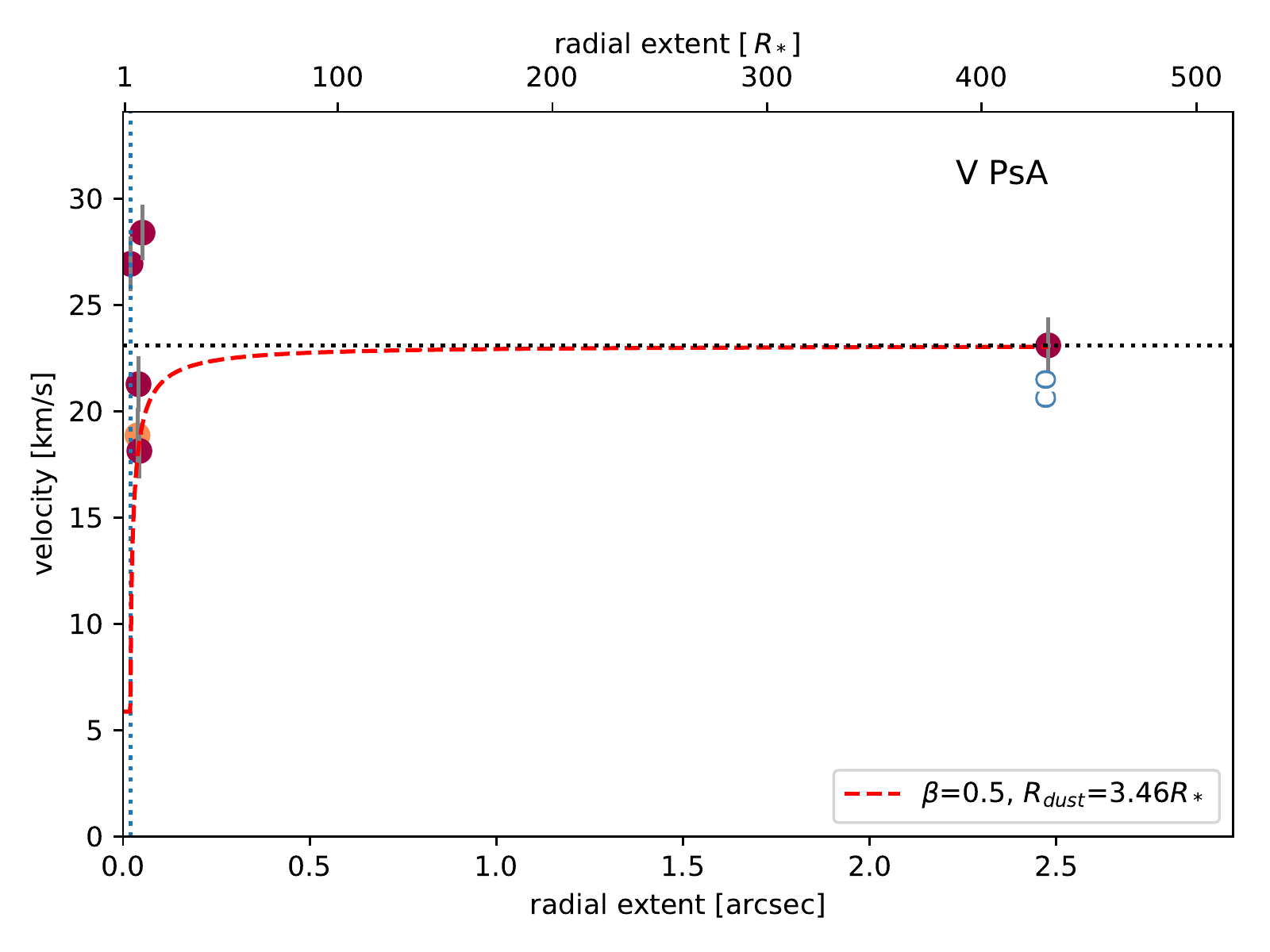}}}
\end{minipage}
\caption{\textbf{Wind kinematics for V~PsA.} See Fig.~\ref{Fig:IRC10529_kinematics}  caption.
Not enough data are available for a reliable determination of the $\beta$ parameter.}
\label{Fig:V_PsA_kinematics}
\end{figure*}

\begin{figure*}[htpb]
\begin{minipage}[t]{.495\textwidth}
        \centerline{\resizebox{\textwidth}{!}{\includegraphics[angle=0]{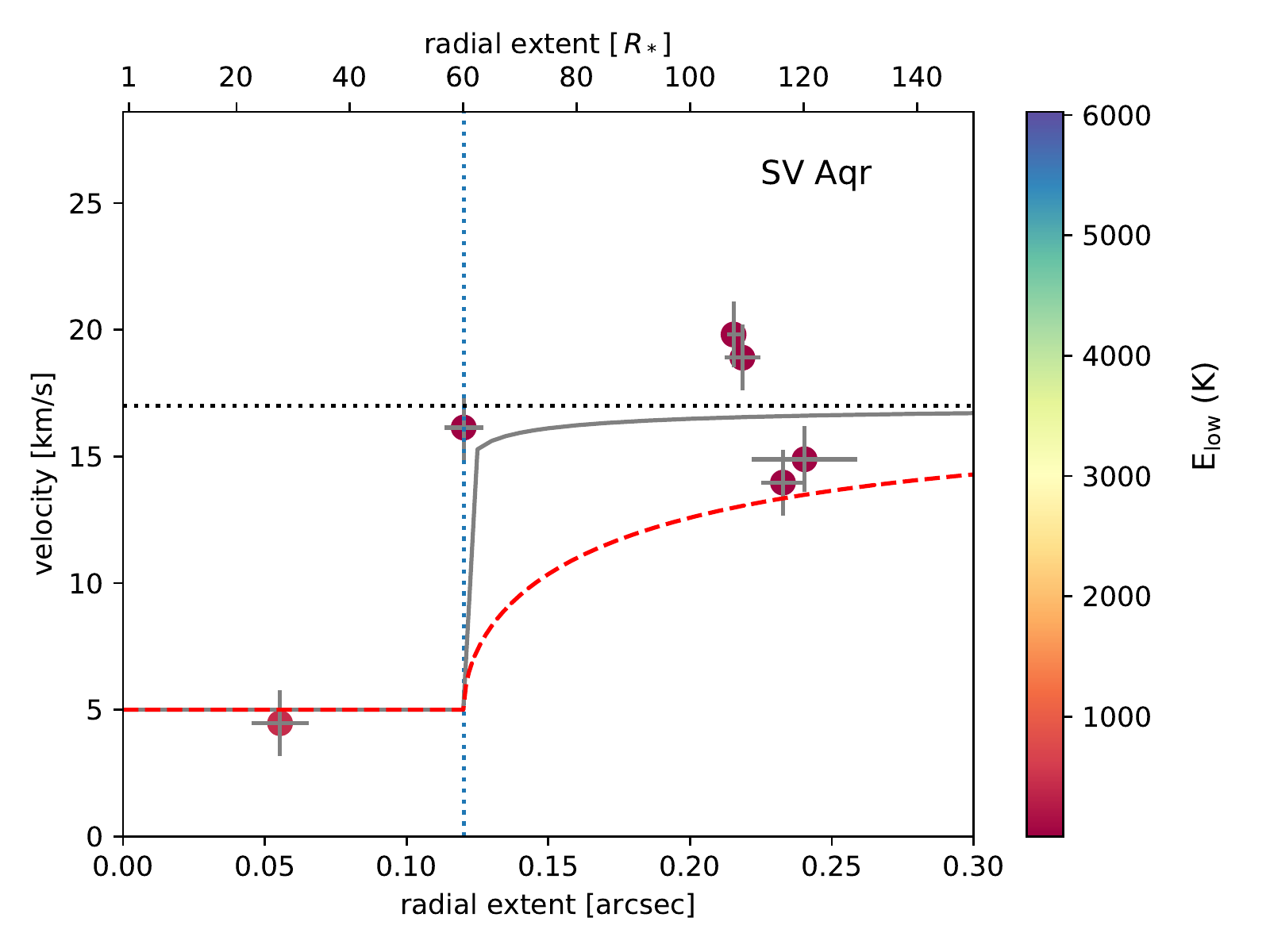}}}
\end{minipage}
    \hfill
\begin{minipage}[t]{.495\textwidth}
        \centerline{\resizebox{\textwidth}{!}{\includegraphics[angle=0]{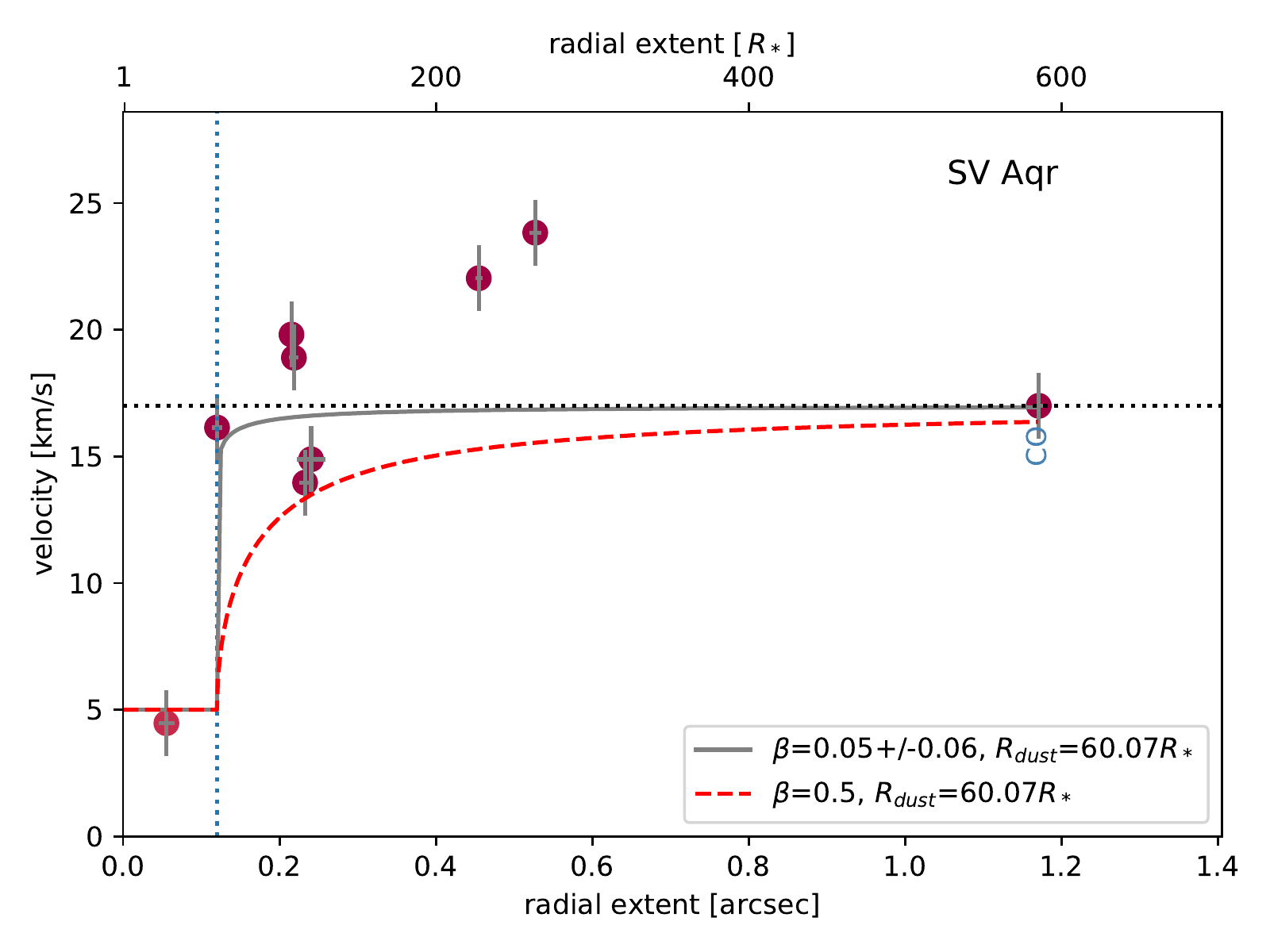}}}
\end{minipage}
\caption{\textbf{Wind kinematics for SV~Aqr.} See Fig.~\ref{Fig:IRC10529_kinematics}  caption. Not enough data are available for a 
reliable determination of the $\beta$ parameter.}
\label{Fig:SV_Aqr_kinematics}
\end{figure*}

\begin{figure*}[htpb]
\begin{minipage}[t]{.495\textwidth}
        \centerline{\resizebox{\textwidth}{!}{\includegraphics[angle=0]{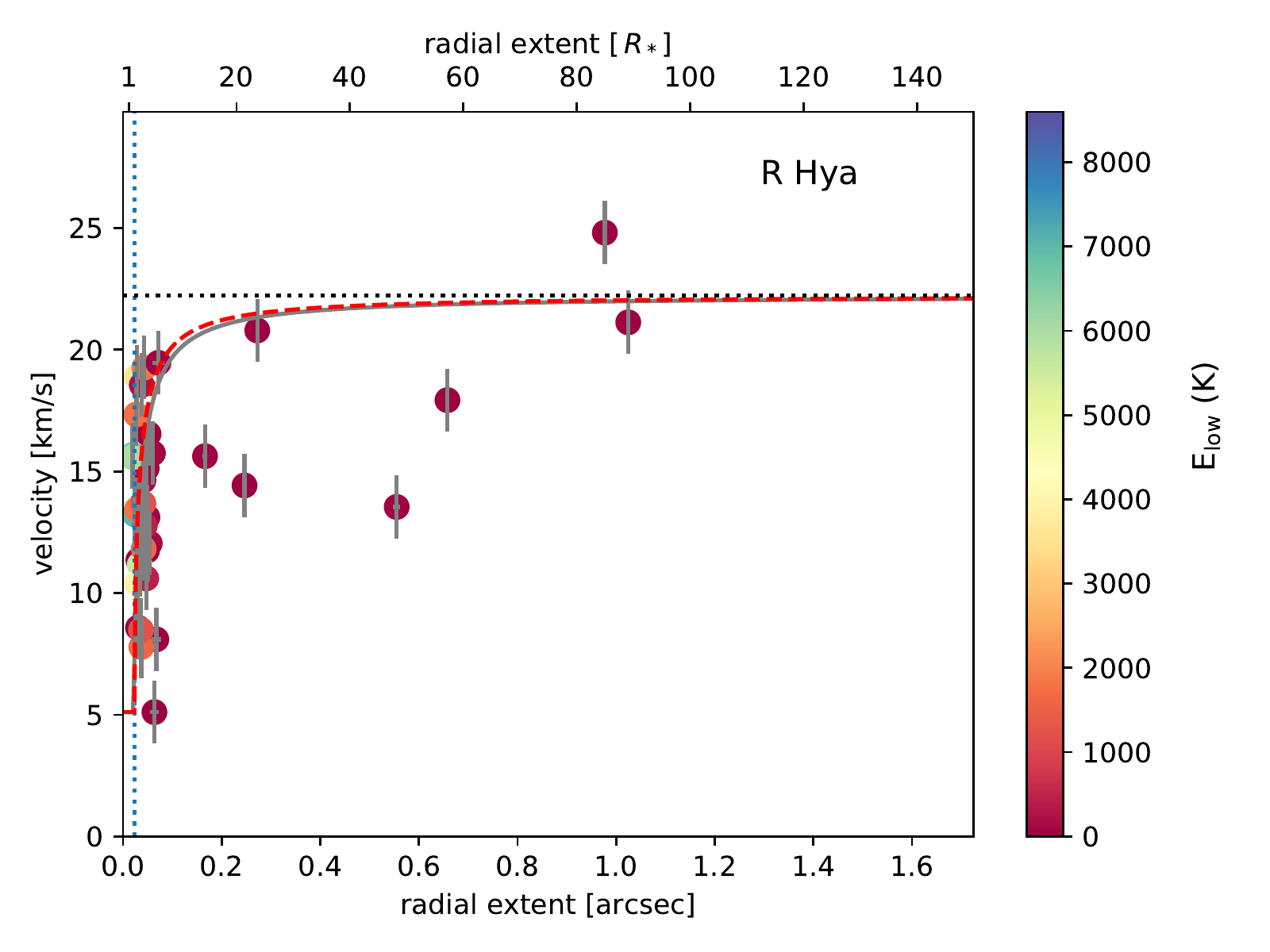}}}
\end{minipage}
    \hfill
\begin{minipage}[t]{.495\textwidth}
        \centerline{\resizebox{\textwidth}{!}{\includegraphics[angle=0]{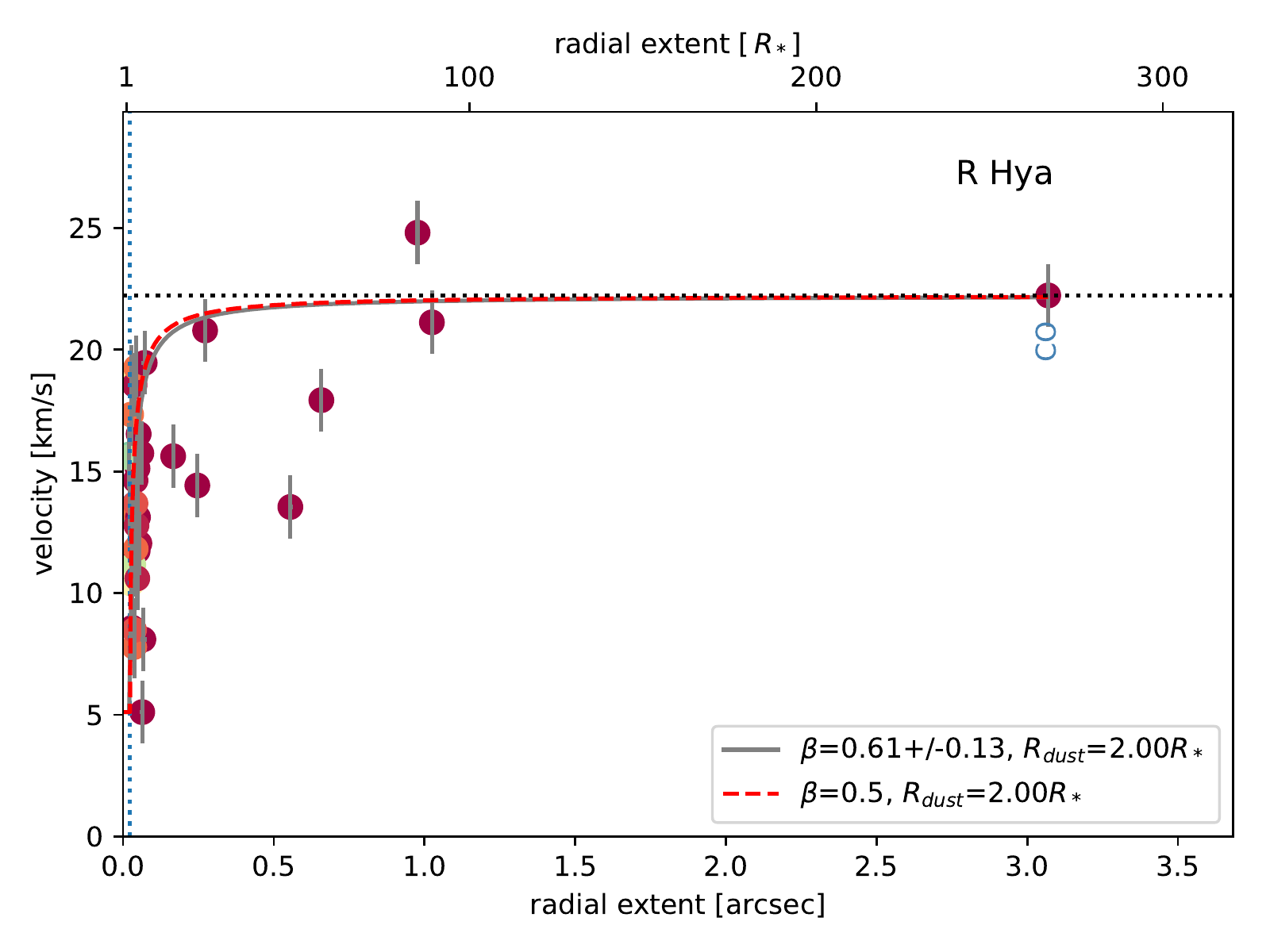}}}
\end{minipage}
\caption{\textbf{Wind kinematics for R~Hya.} See Fig.~\ref{Fig:IRC10529_kinematics}  caption.}
\label{Fig:R_Hya_kinematics}
\end{figure*}

\begin{figure*}[htpb]
\vspace{1ex}
\begin{minipage}[t]{.495\textwidth}
        \centerline{\resizebox{\textwidth}{!}{\includegraphics[angle=0]{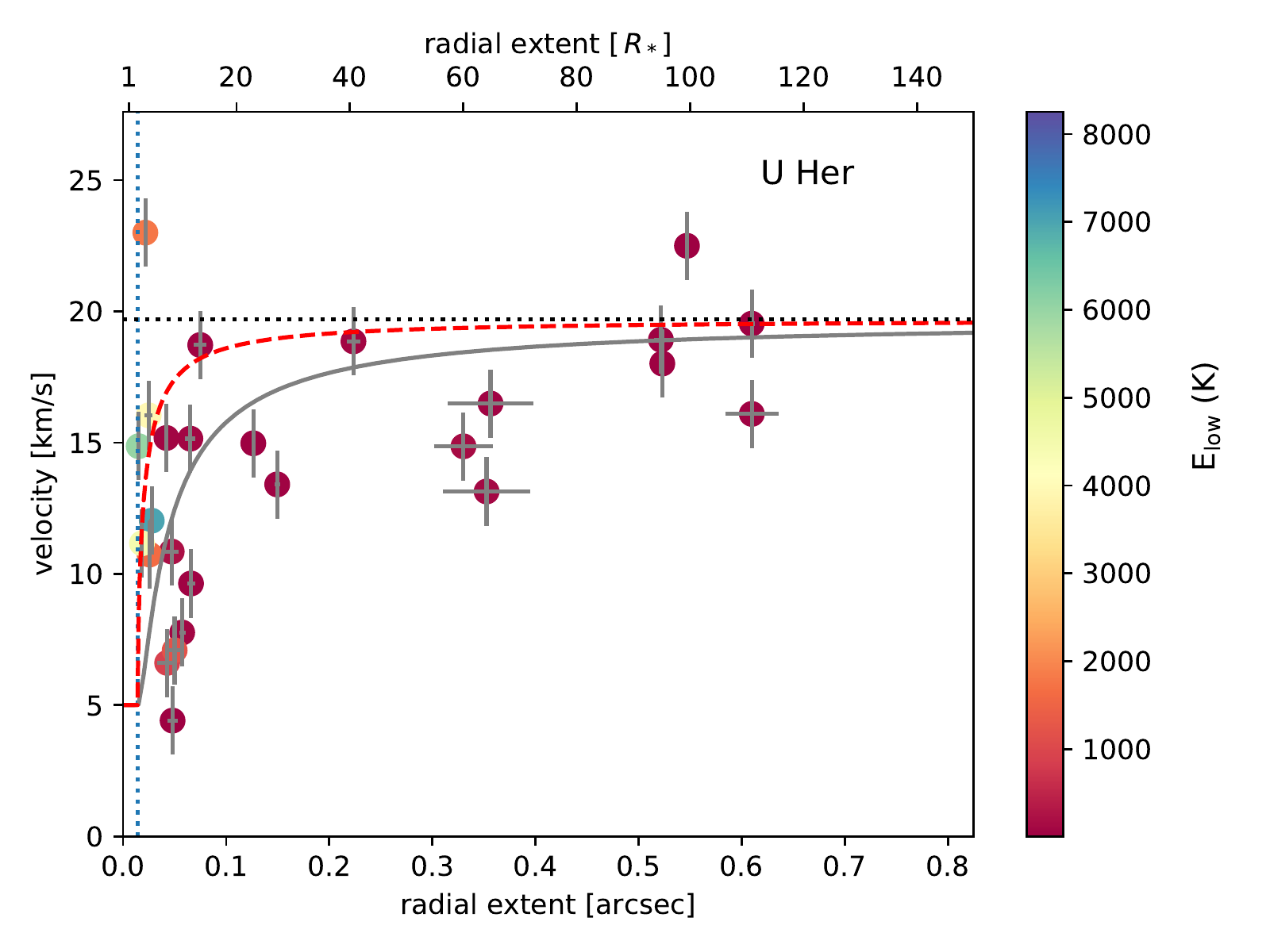}}}
\end{minipage}
    \hfill
\begin{minipage}[t]{.495\textwidth}
        \centerline{\resizebox{\textwidth}{!}{\includegraphics[angle=0]{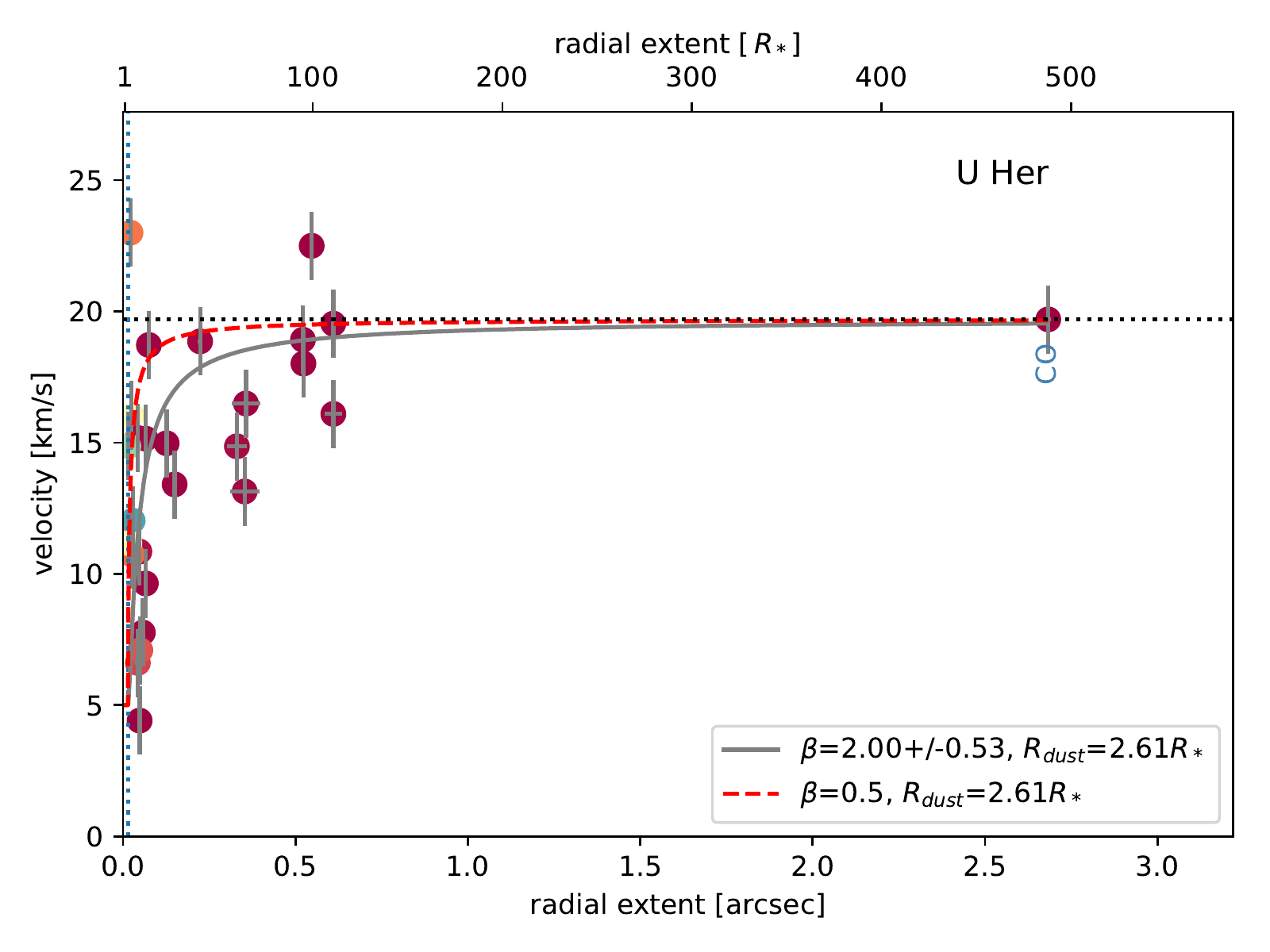}}}
\end{minipage}
\caption{\textbf{Wind kinematics for U~Her.} See Fig.~\ref{Fig:IRC10529_kinematics}  caption.}
\label{Fig:U_Her_kinematics}
\end{figure*}

\begin{figure*}[htpb]
\vspace{1ex}
\begin{minipage}[t]{.495\textwidth}
        \centerline{\resizebox{\textwidth}{!}{\includegraphics[angle=0]{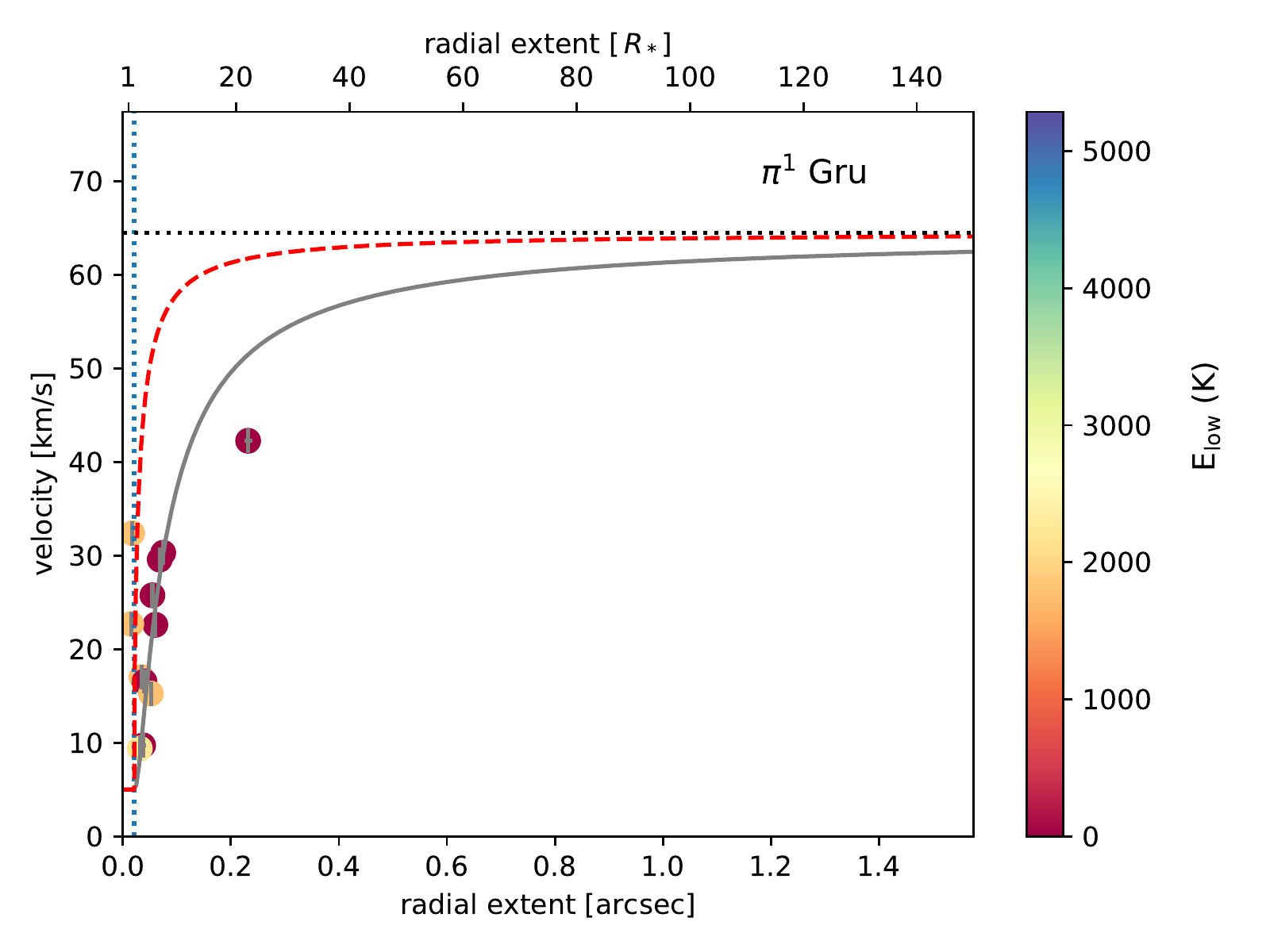}}}
\end{minipage}
    \hfill
\begin{minipage}[t]{.495\textwidth}
        \centerline{\resizebox{\textwidth}{!}{\includegraphics[angle=0]{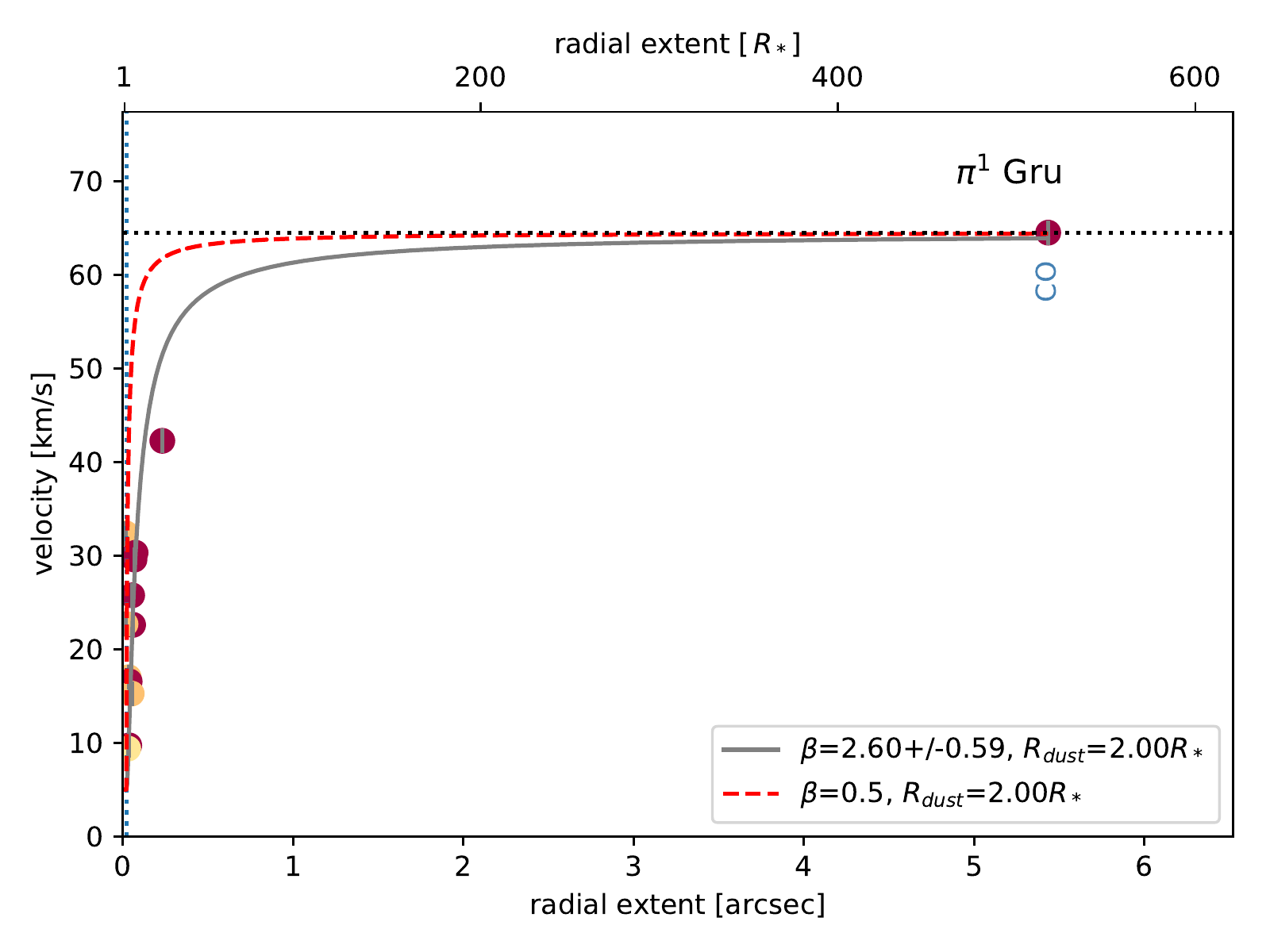}}}
\end{minipage}
\caption{\textbf{Wind kinematics for $\pi^1$~Gru.} See Fig.~\ref{Fig:IRC10529_kinematics}  caption.}
\label{Fig:pi1_Gru_kinematics}
\end{figure*}

\begin{figure*}[htpb]
\vspace{1ex}
\begin{minipage}[t]{.495\textwidth}
        \centerline{\resizebox{\textwidth}{!}{\includegraphics[angle=0]{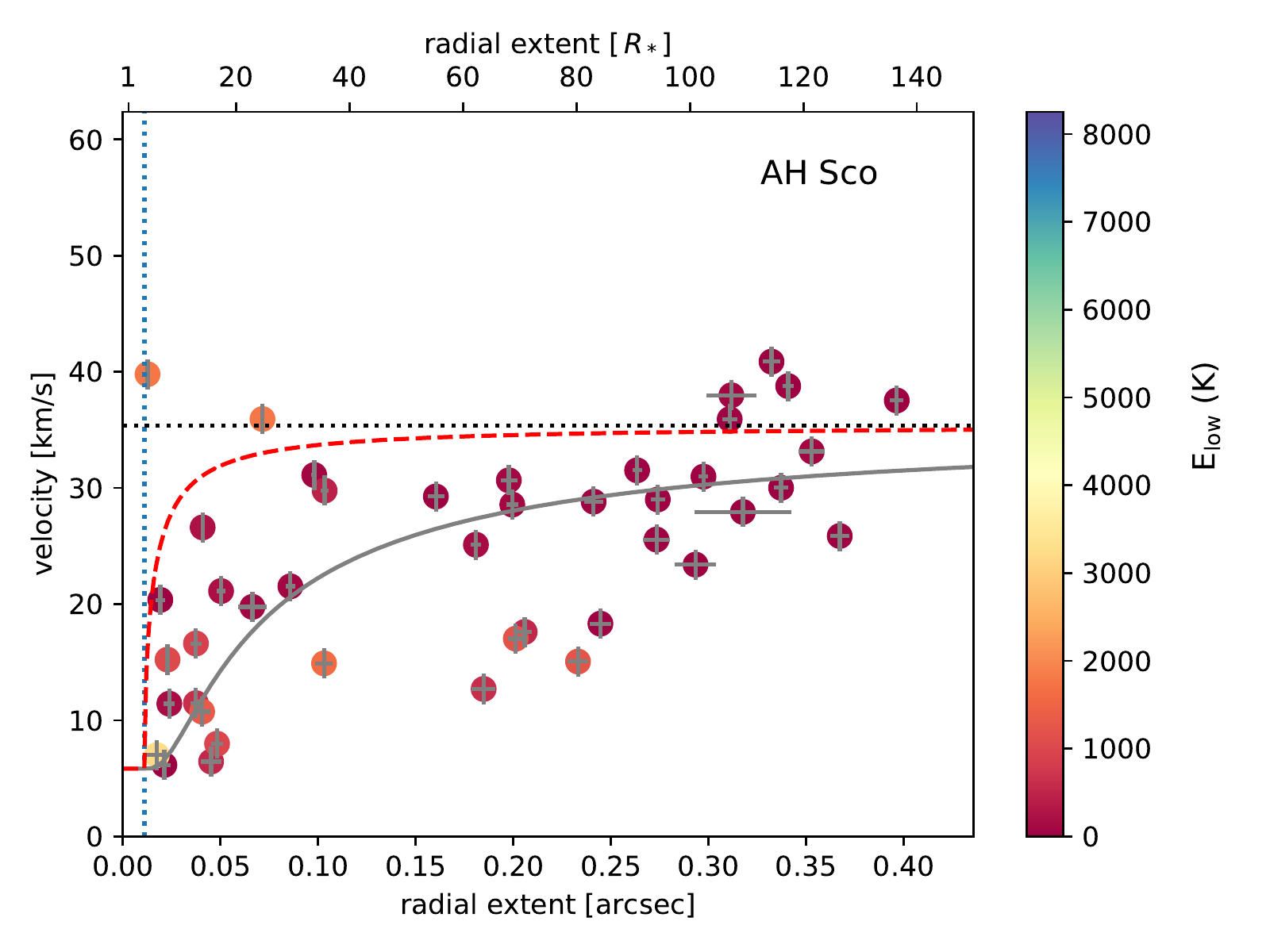}}}
\end{minipage}
    \hfill
\begin{minipage}[t]{.495\textwidth}
        \centerline{\resizebox{\textwidth}{!}{\includegraphics[angle=0]{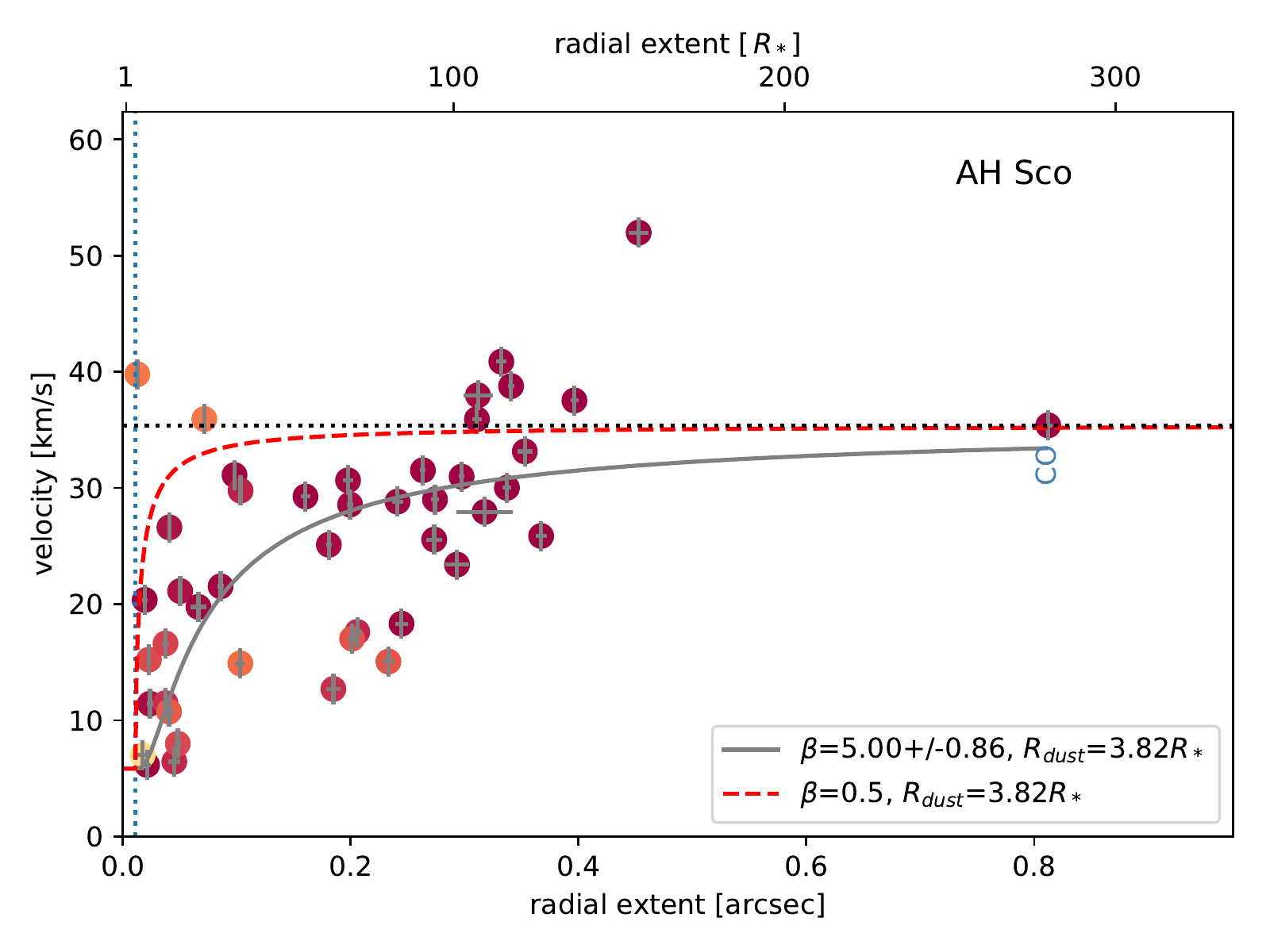}}}
\end{minipage}
\caption{\textbf{Wind kinematics for AH~Sco.} See Fig.~\ref{Fig:IRC10529_kinematics}  caption. Wind velocity profile constructed 
only on the basis of the medium and high spatial resolution data, since the low spatial resolution data still need to be acquired.}
\label{Fig:AH_Sco_kinematics}
\end{figure*}

\begin{figure*}[htpb]
\vspace{1ex}
\begin{minipage}[t]{.495\textwidth}
        \centerline{\resizebox{\textwidth}{!}{\includegraphics[angle=0]{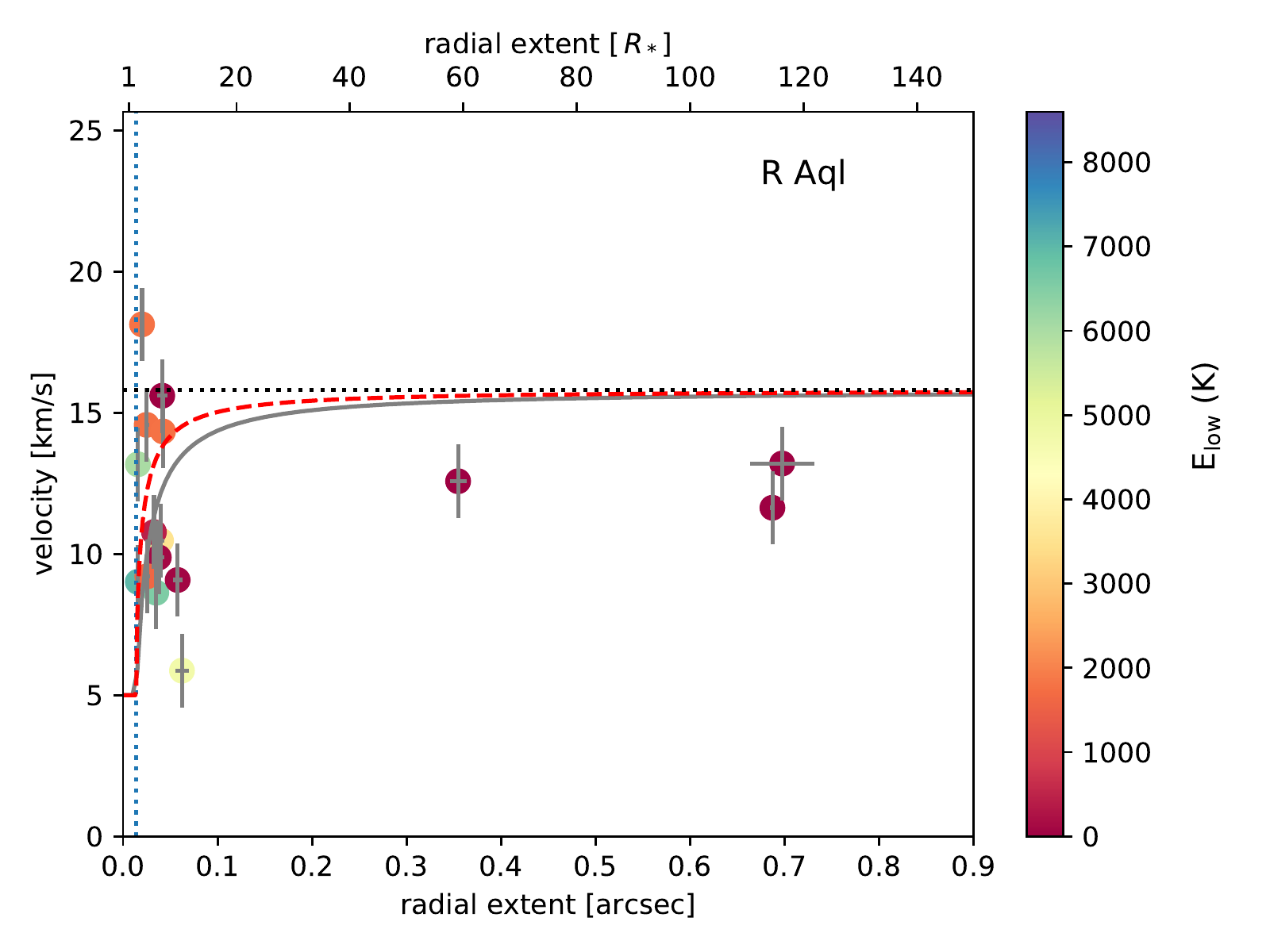}}}
\end{minipage}
    \hfill
\begin{minipage}[t]{.495\textwidth}
        \centerline{\resizebox{\textwidth}{!}{\includegraphics[angle=0]{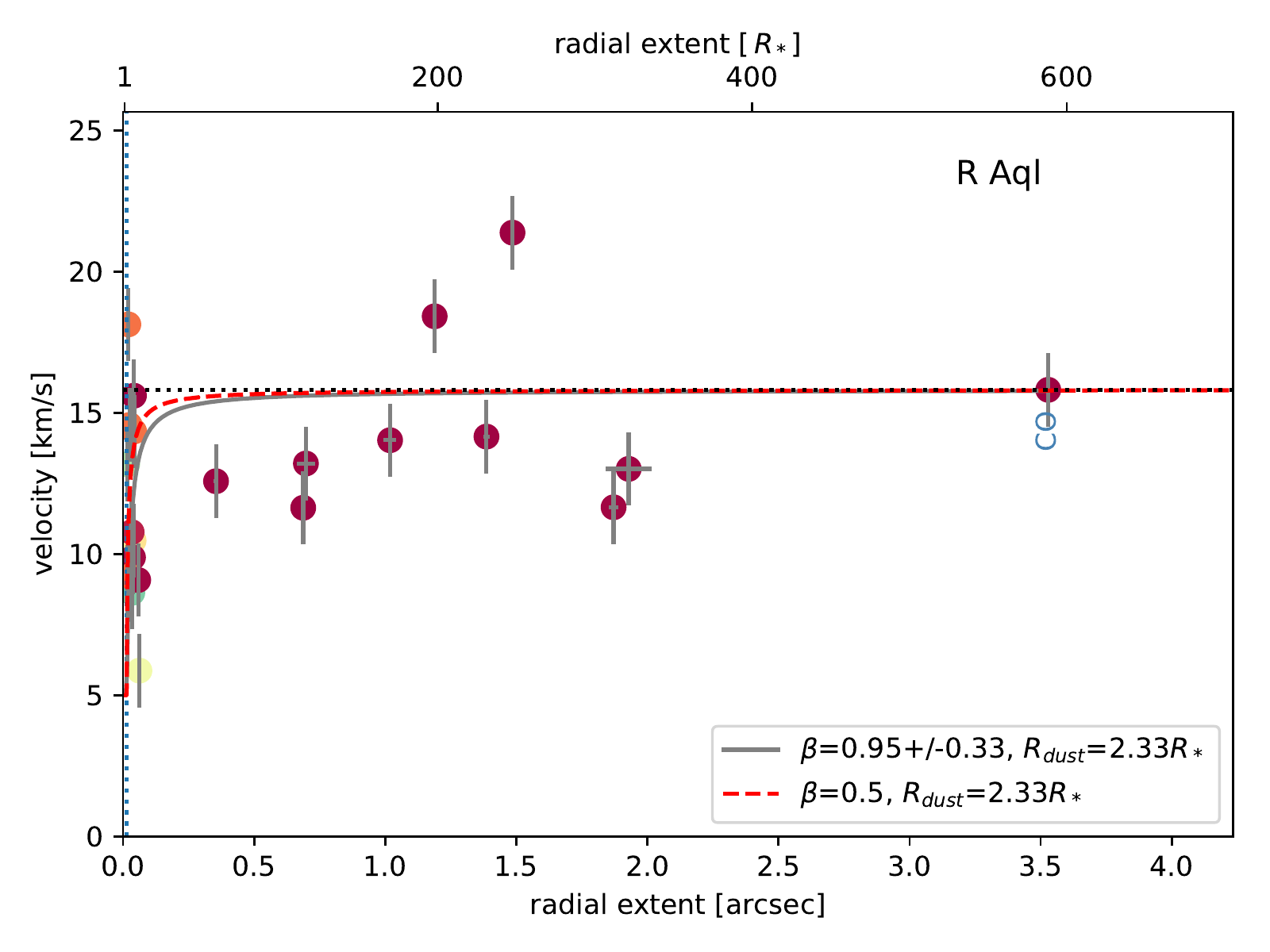}}}
\end{minipage}
\caption{\textbf{Wind kinematics for R~Aql.} See Fig.~\ref{Fig:IRC10529_kinematics}  caption.}
\label{Fig:R_Aql_kinematics}
\end{figure*}

\begin{figure*}[htpb]
\vspace{1ex}
\begin{minipage}[t]{.495\textwidth}
        \centerline{\resizebox{\textwidth}{!}{\includegraphics[angle=0]{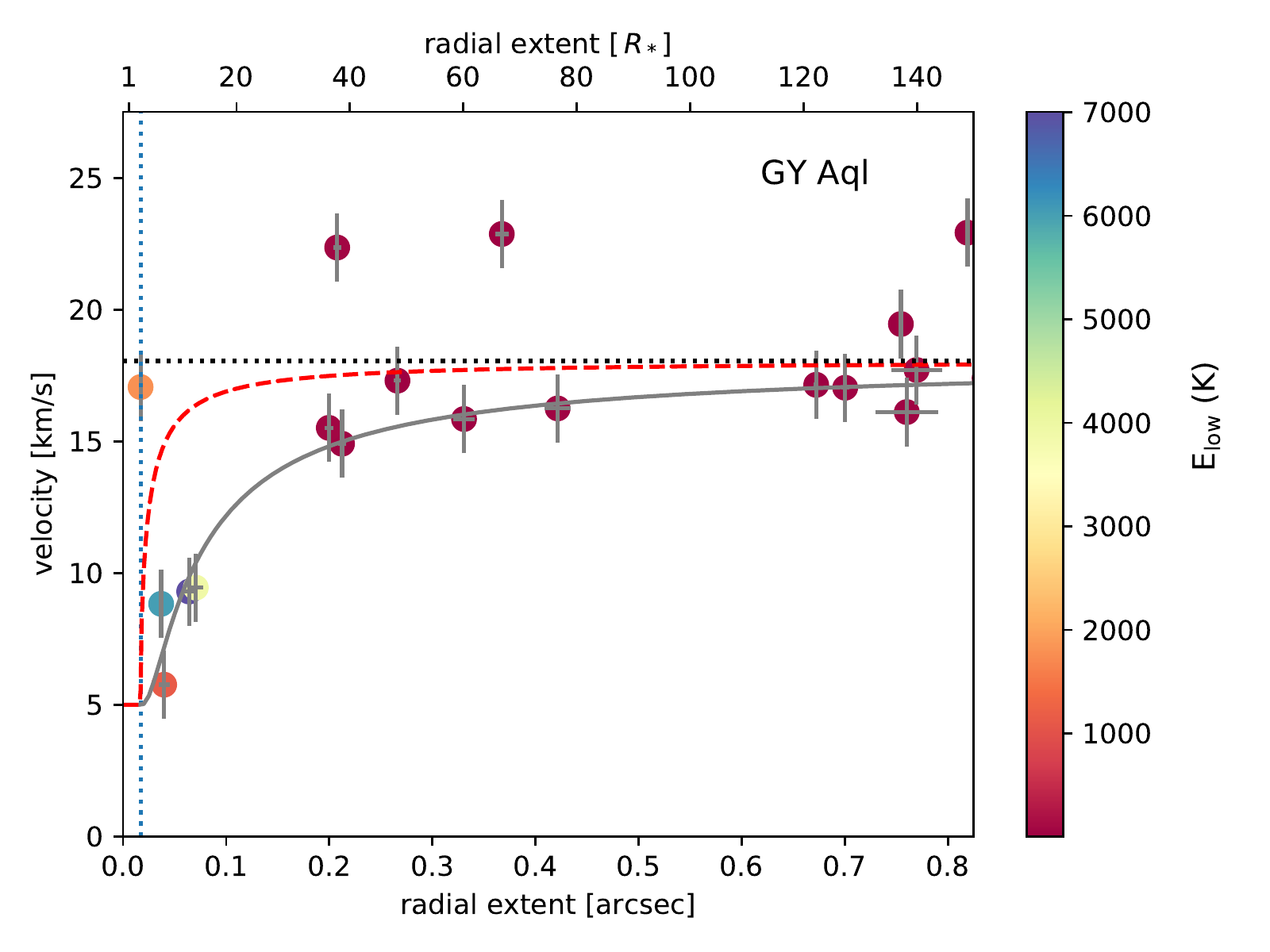}}}
\end{minipage}
    \hfill
\begin{minipage}[t]{.495\textwidth}
        \centerline{\resizebox{\textwidth}{!}{\includegraphics[angle=0]{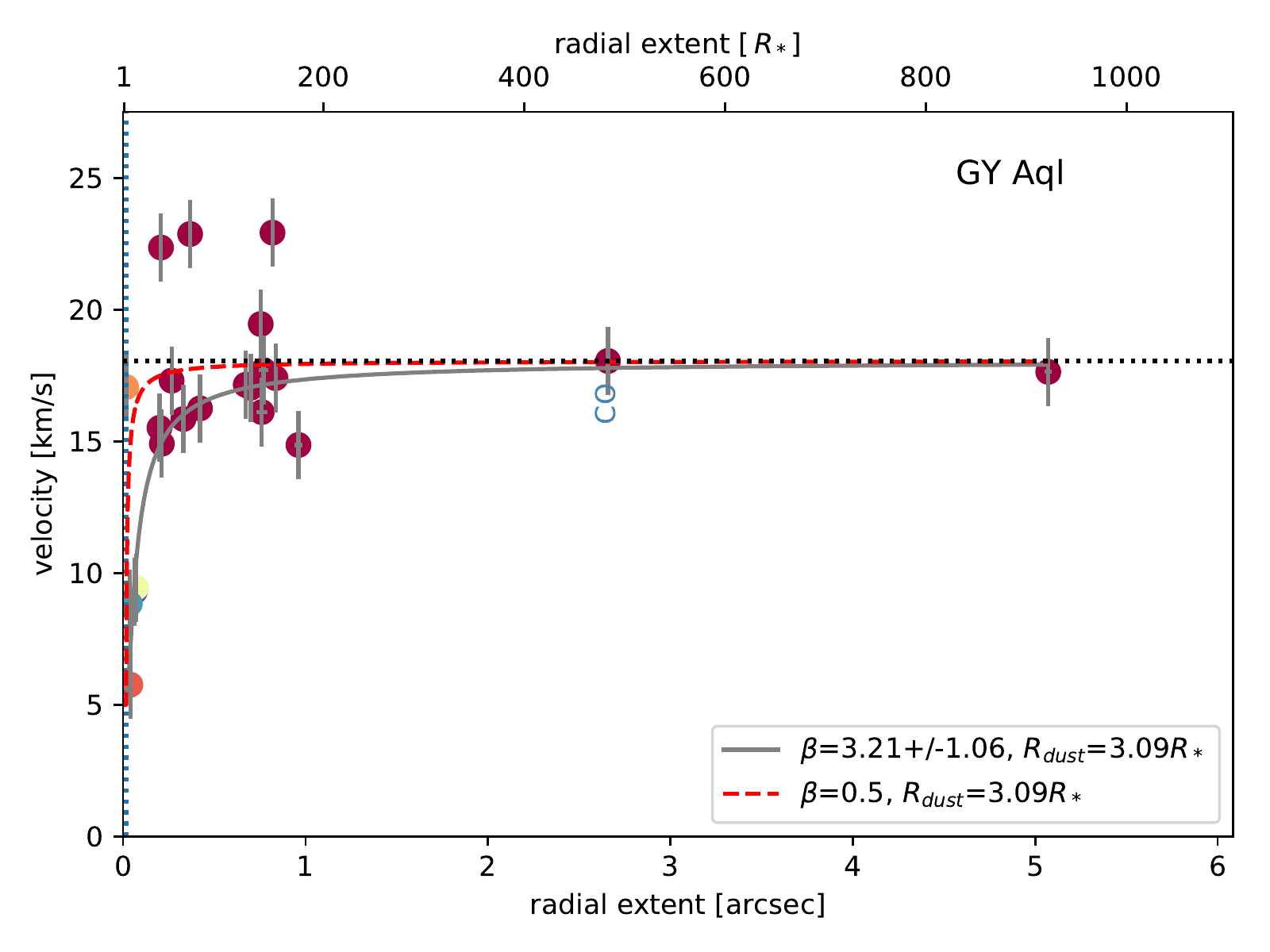}}}
\end{minipage}
\caption{\textbf{Wind kinematics for GY~Aql.} See Fig.~\ref{Fig:IRC10529_kinematics}  caption.}
\label{Fig:GY_Aql_kinematics}
\end{figure*}

\begin{figure*}[htpb]
\vspace{1ex}
\begin{minipage}[t]{.495\textwidth}
        \centerline{\resizebox{\textwidth}{!}{\includegraphics[angle=0]{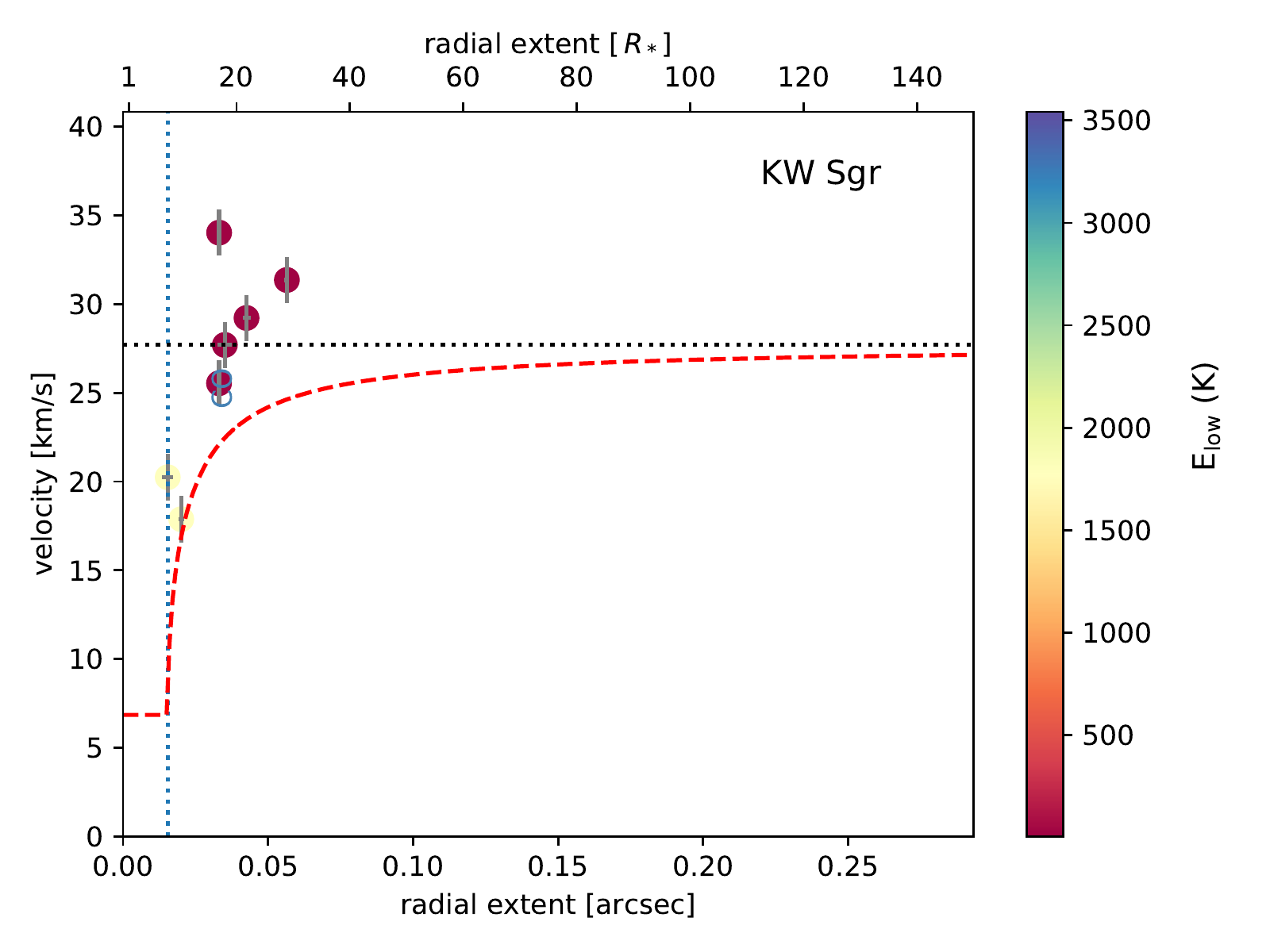}}}
\end{minipage}
    \hfill
\begin{minipage}[t]{.495\textwidth}
        \centerline{\resizebox{\textwidth}{!}{\includegraphics[angle=0]{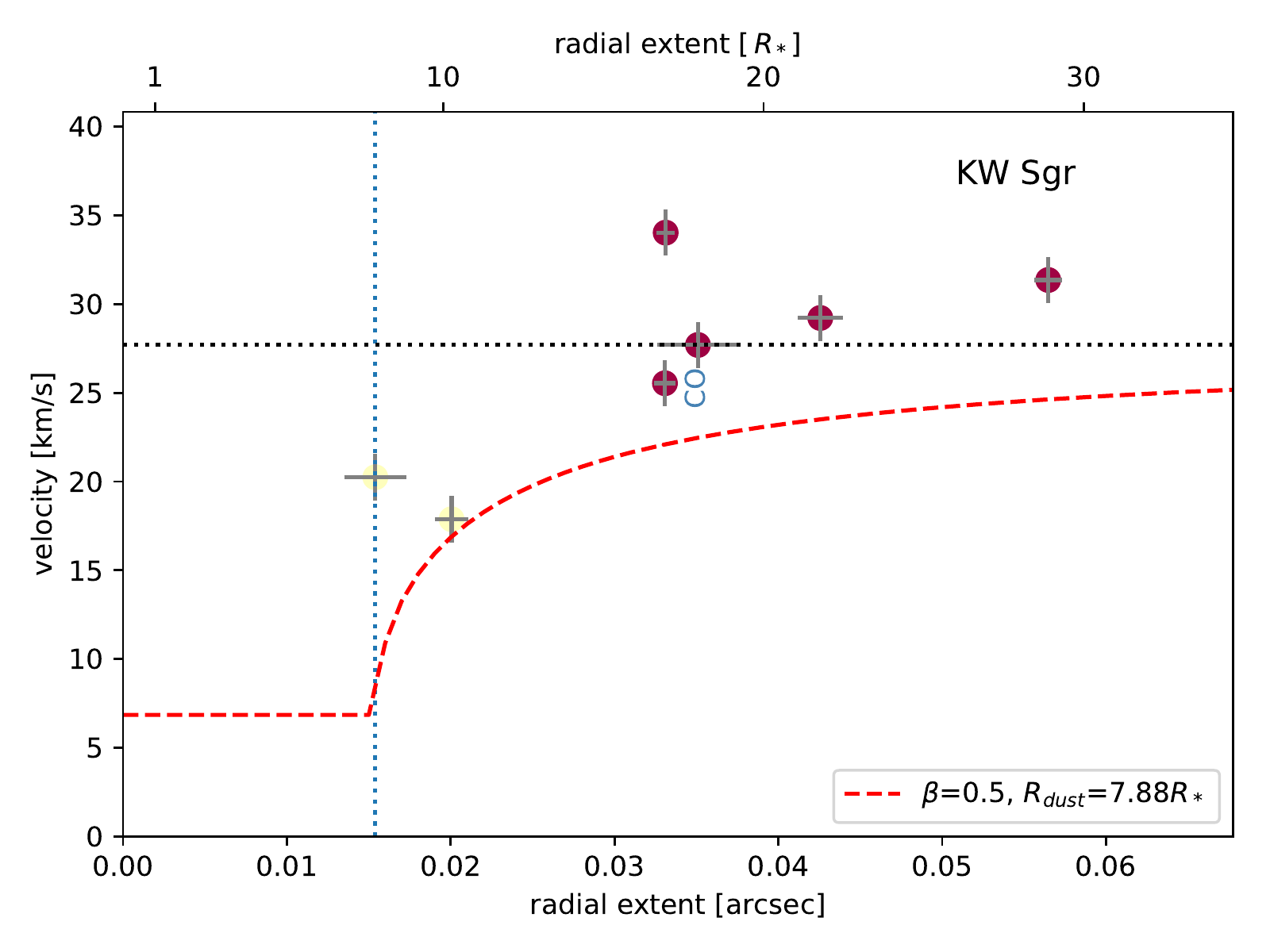}}}
\end{minipage}
\caption{\textbf{Wind kinematics for KW~Sgr.} See Fig.~\ref{Fig:IRC10529_kinematics}  caption. Wind velocity profile constructed 
only on the basis of the medium and high spatial resolution data, since the low spatial resolution data still need to be acquired.
Not enough data are available for a reliable determination of the $\beta$ parameter.  
Since the $^{12}$CO v=0 J=2-1 remains undetected in the medium-resolution data, the CO extent is deduced from the high-resolution data.}
\label{Fig:KW_Sgr_kinematics}
\end{figure*}

\begin{figure*}[htpb]
\vspace{1ex}
\begin{minipage}[t]{.495\textwidth}
        \centerline{\resizebox{\textwidth}{!}{\includegraphics[angle=0]{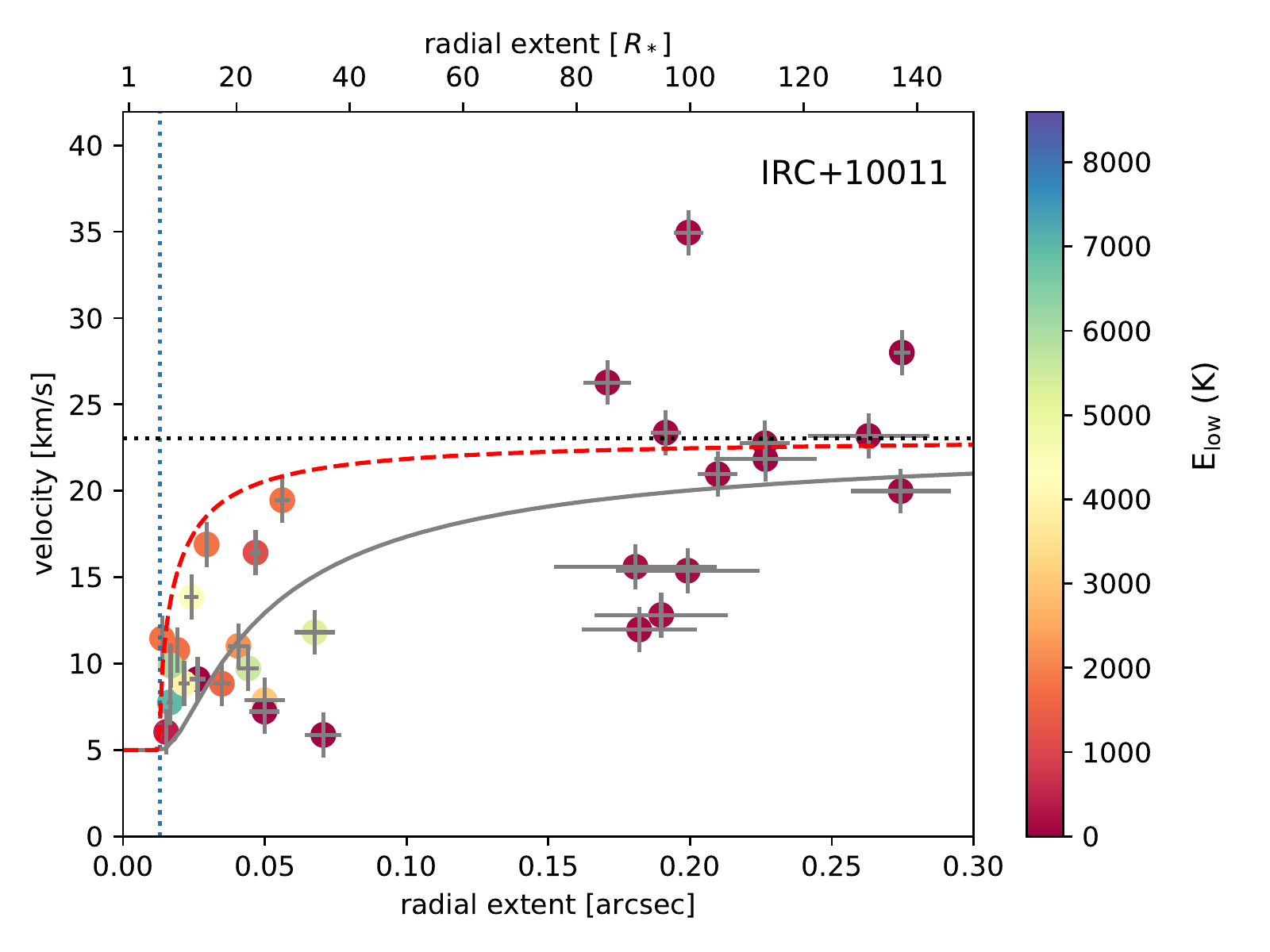}}}
\end{minipage}
    \hfill
\begin{minipage}[t]{.495\textwidth}
        \centerline{\resizebox{\textwidth}{!}{\includegraphics[angle=0]{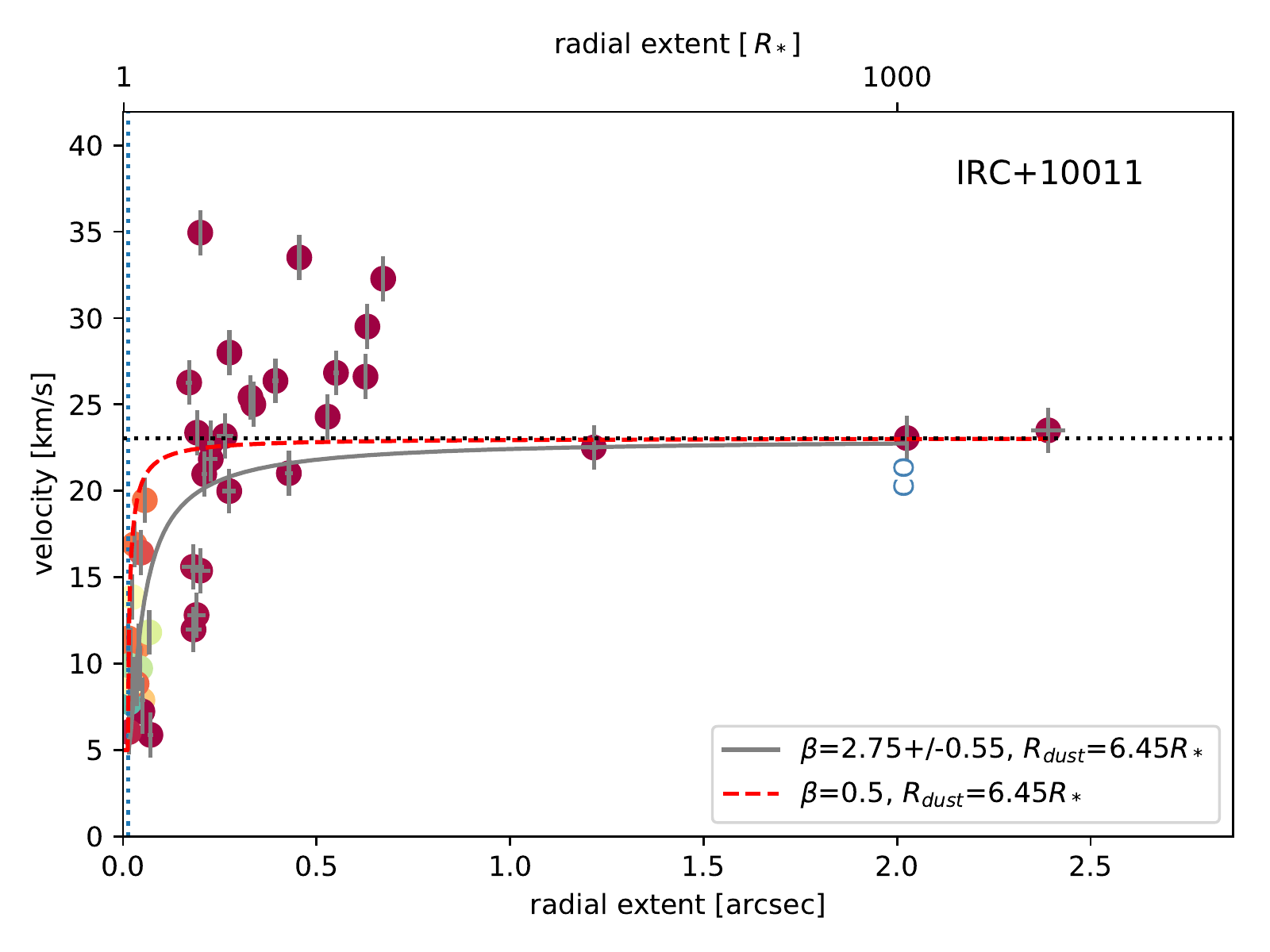}}}
\end{minipage}
\caption{\textbf{Wind kinematics for IRC\,+10011.} See Fig.~\ref{Fig:IRC10529_kinematics}  caption.}
\label{Fig:IRC10011_kinematics}
\end{figure*}

\begin{figure*}[htpb]
\vspace{1ex}
\begin{minipage}[t]{.495\textwidth}
        \centerline{\resizebox{\textwidth}{!}{\includegraphics[angle=0]{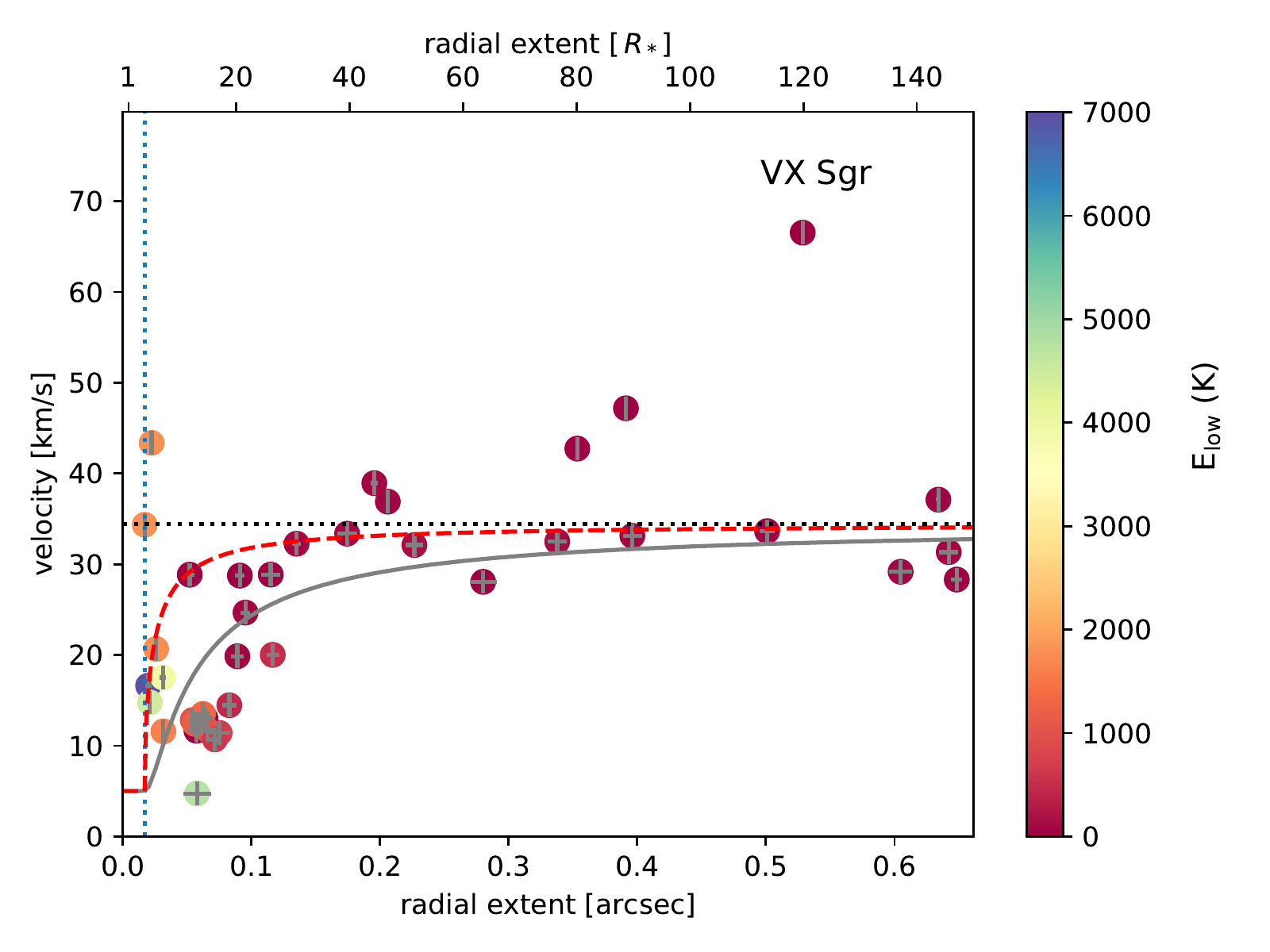}}}
\end{minipage}
    \hfill
\begin{minipage}[t]{.495\textwidth}
        \centerline{\resizebox{\textwidth}{!}{\includegraphics[angle=0]{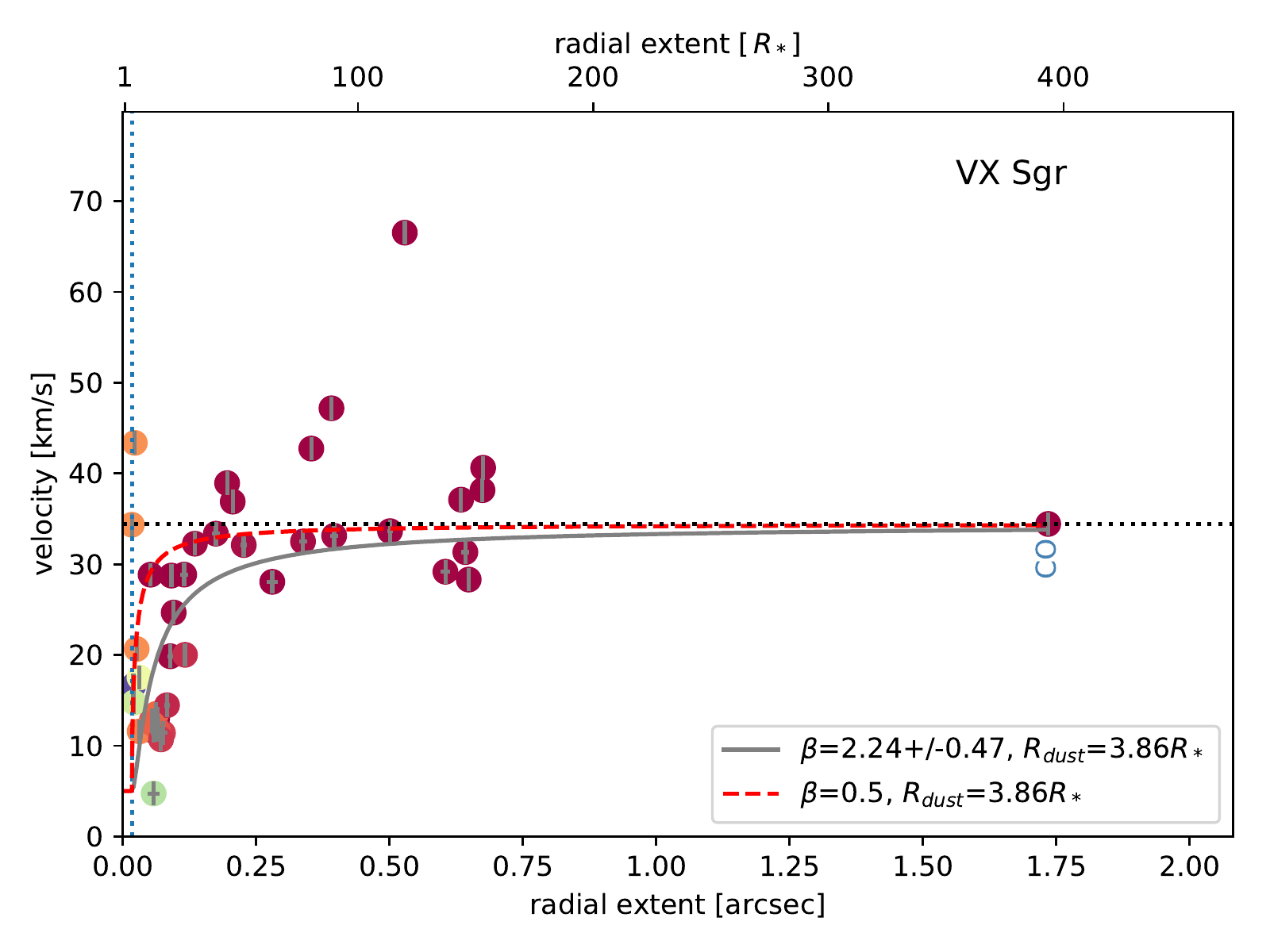}}}
\end{minipage}
\caption{\textbf{Wind kinematics for VX~Sgr.} See Fig.~\ref{Fig:IRC10529_kinematics}    caption.}
\label{Fig:VX_Sgr_kinematics}
\end{figure*}

%  ++++++++++++++++++++++++++++++++++++++++++++++++++++++++++++++++++++++++++++++++++++++++++++++
%  +++++++++++++++++++++++++++  WIND KINEMATICS FOR RW Sco ARBITRIALY ADDED HERE ++++++++++++++++++++
%  ++++++++++++++++++++++++++++++++++++++++++++++++++++++++++++++++++++++++++++++++++++++++++++++

\begin{figure*}[!htpb]
\vspace{1ex}
\begin{minipage}[t]{.495\textwidth}
        \centerline{\resizebox{\textwidth}{!}{\includegraphics[angle=0]{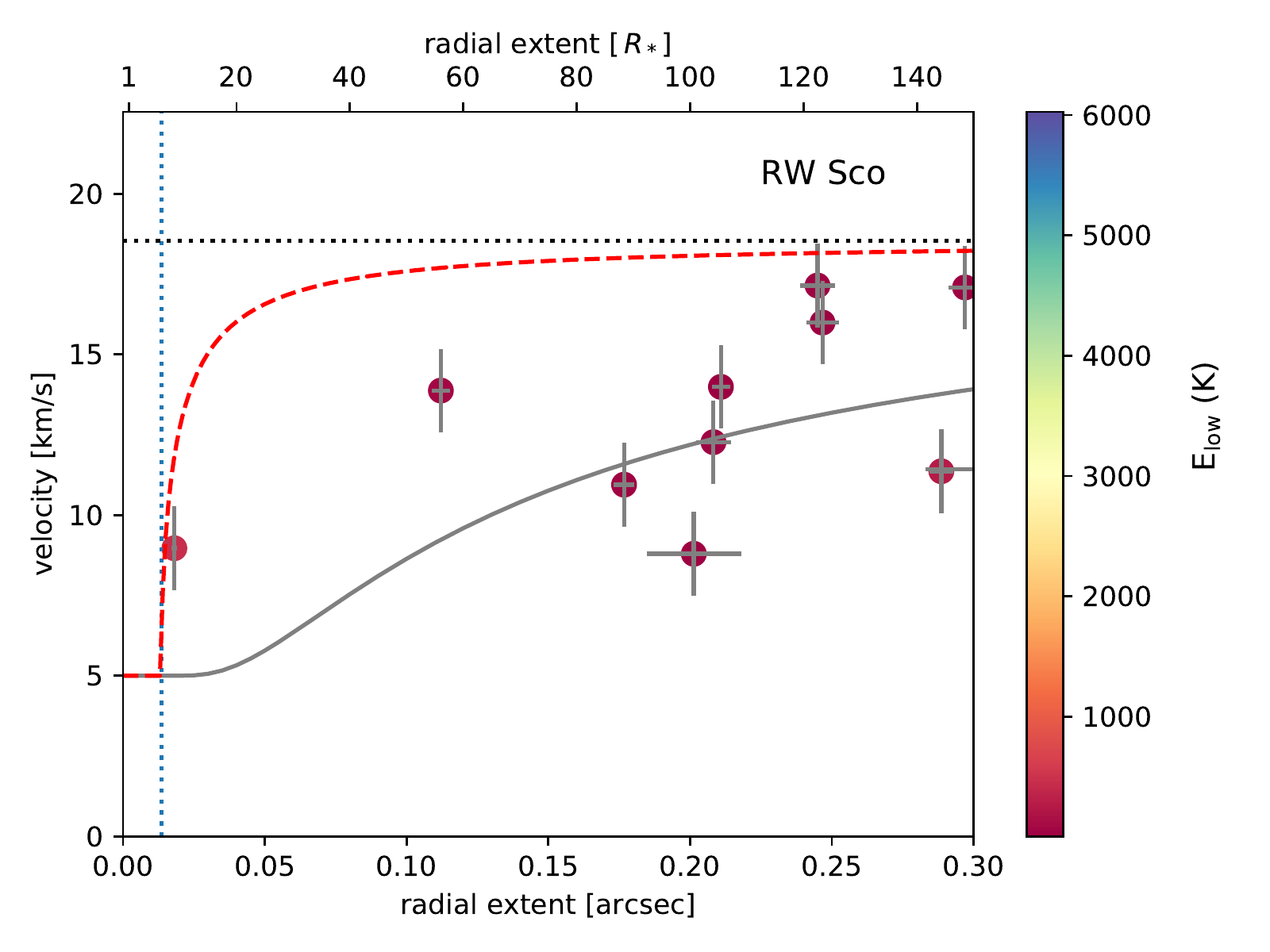}}}
\end{minipage}
    \hfill
\begin{minipage}[t]{.495\textwidth}
        \centerline{\resizebox{\textwidth}{!}{\includegraphics[angle=0]{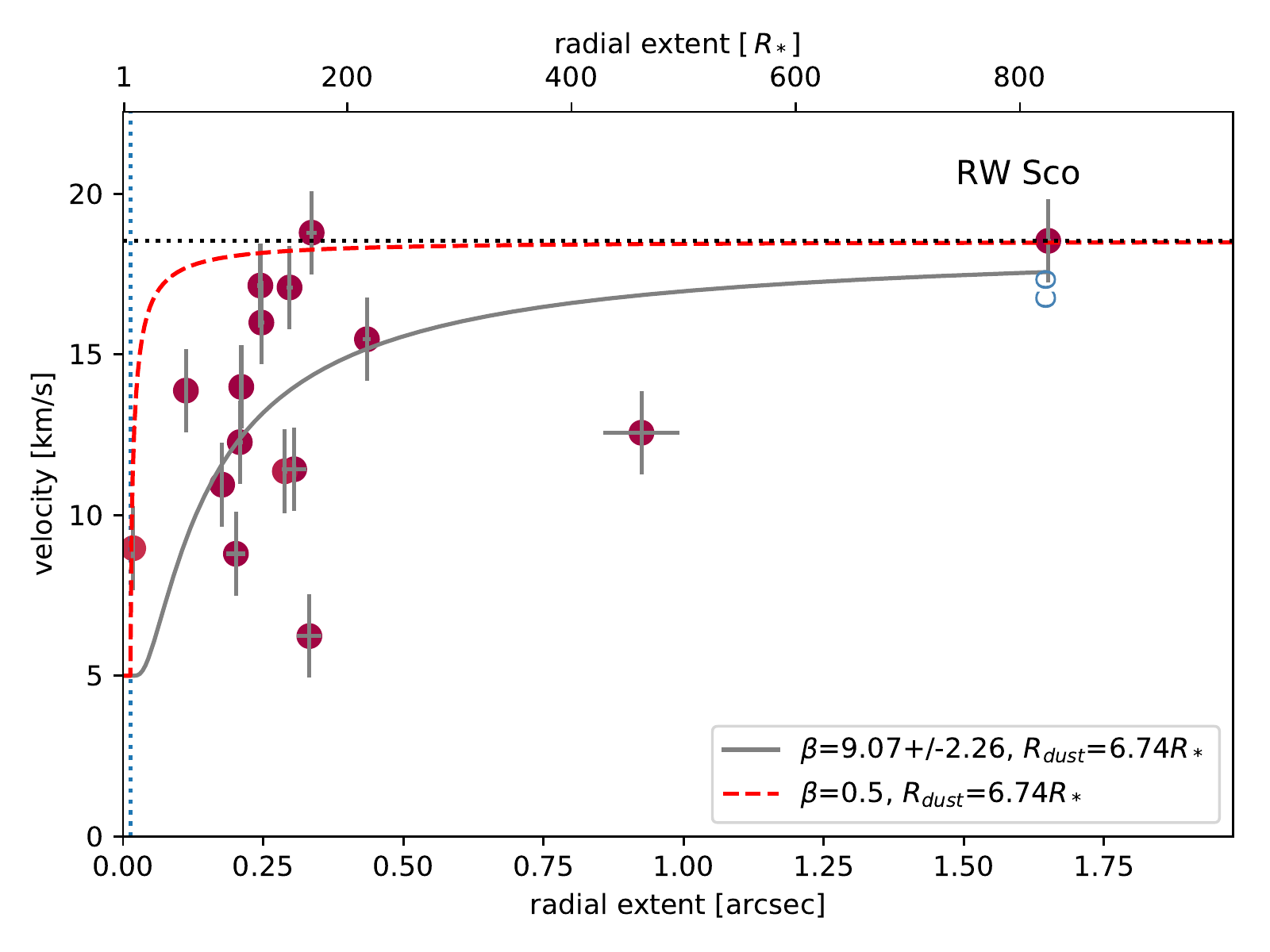}}}
\end{minipage}
\label{Fig:RW_Sco_kinematics}
\end{figure*}
\begin{figure*}[!htpb]
\vspace{1ex}
%\begin{minipage}[t]{.495\textwidth}
%        \centerline{\resizebox{\textwidth}{!}{\includegraphics[angle=0]{figures/IRC-10529_p_extent_restricted_merged_zoom_ad.pdf}}}
%\end{minipage}
    \hfill
%\begin{minipage}[t]{.495\textwidth}
%        \centerline{\resizebox{\textwidth}{!}{\includegraphics[angle=0]{figures/IRC-10529_p_extent_restricted_merged_ad.pdf}}}
%\end{minipage}
\caption{\textbf{Wind kinematics for RW~Sco.} See Fig.~\ref{Fig:IRC10529_kinematics} 
 caption.}
\label{Fig:RW_Sco_kinematics}
\end{figure*}

%  ############################################################################################################ 
%  ##########################################   APPENDIX D  ##################################################### 
%  ############################################################################################################ 
\afterpage{\clearpage}
\newpage
\mbox{}

\section{Determining the terminal wind velocity and the impact of pulsation-induced shocks on the velocity measure}
\label{Sect:terminal_velocity}

In the main paper, we argue that the $^{12}$CO v=0 J=2-1 line in the low resolution {\sc atomium} data can be used to determine the 
terminal wind velocity, and its integrity as a diagnostic is not perturbed by pulsation-induced shocks that occur in the innermost few stellar radii. 
We base our arguments on theoretical simulations of a smooth spherically symmetric wind in which the parameters resemble 
R~Aql (see Table~\ref{Table:targets}).
The level populations and corresponding intensities of CO in the simulations were computed on the assumption of a CO abundance of 
[CO/H$_2$]\,=\,2$\times$10$^{-4}$ with the (3D) non-LTE radiative transfer code 
{\sc magritte} by \citet{2020MNRAS.492.1812D,DeCeuster2020MNRAS.499.5194D} which includes the CMB. 
The parameters for R~Aql resemble the stellar parameters in Table~\ref{Table:targets} in the main text.
The temperature profile is assumed to be similar to that of \citet{Danilovich2017A&A...606A.124D, Danilovich2020}:
$T(r) = T_{\star} (R_{\star}/r)^{0.65}$; and the collisional rates and Einstein $A$ coefficients were taken from the LAMDA database 
\citep{2020Atoms...8...15V, 2010ApJ...718.1062Y, Schoier2005A&A...432..369S}.

\vspace{0.25cm}		
In the first set of models, the wind velocity profile follows the analytic expression of Eq.~\eqref{Eq:velocity} with parameters $v_0\!=\!1$\,km/s, $v_\infty\!=\!12.8$\,km/s, and $\beta = 1, {\rm{or}}~5$. 
Depicted in Fig.~\ref{Fig:theory_vwind} are the  $^{12}$CO v=0 J=2-1 velocity measures as a function of the aperture size.
It is evident that the CO velocities grow when the aperture size increases from small to large scales.
The velocity measure can be larger than the input terminal wind velocity of 12.8\,km/s, owing to the effect of thermal broadening 
($v_{\rm{therm}}\!=\!\sqrt{2 k T/m}$) and turbulent broadening ($v_{\rm{turb}}$\,=\,1.5\,km/s) that is accounted for in the full width 
at half maximum of the Gaussian broadened profile 
(i.e., FWHM\,=\,$2\sqrt{2\ln2}\sigma$, where $\sigma\!=\!\sqrt{v_{\rm{therm}}^2 + v_{\rm{turb}}^2}$\,\,). 
An increase in sensitivity of the observations (and hence a lower noise value) yields a more accurate sampling of the weak wings of the line profile 
where the broadening manifests itself --- particularly in the case of optically thick line profiles \citep{DeBeck2012A&A...539A.108D}.
The blue-wing velocity measure is often smaller than the corresponding red-wing velocity measure, owing to the effect of the blue wing absorption 
\citep{Morris1985A&A...142..107M, Schoenberg1988A&A...195..198S}. 

%  +++++++++++++++++++++++++++++++++++++++++++++++++++++++++++++++++++++++++++++++

\begin{figure}[htp]
	\begin{minipage}[t]{.495\textwidth}
%		\centerline{\resizebox{\textwidth}{!}{\includegraphics[angle=0]{figures/wing_velocities_beta_1.png}}}
				\centerline{\resizebox{\textwidth}{!}{\includegraphics[angle=0]{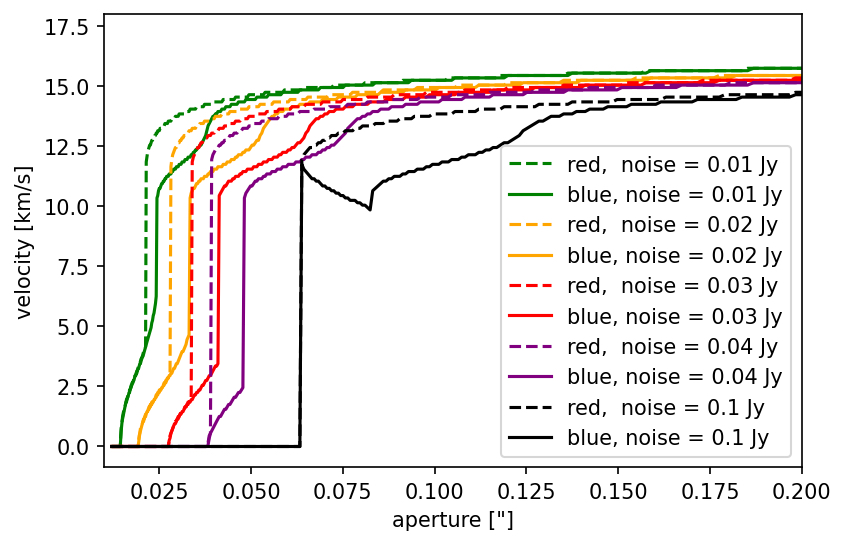}}}

	\end{minipage}
	\hfill
	\begin{minipage}[t]{.495\textwidth}
%		\centerline{\resizebox{\textwidth}{!}{\includegraphics[angle=0]{figures/wing_velocities_beta_5.png}}}
				\centerline{\resizebox{\textwidth}{!}{\includegraphics[angle=0]{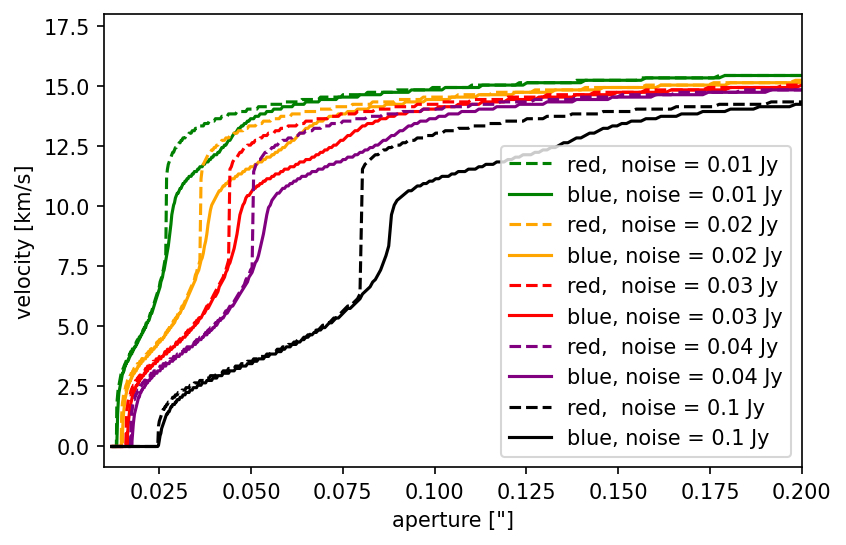}}}
	\end{minipage}
	\caption{\textbf{Change in the $^{12}$CO v=0 J=2-1 velocity measure as a function of the aperture size.} 
The velocity measures are extracted for a range of aperture sizes and noise levels following the same procedure as outlined in Step~2 in Sect.~\ref{Sec:Methodology}. 
The {\it left panel} is constructed for a velocity profile with $\beta$\,=\,1 and the {\it right panel} for $\beta$\,=\,5.
Velocity measures extracted from the red (blue) wing are indicated with `red' (`blue') in the legend in the panels.}
	\label{Fig:theory_vwind}
\end{figure}

%  +++++++++++++++++++++++++++++++++++++++++++++++++++++++++++++++++++++++++++++++

The combined {\sc atomium} data is optimal for establishing whether the increase in CO velocities with aperture size is a general trend.
To date, the data for the three separate spatial resolutions and the combined dataset are available for six stars: R~Hya, $\pi^1$~Gru, 
R~Aql, IRC\,$-$10529, IRC\,+10011, and VX~Sgr, but only $\pi^1$~Gru has been analysed in detail \citep{2020A&A...644A..61H}.
Shown in Fig.~\ref{Fig:terminal_velocities1} is the change in velocities of the $^{12}$CO v=0 J=2-1 
and $^{28}$SiO v=0 J=5-4 lines with aperture size when they are extracted from the combined datasets for the 6~sources 
by following Step~1 and Step~2 in Sect.~\ref{Sec:Methodology}. 
In Fig.~\ref{Fig:terminal_velocities1} we also compare the velocity profiles with the velocity measure extracted from the low-resolution 
{\sc atomium} data for the $^{12}$CO v=0 J=2-1 line [see $v_\infty^{\rm{ com}}$(CO) in column~(3) of Table~\ref{Table:beta}].
For most sources --- except for the blue-shifted velocity in IRC\,+10011, and the red-shifted velocity in R~Hya  --- the CO velocity grows 
with increasing aperture size and reaches a plateau beyond $\sim$100 stellar radii. 
The velocity measure [$v_\infty^{\rm{ com}}$(CO)] derived from the  $^{12}$CO v=0 J=2-1 data at low angular resolution is a good tracer 
for the plateau and the terminal wind velocity, provided the thermal broadening, the turbulent broadening, and the spectral resolution 
of $\sim$1.3\,km/s are accounted for.

%  +++++++++++++++++++++++++++++++++++++++++++++++++++++++++++++++++++++++++++++++

\begin{figure*}[t]
\begin{minipage}[t]{.495\textwidth}
        \centerline{\resizebox{\textwidth}{!}{\includegraphics[angle=0]{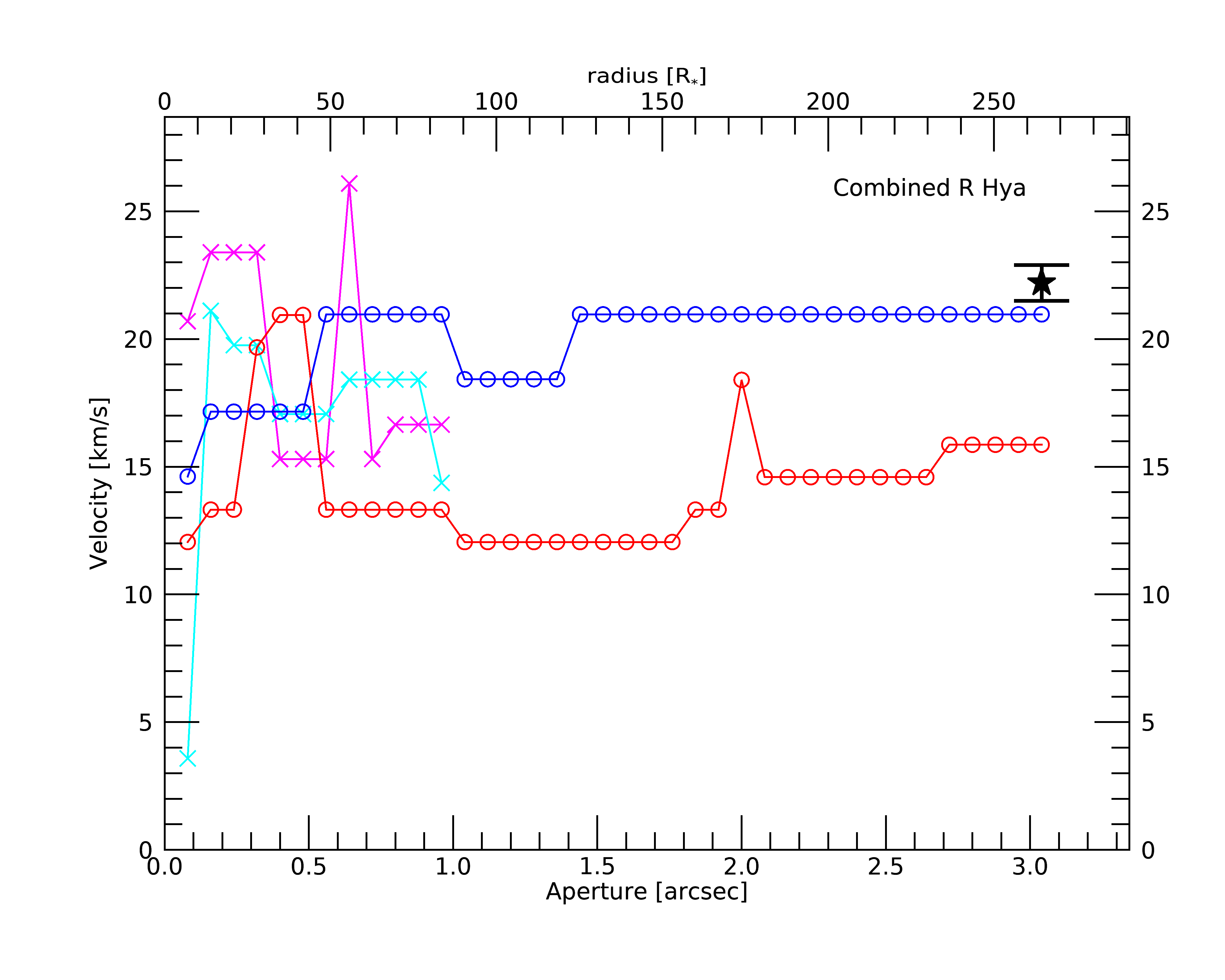}}}
\end{minipage}
    \hfill
\begin{minipage}[t]{.495\textwidth}
        \centerline{\resizebox{\textwidth}{!}{\includegraphics[angle=0]{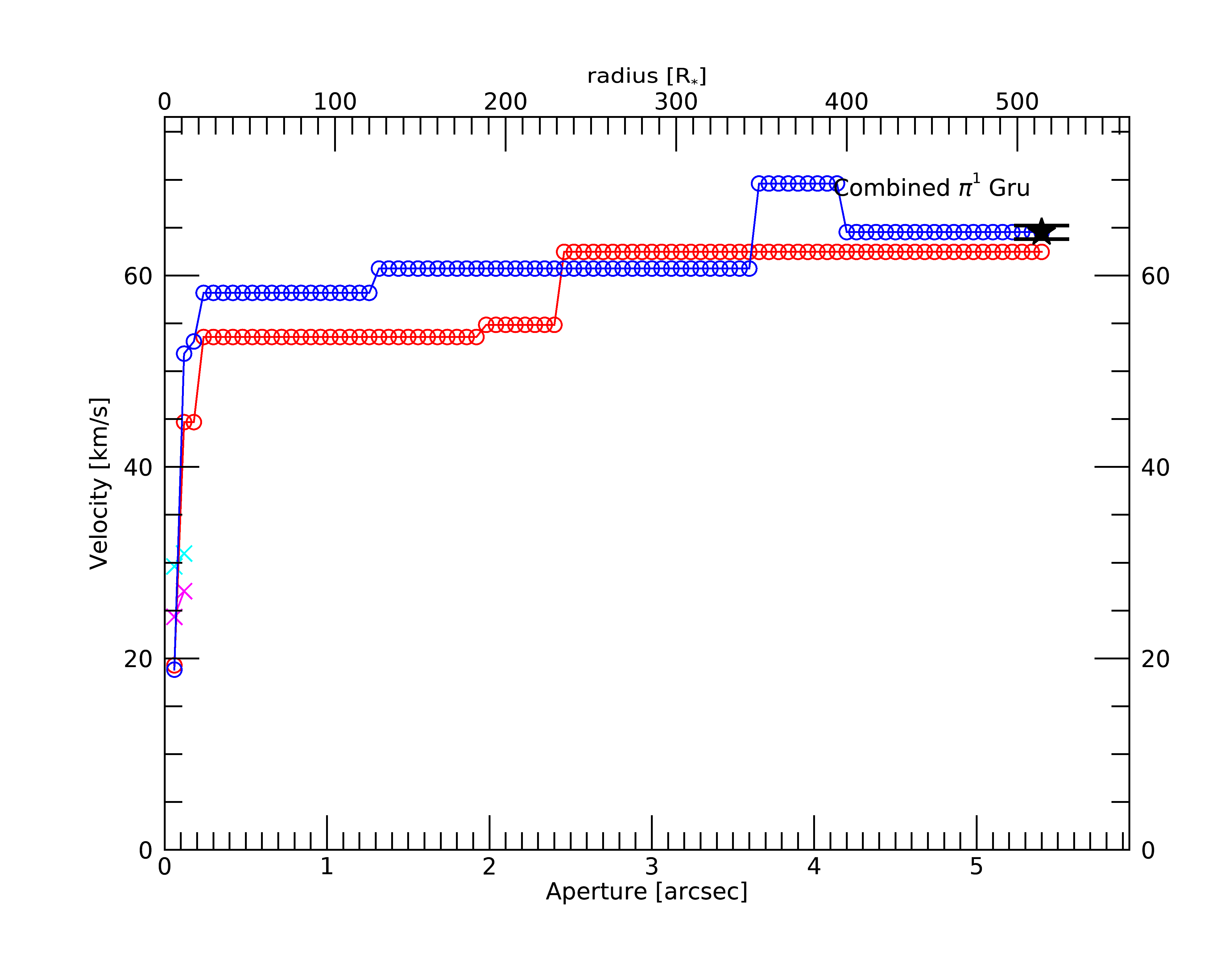}}}
\end{minipage}
\begin{minipage}[t]{.495\textwidth}
	\centerline{\resizebox{\textwidth}{!}{\includegraphics[angle=0]{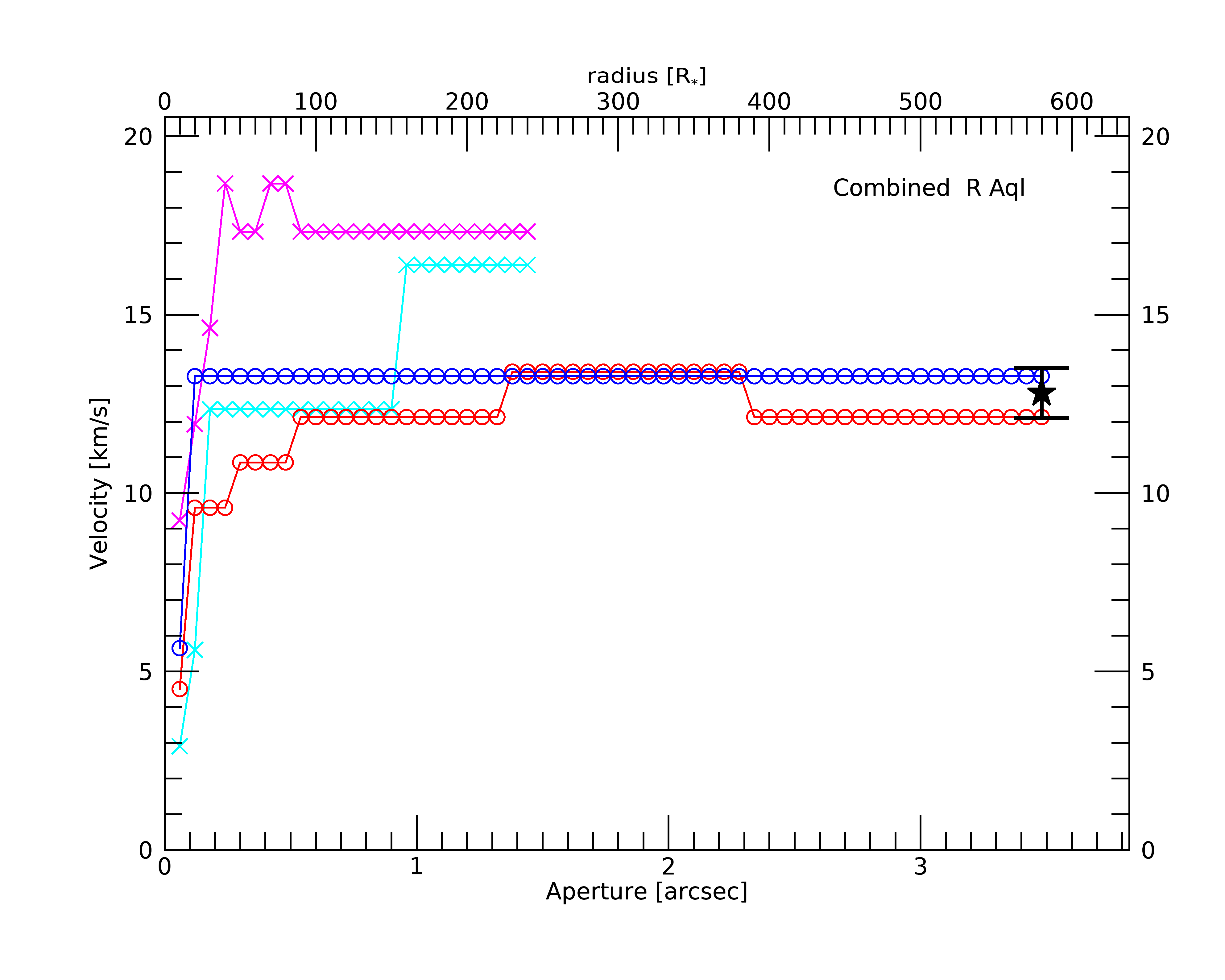}}}
\end{minipage}
\hfill
\begin{minipage}[t]{.495\textwidth}
	\centerline{\resizebox{\textwidth}{!}{\includegraphics[angle=0]{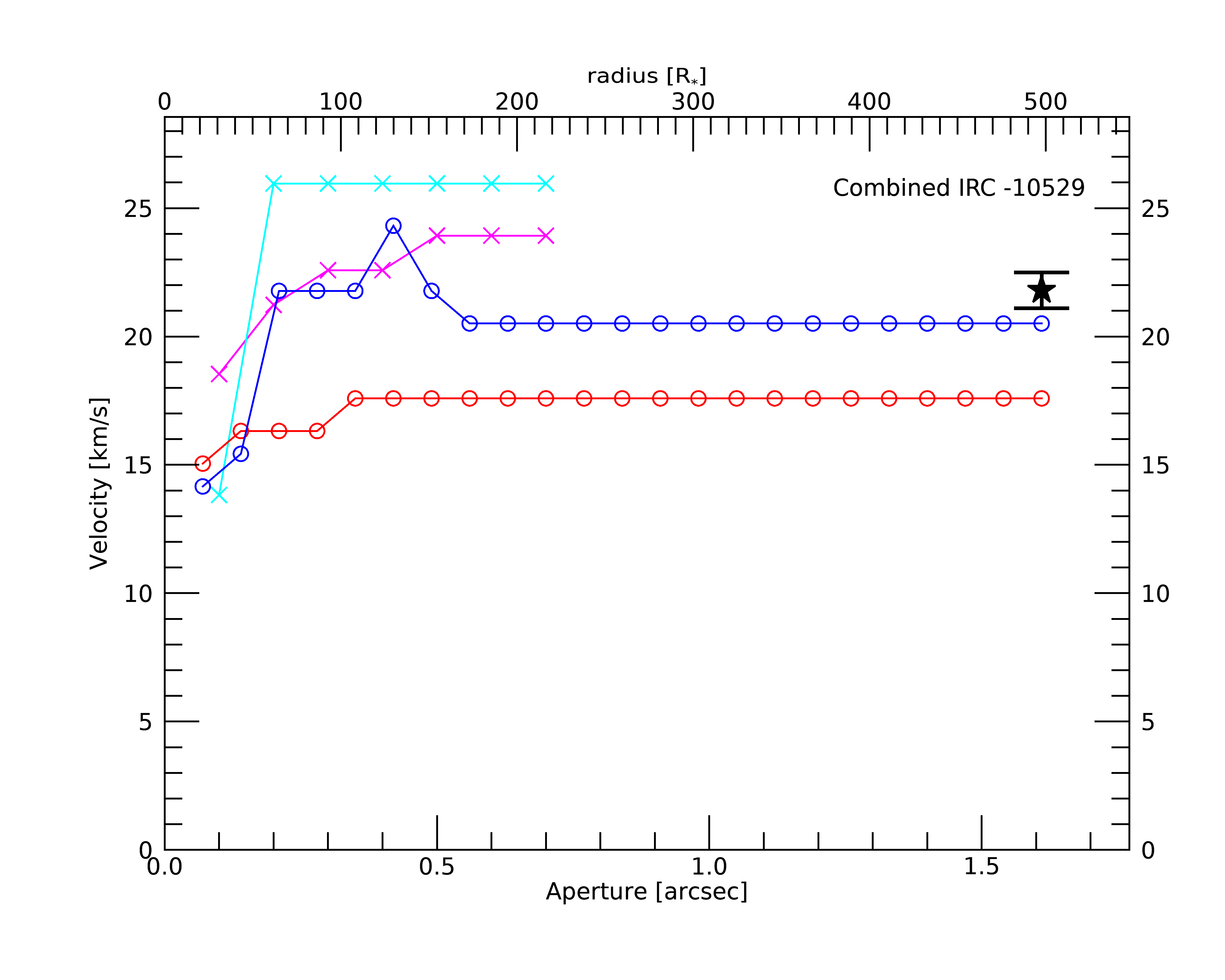}}}
\end{minipage}
\begin{minipage}[t]{.495\textwidth}
	\centerline{\resizebox{\textwidth}{!}{\includegraphics[angle=0]{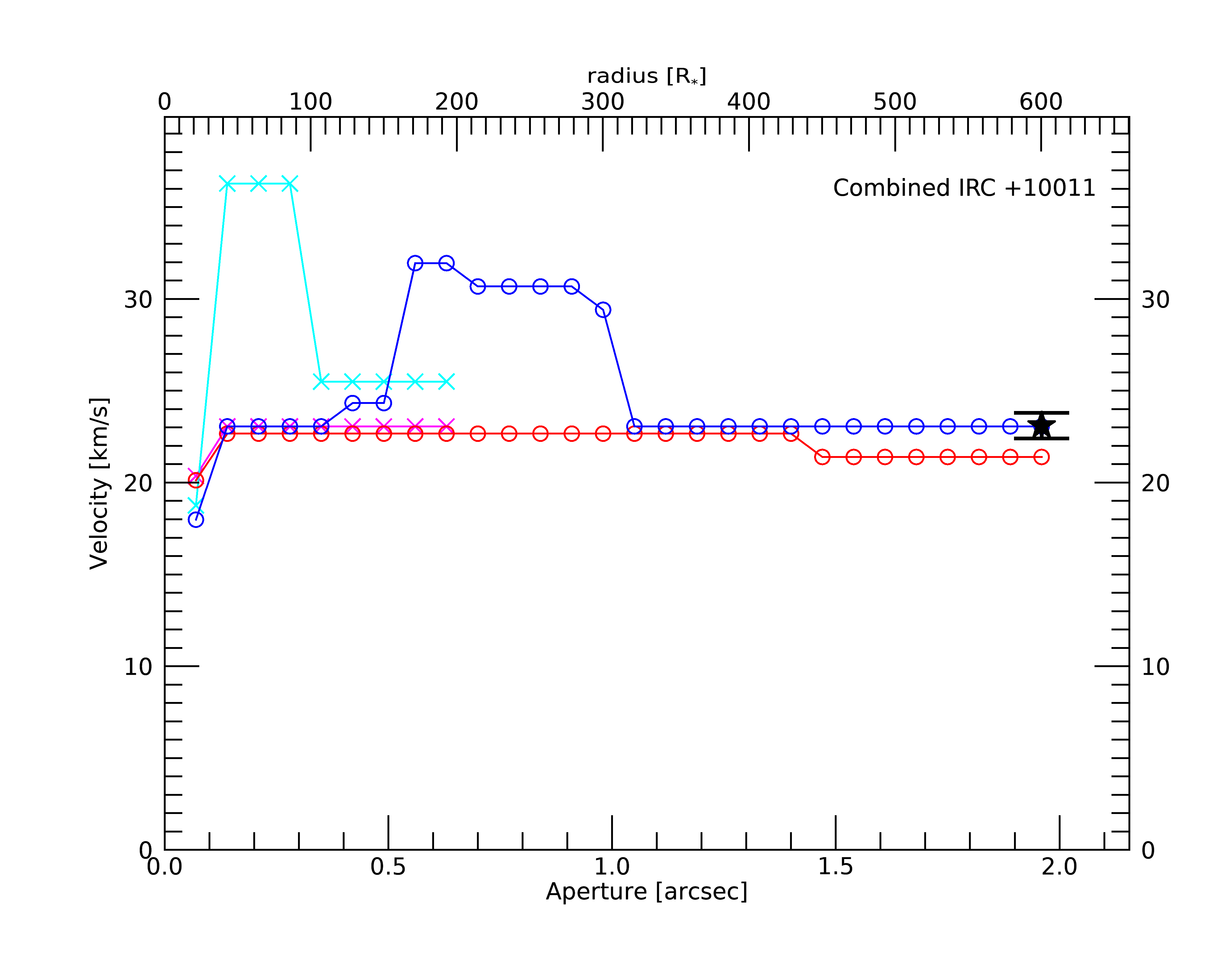}}}
\end{minipage}
\hfill
\begin{minipage}[t]{.495\textwidth}
	\centerline{\resizebox{\textwidth}{!}{\includegraphics[angle=0]{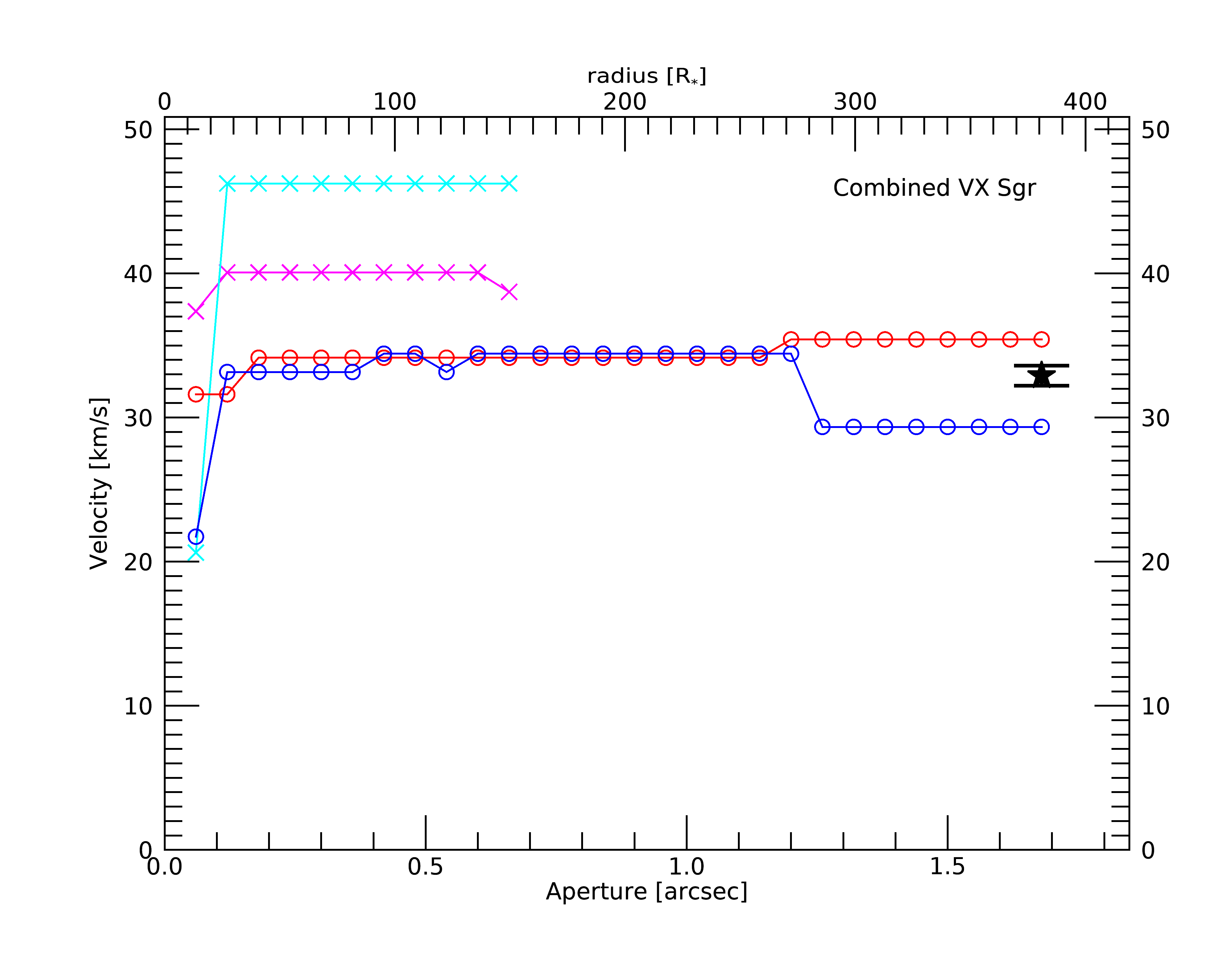}}}
\end{minipage}
\caption{  \textbf{ {\boldmath $^{12}$}CO v=0 J=2-1 and {\boldmath $^{28}$}SiO v=0 J=5-4 wind velocities of R~Hya, 
{\boldmath $\pi^1$}\,Gru, R~Aql, IRC\,$-$10529, IRC\,+10011, and VX~Sgr.}
\hspace{0.125cm} Plots of the blue and red wing velocity of the
$^{12}$CO v=0 J=2-1 line (in {\blue{blue}} and {\red{red}}, respectively) and the $^{28}$SiO v=0 J=5-4 line (in {\cyan{cyan}} and 
{\pink{pink}}, respectively) derived from the {\sc atomium} combined dataset for a range of extraction apertures for R~Hya, 
$\pi^1$~Gru, R~Aql, IRC\,$-$10529, IRC\,+10011, and VX~Sgr, by following Step~1 and Step~2 in Sect.~\ref{Sec:Methodology}.
The black star ({\Large $ \star$}) denotes the $^{12}$CO v=0 J=2-1 velocity extracted from the {\sc atomium} low-resolution data, 
and the error bar denotes the approximate spectral resolution of 1.3~km/s.}
\label{Fig:terminal_velocities1}
\end{figure*}

%  +++++++++++++++++++++++++++++++++++++++++++++++++++++++++++++++++++++++++++++++

\medskip

Understanding how pulsation-induced shocks might impact the velocity measure has a strong bearing on the interpretation 
of the observationally derived wind kinematics discussed in Sect.~\ref{Sec:Interpretation}. 
Analogous to Fig.~\ref{Fig:theory_vwind}, we used the non-LTE radiative transfer code {\sc magritte} 
\citep{2020MNRAS.492.1812D, DeCeuster2020MNRAS.499.5194D}, but this time rather than using the standard beta-law wind profiles from Eq.~\eqref{Eq:velocity}, the wind velocity profile has been modified 
to mimic the effect of pulsation-induced shocks within a 1D wind geometry. 
We used the results of \citet[][see their Fig.~1]{Bladh2019A&A...626A.100B}, which we extrapolated to larger distances from the star by using 
a fit that follows a beta-velocity profile (see left panel in Fig.~\ref{Fig:theory_vwind_shock}).  
The output image is then run through the ALMA simulator tool for setups resembling the compact, medium, and extended configurations. 
The simulated output data are then treated in the same way as the {\sc atomium} data for the extraction of the velocity measure as a function 
of the aperture size by following Step~1 and Step~2 in Sect.~\ref{Sec:Methodology} (see the panels labelled `normal shock' model in
Fig.~\ref{Fig:theory_vwind_shock_traced}). 
Two main conclusions can be drawn from comparing the `no shock' (upper row) and `normal shock' (middle row) panel of $^{12}$CO v=0 J=2-1: 
(1) for the case in which the velocity of the shock amplitude is lower than the terminal wind velocity, it is apparent that the velocity measure
extracted from the compact CO v=0 J=2-1 data is the same for the `no shock' and `normal shock' model, confirming that the compact 
CO v=0 J=2-1 data is a good measure of the terminal wind velocity (if the effect of thermal and turbulent broadening, and the spectral 
resolution of the {\sc atomium} data are accounted for); and
(2) the velocity measures derived from the $^{12}$CO extended configuration data are slightly higher if shocks are accounted for, 
with the shocks manifesting themselves in the faint more extended wings. 

%for (see upper and middle row of  Fig.~\ref{Fig:theory_vwind_shock_traced}, left panels). }}}

%  +++++++++++++++++++++++++++++++++++++++++++++++++++++++++++++++++++++++++++++++
	\begin{figure}[h]
%\vspace{5.0cm}
		\begin{minipage}[t]{.495\textwidth}
			\centerline{\resizebox{\textwidth}{!}{\includegraphics[angle=0]{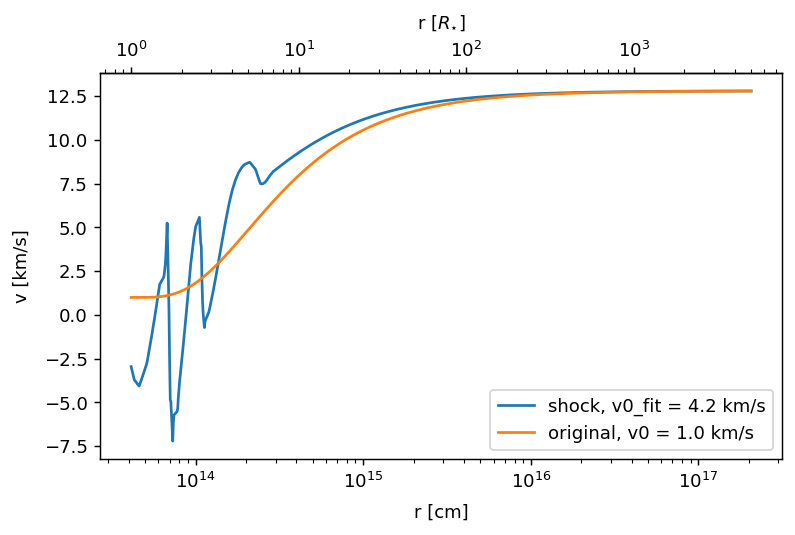}}}
		\end{minipage}
		\hfill
		\begin{minipage}[t]{.495\textwidth}
			\centerline{\resizebox{\textwidth}{!}{\includegraphics[angle=0]{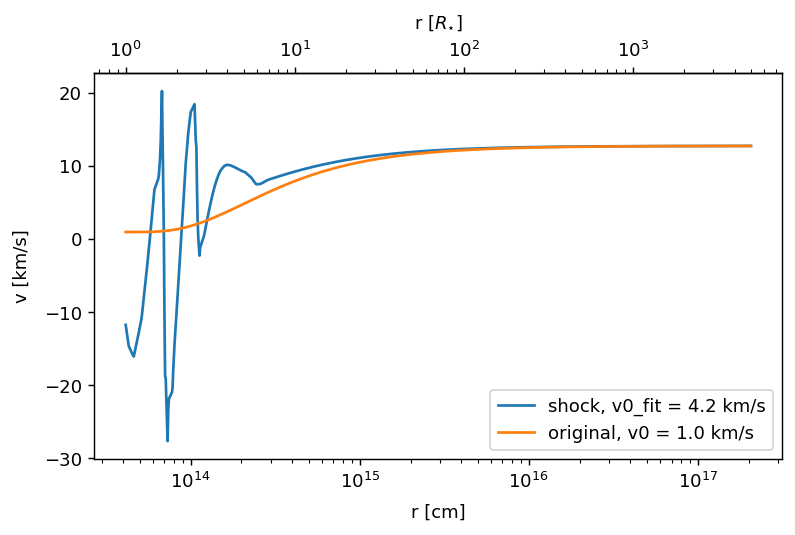}}}
		\end{minipage}
		\caption{ {\bf Wind velocity profile mimicking pulsation-induced shocks just above the stellar atmosphere.} 
				The orange curve shows a beta velocity wind profile for $\beta\!=\!5$ and $v_0\!=\!1$\,km/s. 
				The blue curve in the {\it left panel} is constructed by: 
				(i)~using the shock velocity modelled by \citet[][see their Fig.~1]{Bladh2019A&A...626A.100B} up to $\sim$10\,\Rstar\, followed by; 
				(ii)~a beta velocity profile that is fitted through the velocity points beyond $\sim$8\,\Rstar, with $\beta\!=\!5$ and the fit parameter $v_0$
				which produces a smooth transition from the pulsation-dominated region towards the freely expanding wind region. 
				In the {\it right panel}, the blue curve is constructed in a similar way as in the left panel, but this time the shock velocities modelled by \citet{Bladh2019A&A...626A.100B} are multiplied by a factor of 3.}
		\label{Fig:theory_vwind_shock}
	\end{figure}

%  +++++++++++++++++++++++++++++++++++++++++++++++++++++++++++++++++++++++++++++++	
	\begin{figure}[htp]
	\begin{minipage}[t]{.495\textwidth}
%		\centerline{\resizebox{\textwidth}{!}{\includegraphics[angle=0]{figures/plot_no_shock_CO_rms_sim.png}}}
				\centerline{\resizebox{\textwidth}{!}{\includegraphics[angle=0]{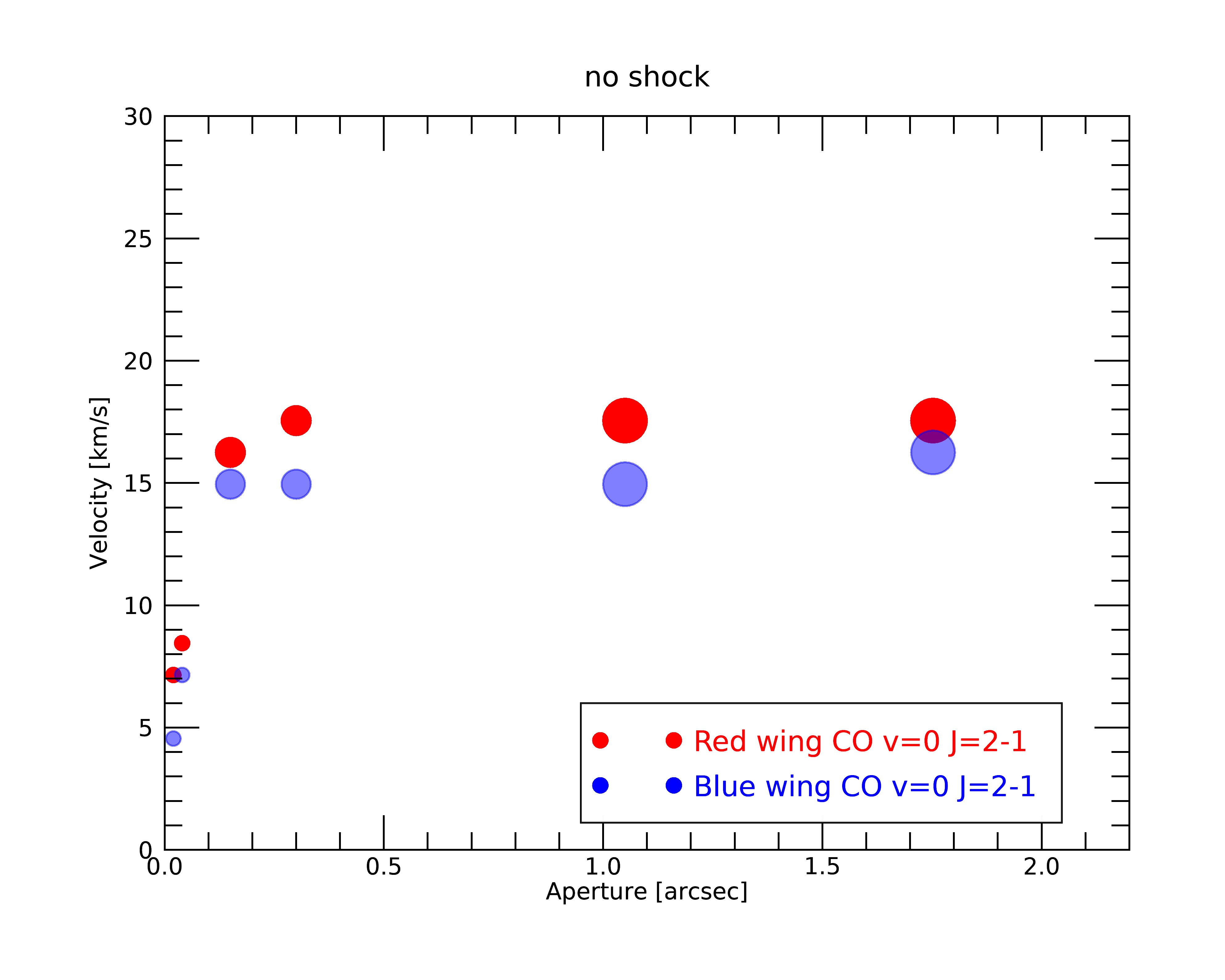}}}
	\end{minipage}
	\hfill
	\begin{minipage}[t]{.495\textwidth}
%		\centerline{\resizebox{\textwidth}{!}{\includegraphics[angle=0]{figures/plot_no_shock_SiO_rms_sim.png}}}
				\centerline{\resizebox{\textwidth}{!}{\includegraphics[angle=0]{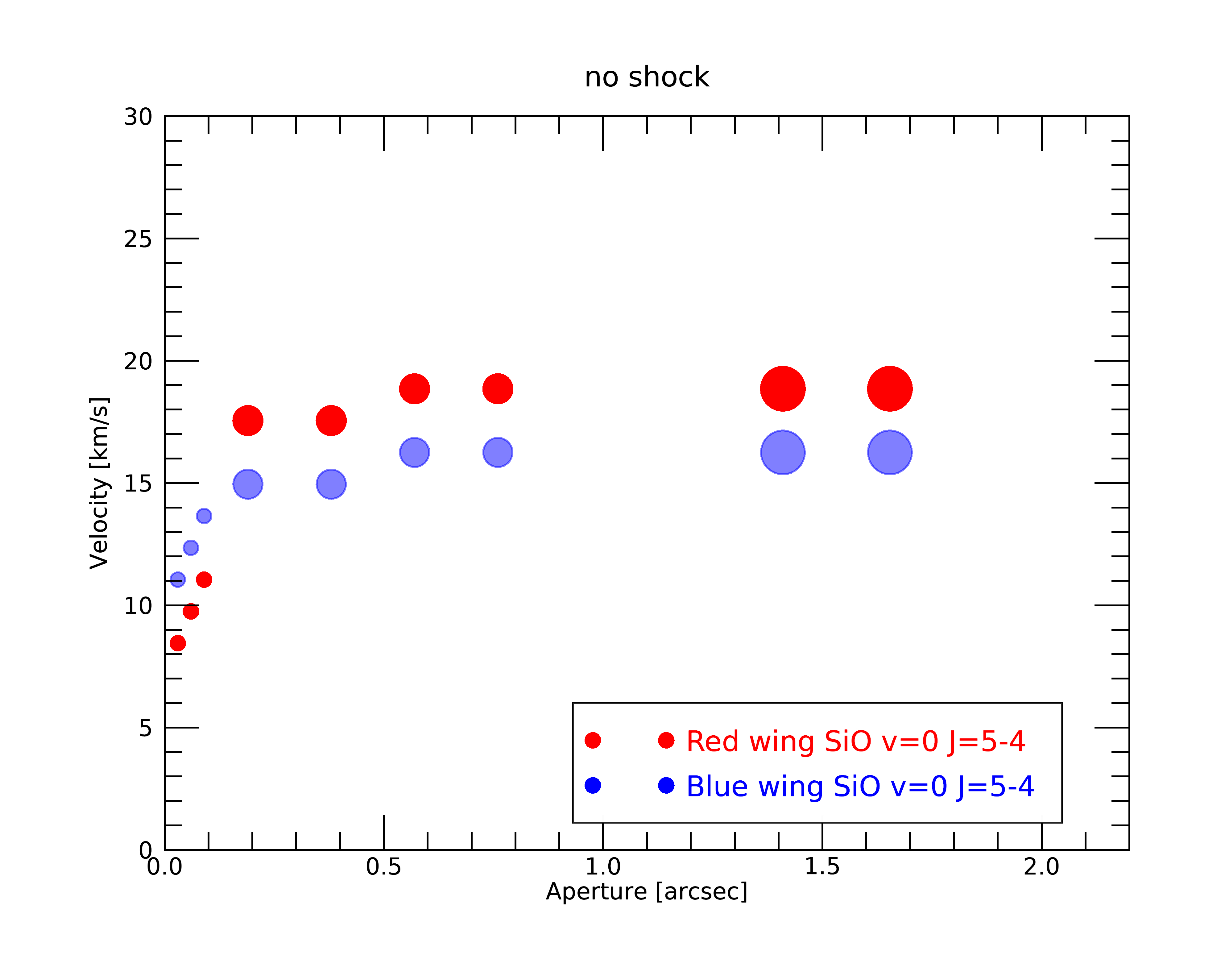}}}
	\end{minipage}
	\begin{minipage}[t]{.495\textwidth}
%		\centerline{\resizebox{\textwidth}{!}{\includegraphics[angle=0]{figures/plot_normal_shock_CO_rms_sim.png}}}
				\centerline{\resizebox{\textwidth}{!}{\includegraphics[angle=0]{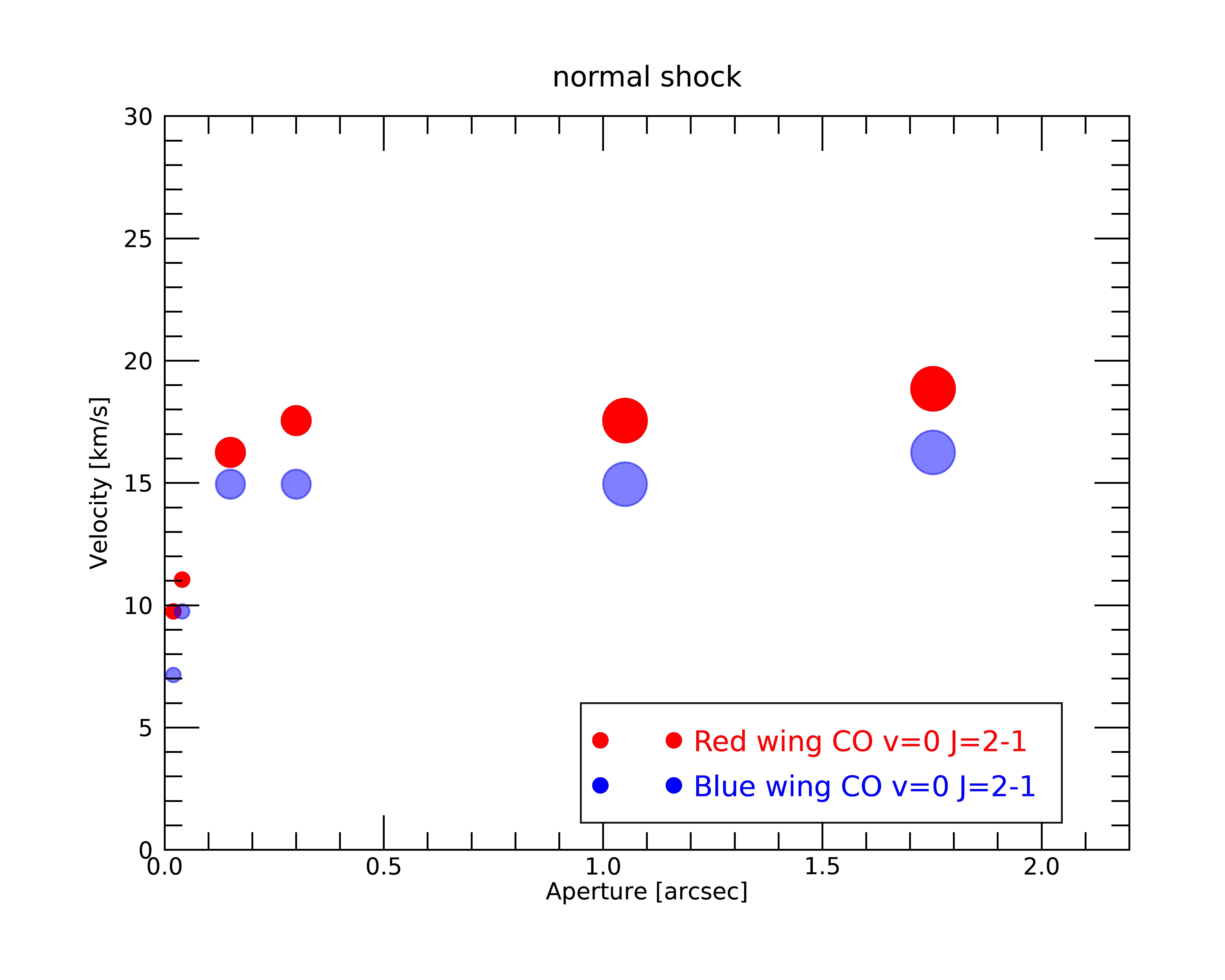}}}
	\end{minipage}
	\hfill
	\begin{minipage}[t]{.495\textwidth}
%		\centerline{\resizebox{\textwidth}{!}{\includegraphics[angle=0]{figures/plot_normal_shock_SiO_rms_sim.png}}}
				\centerline{\resizebox{\textwidth}{!}{\includegraphics[angle=0]{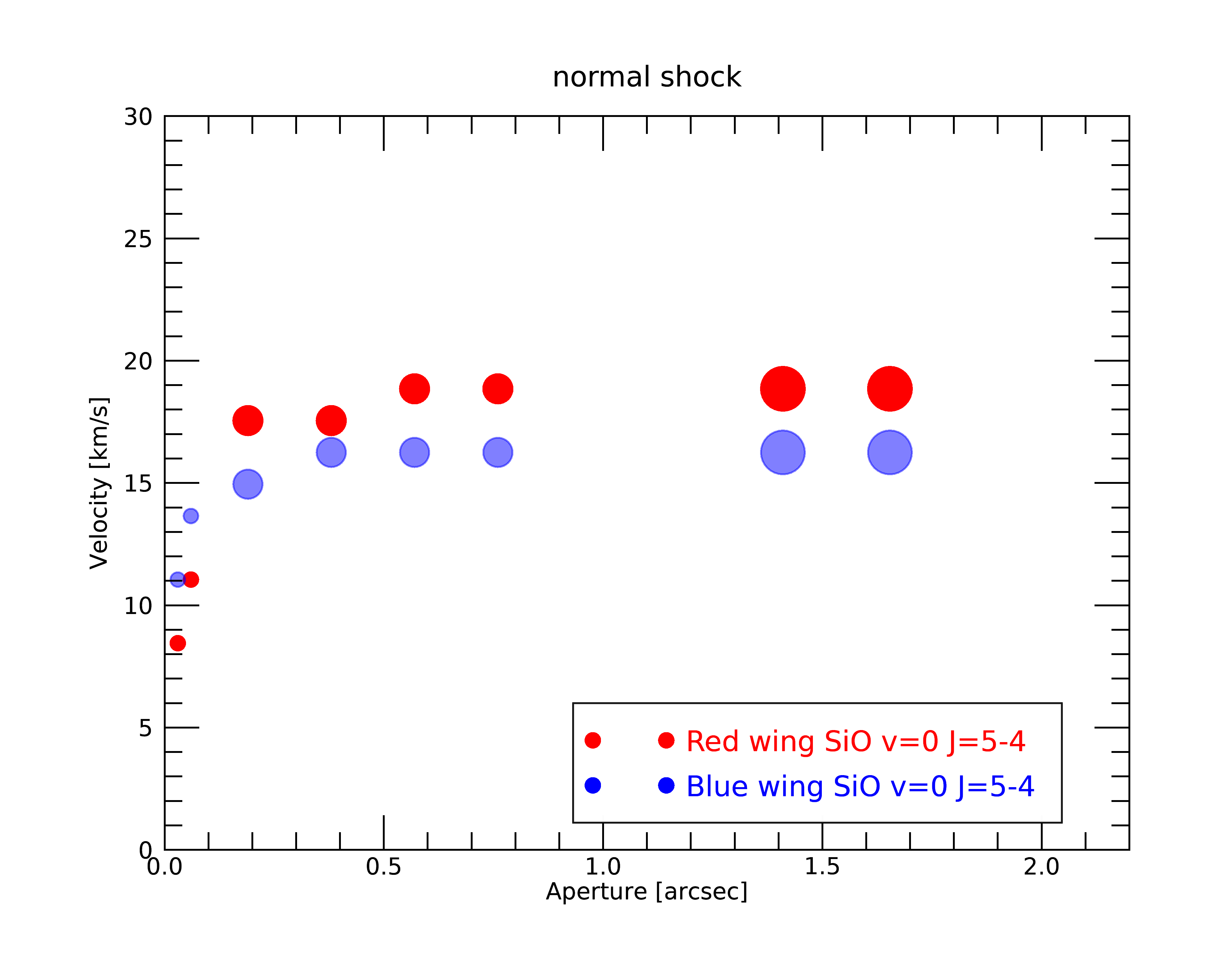}}}

	\end{minipage}
	\begin{minipage}[t]{.495\textwidth}
%		\centerline{\resizebox{\textwidth}{!}{\includegraphics[angle=0]{figures/plot_strong_shock_CO_rms_sim.png}}}
				\centerline{\resizebox{\textwidth}{!}{\includegraphics[angle=0]{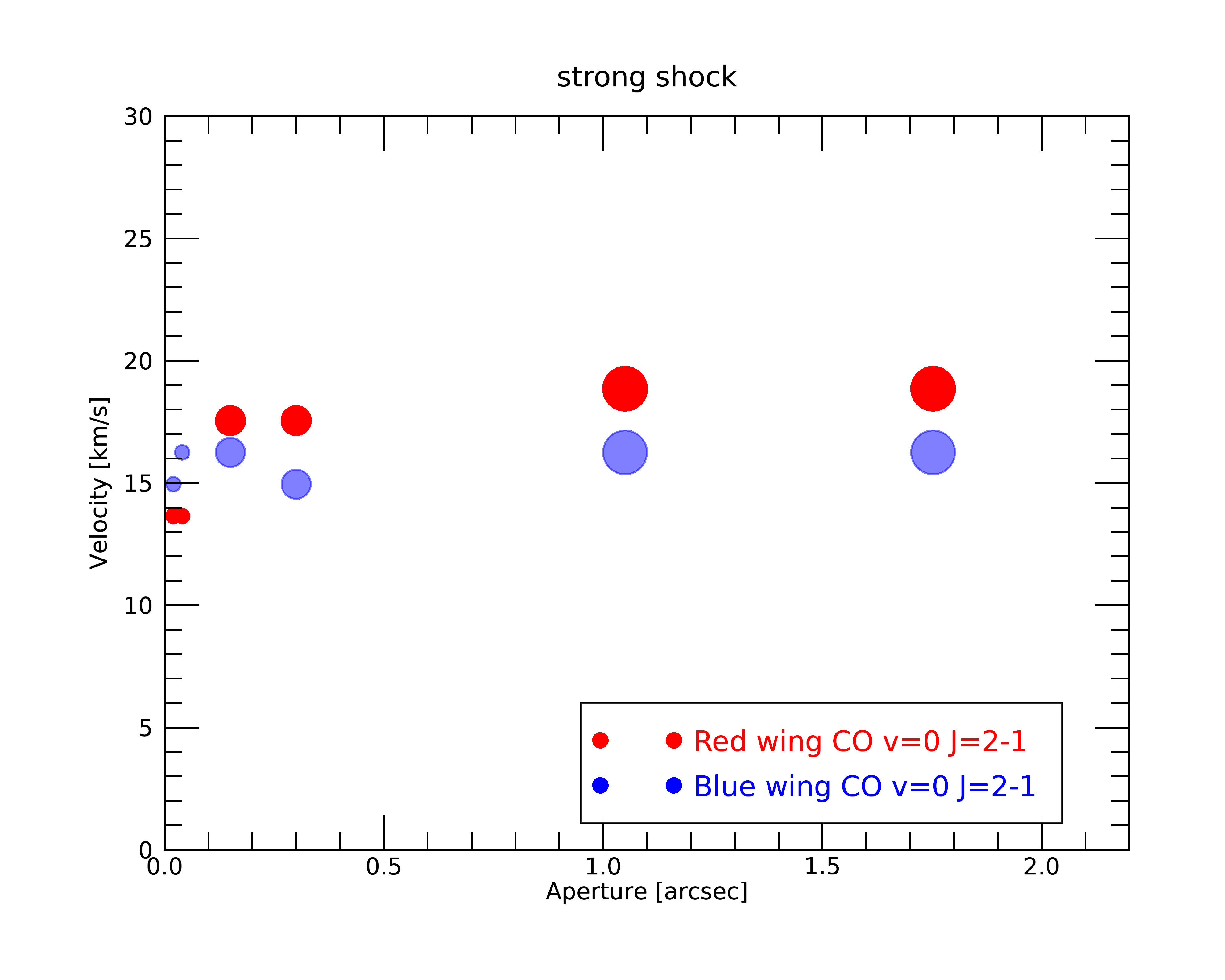}}}
	\end{minipage}
	\hfill
	\begin{minipage}[t]{.495\textwidth}
		\centerline{\resizebox{\textwidth}{!}{\includegraphics[angle=0]{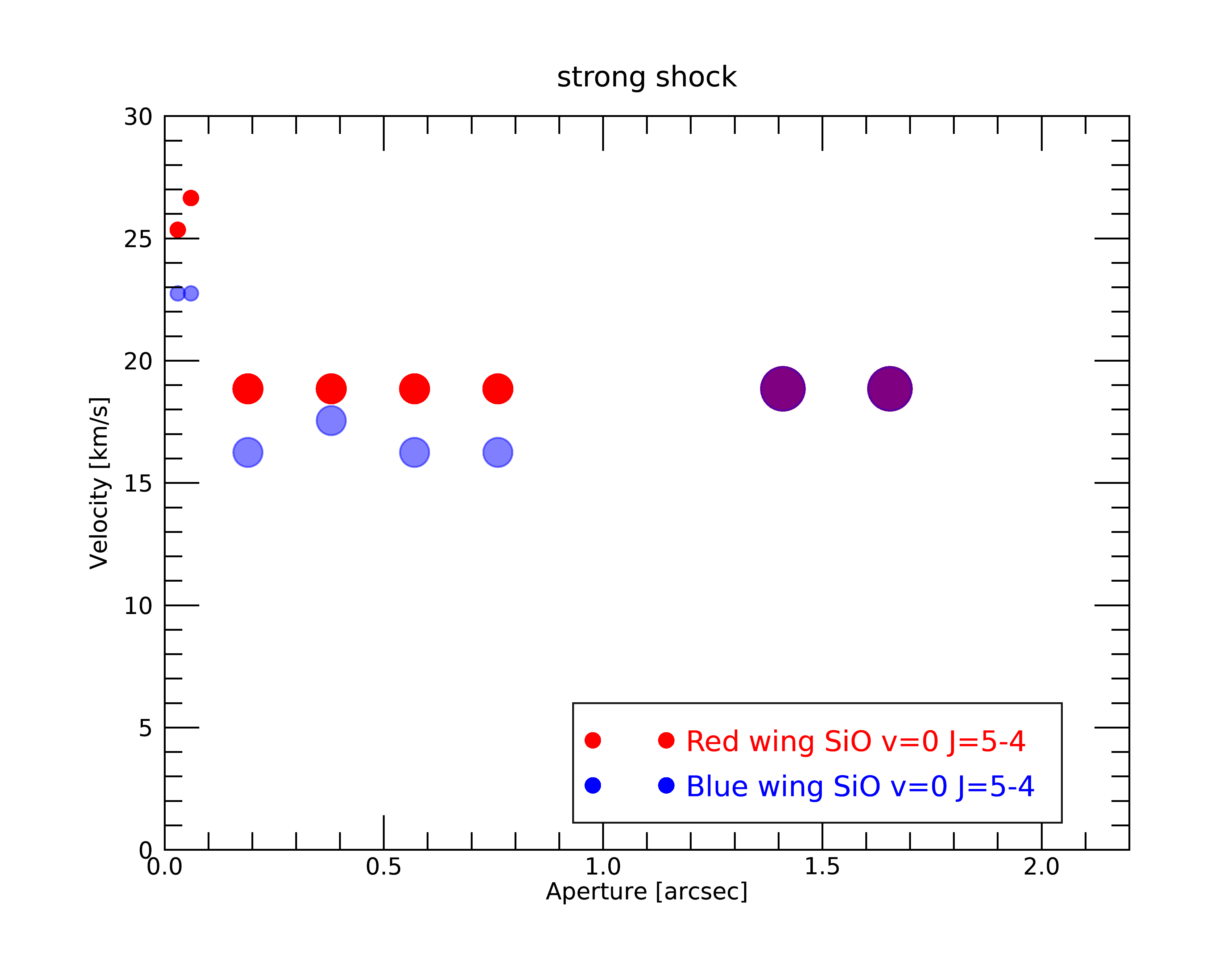}}}
%				\centerline{\resizebox{\textwidth}{!}{\includegraphics[angle=0]{NewFigs/plot_strong_shock_SiO_rms_sim.jpg}}}
%				\centerline{\resizebox{\textwidth}{!}{\includegraphics[angle=0]{NewFigs/plot_strong_shock_SiO_rms_sim_R1.pdf}}}

	\end{minipage}
	\caption{\textbf{Simulated wind velocity measures for $^{12}$CO v=0 J=2=1 (left) and $^{12}$SiO v=0 J=5-4 (right).}
		The upper row depicts the model without shocks, the middle (bottom) row the shocked wind model for the velocity profile shown 
		in the left (right) panel of Fig.~\ref{Fig:theory_vwind_shock}.
		The calculated intensities of $^{12}$CO and $^{12}$SiO account for the different array configurations and extraction apertures 
		in ATOMIUM 	
		(see Sect.~\ref{Sec:Observations} for details).
		The velocity measures for the extended (small dots), medium (medium-sized dots), and compact (large dots) configurations 
		were extracted by following Step 1 and Step 2 in Sect.~\ref{Sec:Methodology}.
		The blue dots representing the blue wing velocities are slightly transparent to allow for visualising the red wing velocities 
		in cases where the red and blue dots coincide.}
	\label{Fig:theory_vwind_shock_traced}
\end{figure}

%  +++++++++++++++++++++++++++++++++++++++++++++++++++++++++++++++++++++++++++++++

\medskip
	 
We also computed the SiO v=0 J=5-4 intensities by following the same procedure as for CO  (see right hand panels in 
Fig.~\ref{Fig:theory_vwind_shock_traced}). 
To account for the depletion of SiO by dust condensation and its potential dissociation, the relative abundance distribution of SiO was 
assumed to follow a Gaussian of the form \citep{Decin2010A&A...516A..69D, Danilovich2014A&A...569A..76D}
	 				\begin{equation}
	 					[{\rm{SiO/H_2}}] = 3 \times 10^{-5} \exp\left(-(r/r_e)^2\right), 
	 				\end{equation}
where the $e$-folding radius $r_e$ ($2.3 \times 10^{16}$\,cm) was determined by following \citet{Gonzalez2003A&A...411..123G}. 
In spite of its lower abundance, the detectable extent of the SiO v=0 J=5-4 line observed in the extended configuration is a factor $\sim$2 larger than 
for CO v=0 J=2-1 (0\farcs09 versus 0\farcs04), establishing the radiative nature of SiO (see Sect.~\ref{Sec:atomium_as_binary}). 
Analogous to CO, the effect of pulsation-induced shocks in the first few stellar radii 
--- where velocities of around 7\,--\,12\,km/s are greater than the local sound speed ---
are observed in the smallest extraction apertures of the extended configuration data. 
As Fig.~\ref{Fig:theory_vwind_shock_traced} demonstrates, observations with spatial resolution better than 
0\farcs150 are a prerequisite for characterizing this complex region in more detail.

\medskip
		
The question still remaining is whether and how physical phenomena yielding a velocity amplitude greater than the terminal wind 
velocity manifest themselves in the velocity measure. 
We therefore take an extreme example and multiply the shock amplitudes from \citet{Bladh2019A&A...626A.100B} by a factor~3 
(see right panel of Fig.~\ref{Fig:theory_vwind_shock}), where the outcome of the velocity measures are shown in the bottom panels of Fig.~\ref{Fig:theory_vwind_shock_traced} and are referred to as the `strong shock' model. 
Here as well, several interesting conclusions can be drawn: 
(1)  the compact configuration data of the $^{12}$CO v=0 J=2-1 line are not affected by the presence of shocks in the inner wind, 
and hence these data are a reliable tracer of the terminal wind velocity given the line broadening and spectral resolution of the data; 
(2)   the extended configuration data of both the $^{12}$CO v=0 J=2-1 and $^{28}$SiO v=0 J=5-4 line bear a signature of the extreme shock 
velocities in the first few stellar radii of the circumstellar envelope, especially in the case of SiO\,; and
(3)   the extreme shock velocities are not traced in both the medium and compact configuration data.
With respect to the latter conclusion, it should be noted that if the rms were three times smaller than the present value,
the weak extended wings could be better captured, thereby allowing the (shock) signature to be traced in the low and medium resolution data, 
although the convolution with the (Gaussian) broadening profile will also be tracked to lower intensity levels and hence yield broader profiles.

\medskip
		
Our current physical understanding of pulsation-induced shocks, however, does not validate the use of the `strong shock model',
because both models and observations indicate complex, non-monotonic velocity fields with relative macroscopic motions
of only some 10\,km/s \citep{2010A&A...514A..35N}. 
Relying then on the simulations for the `normal shock' model, the question that still needs to be addressed is why the observed SiO velocity measures 
can be significantly larger than that of CO as shown, for instance, for R~Aql in Fig.~\ref{Fig:terminal_velocities1}. 
Moreover, the same figure shows a trend in SiO velocity measures that is not captured by any of the shock simulations shown in Fig.~\ref{Fig:theory_vwind_shock_traced} in which the medium and compact configuration data roughly show a constant velocity measure.
Although detailed modelling of the {\sc atomium} wind kinematic profiles is beyond the scope of this paper, the 1D simulations performed here
can guide a thought experiment.  
A binary companion can impact the radial velocity field in a qualitatively similar way as pulsation-induced shocks in the sense that the 
radial velocity can have a wave-like character (see Fig.~\ref{Fig:velocities_binaries}). 
The amplitude of the velocity variations will increase for more massive and closer-in companions, and can attain values well above 
20\,km/s where velocity variations of 40\,--\,60\,km/s are not an exception \citep{Maes}. 
Guided by the outcome of the `strong shock' models, we stipulate that the particular behaviour of the wind kinematics profile in various 
{\sc atomium} sources can be explained by binary interaction. 
The high velocities captured in the SiO measurements can be caused by, for instance, the gravitational well of the companion or the formation 
of an equatorial density enhancement with potentially a Keplerian velocity field similar to the case of L$_2$~Puppis \citep{Kervella2016A&A...596A..92K}.

%  ############################################################################################# 
\afterpage{\clearpage}
\newpage
%\mbox{}
\section{ATOMIUM image cube properties}
 
Each target's `ALMA name' is of the form AH\_Sco. Each Scheduling Block (SB) is then labelled AH\_Sco\_a\_06\_TM1 etc.\ for the extended configuration, 
where (a, b, c, d) denote each frequency combination as in Fig.~\ref{Fig:spw}. 
The mid configuration SBs are labelled similarly ending in TM2. 
The compact configuration data is labelled AH\_Sco\_e\_06\_TM1
etc. where (e, f) denote each frequency combination as in Fig.~\ref{Fig:spw}. 
However, for a few targets in compact, inconsistent capitalisation was used for the target names and thus for the SBs. 
We made these consistent during data processing; the actual observing SB names are also given in Table~\ref{tab:obs_prop}.
The science spw in each SB tuning are numbered 25, 27, 29, 31 in ascending frequency. 
In the concatenated visibility data files, the spw become re-numbered in the order of observing time (thus, differing from target to target).  
For the final cubes that can be retrieved from the {\sc atomium} website, we re-numbered these in frequency order $00, 01, \cdots, 15$.
Tables~\ref{tab:cont} and \ref{tab:cube} list the properties of each continuum and of each cube image, respectfully.
%{\red {Listed in Table~\ref{tab:cont} are the properties of each continuum and of each cube image.}}

%In Table~\ref{tab:cont} and Table~\ref{tab:cube}, we list the properties of each continuum and cube image, respectively.

\onecolumn
 
\longtab{
\begin{landscape}
\begin{longtable}{lllllllll}
\caption{\label{tab:obs_prop} \textbf{Observational properties of the {\sc atomium} project.} }   \\
\hline\hline 
SB & Config & Phase-ref & Phase-ref R.A. & Phase-ref Dec. & Sep.  & PWV & Date              & ASDM \\ 
      &             &                 & (ICRS)              & (ICRS)               & (deg) & (mm)  & (YYYYDDMM) &             \\ 
\hline 
\endfirsthead
%  ------------------------------------------------
\caption{continued.}\\
\hline\hline
SB & Config & Phase-ref & Phase-ref R.A. & Phase-ref Dec. & Sep.    & PWV    & Date                & ASDM \\
      &             &                 &  (ICRS)             &   (ICRS)             &  (deg)   &  (mm)  &  (YYYYMMDD)   &             \\        
\hline
\endhead
\hline
\endfoot
%\hline
%\endlastfoot
AH\_Sco\_a\_06\_TM1     &   extended & J1717-3342 &  17:17:36.02945  &  --33:42:08.8277 &  1.91  &   0.4 & 20190706    & uid\_\_\_A002\_Xde63ab\_X9097 \\   
AH\_Sco\_a\_06\_TM1     &   extended & J1717-3342 &  17:17:36.02945  &  --33:42:08.8277 &  1.91  &   0.4 & 20190706    &    uid\_\_\_A002\_Xde63ab\_X95c6 \\
AH\_Sco\_b\_06\_TM1     &   extended & J1717-3342 &  17:17:36.02945  &  --33:42:08.8277 &  1.91  &   0.4 & 20190706    &    uid\_\_\_A002\_Xde63ab\_X9b95 \\
AH\_Sco\_b\_06\_TM1     &   extended & J1717-3342 &  17:17:36.02945  &  --33:42:08.8277 &  1.91  &   0.8 & 20190707    &    uid\_\_\_A002\_Xde8105\_Xe0b \\
AH\_Sco\_c\_06\_TM1     &   extended & J1717-3342 &  17:17:36.02945  &  --33:42:08.8277 &  1.91  &   2.3 & 20190710    &    uid\_\_\_A002\_Xde9c3e\_X7537 \\
AH\_Sco\_c\_06\_TM1     &   extended & J1717-3342 &  17:17:36.02945  &  --33:42:08.8277 &  1.91  &   2.3 & 20190710    &    uid\_\_\_A002\_Xde9c3e\_X7ad8 \\
AH\_Sco\_c\_06\_TM1     &   extended & J1717-3342 &  17:17:36.02945  &  --33:42:08.8277 &  1.91  &   2.3 & 20190710    &    uid\_\_\_A002\_Xde9c3e\_X7f89 \\
AH\_Sco\_d\_06\_TM1     &   extended & J1717-3342 &  17:17:36.02945  &  --33:42:08.8277 &  1.91  &   1.4 & 20190708    &    uid\_\_\_A002\_Xde8105\_X761c \\
\end{longtable}
\tablefoot{    
`SB' refers to the Scheduling Block, `Config' to the array configuration, `Phase-ref' to the phase reference source, 
`R.A.' and `Dec.' to the right ascension and declination, \hfill\break
\noindent `Sep.' is the angular separation between the target and the phase-reference source, 
`PWV' is the precipitable water vapour at the Date of the observations, and `ASDM' is the ALMA \hfill\break 
\noindent archival name.  The PWV values are for the start of each night and vary (usually by only 10\% or less) during observations. 
U\_Del\_f\_06\_TM1 and KW\_Sgr\_a\_06\_TM2 have not yet \hfill\break
\noindent been fully observed. } \\ 

%\vspace{0.25cm)
\noindent\makebox[\textwidth]
{$^*$~Owing to initial inconsistent capitalisation, SG marked $^*$ were originally named as follows: 
pi1\_gru\_a\_06\_TM1, pi1\_gru\_b\_06\_TM1, vx\_sgr\_a\_06\_TM1, }
\noindent\makebox[\textwidth]
{ \hspace{3.3cm} 
vx\_sgr\_b\_06\_TM1, T\_mic\_a\_06\_TM1, T\_mic\_b\_06\_TM1.  The SG were subsequently renamed as indicated below 
in our data products. }
\hspace{6.cm} 
\end{landscape}
}

%{$^*$: Owing to initial inconsistent capitalisation, the SG marked $^*$ were originally named as follows: 
%pi1\_gru\_a\_06\_TM1, pi1\_gru\_b\_06\_TM1, vx\_sgr\_a\_06\_TM1, }
%\noindent\makebox[\textwidth]
%{ vx\_sgr\_b\_06\_TM1, T\_mic\_a\_06\_TM1, T\_mic\_b\_06\_TM1; we renamed these as given below in our data products.

%\end{appendix}

%\end{document}

%  ################################################################################################################
%  ################################################################################################################
%  ################################################################################################################

%  ################################################################################################################
%  ########################### REVISED TABLE C.2 FROM ANITA ON THURSDAY 9 JULY 2020     ##############################
%  ################################################################################################################

\onecolumn

\longtab{
\begin{landscape}
\begin{longtable}{llllrrrrrllcclrc}
%\begin{table*}[htp]
\caption{  \label{tab:cont}  \textbf{Continuum image properties.} } \\
%\label{tab:cont}
%\begin{tabular}{llllrrrrrrrrrrrr}
\hline\hline
Star &Config.&$b_{\mathrm{maj}}$&$b_{\mathrm{min}}$&$b_{\mathrm{pa}}$&Imsize&MRS &Cont.&$\sigma_{\mathrm{rms}}^{\mathrm{cont}}$& Peak  RA & Peak Dec.&$\sigma_{\mathrm{RA}}^{\mathrm{cont}}$&$\sigma_{\mathrm{Dec}}^{\mathrm{cont}}$&Peak$^{\mathrm{cont}}$&$\sigma_{\mathrm{Peak}}^{\mathrm{cont}}$&Mid Freq.  \\
      &       &\multicolumn{2}{c}{(arcsec)}& (deg)&\multicolumn{2}{c}{(arcsec)} & (GHz) & (mJy)&(ICRS)  & (ICRS) &\multicolumn{2}{c}{(mas)}& \multicolumn{2}{c}{(mJy/bm)} & (GHz) \\
\hline
\endfirsthead
%  -----------------------------------------------
\caption{continued.}\\
\hline\hline
Star &Config.&$b_{\mathrm{maj}}$&$b_{\mathrm{min}}$&$b_{\mathrm{pa}}$&Imsize&MRS &Cont.&$\sigma_{\mathrm{rms}}^{\mathrm{cont}}$& Peak  RA & Peak Dec.&$\sigma_{\mathrm{RA}}^{\mathrm{cont}}$&$\sigma_{\mathrm{Dec}}^{\mathrm{cont}}$&Peak$^{\mathrm{cont}}$&$\sigma_{\mathrm{Peak}}^{\mathrm{cont}}$&Mid Freq\\
      &       &\multicolumn{2}{c}{(arcsec)}& (deg)&\multicolumn{2}{c}{(arcsec)} & (GHz) & (mJy) & (ICRS)  & (ICRS) &\multicolumn{2}{c}{(mas)}& \multicolumn{2}{c}{(mJy/bm)} & (GHz)\\    
\hline
\endhead
\hline
\endfoot
AH\_Sco    & extended  & 0.023 &  0.023 &   70 &    1.0 &   0.5 & 18.61 &   0.011 & 17:11:17.01591&--32:19:30.7643 &  0.1 &  0.1 &  7.48 &  0.04 &    241.78\\
AH\_Sco    & mid       & 0.159 &  0.100 & --79 &    4.0 &   1.7 & 17.08 &   0.014 & 17:11:17.01635&--32:19:30.7669 &  1.2 &  2.6 &  6.48 &  0.13 &    241.78\\
GY\_Aql    & extended  & 0.025 &  0.022 & --56 &    1.0 &   0.4 & 23.81 &   0.023 & 19:50:06.31478&--07:36:52.1890 &  0.1 &  0.1 & 9.00 &  0.04 &    241.75\\
GY\_Aql    & mid       & 0.324 &  0.247 & --70 &   24.0 &   4.0 & 18.03 &   0.026 & 19:50:06.31432&--07:36:52.2006 &  0.7 &  1.0 &  8.53 &  0.06 &    241.75\\
GY\_Aql    & compact   & 1.220 &  0.897 &   64 &   24.0 &   9.5 &  8.69 &   0.040 & 19:50:06.31672&--07:36:52.3182 &  7.0 &  9.5 &  9.45 &  0.16 &    238.44\\
IRC+10011  & extended  & 0.027 &  0.019 &   31 &    1.0 &   0.4 & 23.55 &   0.020 & 01:06:25.98833& +12:35:52.8487 &  0.1 &  0.1 & 11.19 &  0.09 &    241.77\\
IRC+10011  & mid       & 0.112 &  0.100 &   38 &    6.0 &   1.6 & 18.49 &   0.033 & 01:06:25.98838& +12:35:52.8565 &  0.6 &  0.6 & 11.70 &  0.14 &    241.77\\
IRC+10011  & compact   & 0.722 &  0.686 & --59 &   24.0 &   7.4 &  8.53 &   0.051 & 01:06:25.98542& +12:35:52.8578 &  2.2 &  2.2 & 12.44 &  0.09 &    238.43\\
IRC-10529  & extended  & 0.026 &  0.023 & --55 &    1.0 &   0.4 & 23.63 &   0.028 & 20:10:27.87133&--06:16:13.7402 &  0.1 &  0.2 &  7.31 &  0.09 &    241.79\\
IRC-10529  & mid       & 0.146 &  0.113 & --63 &    4.0 &   2.0 & 15.33 &   0.027 & 20:10:27.87259&--06:16:13.7475 &  0.8 &  1.2 &  7.26 &  0.11 &    241.76\\
IRC-10529  & compact   & 0.788 &  0.627 &   76 &   24.0 &   8.9 &  6.97 &   0.052 & 20:10:27.86978&--06:16:13.7251 &  4.4 &  6.3 &  6.25 &  0.10 &    238.48\\
KW\_Sgr    & extended  & 0.022 &  0.020 & --66 &    0.8 &   0.5 & 24.31 &   0.008 & 17:52:00.72819&--28:01:20.5715 &  0.1 &  0.1 &  2.90 &  0.01 &    241.77\\
KW\_Sgr    & mid       & 0.157 &  0.098 & --75 &    4.0 &   2.0 & 22.42 &   0.019 & 17:52:00.72839&--28:01:20.5846 &  0.4 &  1.0 &  2.63 &  0.03 &    241.74\\
pi1\_Gru   & extended  & 0.019 &  0.019 &   60 &    0.6 &   0.4 & 24.21 &   0.015 & 22:22:44.26959&--45:56:53.0065 &  0.2 &  0.2 & 17.79 &  0.04 &    241.79\\
pi1\_Gru   & mid       & 0.248 &  0.235 &   30 &    8.0 &   3.9 & 20.49 &   0.034 & 22:22:44.26654&--45:56:52.9986 &  0.2 &  0.2 & 32.33 &  0.08 &    241.79\\
pi1\_Gru   & compact   & 0.866 &  0.774 & --86 &   24.0 &   9.3 & 10.36 &   0.036 & 22:22:44.26861&--45:56:52.9890 &  0.4 &  0.4 & 31.29 &  0.04 &    238.45\\
RW\_Sco    & extended  & 0.024 &  0.020 & --70 &    1.0 &   0.4 & 24.53 &   0.020 & 17:14:51.68672&--33:25:54.5437 &  0.2 &  0.2 &  5.84 &  0.10 &    241.83\\
RW\_Sco    & mid       & 0.147 &  0.120 & --86 &    4.0 &   1.9 &  6.75 &   0.040 & 17:14:51.68671&--33:25:54.5440 &  0.4 &  0.6 &  5.26 &  0.04 &    242.02\\
RW\_Sco    & compact   & 0.928 &  0.701 &   86 &   24.0 &   9.0 & 10.30 &   0.034 & 17:14:51.68927&--33:25:54.5042 &  2.5 &  4.2 &  3.37 &  0.03 &    238.52\\
R\_Aql     & extended  & 0.024 &  0.022 & --13 &    1.0 &   0.4 & 22.63 &   0.008 & 19:06:22.25672& +08:13:46.6778 &  0.1 &  0.1 & 17.02 &  0.03 &    241.74\\
R\_Aql     & mid       & 0.306 &  0.238 & --54 &    8.0 &   3.8 & 20.31 &   0.030 & 19:06:22.25564& +08:13:46.7063 &  0.8 &  1.0 & 18.04 &  0.13 &    241.74\\
R\_Aql     & compact   & 0.764 &  0.648 &   83 &   24.0 &   7.7 & 10.25 &   0.042 & 19:06:22.26051& +08:13:46.6697 &  1.5 &  2.1 & 15.91 &  0.09 &    238.43\\
R\_Hya     & extended  & 0.034 &  0.025 &   67 &    1.0 &   0.6 & 23.55 &   0.057 & 13:29:42.70211&--23:16:52.5146 &  0.3 &  0.4 & 41.86 &  0.27 &    241.78\\
R\_Hya     & mid       & 0.256 &  0.223 &   70 &    8.0 &   3.5 & 19.02 &   0.028 & 13:29:42.70465&--23:16:52.5318 &  0.2 &  0.2 & 54.44 &  0.10 &    241.82\\
R\_Hya     & compact   & 0.830 &  0.600 &   79 &   24.0 &   8.7 & 10.09 &   0.051 & 13:29:42.70448&--23:16:52.5536 &  0.2 &  0.4 & 65.55 &  0.06 &    238.47\\
SV\_Aqr    & extended  & 0.022 &  0.021 &   43 &    1.0 &   0.4 & 28.96 &   0.009 & 23:22:45.40025&--10:49:00.1874 &  0.2 &  0.2 &  1.43 &  0.02 &    241.78\\
SV\_Aqr    & mid       & 0.124 &  0.104 & --75 &    8.0 &   1.6 & 27.55 &   0.023 & 23:22:45.39878&--10:49:00.1789 &  0.6 &  0.7 &  2.17 &  0.03 &    241.77\\
SV\_Aqr    & compact   & 0.886 &  0.747 &   74 &   24.0 &   9.8 & 10.92 &   0.038 & 23:22:45.39676&--10:49:00.2442 &  7.9 &  9.3 &  1.35 &  0.03 &    238.46\\
S\_Pav     & extended  & 0.025 &  0.020 & --13 &    1.0 &   0.4 & 22.35 &   0.010 & 19:55:14.00546&--59:11:45.1943 &  0.1 &  0.1 & 21.75 &  0.04 &    241.79\\
S\_Pav     & mid       & 0.304 &  0.234 &   56 &    8.0 &   3.3 & 20.20 &   0.022 & 19:55:14.00227&--59:11:45.1462 &  0.2 &  0.2 & 31.04 &  0.04 &    241.89\\
S\_Pav     & compact   & 1.026 &  0.983 & --56 &   24.0 &   8.7 & 10.22 &   0.051 & 19:55:13.99589&--59:11:45.0735 &  1.7 &  1.8 & 27.24 &  0.11 &    238.48\\
T\_Mic     & extended  & 0.024 &  0.021 & --73 &    1.0 &   0.4 & 24.94 &   0.013 & 20:27:55.17974&--28:15:39.5529 &  0.1 &  0.1 & 23.00 &  0.07 &    241.75\\
T\_Mic     & mid       & 0.268 &  0.225 & --89 &    8.0 &   4.0 & 19.36 &   0.025 & 20:27:55.17968&--28:15:39.5631 &  0.1 &  0.1 & 30.14 &  0.03 &    241.75\\
T\_Mic     & compact   & 1.047 &  0.730 & --79 &   24.0 &   9.3 & 10.99 &   0.059 & 20:27:55.18152&--28:15:39.4732 &  0.8 &  1.6 & 26.39 &  0.08 &    238.45\\
U\_Del     & extended  & 0.030 &  0.021 & --25 &    1.0 &   0.4 & 25.44 &   0.010 & 20:45:28.25002& +18:05:23.9761 &  0.0 &  0.0 &  6.49 &  0.02 &    241.78\\
U\_Del     & mid           & 0.316 &  0.235 & --33 &    8.0 &   3.3 & 14.84 &   0.028 & 20:45:28.24967& +18:05:23.9930 &  1.4 &  1.2 &  7.25 &  0.06 &    241.68\\
U\_Del     & compact   & 1.165 &  1.013 &   33 &   24.0 &   9.0 & 11.00 &   0.048 & 20:45:28.25138& +18:05:23.9726 &  4.6 &  4.0 &  7.36 &  0.06 &    238.44\\
U\_Her     & extended  & 0.024 &  0.018 &    8 &    0.6 &   0.4 & 23.33 &   0.013 & 16:25:47.45136& +18:53:32.6663 &  0.1 &  0.1 & 11.60 &  0.08 &    241.79\\
U\_Her     & mid       & 0.267 &  0.195 & --33 &    8.0 &   2.2 & 16.83 &   0.048 & 16:25:47.45134& +18:53:32.7012 &  2.3 &  1.9 & 14.77 &  0.18 &    241.79\\
U\_Her     & compact   & 0.997 &  0.843 &   26 &   24.0 &   9.7 &  9.75 &   0.054 & 16:25:47.45145& +18:53:32.6428 &  1.2 &  1.0 & 17.29 &  0.05 &    238.48\\
VX\_Sgr    & extended  & 0.028 &  0.020 &   89 &    0.6 &   0.4 & 18.16 &   0.019 & 18:08:04.04604&--22:13:26.6209 &  0.1 &  0.1 & 14.58 &  0.08 &    241.77\\
VX\_Sgr    & mid       & 0.162 &  0.095 & --75 &    4.0 &   1.5 & 16.11 &   0.030 & 18:08:04.04466&--22:13:26.6121 &  0.2 &  0.4 & 16.34 &  0.08 &    241.77\\
VX\_Sgr    & compact   & 1.127 &  0.809 &   79 &   36.0 &  10.0 &  8.05 &   0.039 & 18:08:04.04934&--22:13:26.6426 &  1.6 &  2.8 & 15.96 &  0.08 &    238.46\\
V\_PsA     & extended  & 0.023 &  0.021 & --77 &    1.0 &   0.4 & 24.78 &   0.009 & 22:55:19.72280&--29:36:45.0384 &  0.1 &  0.1 &  8.93 &  0.03 &    241.78\\
V\_PsA     & mid       & 0.283 &  0.229 &   85 &    8.0 &   4.0 & 20.99 &   0.020 & 22:55:19.72033&--29:36:45.0298 &  0.2 &  0.4 &  9.35 &  0.03 &    241.67\\
V\_PsA     & compact   & 0.995 &  0.753 &   87 &   24.0 &   9.0 & 11.28 &   0.030 & 22:55:19.72043&--29:36:45.0559 &  1.4 &  2.4 &  8.67 &  0.04 &    238.45\\
W\_Aql     & extended  & 0.024 &  0.021 & --47 &    1.0 &   0.4 & 24.22 &   0.005 & 19:15:23.37809&--07:02:50.3306 &  0.1 &  0.1 &  6.53 &  0.04 &    241.80\\
W\_Aql     & mid       & 0.351 &  0.223 & --68 &    8.0 &   3.9 & 18.75 &   0.030 & 19:15:23.37954&--07:02:50.3165 &  2.4 &  4.0 &  5.63 &  0.12 &    241.69\\
W\_Aql     & compact   & 0.920 &  0.667 &   76 &   24.0 &   8.9 &  6.61 &   0.056 & 19:15:23.38051&--07:02:50.3096 &  5.6 &  8.7 &  7.62 &  0.14 &    238.49\\
\hline
%\end{tabular}\\
%\end{table*}
\end{longtable}
\tablefoot{ 
These are taken from the image after the optimum number of rounds of 
self-calibration. Config. is the array combination, determining $b_\text{maj}$, $b_\text{min}$ and $b_\text{PA}$, 
the major and minor axis and the position angle of the synthesized beam, respectively.  
Imsize is the image size and MRS is the maximum recoverable angular scale.
Cont. is the total line-free
  bandwidth, spread over all tunings. The continuum $\sigma_{\mathrm{rms}}^{\rm{cont}}$ noise is measured in a region 
of $\sim$10\% the total image area clear of the emission, in the images without primary beam correction. 
Peak RA and Peak Dec. are the position of a 2D Gaussian component fitted
to the peak, fitting uncertainties $\sigma_{\mathrm{RA}}^{\mathrm{cont}}$, $\sigma_{\mathrm{Dec}}^{\mathrm{cont}}$. 
Peak$^{\mathrm{cont}}$ and $\sigma_{\mathrm{Peak}}^{\mathrm{cont}}$ are the peak flux density and stochiastic uncertainty, and Mid Freq 
is the approximate mid point of the line-free coverage.}  \\
\end{landscape}
}

%  ################################################################################################################
%  ################################################################################################################
%  ################################################################################################################

%  ################################################################################################################
%  ########################### REVISED TABLE C.3 FROM ANITA ON THURSDAY 9 JULY 2020     ##############################
%  ################################################################################################################
\newpage

\setlength{\LTcapwidth}{\linewidth}
\longtab{
%\begin{landscape}
\begin{longtable}{lllllllrrr}
\caption{\label{tab:cube} \textbf{Image cube properties.} }      \\
\hline 
\multicolumn{1}{l}{Star}  &  Configuration  &Cube & Low    &  High   & $b_\text{maj}$       & $b_\text{min}$  & $b_\text{PA}$ 
& Imsize                          & $\sigma_{\mathrm{rms}}$         \\
                                       &                          & No.   & (GHz)  & (GHz)  & (arcsec)                  & (arcsec)              &(deg)                  
& (arcsec)                       & (mJy)    \\
\hline 
\endfirsthead
\caption{continued.}\\
\hline\hline
Star &  Configuration  &Cube      & Low    &  High   & $b_\text{maj}$ & $b_\text{min}$  & $b_\text{PA}$ &Imsize & $\sigma_{\mathrm{rms}}$\\
        &          &No.      & (GHz)  & (GHz)   & (arcsec)& (arcsec)&(deg)& (arcsec)& (mJy)\\
\hline
\endhead
\hline
\endfoot
\hline
AH\_Sco   & extended  & 00  & 213.865 & 215.738 & 0.024  & 0.022 &    0 &    2.0 &   0.4\\
AH\_Sco   & extended  & 01  & 216.065 & 217.938 & 0.025  & 0.023 &    5 &    2.0 &   0.6\\
AH\_Sco   & extended  & 02  & 220.265 & 222.138 & 0.027  & 0.023 &   86 &    2.0 &   0.6\\
AH\_Sco   & extended  & 03  & 223.660 & 225.533 & 0.026  & 0.024 &   82 &    2.0 &   0.5\\
AH\_Sco   & extended  & 04  & 227.263 & 229.136 & 0.022  & 0.021 & --12 &    2.0 &   0.5\\
AH\_Sco   & extended  & 05  & 229.618 & 231.491 & 0.022  & 0.020 & --21 &    2.0 &   0.4\\
AH\_Sco   & extended  & 06  & 235.468 & 237.341 & 0.025  & 0.021 & --84 &    2.0 &   0.5\\
AH\_Sco   & extended  & 07  & 239.187 & 240.122 & 0.026  & 0.023 &   74 &    2.0 &   0.8\\
\hline
\end{longtable}
\tablefoot{
Low and High are the minimum and maximum observed frequencies in the cube. 
The parameters $b_\text{maj}$, $b_\text{min}$ and $b_\text{PA}$ are the major and minor axis and the position angle of the synthesized beam, respectively.  
The noise $\sigma_{\mathrm{rms}}$ is measured from a selection of emission-free channels in the cube without the primary beam correction. 
U~Del compact configuration cubes 08, 09, 12 and 13, and KW~Sgr mid configuration cubes 00, 01, 04 and 05 have not yet been fully observed.}  \\
%\end{landscape}
}

%  ################################################################################################################
%  ################################################################################################################
%  ################################################################################################################

\end{appendix}

%\bibliography{overview_v4}{}

%\bibliography{ATOMIUM(overview_v4)CAG_23April2020}{}

%\bibliography{ATOMIUM_overview}{}
%\bibliographystyle{aasjournal}



%  ##################################################################################################
%  ##################################################################################################
%  ##################################################################################################
\end{document}